\documentclass[11pt,twoside]{mitthesis}
\usepackage{lgrind}
\usepackage{arydshl}
\usepackage{multirow}
\usepackage{supertabular}
\usepackage{graphicx}
\newcommand{\eq}[1]{(\ref{#1})}
\newcommand{\bun}{\hat{\mathbf{b}}}
\newcommand{\eun}{\hat{\mathbf{e}}}
\newcommand{\phiave}{\langle \phi \rangle}
\newcommand{\phiwig}{\widetilde{\phi}}
\newcommand{\Phiwig}{\widetilde{\Phi}}
\newcommand{\br}{\mathbf{r}}
\newcommand{\bv}{\mathbf{v}}
\newcommand{\bR}{\mathbf{R}}
\newcommand{\bJ}{\mathbf{J}}
\newcommand{\bB}{\mathbf{B}}
\newcommand{\bV}{\mathbf{V}}
\newcommand{\matrixtop}[1]{\buildrel\leftrightarrow\over{#1}}
\newcommand{\matI}{\matrixtop{\mathbf{I}}}
\newcommand{\fwig}{\widetilde{f}}
\newcommand{\nablaave}{\overline{\nabla}}
\newcommand{\ssim}{ {\scriptstyle {{_{\displaystyle <}}\atop{\displaystyle \sim}}} }
\newcommand{\gsim}{ {\scriptstyle {{_{\displaystyle >}}\atop{\displaystyle \sim}}} }
\newcommand{\dotcross}{ \raise 0.65ex\hbox{${\scriptstyle {{_{\displaystyle \cdot}}\atop\times}}$} }
\newcommand{\crossdot}{ \raise 0.5ex\hbox{${\scriptstyle {{_\times}\atop{\displaystyle \cdot}}}$} }
\newcommand{\rhobf}{\mbox{\boldmath$\rho$}}
\newcommand{\pibf}{\mbox{\boldmath$\pi$}}
\newcommand{\kappabf}{\mbox{\boldmath$\kappa$}}
\newcommand{\zetabf}{\mbox{\boldmath$\zeta$}}
\newcommand{\zun}{\hat{\zetabf}}
\newcommand{\Gammabf}{\mbox{\boldmath$\Gamma$}}
\newcommand{\thetabf}{\mbox{\boldmath$\theta$}}
\newcommand{\ubrace}[2]{ \mathop{\underbrace{#1}}_{\displaystyle #2} }

\begin{document}

%
%
%
%
%
%
%
%

\title{Extension of gyrokinetics to transport time scales}

\author{F\'elix Ignacio Parra D\'iaz}
       \prevdegrees{Aeronautical Engineer, Universidad Polit\'ecnica de Madrid (2004) \\
                    S.M., Massachusetts Institute of Technology (2007)}
\department{Department of Aeronautics and Astronautics}

\degree{Doctor of Philosophy}

\degreemonth{June}
\degreeyear{2009}
\thesisdate{May 21, 2009}



\supervisor{Peter J. Catto}{Senior Research Scientist}

\thesischairman{Manuel Mart\'inez-S\'anchez}{Full Professor}

\thesismember{Jeffrey P. Freidberg}{Full Professor}

\chairman{David L. Darmofal}{Associate Department Head}{Chair, Committee on Graduate Students}

\maketitle



\cleardoublepage
\setcounter{savepage}{\thepage}
\begin{abstractpage}
%
%
%

In the last decade, gyrokinetic simulations have greatly improved
our theoretical understanding of turbulent transport in fusion
devices. Most gyrokinetic models in use are $\delta f$ simulations
in which the slowly varying radial profiles of density and
temperature are assumed to be constant for turbulence saturation
times, and only the turbulent electromagnetic fluctuations are
calculated. Due to the success of these models, new massive
simulations are being built to self-consistently determine the
radial profiles of density and temperature. However, these new
codes have failed to realize that modern gyrokinetic formulations,
composed of a gyrokinetic Fokker-Planck equation and a gyrokinetic
quasineutrality equation, are only valid for $\delta f$
simulations that do not reach the longer transport time scales
necessary to evolve radial profiles. In tokamaks, due to
axisymmetry, the evolution of the axisymmetric radial electric
field is a challenging problem requiring substantial modifications
to gyrokinetic treatments. The radial electric field, closely
related to plasma flow, is known to have a considerable impact on
turbulence saturation, and any self-consistent global simulation
of turbulent transport needs an accurate procedure to determine
it. In this thesis, I study the effect of turbulence on the global
electric field and plasma flows. By studying the current
conservation equation, or vorticity equation, I prove that the
long wavelength, axisymmetric flow must remain neoclassical and I
show that the tokamak is intrinsically ambipolar, i.e., the radial
current is zero to a very high order for any long wavelength
radial electric field. Intrinsic ambipolarity is the origin of the
problems with the modern gyrokinetic approach since the lower
order gyrokinetic quasineutrality (if properly evaluated) is
effectively independent of the radial electric field. I propose a
new gyrokinetic formalism in which, instead of a quasineutrality
equation, a current conservation equation or vorticity equation is
solved. The vorticity equation makes the time scales in the
problem explicit and shows that the radial electric field is
determined by the conservation of toroidal angular momentum.

\end{abstractpage}


\cleardoublepage

\section*{Acknowledgments}

First and foremost, I am deeply grateful to my wife Violeta. She is probably the
only person that can truly appreciate how much effort I had to dedicate to this thesis. Without
her unrelenting support I would have not been able to achieve my doctoral degree. 

I am also indebted to my parents Nieves and Ignacio, and my brother Braulio. They were able 
to support me despite the distance.

This thesis would have not been possible without Peter Catto's judicious tutoring and 
Bill Dorland's support. Peter has taught me all I know about kinetic theory in plasma 
physics, led my work through the rough patches and, what is more important, laid down
clearly with his example the path I want to follow in my future career. Bill made possible
this thesis with his support and warm friendship.

I sincerely thank the members of my committee, Jeffrey Freidberg and Manuel Mart\'inez-S\'anchez, and the 
readers, Ian Hutchinson and Jes\'us Ramos, for being so patient with the initial write-up. Their 
questions and suggestions helped me improve the presentation of this thesis.

Finally, I am grateful to my office mates Grisha, Antoine, Susan and Matt, and to my friends 
Arturo, Carolina, Luis, Kate, Roberto, Ana, Joaquim, Joel, Chantal, Tanya, Andrew, Robyn, 
Michael, Kyle, Alex and Steve. I could not have asked for a more enjoyable and supporting 
group of friends.


\pagestyle{plain}
\tableofcontents
\newpage
\listoffigures
\newpage
\listoftables

\chapter*{Nomenclature}

\begin{supertabular}{l p{0.8\textwidth} }
\multicolumn{2}{l}{\textbf{Miscellaneous}} \\
$\overline{(\ldots)}$ & Gyroaverage holding $\br$, $v_{||}$, $v_\bot$ and $t$ fixed. \\
$\langle \ldots \rangle$ & Gyroaverage holding $\bR$, $E$, $\mu$ and $t$ fixed. \\
$\langle \ldots \rangle_\psi$ & In tokamaks, flux surface average. \\
$\nablaave$ & Gradient holding $E_0$, $\mu_0$, $\varphi_0$ and $t$ fixed. \\
\multicolumn{2}{l}{\textbf{Greek letters}} \\
$\psi$ & In tokamaks, poloidal magnetic field flux, radial coordinate. \\
$\theta$ & In tokamaks, poloidal angle; in $\theta$-pinches, azimuthal angle. \\
$\hat{\thetabf}$ & In $\theta$-pinches, unit vector in the azimuthal direction. \\
$\zeta$, $\zun$ & In tokamaks, toroidal angle and unit vector in the toroidal direction. \\
$\epsilon$ & In tokamaks, inverse aspect ratio $a/R$. \\
$\rho_e$, $\rho_i$ & Electron and ion gyroradii, $mcv_e/eB$ and $Mcv_i/ZeB$. \\
$\Omega_e$, $\Omega_i$ & Electron and ion gyrofrequencies, $eB/mc$ and $ZeB/Mc$. \\
$\delta_e$, $\delta_i$ & Expansion parameters $\delta_e = \rho_e/L \ll \delta_i = \rho_i/L \ll 1$. \\
$\lambda_D$ & Debye length. \\
$\nu_{ii}$, $\nu_{ie}$, $\nu_{ee}$, $\nu_{ei}$ & Ion-ion, ion-electron, electron-electron and electron-ion Braginskii collision frequencies. \\
$\omega_\ast$ & Drift wave frequency. \\
$\omega_\ast^{n,T}$ & Drift wave frequency in gyrokinetic equation \eq{deltaf_FP} dependent on $\nabla n_i$ and $\nabla T_i$. \\
$\phi$ & Electrostatic potential. \\
$\phiave$ & Electrostatic potential averaged in a gyromotion [see \eq{phiave}]. \\
$\phiwig$ & Difference between the potential seen by the particle and the potential averaged in a gyromotion [see \eq{phiwig}]. \\
$\Phiwig$ & Indefinite integral of $\phiwig$ with vanishing gyroaverage [see \eq{Phiwig}]. \\
$\mu$ & Gyrokinetic magnetic moment defined to be an adiabatic invariant to higher order [see \eq{mutotal_def}]. \\
$\mu_g$ & Gyrokinetic magnetic moment in which the explicit dependence on the potential has been subtracted. \\
$\mu_0$ & Lowest order magnetic moment of the particle $v_\bot^2/2B$. \\
$\mu_1$ & First order correction to the gyrokinetic magnetic moment [see \eq{correc_mu_o1}]. \\
$\mu_{10}$ & First order correction to the gyrokinetic magnetic moment in which the explicit dependence on the potential has been subtracted [see \eq{mu10}]. \\
$\varphi$ & Gyrokinetic gyrophase in which the fast time variation has been averaged out [see \eq{varphitotal_def}]. \\
$\varphi_0$ & Lowest order gyrophase of the particle [see \eq{gyrophase_def}]. \\
$\varphi_1$ & First order correction to the gyrokinetic gyrophase [see \eq{correc_varphi_o1}]. \\
$\varphi_{10}$ & First order correction to the gyrokinetic gyrophase in which the explicit dependence on the potential has been subtracted [see \eq{varphi10}]. \\
$\matrixtop{\pibf}_i$ & Ion viscosity. Its definition includes the Reynolds stress because the average velocity has not been subtracted [see \eq{pi_def}]. \\
$\pibf_{ig||}$ & Vector that gives the transport of parallel ion momentum by the $E\times B$ and magnetic drifts and the finite gyroradius drift $\tilde{\bv}_1$ [see \eq{pig_par}]. \\
$\matrixtop{\pibf}_{ig\times}$ & Tensor that gives the transport of perpendicular ion momentum by the parallel velocity, the $E\times B$ and magnetic drifts and the finite gyroradius drift $\tilde{\bv}_1$ [see \eq{pig_perp}]. \\
$\matrixtop{\pibf}_{iG}$ & Effective viscosity for gyrokinetic vorticity equation \eq{vort_type2}. \\
$\varpi$ & Vorticity [see \eq{vort_def}]. \\
$\varpi_G$ & Gyrokinetic ``vorticity" [see \eq{vortGKdef}]. \\
$\varpi_G^{(2)}$ & Higher order gyrokinetic ``vorticity" [see \eq{vortGKdef_o2}]. \\
\multicolumn{2}{l}{\textbf{Roman letters}} \\
$a$ & In tokamaks, minor radius. \\
$\bB$, $B$, $\bun$ & Magnetic field, magnetic field magnitude, and unit vector parallel to the magnetic field. \\
$B_p$ & In tokamaks, magnitude of the poloidal component of the magnetic field. \\
$c$ & Speed of light. \\
$D_{gB}$ & GyroBohm diffusion coefficient $\delta_i \rho_i v_i$. \\
$e$ & Electron charge magnitude. \\
$\eun_1$, $\eun_2$ & Orthonormal vectors such that $\eun_1 \times \eun_2 = \bun$ used to define the gyrophase $\varphi_0$ [see \eq{gyrophase_def}]. \\
$E$ & Gyrokinetic kinetic energy in which the fast time variation has been averaged out [see \eq{Etotal_def}]. \\
$E_0$ & Kinetic energy of the particle $v^2/2$. \\
$E_1$ & First order correction to the gyrokinetic kinetic energy [see \eq{correc_E_o1}]. \\
$E_2$ & Second order correction to the gyrokinetic kinetic energy [see \eq{correc_E_o2}]. \\
$\mathbf{E}$ & Electric field. \\
$f_e$ & Electron distribution function, dependent on $\br$, $E_0$, $\mu_0$ and $\varphi_0$. \\
$\overline{f}_e$ & Gyrophase independent piece of the electron distribution function. \\
$f_e - \overline{f}_e$ & Gyrophase dependent piece of the electron distribution function. \\
$f_i$ & Ion distribution function, usually dependent on the gyrokinetic variables $\bR$, $E$ and $\mu$. \\
$\fwig_i$ & Small piece of the ion distribution function that depends on the gyrokinetic gyrophase responsible for classical diffusion [see \eq{fwig_sol}]. \\
$f_{ig}$ & Distribution function found by replacing the gyrokinetic variables $\bR$, $E$ and $\mu$ in $f_i (\bR, E, \mu, t)$ by $\bR_g$, $E_0$ and $\mu_0$ [see \eq{fi1}]. \\
$f_{iG}$ & Distribution function found by replacing the gyrokinetic variables $\bR$, $E$ and $\mu$ in $f_i (\bR, E, \mu, t)$ by $\bR_g$, $E_0$ and $\mu_g$. \\
$f_{i0}$ & Distribution function found by replacing the gyrokinetic variables $\bR$, $E$ and $\mu$ in $f_i (\bR, E, \mu, t)$ by $\br$, $E_0$ and $\mu_0$. \\
$\overline{f}_i$ & Gyrophase independent piece of the ion distribution function when written in physical phase space variables $\br$, $E_0$, $\mu_0$ and $\varphi_0$. \\
$f_i - \overline{f}_i$ &  Gyrophase dependent piece of the ion distribution function when written in physical phase space variables $\br$, $E_0$, $\mu_0$ and $\varphi_0$. \\
$f_{Me}$, $f_{Mi}$ & Lowest order electron and ion distribution functions, assumed to be stationary Maxwellians.\\
$\mathbf{F}_{ei}$ & Electron collisional momentum exchange with ions. \\
$\mathbf{F}_{iB}$ & Change in the perpendicular momentum of the gyromotion due to variations in the magnetic field strength [see \eq{force_B}]. \\
$\mathbf{F}_{iB}^{(2)}$ & Higher order version of $\mathbf{F}_{iB}$ [see \eq{force_B_o2}]. \\
$\mathbf{F}_{iC}$ & Force due to finite gyroradius effects on collisions [see \eq{FiC}]. \\
$\tilde{F}_{iE}$ & Change in the parallel momentum due to the short wavelength components of the electrostatic potential [see \eq{force_E}]. \\
$\tilde{F}_{iE}^{(2)}$ & Higher order version of $\tilde{F}_{iE}$ [see \eq{force_E_o2}]. \\
$h_i$ & Correction to the Maxwellian for ions. \\
$h_{i1}$, $h_{i1}^\mathrm{tb}$, $h_{i1}^\mathrm{nc}$ & First order correction to the Maxwellian in $\delta_i$, decomposed into two pieces: the piece due to turbulence and the neoclassical contribution. \\
$h_{i2}$, $h_{i2}^\mathrm{tb}$, $h_{i2}^\mathrm{nc}$ & Second order correction to the Maxwellian in $\delta_i$, decomposed into two pieces: the piece due to turbulence and the neoclassical contribution. \\
$I$ & Function $R \bB \cdot \zun$; it only depends on $\psi$ to lowest order. \\
$\matI$ & Unit matrix. \\
$J$, $J_u$ & Jacobian of the gyrokinetic transformation [see \eq{result_Jacob} and \eq{Jacob_u}].\\
$\bJ$ & Current density. \\
$\bJ_d$ & Current density due to magnetic drifts [see \eq{J_drift_M}]. \\
$\bJ_{gd}$ & Current density due to magnetic drifts calculated integrating $\bv_{M0} f_{ig}$ over velocity space instead of $\bv_{M0} f_i$ [see \eq{Jg_drift_M}]. \\
$J_{g||}^{(2)}$ & Parallel current density calculated integrating $v_{||} f_{ig}$ over velocity space instead of $v_{||} f_i$ [see \eq{Jgpar_o2}]. The superindex (2) emphasizes that $J_{||} = \bJ \cdot \bun$ and $J_{g||}^{(2)}$ differ only in higher order terms. \\
$\tilde{\bJ}_i$ & Polarization current density in gyrokinetic vorticity equation \eq{vort_type1}. \\
$\tilde{\bJ}_{i\phi}$ & Polarization current density in gyrokinetic vorticity equation \eq{vort_type2}. \\
$\tilde{\bJ}_{i\phi}^{(2)}$ & Polarization current density in gyrokinetic vorticity equation \eq{vort_type2_delta2}. \\
$k_{||}$, $k_\bot$ & Wavenumbers parallel and perpendicular to the magnetic field. \\
$L$ & Characteristic length in the problem. \\
$m$, $M$ & Electron and ion masses. \\
$n_e$, $n_i$ & Electron and ion densities. \\
$n_{ip}$ & Gyrokinetic polarization density defined in \eq{polar_n}. \\
$n_{ip}^{(2)}$ & Higher order gyrokinetic polarization density defined in \eq{nip_o2}. \\
$p_e$, $p_i$ & Electron and ion ``pressures." They are not the usual definitions because the average velocity is not subtracted. \\
$p_{e||}$, $p_{e\bot}$ & Electron parallel and perpendicular ``pressures." They are not the usual definitions because the average velocity is not subtracted. \\
$p_{i||}$, $p_{i\bot}$ & Ion parallel and perpendicular ``pressures." They are not the usual definitions because the average velocity is not subtracted. \\
$p_{ig||}$, $p_{ig\bot}$ & Parallel and perpendicular ``pressures" calculated integrating $Mv_{||}^2 f_{ig}$ and $(Mv_\bot^2/2) f_{ig}$ over velocity space instead of $Mv_{||}^2 f_i$ and $(Mv_\bot^2/2) f_i$. \\
$\matrixtop{\mathbf{P}}_i$ & Total ion stress tensor, including contributions due to the average ion velocity. \\
$q$ & Safety factor. \\
$\br$ & Position of the particle. \\
$r$ & In $\theta$-pinches, radial coordinate. \\
$\hat{\br}$ & In $\theta$-pinches, unit vector in the radial direction. \\
$R$ & In tokamaks, the distance between the axis of symmetry and the position of the particle, also used as major radius in estimates of order of magnitude. \\
$\bR$ & Position of the gyrocenter [see \eq{RGK_def}]. \\
$\bR_g$ & Position of the guiding center $\br + \Omega_i^{-1} \bv \times \bun$. \\
$\bR_1$ & First order correction to the gyrocenter position $\Omega_i^{-1} \bv \times \bun$. \\
$\bR_2$ & Second order correction to the gyrocenter position [see \eq{R_1_main}]. \\
$T_e$, $T_i$ & Electron and ion temperatures. \\
$u$ & Velocity of the gyrocenter parallel to the magnetic field [see \eq{par_drift}]. \\
$u_g$ & Parallel velocity found by replacing the gyrokinetic variables $\bR$, $E$ and $\mu$ in $u(\bR, E, \mu)$ by $\bR_g$, $E_0$ and $\mu_g$. [see \eq{ug_o2}]. \\
$\bv$ & Velocity of the particle. \\
$v_{||}$ & Velocity component parallel to the magnetic field. \\
$v_{||0}$ & Velocity parallel to the magnetic field with finite gyroradius modifications [see \eq{vpar0}]. \\
$\bv_\bot$, $v_\bot$ & Velocity perpendicular to the magnetic field and its magnitude. \\
$v_e$, $v_i$ & Electron and ion thermal velocities, $\sqrt{2T_e/m}$ and $\sqrt{2T_i/M}$. \\
$\bv_E$ & Gyrokinetic $E \times B$ drift [see \eq{drift_E}]. \\
$\bv_{E0}$ & Lowest order gyrokinetic $E \times B$ drift [see \eq{v_E0}]. \\
$\bv_d$ & Gyrokinetic drift composed of $\bv_E$ and $\bv_M$. \\
$\bv_{de}$ & Electron drift [see \eq{drift_electrons}]. \\
$\bv_{dg}$ & Drift found by replacing the gyrokinetic variables $\bR$, $E$ and $\mu$ in $\bv_d (\bR, E, \mu)$ by $\bR_g$, $E_0$ and $\mu_g$ [see \eq{vdg_o2}]. \\
$\bv_M$ & Gyrokinetic magnetic drift [see \eq{drift_M}]. \\
$\bv_{M0}$ & Standard magnetic drift [see \eq{v_M0}]. \\
$\tilde{\bv}_1$ & Drift due to finite gyroradius effects [see \eq{vtilde1}]. \\
$V^\prime$ & In tokamaks, flux surface volume element $dV/d\psi$. \\
$\bV_e$, $\bV_i$ & Electron and ion average velocities. \\
$\bV_{ig}$ & Ion average velocity calculated integrating $\bv f_{ig}$ over velocity space instead of $\bv f_i$. \\
$V_{ig||}^{(2)}$ & Ion parallel average velocity calculated integrating $v_{||} f_{ig}$ over velocity space instead of $v_{||} f_i$ [see \eq{Vigpar_o2}]. The superindex (2) emphasizes that $V_{i||} = \bV_i \cdot \bun$ and $V_{ig||}^{(2)}$ differ only in higher order terms. \\
$\bV_{iC}$ & Ion average velocity due to finite gyroradius effects on collisions [see \eq{nViC}]. \\
$\bV_{iE}$, $\bV_{igd}$ & Ion average velocities due to the gyrokinetic $E\times B$ and magnetic drifts [see \eq{VE_i} and \eq{VM_i}]. \\
$\tilde{\bV}_i$ & Ion average velocity due to finite gyroradius contribution $\tilde{\bv}_1$ [see \eq{Vtilde_i}]. \\
$V_{i||}^\mathrm{nc}$ & Neoclassical ion parallel velocity. \\
$Z$ & Ion charge number. \\
\end{supertabular}

\chapter{Introduction \label{chap_intro}}

Magnetic confinement is the most promising concept for production
of fusion energy and the tokamak is the best candidate among all
the possible magnetic confinement devices. However, transport of
particles, energy and momentum is still not well understood in
tokamaks. The transport is mainly turbulent, and modelling and
predicting how it evolves is necessary to build a viable reactor.

The understanding of turbulent transport in tokamaks has greatly
improved in the last decade, mainly due to more comprehensive
simulations \cite{dorland00, dannert05, candy03, chen03, dimits96,
lin02}. These simulations employ the gyrokinetic formalism to
shorten the computational time. Gyrokinetics is a sophisticated
asymptotic method that keeps finite gyroradius effects without
solving on the gyrofrequency time scale -- a time too short to be
of interest in turbulence. Unfortunately, modern formulations of
gyrokinetics are still only valid for times shorter than the
energy diffusion time or transport time scale. As simulations try
to reach longer time scales, the gyrokinetic formalism needs to be
extended. This thesis identifies the shortcomings of gyrokinetics
at long transport times and solves one of the most pressing
issues, namely, the calculation of the axisymmetric radial
electric field, and thereby, the transport of momentum.

In this introduction, first I will review the characteristics of
turbulence in tokamaks in section \ref{sect_turbulence}. This
review is followed by a brief history of gyrokinetic simulations
in section \ref{sect_history}, with emphasis on new developments.
The new codes being built will require a new formulation of
gyrokinetics valid for longer time scales. In section
\ref{sect_ini_discussion}, I will close the introduction by
discussing the requirements for future gyrokinetic formalisms and
summarizing the rest of the thesis.

\section{Turbulence in tokamaks \label{sect_turbulence}}

Currently, the main part of the turbulence in the core is believed
to be driven by drift waves. These waves propagate in the plasma
perpendicularly to density and temperature gradients. They become
unstable in tokamaks due to the curvature in the magnetic field
and other inhomogeneities. There are several modes that are
considered important, but the two of most interest here are the
Ion Temperature Gradient mode (ITG) \cite{biglari89} and the
Trapped Electron Mode (TEM) \cite{romanelli90}.

These modes have frequencies much smaller than the ion
gyrofrequency. They are unstable at short wavelengths -- the
fastest growing mode wavelengths are on the order of the ion
gyroradius because shorter wavelengths are stabilized by finite
gyroradius effects. The measurements in tokamaks suggest
turbulence correlation lengths on the order of five to ten
gyroradii \cite{fonck93}, which agrees with this idea. The
anisotropy induced by the magnetic field is reflected in the
spatial structure of these modes. The wavelength along the
magnetic field, on the order of the characteristic size of the
device, is much longer than the perpendicular wavelengths.

The ITG and TEM instabilities provide energy for the turbulence at
short wavelengths. By nonlinear beating, part of the turbulence
energy is deposited in a radial mode known as zonal flow
\cite{rosenbluth98, hinton99, diamond05}. The zonal flow is a
radial structure in the radial electric field that gives rise to a
sheared poloidal and toroidal $E \times B$ flow. It is a robust
mode because it does not have any parallel electric field and
hence electrons cannot shield it or Landau damp it. Then, any
energy deposited in the zonal flow will remain there, leading to a
rapid nonlinear growth of this mode. It has an impact on
turbulence dynamics because the velocity shear decorrelates the
turbulence at the shorter wavelengths. The statistical equilibrium
of the turbulence is determined by the feedback between zonal flow
and short wavelength fluctuations. The effect of zonal flow is so
important that it can suppress the turbulence when the instability
is not too strong \cite{dimits00}.

The radial electric field is crucial in the saturation of
turbulence. Its short wavelength radial structure is the zonal
flow, whose importance in turbulence dynamics has already been
discussed. It is also quite clear that it plays an important role
in the pedestal of high confinement or H-mode plasmas in tokamaks
\cite{connor00}, where the shear in the macroscopic radial
electric field becomes large. Experimentally, it is observed that
the radial electric field shear increases before the turbulent
fluctuations are suppressed, radial transport is quenched and the
gradient of density increases to form the pedestal \cite{moyer95}.
Both zonal flow and transport barriers highlight the importance of
the calculation of the radial electric field in any turbulence
simulation. In this thesis, I will show that the traditional
gyrokinetic approach is unable to provide the correct long
wavelength axisymmetric radial electric field. This problem has
gone undetected up until now because it is only noticeable at long
time scales.

To summarize, the turbulence in a tokamak is characterized by
electromagnetic fluctuations with wavelengths as small as the ion
gyroradius. On the other hand, the frequency of these fluctuations
is much smaller than the ion gyrofrequency, making the timescales
of gyromotion and turbulence so disparate that both can be treated
independently. It is this scale separation that gyrokinetics
exploits by ``averaging out" the gyromotion, while keeping the
finite gyroradius effects. However, as formulated, the traditional
gyrokinetic model does not contain enough physics to provide the
self-consistent long wavelength axisymmetric radial electric
field. Since the radial electric field affects turbulent
transport, we need to extend the gyrokinetic formalism to
calculate it.

\section{Gyrokinetics: history and current challenges \label{sect_history}}

The gyrokinetic model is a more suitable way of writing the
Fokker-Planck equation for low frequencies and short perpendicular
wavelengths. The idea of gyrokinetics is to define new variables
to replace the position $\br$ and velocity $\bv$ of the particle.
The gyrokinetic variables are constructed such that the the
gyromotion is decoupled from the slowly varying electromagnetic
fluctuations. This approach is especially convenient in turbulence
simulation because retaining the gyromotion is unnecessary. The
gyrokinetic variables are the appropriate variables to solve the
problem since they retain wavelengths on the order of the ion
gyroradius and ignore the high frequencies.

Gyrokinetics had its roots in reduced kinetic techniques used to
analyze stability problems with finite gyroradius effects. The
early works of Rutherford and Frieman \cite{rutherford68} and
Taylor and Hastie \cite{taylor68} treated small perpendicular
wavelengths in stability calculations for general magnetic field
geometries by using an eikonal approximation. Years later, Catto
\cite{catto78, catto81} formulated the gyrokinetic approach by
introducing the gyrokinetic change of variables.

The gyrokinetic formulation eventually evolved to a nonlinear
model. Frieman and Chen \cite{frieman82} developed a nonlinear
theory for perturbations of small amplitude over the distribution
function in general magnetic field geometry. Their work was
extended later for a full distribution function in a slab geometry
by Dubin \emph{et al} \cite{dubin83}. Hahm \emph{et al} extended
the work of Dubin \emph{et al} to electromagnetic perturbations
\cite{hahm88}, and toroidal geometry \cite{hahm88b}. These
nonlinear gyrokinetic equations were found employing a Hamiltonian
formulation and Lie transforms \cite{littlejohn81, brizard07}.

Based on these seminal nonlinear models, Lee developed the first
gyrokinetic code for investigation of the drift wave turbulence
\cite{lee83, lee87}. This first approach was a primitive
Particle-In-Cell (PIC) $\delta f$ model. The $\delta f$ models
avoid solving for the full distribution function, which would
require much computational time. Instead, the distribution
function is assumed to be Maxwellian to lowest order, and a
nonlinear equation for $\delta f$ retaining small fluctuations is
solved. The assumption is that the time it takes the turbulence to
saturate is much shorter than the diffusion time. Then, the
density and temperature profiles are given as an input and do not
change in time. In these codes, the turbulent fluctuations evolve
and saturate, and from their saturated value we can calculate the
radial particle and heat fluxes. The modern, less noisy $\delta f$
models originate in the ideas put forth by Kotschenreuther in
\cite{kotschen88}.

Several $\delta f$ codes, both continuum, like GS2
\cite{dorland00}, GENE \cite{dannert05} and GYRO \cite{candy03},
and PIC, like GEM \cite{chen03}, PG3EQ \cite{dimits96} and GTC
\cite{lin02}, have been developed and benchmarked. It is based on
these codes that most of the recent advances in tokamak turbulence
theory have occurred.

As it was already pointed out, the $\delta f$ codes are only
useful to compute the particle and heat fluxes once profiles for
density and temperature are given. It is necessary to develop a
new generation of models capable of self-consistently calculating
and evolving those profiles. It is not obvious that it can be done
with the current gyrokinetic formalism. For the $\delta f$ models
it was enough to run the codes until the turbulence had saturated,
but in order to let the profiles relax to their equilibrium, runs
on the order of the transport time scale are needed. This
extension is both a costly numerical task and an unsolved physical
problem. Models that reach transport time scales need to take into
account phenomena that were negligible when looking for the
turbulence saturation. In gyrokinetics, corrections to the
velocity of the particles small in a ion gyroradius over scale
length are neglected. However, as run times become longer, these
terms must be retained since a small velocity correction gives a
considerable contribution to the total particle motion.

In recent years, several groups have begun to build codes that
evolve the full distribution function, without splitting it into a
slowly varying Maxwellian and a fast, fluctuating piece. These
simulations, known as full $f$ models, are employing the
traditional gyrokinetic formulation. In this thesis, I will argue
that this approach is inadequate since it is unable to solve for
the self-consistent radial electric field that is crucial for the
turbulence.

Before getting into details, I will briefly review the four main
efforts in this field: GYSELA \cite{grandgirard06}, ELMFIRE
\cite{heikkinen08}, XGC \cite{chang08} and TEMPEST \cite{xu07}.
All these models are electrostatic. ELMFIRE and XGC are PIC
simulations, and GYSELA and TEMPEST are continuum codes. GYSELA
and XGC calculate the full ion distribution function, but they
adopt a fluid model for electrons that assumes an adiabatic
response along the magnetic field lines. ELMFIRE and TEMPEST solve
kinetically for both ions and electrons. Importantly, all four
models find the electrostatic potential from a gyrokinetic
Poisson's equation \cite{lee87}. This gyrokinetic Poisson's
equation just imposes that the ion and electron density must be
equal. It looks like Poisson's equation because there is a piece
of the ion density, known as polarization density, that can be
written explicitly as a Laplacian of the potential. Regardless of
its appearance, the gyrokinetic Poisson's equation is no more than
a lower order quasineutrality condition. It is lower order because
the density is calculated from the gyrokinetic equation in which
higher order terms have been neglected. This is the most
problematic part of these models, as I will demonstrate in this
thesis. Interestingly, GYSELA, ELMFIRE and XGC have reported an
extreme sensitivity to the initialization.

Operationally, the polarization density depends on the velocity
space derivatives of the distribution function (the polarization
density is presented in section~\ref{sect_GKpoisson}). It is
difficult to evaluate directly. In GYSELA, XGC and TEMPEST, the
wavelengths are taken to be longer than the ion gyroradius and the
distribution function is assumed to be close to a Maxwellian to
obtain a simplified expression. In an attempt to circumvent this
problem, ELMFIRE employs the gyrokinetic variables proposed by
Sosenko \emph{et al} \cite{sosenko01}. These variables include a
polarization drift that largely removes the polarization density.
With the polarization velocity, it is possible to use implicit
numerical schemes that give the dependence of the ion density with
the electrostatic potential.

To summarize, gyrokinetic modelling has been successfully used for
studying turbulence in the past decade. Codes based on $\delta f$
formulations, especially the continuum ones, have provided
valuable insights into tokamak anomalous transport. Currently,
there is an interest in extending these simulations to transport
timescales, and that requires careful evaluation of both physical
and numerical issues. As a result, several groups are building and
testing full $f$ simulations. In these codes, solving for the
axisymmetric radial electric field is crucial because it
determines the poloidal and toroidal flows, and those flows
strongly affect and, near marginality, control the turbulence
level. Unfortunately, the full $f$ community has failed to realize
that a straightforward extension of the equations valid for
$\delta f$ codes are unable to provide the long wavelength radial
electric field. The objective of this thesis is exposing this
problem and proposing a solution.

\section{Calculating the radial electric field \label{sect_ini_discussion}}

There are several problems that a gyrokinetic formulation has to
face before it is satisfactory for long time scales. The main
issue is the missing higher order terms in the gyrokinetic
Fokker-Planck equation. The transport of particles, momentum and
energy from one flux surface to the next is slow compared to the
typical turnover time of turbulent eddies, the characteristic time
scale for the traditional gyrokinetic formulation. To see this,
recall that the typical structures in the turbulence are of the
size of the ion gyroradius $\rho_i = Mcv_i/ZeB$, with $v_i =
\sqrt{2T_i/M}$ the ion thermal velocity, $Ze$, $M$ and $T_i$ the
ion charge, mass and temperature, $B$ the magnetic field
magnitude, and $e$ and $c$ the electron charge magnitude and the
speed of light. Then, eddies are of ion gyroradius size, requiring
many eddies -- and hence many eddy turnover times -- for a
particle to diffuse out of the tokamak.

The gyrokinetic Fokker-Planck equation is derived to an order
adequate for simulation of turbulence saturation, i.e., for time
scales on the order of the eddy turnover time. This equation is
too low of an order for transport time scales because flows and
fluxes that were neglected as small now have enough time to
contribute to the motion of the particles. In other words, a
higher order distribution function and hence a higher order
Fokker-Planck equation are required. For this reason, the
extension of gyrokinetics to transport time scales must draw from
the experience developed in neoclassical theory \cite{hinton76,
helander02bk}. Not only can neoclassical transport compete with
the turbulent fluxes in some limited cases, but the tools and
techniques developed in neoclassical theory become extremely
useful because they require only a lower order distribution
function to determine higher order radial fluxes of particles,
energy and momentum. The application of neoclassical tools in
gyrokinetic simulations is described in \cite{catto08} and
extended herein.

The physics in which the modern gyrokinetic formulation is
especially flawed is the calculation of the long wavelength radial
electric field. In this case, the comparison between neoclassical
theory and gyrokinetics is striking. In neoclassical theory, the
tokamak is intrinsically ambipolar due to its axisymmetry
\cite{kovrizhnykh69, rutherford70}, i.e., the plasma remains
quasineutral for any value of the radial electric field unless the
distribution function is known to higher order than second in an
expansion on the ion gyroradius over the scale length. The reason
for this is that the radial electric field is related to the
toroidal velocity through the $E\times B$ drift. Due to
axisymmetry, the evolution of the toroidal velocity only depends
on the small off-diagonal terms of the viscosity, making
impossible the self-consistent calculation of the radial electric
field unless the proper off-diagonal terms are included. The
distribution function required to directly obtain the viscosity is
higher order than second; the order at which intrinsic
ambipolarity is maintained. The axisymmetric radial electric field
has only been recently found in the Pfirsch-Schl\"{u}ter regime
\cite{catto05a, catto05b, catto05c, wong07}, and there has been
some incomplete work on the banana regime for high aspect ratio
tokamaks \cite{rosenbluth71, wong05}.

In gyrokinetics, however, the electric field is found from a lower
order gyrokinetic quasineutrality equation \cite{dubin83, lee83}
rather than from the transport of toroidal angular momentum.
Implicitly, it is assumed that the tokamak is not intrinsically
ambipolar in the presence of turbulence. In this thesis, I prove
that even turbulent tokamaks are intrinsically ambipolar in the
gyrokinetic ordering. Consequently, if the radial electric field
is to be retrieved from a quasineutrality equation, the
distribution function must be found to a hopelessly high order.
The physics that determine the radial electric field, namely, the
transport of angular momentum, enters the quasineutrality
condition only in higher order terms, making the gyrokinetic
quasineutrality equation inadequate for the calculation.

In this thesis, I pay special attention to the evolution of the
long wavelength axisymmetric radial electric field in the presence
of drift wave turbulence. Employing a current conservation
equation or vorticity equation, I assess the feasibility of
different methods to find the long wavelength axisymmetric radial
electric field. Each method requires the ion Fokker-Planck
equation to a different order in $\delta_i = \rho_i/L \ll 1$ and
$B/B_p \gsim 1$, with $\rho_i$ the ion gyroradius, $L$ a
characteristic size in the machine, typically the minor radius
$a$, $B$ the magnitude of the magnetic field and $B_p$ the
magnitude of its poloidal component. The different methods
explored in this thesis are summarized in
table~\ref{table_methods}. In this table, I give the chapter in
which the method is presented and the order of magnitude to which
the ion distribution function must be known compared to the zeroth
order distribution function $f_{Mi}$. To better explain the
classification by required order of magnitude of $f_i$, I present
here a very simplified heuristic study of the first method in
table~\ref{table_methods}. This method, known as the gyrokinetic
quasineutrality equation, will be rigourously described at the end
of chapter~\ref{chap_gyrokinetics}. In the gyrokinetic
quasineutrality equation, used in modern gyrokinetics, the
electric field is adjusted so that the ion and electron densities
satisfy $Zen_i = en_e$, and the densities are calculated by direct
integration of the distribution functions, i.e., $n_i = \int
d^3v\, f_i$ and $n_e = \int d^3v\, f_e$. The ion and electron
Fokker-Planck equations used to solve for the distribution
functions $f_i$ and $f_e$ are approximate. In
chapter~\ref{chap_gyrokinetics}, I will describe in more detail
the formalism to obtain these approximate equations. For now, it
is enough to consider a heuristic form of the long wavelength
limit of these equations in which only the motion of the guiding
center $\bR$ is considered. Then, schematically the Fokker-Planck
equations for ions and electrons are
\begin{equation}
\frac{\partial f_i}{\partial t} + \dot{\bR}_i \cdot \nabla f_i +
\ldots = C_{i,\mathrm{eff}} \{ f_i, f_e \} \label{sym_FPeq_ion}
\end{equation}

\noindent and
\begin{equation}
\frac{\partial f_e}{\partial t} + \dot{\bR}_e \cdot \nabla f_e +
\ldots = C_{e,\mathrm{eff}} \{ f_e, f_i \}, \label{sym_FPeq_elect}
\end{equation}

\begin{table}
\begin{center}
\renewcommand{\arraystretch}{1.5}
\begin{tabular}{|l|l|c|c|}
\hline \multicolumn{2}{|c|}{Method} & Order of $f_i$ & Chapter \\
\hline \hline \multicolumn{2}{|l|}{Gyrokinetic quasineutrality
equation} & $\delta_i^4 f_{Mi}$ & \ref{chap_gyrokinetics} \\
\hline \multirow{3}{0.2\textwidth}{Radial transport of toroidal
angular momentum} & Evaluated directly from $f_i$ & $\delta_i^3
f_{Mi}$ & \ref{chap_intrinsic} \\ \cline{2-4} & Moment
equation & $\delta_i^2 f_{Mi}$ & \ref{chap_angularmomentum} \\
\cline{2-4} & Moment equation and $B/B_p \gg 1$ & $(B/B_p)
\delta_i^2 f_{Mi}$ & \ref{chap_angularmomentum} \\ \hline
\end{tabular}
\end{center}

\caption[Different methods to obtain the radial electric
field]{Comparison of different methods to obtain the long
wavelength axisymmetric radial electric field.}
\label{table_methods}
\end{table}

\noindent where $\bR_i$ and $\bR_e$ are the ion and electron
guiding center positions, $\dot{\bR}_i = v_{||i} \bun + \bv_{di} +
\ldots$ and $\dot{\bR}_e = v_{||e} \bun + \bv_{de} + \ldots$ are
the drifts of those guiding centers, and $C_{i,\mathrm{eff}}$ and
$C_{e,\mathrm{eff}}$ are the effective collision operators. The
lower order drifts are the parallel velocities $v_{||i} \sim v_i$
and $v_{||e} \sim v_e$, and the perpendicular drifts $\bv_{di} =
\bv_{Mi} + \bv_E \sim \delta_i v_i$ and $\bv_{de} = \bv_{Me} +
\bv_E \sim \delta_e v_e \sim \delta_i v_i$, with $\bv_{Mi}$ and
$\bv_{Me}$ the magnetic drifts, $\bv_E = (c/B) \mathbf{E} \times
\bun$ the $E\times B$ drift and $\mathbf{E}$ the electric field.
The ion drifts are expanded in the small parameter $\delta_i =
\rho_i/L \ll 1$, and the electron drifts are expanded in the small
parameter $\delta_e = \rho_e/L \sim \delta_i \sqrt{m/M} \ll
\delta_i$, with $\rho_e = mcv_e/eB$ the electron gyroradius, $v_e
= \sqrt{2T_e/m}$ the electron thermal speed, and $T_e \sim T_i$
and $m$ the electron temperature and mass. The next order
corrections in $\delta_i$ to $\dot{\bR}_i$ may be written as
$\dot{\bR}_i = v_{||i} \bun + \bv_{di} + \dot{\bR}_i^{(2)} +
\dot{\bR}_i^{(3)} + \ldots$, with $\dot{\bR}_i^{(2)} =
O(\delta_i^2 v_i)$, $\dot{\bR}_i^{(3)} = O(\delta_i^3 v_i)$... The
operators $C_{i,\mathrm{eff}}$ and $C_{e,\mathrm{eff}}$ are
asymptotic expansions in $\delta_i$ as well. The next order
corrections for $\dot{\bR}_e$ are small in the parameter $\delta_e
\ll \delta_i$. For this heuristic introduction, I will drop the
next order corrections to $\dot{\bR}_e$ exploiting the scale
separation $\delta_e \ll \delta_i$ to find $\dot{\bR}_e \simeq
v_{||e} \bun + \bv_{de}$. This is not rigorous, but it does not
change the final result and simplifies the derivation. Under all
these assumptions, the ion and electron distribution functions can
be solved for perturbatively, giving $f_i = f_i^{(0)} + f_i^{(1)}
+ f_i^{(2)} + \ldots$ and $f_e = f_e^{(0)} + f_e^{(1)} + f_e^{(2)}
+ \ldots$, with $f_i^{(0)} = f_{Mi}$, $f_i^{(1)} = O(\delta_i
f_{Mi})$, $f_i^{(2)} = O(\delta_i^2 f_{Mi})$... and $f_e^{(0)} =
f_{Me}$, $f_e^{(1)} = O(\delta_i f_{Me})$, $f_e^{(2)} =
O(\delta_i^2 f_{Me})$...

In this thesis I prove that to find the long wavelength radial
electric field in axisymmetric configurations using the
gyrokinetic quasineutrality equation, it is necessary to solve the
Fokker-Planck equations to fourth order because equations
\eq{sym_FPeq_ion} and \eq{sym_FPeq_elect} satisfy the condition
\begin{eqnarray}
\frac{\partial}{\partial t} \langle Zen_i - en_e \rangle_\psi
\equiv - \left \langle \nabla \cdot \left [ \int d^3v \left ( Ze
f_i \dot{\bR}_i - e f_e \dot{\bR}_e \right ) \right ] \right
\rangle_\psi \nonumber \\ + \left \langle \int d^3v \left ( Ze
C_{i,\mathrm{eff}} \{f_i, f_e\} + e C_{e,\mathrm{eff}} \{f_e,
f_i\} \right ) \right \rangle_\psi = O(\delta_i^4 en_ev_i/L).
\label{sym_condition3order}
\end{eqnarray}

\noindent Here, $\langle \ldots \rangle_\psi$ is the flux surface
average. For now, it is only important to know that $\langle
\ldots \rangle_\psi$ makes the non-axisymmetric pieces vanish [see
chapter~\ref{chap_intrinsic}]. Since the axisymmetric radial
electric field adjusts so that the axisymmetric pieces of the ion
and electron densities $\langle n_i \rangle_\psi$ and $\langle n_e
\rangle_\psi$ satisfy quasineutrality, equation
\eq{sym_condition3order} requires that terms of order $\delta_i^4
f_{Mi} v_i/L$ be kept in equations \eq{sym_FPeq_ion} and
\eq{sym_FPeq_elect} to obtain the self-consistent radial electric
field. Gyrokinetic codes solve a Fokker-Planck equation only
through $O(\delta_i f_{Mi} v_i/L)$, leaving the radial electric
field as a free parameter in the best case (intrinsic
ambipolarity), or finding an unphysical result in the worst
scenario. From an ion Fokker-Planck equation of order $\delta_i^4
f_{Mi} v_i/L$ it is possible in principle (but not in practice) to
obtain a distribution function good to order $\delta_i^4 f_{Mi}$.
For this reason, in table~\ref{table_methods} the order to which
the distribution function is required is $\delta_i^4 f_{Mi}$.
Interestingly, equation \eq{sym_condition3order} simplifies
considerably because of the flux surface average, giving
\begin{eqnarray}
\frac{\partial}{\partial t} \langle Zen_i - en_e \rangle_\psi
\equiv - Ze \left \langle \nabla \cdot \left [ \int d^3v \left (
f_i^{(2)} \dot{\bR}_i^{(2)} + f_i^{(1)} \dot{\bR}_i^{(3)} +
f_i^{(0)} \dot{\bR}_i^{(4)} \right ) \right ] \right \rangle_\psi
\nonumber \\ + \left \langle \int d^3v \left ( Ze
C^{(4)}_{i,\mathrm{eff}} \{ f_i^{(1)}, f_i^{(2)} \} - e
C^{(4)}_{e,\mathrm{eff}} \{ f_i^{(1)}, f_i^{(2)} \} \right )
\right \rangle_\psi = O(\delta_i^4 en_ev_i/L).
\label{sym_condition3order_2}
\end{eqnarray}

\noindent Notice that, after flux surface averaging and
integrating over velocity space, the difference of the fourth
order pieces of the collision operators only depends on
$f_i^{(1)}$ and $f_i^{(2)}$, and the terms
\begin{equation}
\langle \nabla \cdot ( J_{||}^{(4)} \bun ) \rangle_\psi \equiv
\left \langle \nabla \cdot \left [ \int d^3v\, \left ( Ze
f_i^{(4)} v_{||i} \bun - e f_e^{(4)} v_{||e} \bun \right ) \right
] \right \rangle_\psi = 0 \label{sym_parcurr}
\end{equation}

\noindent and
\begin{equation}
\langle \nabla \cdot \bJ_d^{(3)} \rangle_\psi \equiv \left \langle
\nabla \cdot \left [ \int d^3v\, \left ( Ze f_i^{(3)} \bv_{di} - e
f_e^{(3)} \bv_{de} \right ) \right ] \right \rangle_\psi = 0
\label{sym_driftcurr}
\end{equation}

\noindent exactly vanish in axisymmetric configurations. The term
\eq{sym_parcurr} is the contribution of the parallel current
density $J_{||}$ to the radial current. Since the radial current
is perpendicular, this contribution is obviously zero. The term
\eq{sym_driftcurr} is more subtle. It is the contribution of the
current density $\bJ_d$ due to the magnetic drifts. This
contribution vanishes in axisymmetric configurations because to
lowest order the net radial displacement due to magnetic drifts is
zero. In chapter~\ref{chap_intrinsic}, I will show that the radial
component of the current density $\bJ_d$ is related to the
parallel and perpendicular pressures $p_{||}$ and $p_\bot$ in the
momentum conservation equation, and these pressures finally do not
enter in the calculation of the toroidal rotation, the quantity
that determines the radial electric field.

Importantly, according to equation \eq{sym_condition3order_2}, a
distribution function $f_i$ good to $O(\delta_i^2 f_{Mi})$ is
enough to calculate the radial electric field although we need the
higher order corrections $\dot{\bR}_i^{(2)}$, $\dot{\bR}_i^{(3)}$
and $\dot{\bR}_i^{(4)}$; terms never employed in gyrokinetic
codes. In this thesis, I exploit the fact that only the second
order correction of $f_i$ is needed to obtain the long wavelength
axisymmetric electric field. Moreover, I will not need to compute
the higher order terms $\dot{\bR}_i^{(2)}$, $\dot{\bR}_i^{(3)}$
and $\dot{\bR}_i^{(4)}$ explicitly. It is possible to circumvent
this calculation by employing the radial transport of toroidal
angular momentum, introduced in chapter~\ref{chap_intrinsic}. By
doing so we gain two orders in $\delta_i$, i.e., we make explicit
that according to \eq{sym_condition3order_2} we only need an ion
distribution function good to order $\delta_i^2 f_{Mi}$. It is not
necessary to obtain the higher order corrections to $\dot{\bR}_i$
because the toroidal angular momentum conservation equation is
obtained from a full Fokker-Planck equation where no approximation
for small $\delta_i$ has been made. Importantly, the procedure
used to evaluate the radial transport of toroidal angular momentum
makes a considerable difference. As given in
table~\ref{table_methods}, direct evaluation from the ion
distribution function requires the ion distribution function to
$O(\delta_i^3 f_{Mi})$, whereas the moment approach presented in
chapter~\ref{chap_angularmomentum} only requires a distribution
function good to $O(\delta_i^2 f_{Mi})$. The moment approach works
because the radial transport of toroidal angular momentum depends
only on the third order gyrophase dependent piece of the ion
distribution function. The third order gyrophase dependent piece
of the ion distribution function can be expressed as a function of
the second order piece by using the full Fokker-Planck equation,
and the moment approach is a simple way to write that relation.

The second order distribution function is still an order higher
than usual gyrokinetic codes are built for, but there is a
possible simplification listed last in table~\ref{table_methods}.
The idea is exploiting the usually largish parameter $B/B_p \sim
10$. The new method, described at the end of
chapter~\ref{chap_angularmomentum}, is advantageous because
conventional gyrokinetic Fokker-Planck equations can provide, with
only a few modifications in the implementation, the ion
distribution function to order $(B/B_p) \delta_i^2 f_{Mi}$, high
enough order to self-consistently determine the long wavelength
axisymmetric radial electric field for $B/B_p \gg 1$.

Along with the long wavelength axisymmetric radial electric field,
I investigate the evolution of the axisymmetric flows in drift
wave turbulence. Under the assumptions explained at the beginning
of chapter~\ref{chap_gyrovorticity}, the long wavelength
axisymmetric flows remain neoclassical even in turbulent tokamaks.
Moreover, it is possible to prove that in the modern gyrokinetic
formalism the axisymmetric components of the flows and the radial
electric field with radial wavelengths above $\sqrt{\rho_i L}$ are
unreliable due to intrinsic ambipolarity. This result is related
to the unrealistic higher order distribution functions needed to
obtain the long wavelength axisymmetric radial electric field from
the quasineutrality equation.

The rest of the thesis is organized as follows. In
chapter~\ref{chap_intrinsic}, the calculation of the radial
electric field is formulated in terms of a current conservation
equation or vorticity equation. I show that, for the axisymmetric
radial electric field, the vorticity equation reduces to the
radial transport of toroidal angular momentum. The vorticity
equation must be evaluated in the presence of turbulence, and the
natural formulation for drift wave turbulence is the gyrokinetic
formalism, presented in chapter~\ref{chap_gyrokinetics}. In
chapter~\ref{chap_gyrovorticity}, the gyrokinetic formulation is
applied to the vorticity equation. The resulting equation makes
explicit the time scales involved in the evolution of the
axisymmetric flows and the axisymmetric radial electric field.
With this formulation, turbulent tokamaks are proven to be
intrinsically ambipolar to the same order as neoclassical theory.
In chapter~\ref{chap_angularmomentum}, I present a different
approach based on the radial transport of toroidal angular
momentum that only requires an $O(\delta_i^2 f_{Mi})$ distribution
function, and by exploiting an expansion in $B/B_p \gg 1$, I
formulate a relatively simple model capable of self-consistently
evolving the axisymmetric radial electric field in the core of a
tokamak. By employing an example with simplified geometry in
chapters~\ref{chap_gyrovorticity} and \ref{chap_angularmomentum},
I illustrate the problems that arise from the use of the
gyrokinetic quasineutrality equation, and how a new approach can
solve them. Finally, in chapter~\ref{chap_conclusion}, I summarize
the findings in this thesis, and I describe a program to gradually
implement a new gyrokinetic formulation in current codes that can
solve for the axisymmetric radial electric field.

\chapter{Vorticity and intrinsic ambipolarity \label{chap_intrinsic}}

In this chapter, I study the quasineutrality equation in an
axisymmetric configuration. The time derivative of
quasineutrality, also known as vorticity equation, makes the time
scales in the problem explicit. With this equation, it is possible
to show that the radial current is zero to a very high order
independently of the axisymmetric, long wavelength radial electric
field, i.e., both ions and electrons drift radially in an
intrinsically ambipolar manner even in the presence of turbulence.
Moreover, if the radial current is calculated to high enough
order, I can also show that forcing it to vanish is equivalent to
solving the toroidal angular momentum conservation equation.

The chapter is organized as follows. In
section~\ref{sect_orderings1}, I explain and justify the
assumptions necessary to simplify the problem. In
section~\ref{sect_FPvorticity}, I derive a vorticity equation from
the full Fokker-Planck equation. This equation cannot be usefully
implemented as it is written in this chapter because some of the
terms are difficult to evaluate. This issue will be addressed in
chapter~\ref{chap_gyrovorticity}. However, this vorticity equation
is useful because it makes the study of the evolution of the
radial electric field easier. In
section~\ref{sect_fluxsurfaceaverage}, I flux surface average the
vorticity equation to determine the radial electric field. The
radial electric field adjusts so that the total radial current in
the tokamak vanishes, and the flux surface averaged vorticity
equation is equivalent to imposing that the total radial current
is zero. I show here that the flux surface averaged vorticity
equation is the conservation equation for toroidal angular
momentum. Then, by estimating the size of the term that contains
the transport of angular momentum in the vorticity equation, I can
argue that the radial current is zero to a very high order and
hence the plasma is intrinsically ambipolar to the same order as
neoclassical theory.

\section{Orderings and assumptions \label{sect_orderings1}}

To simplify the calculations, and for the rest of this thesis, I
assume that the electric field is electrostatic, i.e., $\mathbf{E}
= - \nabla \phi$, with $\phi$ the electrostatic potential. In
general, the electromagnetic turbulent fluctuations are important
and should be considered \cite{kim93, snyder01}. However, keeping
electromagnetic effects would obscure derivations that are already
quite involved. Furthermore, in the Coulomb gauge, the
axisymmetric radial electric field is purely derived from the
potential. Thus, calculating the axisymmetric radial electric
field is fundamentally an electrostatic problem. The electrostatic
formulation presented in this thesis will offer a solution that
can be extended later to electromagnetic turbulence.

\begin{figure}

\includegraphics[width = \textwidth]{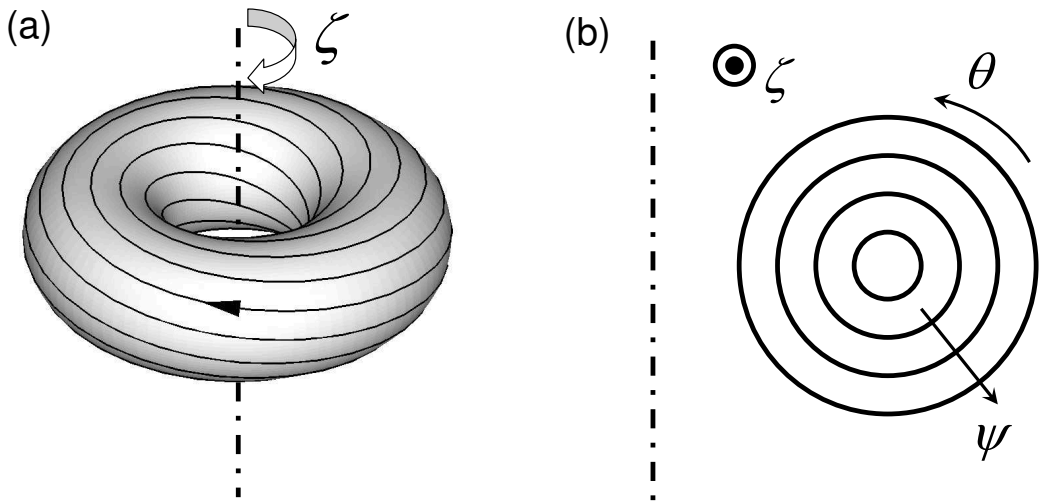}

\vspace{-4cm} \caption[Magnetic coordinates in tokamaks]{Magnetic
coordinates in tokamaks. (a) Three-dimensional view of the tokamak
where the magnetic field lines are the solid black lines and the
axis of symmetry is the chain-dot line. (b) Poloidal plane where
the flux surfaces are schematically represented as concentric
circles. Here, $\zeta$ points out of the paper, $\psi$ labels
different flux surfaces, and $\theta$ defines the position within
the flux surface.} \label{fig_tokamakgeom}
\end{figure}

To be consistent with the electrostatic electric field, the
magnetic field $\bB$ is assumed to be constant in time. In
addition, it has a characteristic length of variation much larger
than the ion gyroradius. I use an axisymmetric magnetic field,
\begin{equation}
\bB = I \nabla \zeta + \nabla \zeta \times \nabla \psi,
\label{B_def}
\end{equation}

\noindent with $\psi$ and $\zeta$ the magnetic flux and toroidal
angle coordinates. The vector $\nabla \zeta = \zun / R$ with
$\zun$ the unit vector in the toroidal direction and $R$ the
radial distance to the symmetry axis of the torus. I use a
poloidal angle $\theta$ as the third coordinate, and employ the
unit vector $\bun = \mathbf{B}/B$ with $B = |\mathbf{B}|$. The
coordinates $\psi$, $\zeta$ and $\theta$ are shown in
figure~\ref{fig_tokamakgeom}. The toroidal magnetic field,
$B_\zeta = I/R$, is determined by the function $I$ that only
depends on the radial variable $\psi$ to zeroth order.

The results in this thesis are expansions in the small parameter
$\delta_i = \rho_i/L \ll 1$, with $L$ the characteristic length in
the problem. Collisions and non-neutral effects must be ordered
with respect to $\delta_i$. To do so, I will use the
characteristic tokamak values given in table~\ref{table_numbers}.
In this table, the minor and major radii and the magnetic field
strength are taken from \cite{hutchinson94} for Alcator C-Mod and
\cite{luxon02} for DIII-D. The electron density $n_e$ and the
electron temperature $T_e$ are given in \cite{hutchinson94} for
Alcator C-Mod, and in \cite{groebner90, gohil06} for DIII-D. The
ion temperature $T_i$ is difficult to measure, but in general can
be assumed to be of the order of the electron temperature $T_e$.
The average ion velocity $V_i$ is taken from \cite{greenwald05,
mcdermott09} for Alcator C-Mod, and from \cite{groebner90,
degrassie07} for DIII-D. The average velocities in DIII-D depend
strongly on the neutral beam injection, ranging from 20~km/s when
the neutral beams are turned off or their net momentum input is
zero, to 200~km/s in cases that the beams drive large rotation.
Alcator C-Mod, on the other hand, does not have neutral beam
injection and tends to rotate at lower speeds. The rest of the
quantities in table~\ref{table_numbers} are calculated from the
measured values. These quantities are the electron and ion thermal
velocities $v_e = \sqrt{2T_e/m}$ and $v_i = \sqrt{2T_i/M}$; the
electron and ion gyrofrequencies $\Omega_e = eB/mc$ and $\Omega_i
= ZeB/Mc$; the electron and ion gyroradii, $\rho_e = v_e/\Omega_e$
and $\rho_i = v_i / \Omega_i$; the plasma frequency $\omega_p =
\sqrt{4\pi e^2 n_e/m}$ and Debye length $\lambda_D =
v_e/\omega_p$; the electron-electron, electron-ion, ion-electron
and ion-ion Coulomb collision frequencies $\nu_{ee} =
(4\sqrt{2\pi}/3) \times (e^4 n_e \ln \Lambda/\sqrt{m} T_e^{3/2})$,
$\nu_{ei} = Z \nu_{ee}$, $\nu_{ie} = (2Z^2 m/M) \nu_{ee}$ and
$\nu_{ii} = (4\sqrt{\pi}/3) \times (Z^3 e^4 n_e \ln \Lambda/
\sqrt{M} T_i^{3/2})$, and the corresponding mean free paths
$\lambda_{ee} = v_e/\nu_{ee}$, $\lambda_{ei} = v_e/\nu_{ei}$,
$\lambda_{ie} = v_i/\nu_{ie}$ and $\lambda_{ii} = v_i/\nu_{ii}$.

\begin{table}
\begin{center}
\renewcommand{\arraystretch}{1.5}
\begin{tabular}{|l|c|c|c|c|}
\cline{2-5} \multicolumn{1}{l|}{ } & \multicolumn{2}{c|} {Alcator C-Mod} & \multicolumn{2}{c|}{DIII-D} \\
\cline{2-5} \multicolumn{1}{l|}{ } & Core & Separatrix & Core & Separatrix \\
\hline $B$ $(T)$ & $5$ & $5$ & $2$ & $2$ \\
$n_e$ (m$^{-3}$) & $3\times 10^{20}$ & $10^{20}$ & $3\times 10^{19}$ & $10^{19}$ \\
$T_e \sim T_i$ (keV) & $2$ & $0.1$ & $2$ & $0.1$  \\
$V_i$ (km/s) & $20$ & $20$ & $20 - 200$ & $20$ \\
\hline $v_e$ (km/s) & $27000$ & $5900$ & $27000$ & $5900$ \\
$v_i$ (km/s) & $620$ & $140$ & $620$ & $140$ \\
\hline $\Omega_e$ (GHz) & 880 & 880 & 350 & 350 \\
$\Omega_i$ (GHz) & $0.48$ & $0.48$ & $0.19$ & $0.19$ \\
$\rho_e$ ($\mu$m) & $31$ & $6.7$ & $77$ & $17$ \\
$\rho_i$ ($\mu$m) & $920$ & $290$ & $3300$ & $740$ \\
\hline $\omega_p$ (GHz) & $980$ & $560$ & $310$ & $180$ \\
$\lambda_D$ ($\mu$m) & $28$ & $11$ & $87$ & $33$ \\
\hline $\nu_{ee} \sim \nu_{ei}$ (kHz) & $150$ & $3600$ & $16$ & $400$ \\
$\nu_{ii}$ (kHz) & $2.4$ & $60$ & $0.26$ & $6.5$ \\
$\nu_{ie}$ (kHz) & $0.16$ & $4.0$ & $0.017$ & $0.44$ \\
$\lambda_{ii} \sim \lambda_{ei} \sim \lambda_{ee}$ (m) & $260$ & $2.3$ & $2400$ & $22$ \\
$\lambda_{ie}$ (m) & $3800$ & $35$ & $35000$ & $320$ \\
\hline Minor radius $a$ (m) & \multicolumn{2}{|c|}{$0.21$} & \multicolumn{2}{c|}{$0.67$ } \\
Major radius $R$ (m) & \multicolumn{2}{|c|}{$0.67$} & \multicolumn{2}{c|}{$1.66$ } \\
\hline
\end{tabular}
\end{center}

\caption[Typical numbers for Alcator C-Mod and DIII-D]{Typical
numbers for Alcator C-Mod \cite{hutchinson94, greenwald05,
mcdermott09} and DIII-D \cite{groebner90, luxon02, gohil06,
degrassie07}.} \label{table_numbers}
\end{table}

In table~\ref{table_numbers}, $V_i/v_i \sim 0.3$ for shots with
neutral beam injection, $V_i/v_i \sim 0.1$ in the pedestal region,
and $V_i/v_i \sim 0.03$ in the core in the absence of neutral beam
injection. In general, in the core, $V_i \sim V_e < v_i$ can be
assumed. Moreover, in the absence of neutral beam injection, the
average ion velocity is comparable to $V_i \sim \delta_i v_i$,
with $\delta_i = \rho_i/L \sim 5 \times 10^{-3} \ll 1$ and $L$ the
characteristic length in the problem, in this case the minor
radius $a$. Ordering $V_i \sim \delta_i v_i \sim \delta_e v_e \sim
V_e$, known as the low flow or drift ordering, is then justified.
This ordering allows the electric field to compete with the
pressure gradient by making the $E\times B$ flow, $(cn_i/B)
\mathbf{E} \times \bun$, and diamagnetic flow, $(c/ZeB) \bun
\times \nabla p_i$, comparable. According to the drift ordering,
the electric field is $\mathbf{E} = - \nabla \phi \sim T_e/eL$,
giving a total electrostatic potential drop across the core of
order $T_e/e$. The cases with neutral beam injection can be
recovered by employing the drift ordering and then sub-expanding
in $eL|\nabla \phi|/T_e \gg 1$, giving a velocity $V_i \sim
(eL|\nabla \phi|/T_e) \delta_i v_i \gg \delta_i v_i$. In the
pedestal, on the other hand, the gradients are large, making $L
\ll a$, and the average velocity is closer to the thermal speed,
$V_i/v_i \sim 0.1$. Interestingly, it is possible to find an
ordering similar to the low flow ordering because the pressure
gradient and the electric field must be allowed to compete
\cite{kagan08}. In this thesis, I focus in the core, where the
drift ordering $V_i \sim \delta_i v_i \sim \delta_e v_e \sim V_e$
is valid, but it is possible that the formalism I will develop
could be extended to the pedestal region.

To evaluate the importance of non-neutral effects, I need to
compare the turbulence frequencies and wavelengths with the plasma
frequency $\omega_p$ and the Debye length $\lambda_D$. I am
interested in the long wavelength, axisymmetric radial electric
field and its evolution in the presence of drift wave turbulence.
In section~\ref{sect_turbulence}, I explained that the drift wave
turbulence spectrum extends from the minor radius to the ion
gyroradius. Theoretical studies \cite{dorland00, jenko00} suggest
that the turbulent spectrum can also reach wavelengths of the
order of the electron gyroradius. Since my research focuses on the
long wavelength part of the electric field, I restrict myself to
perpendicular wavelengths between the ion gyroradius and the minor
radius. In this range, the Debye length is small,
$\lambda_D/\rho_i \sim 0.03 \ll 1$. Moreover, the typical
frequencies are of the order of the drift wave frequency
$\omega_\ast = k_\bot c T_e/eBL_n \sim k_\bot \rho_i v_i/L \ssim
v_i/L \sim \delta_i \Omega_i \ll \Omega_i$, with $L_n = |\nabla
\ln n_e|^{-1}$. The plasma frequency is then very high,
$\omega_\ast/\omega_p \sim 3\times 10^{-6}$, and the plasma may be
assumed quasineutral. There are exceptions in which non-neutral
effects have to be considered. I will point them out as they
appear, but in general I will ignore these corrections to simplify
the derivation. They are easy to implement and they could be added
in the future with the electromagnetic corrections.

Collisional mean free paths range from very long in the center of
the tokamak, $R/\lambda_{ii} \sim R/\lambda_{ee} \sim
R/\lambda_{ei} \sim 10^{-3} \sim \delta_i$, to comparable to the
major radius in the pedestal, $R/\lambda_{ii} \sim R/\lambda_{ee}
\sim R/\lambda_{ei} \sim 0.01 - 0.1$. Notice that the mean free
path is compared to the typical length along the magnetic field,
proportional to the major radius $R$. I will order collisions as
$\nu_{ie} \ll \nu_{ii} \sim v_i/L$ and $\nu_{ee} \sim \nu_{ei}
\sim v_e/L$, and the low collisionality case can be obtained by
then sub-expanding in the small parameter $qR/\lambda_{ii} \sim
qR/\lambda_{ee} \sim qR/\lambda_{ei} \ll 1$, where $q \gsim 1$ is
the safety factor. In this manner, collisions are kept in the
derivation. This collisional ordering implies that particles are
confined long enough to become a Maxwellian to lowest order. Then,
the lowest order ion and electron distribution functions are the
stationary Maxwellians $f_{Mi}$ and $f_{Me}$. They are assumed to
be stationary to be consistent with the drift ordering.

To summarize, electromagnetic effects and non-neutral effects are
dropped. The typical frequency is assumed to be $\omega_\ast \ssim
v_i/L$. Collisions are ordered as $\nu_{ie} \ll \nu_{ii} \sim
v_i/L$ and $\nu_{ee} \sim \nu_{ei} \sim v_e/L$. The ion and
electron distribution functions are stationary Maxwellians,
$f_{Mi}$ and $f_{Me}$, to lowest order. Since $L$ is the
characteristic length in the problem, $\nabla f_i \sim f_{Mi}/L$
and $\nabla f_{Me} \sim f_{Me}/L$. The average velocities are
ordered as small in $\delta_i$, $V_i \sim \delta_i v_i \sim
\delta_e v_e \sim V_e$, and consequently the electric field is
$O(T_e/eL)$. In chapters~\ref{chap_gyrokinetics} and
\ref{chap_gyrovorticity}, I will extend these assumptions to the
shorter turbulent wavelengths. For this chapter, intended as an
introduction to the properties of the long wavelength axisymmetric
radial electric field, it will be enough to consider only the
longer wavelengths, of order $L$. Thus, in this chapter, the
gradients are $\nabla \sim 1/L$.

Finally, in the derivation it will be useful to split the velocity
of the particles into components parallel and perpendicular to the
magnetic field, with $v_{||} = \bv \cdot \bun$ the parallel
component and $\bv_\bot = \bv - v_{||}\bun$ the perpendicular. The
perpendicular velocity is determined by its magnitude $v_\bot =
|\bv_\bot|$ and the gyrophase $\varphi_0$, defined such that
\begin{equation}
\bv_\bot = v_\bot ( \eun_1 \cos \varphi_0 + \eun_2 \sin \varphi_0
), \label{gyrophase_def}
\end{equation}

\noindent where the unit vectors $\bun (\br)$, $\eun_1 (\br)$ and
$\eun_2 (\br)$ are an orthonormal system such that $\eun_1 \times
\eun_2 = \bun$. Notice that $\eun_1$ and $\eun_2$ depend on the
position $\br$ because $\bun$ depends on $\br$. For a general
magnetic field, $\eun_1$ and $\eun_2$ may be chosen to be the
normal $\eun_1 = \bun \cdot \nabla \bun / |\bun \cdot \nabla
\bun|$ and binormal $\eun_2 = \bun \times \eun_1$ of the magnetic
field line. In a tokamak, they could be defined as $\eun_1 =
\nabla \psi/|\nabla \psi|$ and $\eun_2 = (\bun \times \nabla
\psi)/|\nabla \psi|$.

The distinction between the gyrophase independent and dependent
pieces of the distribution function will be important. I will
denote the gyroaverage, or average over the gyrophase $\varphi_0$,
holding $\br$, $v_{||}$, $v_\bot$ and $t$ fixed, by
$\overline{(\ldots)}$. It is important which variables are held
fixed because when the gyrokinetic variables are defined, they
will have their own distinct gyroaverage. Notice that this average
is not weighted with the distribution function, i.e., $\int d^3v f
\overline{Q} \neq \int d^3v f Q$, where $f$ is the distribution
function and $Q(\varphi_0)$ is some given function of the
gyrophase $\varphi_0$.

\section{Vorticity equation \label{sect_FPvorticity}}

To obtain the electrostatic potential and build in the
quasineutrality condition, but also make explicit the time scales
that enter the problem, I work with the current conservation or
vorticity equation,
\begin{equation}
\nabla \cdot \bJ = 0, \label{cserv_J}
\end{equation}

\noindent where $\bJ = Zen_i\bV_i - en_e \bV_e$ is the current
density, and $n_i = \int d^3v\, f_i$, $n_e = \int d^3v\, f_e$,
$n_i \bV_i =  \int d^3v\, \bv f_i$ and $n_e \bV_e =  \int d^3v\,
\bv f_e$ are the ion and electron densities, and the ion and
electron average flows. The functions $f_i$ and $f_e$ are the ion
and electron distribution functions, respectively. The parallel
current $J_{||} = \bJ \cdot \bun$ can be obtained to the requisite
order by integrating over the ion and electron distribution
functions as discussed in more detail in
chapter~\ref{chap_gyrovorticity}.

The perpendicular current $\bJ_\bot = \bJ - J_{||}\bun$ is given
by the perpendicular component of the total momentum conservation
equation. The total momentum conservation equation is
\begin{equation}
\frac{\partial}{\partial t} (n_i M \bV_i) + \nabla \cdot \left [
\int d^3 v \; ( M f_i + m f_e ) \bv \bv \right ] = \frac{1}{c} \bJ
\times \bB. \label{total_momentum}
\end{equation}

\noindent I neglect the inertia of electrons because their mass is
much smaller than the mass of the ions. The total stress tensor
can be rewritten as
\begin{equation}
\int d^3 v \; ( M f_i + m f_e ) \bv \bv = p_\bot (\matI - \bun
\bun) + p_{||} \bun \bun + \matrixtop{\pibf}_i,
\end{equation}

\noindent where $p_\bot = p_{i\bot} + p_{e\bot} = \int d^3 v\; (M
f_i + m f_e) v_\bot^2/2$ is the total perpendicular ``pressure",
$p_{||} = p_{i||} + p_{e||} = \int d^3 v\; (M f_i + m f_e)
v_{||}^2$ is the total parallel ``pressure", and
\begin{equation}
\matrixtop{\pibf}_i = M \int d^3 v \; f_i ( \mathbf{v} \mathbf{v}
- \overline{\bv \bv} ) = M \int d^3 v \; f_i \left [ \bv \bv -
\frac{v_\bot^2}{2} ( \matI - \bun \bun ) - v_{||}^2 \bun \bun
\right ] \label{pi_def}
\end{equation}

\noindent is the ion ``viscosity." The electron viscosity is
neglected because it is $m/M$ smaller, as I will prove in
chapter~\ref{chap_gyrovorticity}. Here, $\matI$ is the unit dyad,
and $\overline{\bv\bv} = (v_\bot^2/2)(\matI - \bun\bun) + v_{||}^2
\bun \bun$ is the gyroaverage of $\bv\bv$ holding $\br$, $v_{||}$,
$v_\bot$ and $t$ fixed. The definitions of the ``pressures"
$p_\bot$ and $p_{||}$ and the ``viscosity" $\matrixtop{\pibf}_i$
differ from the usual in that the average velocity is not
subtracted. The usual perpendicular and parallel pressures are
$p_\bot^\prime = p_\bot - n_i M V_{i\bot}^2/2$ and $p_{||}^\prime
= p_{||} - n_i M V_{i||}^2$, and the usual viscosity is
$\matrixtop{\pibf}_i^\prime = \matrixtop{\pibf}_i - n_i M [ \bV_i
\bV_i - (V_{i\bot}^2/2) (\matI - \bun \bun) - V_{i||}^2 \bun \bun
]$. Notice that $\matrixtop{\pibf}_i$ contains the turbulent
Reynolds stress. In the drift ordering, the pressures and
viscosity used here are more convenient than the more common
definitions.

Obtaining the perpendicular current $\mathbf{J}_\bot$ from
\eq{total_momentum}, substituting it into \eq{cserv_J} and
employing
\begin{equation}
\frac{c}{B} \bun \times \nabla p_\bot = - \nabla \times \left (
\frac{cp_\bot}{B} \bun \right ) + \frac{cp_\bot}{B^2} \bun \times
\nabla B + \frac{cp_\bot}{B} \nabla \times \bun
\end{equation}

\noindent and
\begin{equation}
\nabla \times \bun = \bun \bun \cdot \nabla \times \bun + \bun
\times \kappabf, \label{id_kappa}
\end{equation}

\noindent with $\kappabf = \bun \cdot \nabla \bun$ the curvature
of the magnetic field lines, gives the vorticity equation
\begin{equation}
\frac{\partial \varpi}{\partial t} = \nabla \cdot \left [ J_{||}
\bun + \bJ_d + \frac{c}{B} \bun \times (\nabla \cdot
\matrixtop{\pibf}_i) \right ], \label{vorticity}
\end{equation}

\noindent where
\begin{equation}
\bJ_d = \frac{cp_\bot}{B} \bun \bun \cdot \nabla \times \bun  +
\frac{cp_\bot}{B^2} \bun \times \nabla B + \frac{cp_{||}}{B} \bun
\times \kappabf \label{J_drift_M}
\end{equation}

\noindent is the current due to the magnetic drifts and
\begin{equation}
\varpi = \nabla \cdot \left ( \frac{Zen_i}{\Omega_i} \bV_i \times
\bun \right ) \label{vort_def}
\end{equation}

\noindent is the ``vorticity", with $\Omega_i = ZeB/Mc$ the ion
gyrofrequency. The quantity $\varpi$ has dimensions of charge
density, but it is traditionally called vorticity because for
constant magnetic fields it is proportional to the parallel
component of the curl of the ion flow, $\varpi \propto \bun \cdot
\nabla \times ( n_i \bV_i )$.

Vorticity equations like equation \eq{vorticity} have been used in
the past to solve for the electrostatic potential in fluid models
for turbulence \cite{zeiler97, simakov03}. The vorticity $\varpi$
is evolved in time, and the potential can be obtained from
$\varpi$ by employing the lowest order result
\begin{equation}
\varpi \simeq \nabla \cdot \left [ \frac{Zecn_i}{B\Omega_i}
\nabla_\bot \phi + \frac{c}{B\Omega_i} ( \nabla \cdot
\matrixtop{\mathbf{P}}_i )_\bot \right ]. \label{vort_1o}
\end{equation}

\noindent To find this equation, I substitute in \eq{vort_def} the
lowest order perpendicular flow
\begin{equation}
n_i \bV_{i\bot} \simeq \frac{cn_i}{B} \bun \times \nabla \phi +
\frac{c}{ZeB} \bun \times ( \nabla \cdot \matrixtop{\mathbf{P}}_i
), \label{Vibot_lowest}
\end{equation}

\noindent where $\matrixtop{\mathbf{P}}_i = \int d^3v\, M\bv\bv
f_i$. The lowest order perpendicular ion flow \eq{Vibot_lowest} is
obtained from the conservation of ion momentum in the same way
that the perpendicular current density $\bJ_\bot$ was found from
the total momentum equation \eq{total_momentum}. In the lowest
order result \eq{Vibot_lowest}, I have neglected the ion inertia,
$\partial (n_i M \bV_i)/\partial t \sim \delta_i p_i/L$, and the
ion-electron friction force, $\mathbf{F}_{ei} \sim \sqrt{m/M}
\delta_i p_i/L$.

\section{Radial electric field in the vorticity equation \label{sect_fluxsurfaceaverage}}

In chapter~\ref{chap_gyrovorticity}, I will show that in general
$\nabla \cdot \bJ_d$ dominates or at least is comparable to the
other terms in \eq{vorticity}. However, in axisymmetric
configurations, the physics that determines the radial electric
field is an exception in that $\nabla \cdot \bJ_d$ no longer
dominates and only the viscosity term $(c/B) \bun \times (\nabla
\cdot \matrixtop{\pibf}_i)$ matters.

The radial electric field adjusts so that the radial current and
hence the flux surface average of the vorticity equation vanish.
To see that the flux surface average of the vorticity equation is
equivalent to forcing the radial current to vanish, recall that
the vorticity is $\nabla \cdot \bJ = 0$, and the flux surface
average of this equation gives $\langle \bJ \cdot \nabla \psi
\rangle_\psi = 0$, i.e., the total radial current out of a flux
surface must be zero. The flux surface average of equation
\eq{vorticity} is
\begin{equation}
\frac{\partial}{\partial t} \langle \varpi \rangle_\psi =
\frac{1}{V^\prime} \frac{\partial}{\partial \psi} V^\prime \left
\langle \bJ_d \cdot \nabla \psi - \frac{cI}{B} (\nabla \cdot
\matrixtop{\pibf}_i) \cdot \bun \right \rangle_\psi +
\frac{1}{V^\prime} \frac{\partial^2}{\partial \psi^2} V^\prime
\langle cR \zun \cdot \matrixtop{\pibf}_i \cdot \nabla \psi
\rangle_\psi, \label{vort_surfave}
\end{equation}

\noindent where $\langle \ldots \rangle_\psi = (V^\prime)^{-1}
\int d\theta\, d\zeta (\ldots)/(\bB \cdot \nabla \theta)$ is the
flux surface average and $V^\prime \equiv dV/d\psi = \int
d\theta\, d\zeta (\bB \cdot \nabla \theta)^{-1}$ is the flux
surface volume element. To simplify, I have used
\begin{equation}
\langle \nabla \cdot (\ldots) \rangle_\psi = \frac{1}{V^\prime}
\frac{\partial}{\partial \psi} V^\prime \langle \nabla \psi \cdot
(\ldots) \rangle_\psi, \label{avepsi_div}
\end{equation}
\begin{equation}
\bun \times \nabla \psi = I \bun - RB \zun \label{bxgradpsi}
\end{equation}

\noindent and $R (\nabla \cdot \matrixtop{\pibf}_i) \cdot \zun =
\nabla \cdot (R \matrixtop{\pibf}_i \cdot \zun)$. Equations
\eq{avepsi_div} and \eq{bxgradpsi} are obtained from the
definition of flux surface average and equation \eq{B_def},
respectively. The flux surface average of $\bJ_d \cdot \nabla
\psi$ is conveniently rewritten using \eq{id_kappa}, $(\nabla
\times \bun) \cdot \nabla \psi = \nabla \cdot ( \bun \times \nabla
\psi )$ and \eq{bxgradpsi} to find
\begin{equation}
\bJ_d \cdot \nabla \psi = \frac{c p_{||}}{B} \nabla \cdot ( I\bun)
- \frac{cI p_\bot}{B^2} \bun \cdot \nabla B,
\end{equation}

\noindent where I use that $\nabla \cdot (RB\zun) = 0 = \zun \cdot
\nabla B$ due to axisymmetry. The flux surface average of this
expression is
\begin{equation}
\langle \bJ_d \cdot \nabla \psi \rangle_\psi = - \left \langle
\frac{cI}{B} [ \bun \cdot \nabla p_{||} + (p_{||} - p_\bot) \nabla
\cdot \bun ] \right \rangle_\psi,
\end{equation}

\noindent where I have integrated by parts and used $\bun \cdot
\nabla \ln B = - \nabla \cdot \bun$. Substituting this result into
equation \eq{vort_surfave}, using the parallel component of
\eq{total_momentum} to write
\begin{equation}
\bun \cdot \nabla p_{||} + ( p_{||} - p_\bot ) \nabla \cdot \bun +
(\nabla \cdot \matrixtop{\pibf}_i) \cdot \bun = -
\frac{\partial}{\partial t} M n_i \mathbf{V}_i \cdot \bun
\end{equation}

\noindent and employing
\begin{equation}
\langle \varpi \rangle_\psi - \frac{1}{V^\prime}
\frac{\partial}{\partial \psi} V^\prime \left \langle \frac{Zen_i
I}{\Omega_i} \bV_i \cdot \bun \right \rangle_\psi = -
\frac{1}{V^\prime} \frac{\partial}{\partial \psi} V^\prime \langle
c n_i R M \bV_i \cdot \zun \rangle_\psi
\end{equation}

\noindent gives the conservation equation for toroidal angular
momentum
\begin{equation}
\frac{\partial}{\partial t} \langle n_i R M \bV_i \cdot \zun
\rangle_\psi = - \frac{1}{V^\prime} \frac{\partial}{\partial \psi}
V^\prime \langle R \zun \cdot \matrixtop{\pibf}_i \cdot \nabla
\psi \rangle_\psi, \label{toro_angmom}
\end{equation}

\noindent where I have integrated once in $\psi$ assuming that
there are no sources or sinks of momentum. Equation
\eq{toro_angmom} shows that setting the total radial current to
zero is equivalent to the toroidal angular momentum conservation
equation. Equation \eq{toro_angmom} includes both turbulent and
neoclassical effects. In a model in which the transport time scale
is not reached, as is the usual case in gyrokinetics, there is not
enough time for the angular momentum to diffuse from one flux
surface to the next, keeping then the long wavelength toroidal
velocity constant and equal to its initial value. Consequently,
the long wavelength radial electric field, related to the toroidal
velocity by the $E\times B$ velocity, must not evolve and must be
determined by the initial condition. The vorticity equation makes
this fact explicit by including the radial current density $(c/B)
[\bun \times (\nabla \cdot \matrixtop{\pibf}_i)] \cdot \nabla
\psi$.

Equation \eq{toro_angmom} can be generalized by flux surface
averaging the toroidal angular momentum conservation equation with
the charge density and full current density retained. Writing the
electromagnetic force as the Maxwell stress gives
\begin{equation}
\frac{\partial}{\partial t} \langle n_i R M \bV_i \cdot \zun
\rangle_\psi = - \frac{1}{V^\prime} \frac{\partial}{\partial \psi}
V^\prime \left \langle R \zun \cdot (\matrixtop{\pibf}_i +
\matrixtop{\pibf}_e) \cdot \nabla \psi - \frac{1}{4\pi} R \zun
\cdot ( \mathbf{E} \mathbf{E} + \bB \bB ) \cdot \nabla \psi \right
\rangle_\psi. \label{toro_angmom_2}
\end{equation}

\noindent In equation~\eq{toro_angmom}, the components of the
toroidal angular momentum transport due to the electron viscosity
and the Maxwell stress are neglected. In subsequent chapters I
will argue that the ion viscosity term $\langle R \zun \cdot
\matrixtop{\pibf}_i \cdot \nabla \psi \rangle_\psi$ is of order
$\delta_i^3 p_i R |\nabla \psi|$. The other terms must be compared
to this order of magnitude estimate to asses their importance. In
chapter~\ref{chap_gyrovorticity}, I will show that
$\matrixtop{\pibf}_e \sim \delta_e^2 p_e \sim (m/M) \delta_i^2
p_i$, negligible compared to $\matrixtop{\pibf}_i$ because $m/M =
5 \times 10^{-4} \ll \delta_i$. The magnetic portion of the
Maxwell stress vanishes because the calculation is electrostatic
and the magnetic field is unperturbed and given by \eq{B_def}; in
particular, it satisfies $\bB \cdot \nabla \psi = 0$. The electric
part of the Maxwell stress is a non-neutral contribution. Using
$\mathbf{E} = - \nabla \phi \sim T_e/eL$, I find $R \zun \cdot
\mathbf{E} \mathbf{E} \cdot \nabla \psi / 4 \pi \sim
(\lambda_D/L)^2 p_e R |\nabla \psi|$, with $\lambda_D = \sqrt{T_e/
4\pi e^2 n_e}$ the Debye length. This is one of the instances when
the non-neutral effects may contribute to the final result. In
both Alcator C-Mod and DIII-D, the non-neutral transport of
toroidal angular momentum $R \zun \cdot \mathbf{E} \mathbf{E}
\cdot \nabla \psi / 4 \pi$ is comparable to the viscosity
contribution, albeit somewhat smaller. For the rest of the thesis,
I will drop the non-neutral contribution to simplify the
presentation, but it is important to remark that it could be
easily added if necessary.

The fact that $\langle \bJ \cdot \nabla \psi \rangle_\psi = 0$ is
equivalent to equation \eq{toro_angmom} means that whether a
configuration is intrinsically ambipolar or not depends then on
the size of the radial current density $(c/B) [\bun \times (\nabla
\cdot \matrixtop{\pibf}_i)] \cdot \nabla \psi$. If this current
density is small, as I will prove it to be in
chapter~\ref{chap_gyrovorticity}, the radial current is
effectively zero. Then, if the plasma is quasineutral initially,
it stays so independently of the radial electric field, that is,
the configuration remains intrinsically ambipolar.

I will finish this chapter with some simple estimates that I will
prove in chapters~\ref{chap_gyrovorticity} and
\ref{chap_angularmomentum}. In neoclassical calculations of the
radial electric field \cite{catto05a, catto05b, wong07,
rosenbluth71, wong05}, the flux surface average $\langle R \zun
\cdot \matrixtop{\pibf}_i \cdot \nabla \psi \rangle_\psi$ is of
order $\delta_i^3 p_i R |\nabla \psi|$. The radial current density
$(c/B) [\bun \times (\nabla \cdot \matrixtop{\pibf}_i)]\cdot
\nabla \psi$ associated with this piece of the viscosity is tiny,
of order $\delta_i^4 e n_e v_i |\nabla \psi|$. The estimate from
neoclassical theory is not necessarily applicable to turbulent
transport of toroidal angular momentum, but it is suggestive.
Indeed, the same order of magnitude is recovered if the transport
of toroidal angular momentum is at the gyroBohm level and the
average velocities in the tokamak are small compared to the ion
thermal velocity by $\delta_i$, i.e., $V_i \sim \delta_i v_i$. The
gyroBohm diffusion coefficient is obtained from turbulence
fluctuations employing a simple random walk argument as follows.
In chapter~\ref{chap_intro}, I pointed out that the typical eddy
size is $\Delta_\mathrm{eddy} \sim \rho_i$. The eddy turnover time
is determined by the average velocity $V_i \sim \delta_i v_i$,
giving $\tau_\mathrm{eddy} \sim \Delta_\mathrm{eddy} / V_i \sim
L/v_i$. With these estimates, the gyroBohm diffusion coefficient
is calculated to be $D_{gB} =
\Delta_\mathrm{eddy}^2/\tau_\mathrm{eddy} \sim \delta_i \rho_i
v_i$. Multiplying this diffusivity by the macroscopic gradient of
momentum, $\nabla ( n_i M V_{i\zeta} ) \sim \delta_i n_e M v_i/L$,
I find the same result as the neoclassical calculation, namely,
the transport of momentum is $\delta_i^3 p_i$ and the associated
radial current density is $\delta_i^4 e n_e v_i$. In
chapter~\ref{chap_angularmomentum}, I will argue in favor of this
estimate more strongly. In any case, in current gyrokinetic
formalisms, the flux surface averaged radial current should remain
equal to zero independently of the long wavelength radial electric
field.

\chapter{Derivation of gyrokinetics \label{chap_gyrokinetics}}

Gyrokinetics is a kinetic formalism that is especially adapted to
model drift wave turbulence. By defining more convenient variables
to replace the position $\br$ and velocity $\bv$, it ``averages
out" the fast gyromotion time scales and keeps finite gyroradius
effects, permitting perpendicular wavelengths on the order of the
ion gyroradius. Currently, there is no other model that allows
simulation of short wavelength turbulence in magnetized plasmas in
reasonable computational times.

In this chapter, I derive the gyrokinetic Fokker-Planck equation
and the gyrokinetic quasineutrality equation. The gyrokinetic
Fokker-Planck equation models the response of the plasma to
fluctuations in the electrostatic potential with wavelengths that
may be as small as the ion gyroradius. The gyrokinetic
quasineutrality equation, or gyrokinetic Poisson's equation, is
the equation used so far to determine the self-consistent
electrostatic potential. This equation is given here for
completeness, but in chapter~\ref{chap_gyrovorticity} I will show
that it is flawed at long wavelengths.

The rest of the chapter is organized as follows. In section
\ref{sect_GKorderings}, I lay out the assumptions I need to derive
the gyrokinetic equation. In section~\ref{sect_GKvariables}, I
find the gyrokinetic variables with a new nonlinear approach,
based on the linear derivation of \cite{leecatto83, bernstein85}.
I have already published this derivation in \cite{parra08}. In
section~\ref{sect_GKFPequation}, I give the Fokker-Planck equation
in these new variables. This gyrokinetic Fokker-Planck equation
does not have the fast gyromotion scale, yet it retains finite
gyroradius effects, as desired. I finish by deriving the
traditional gyrokinetic quasineutrality in section
\ref{sect_GKpoisson}. The algebraic details of the calculation are
relegated to Appendices~\ref{app_GKvar} - \ref{app_gyrodepend_kL}.

\section{Orderings \label{sect_GKorderings}}

The assumptions are the same as in section~\ref{sect_orderings1}.
The characteristic frequency of the processes of interest is
assumed to be the drift wave frequency $\omega \sim \omega_\ast =
k_\bot c T_e/e B L_n \sim k_\bot \rho_i v_i/L$, with $L_n^{-1} =
|\nabla \ln n_e|$. To treat arbitrary collisionality, the ion-ion
collision frequency $\nu_{ii}$ is assumed to be of the order of
the transit time of ions, $\nu_{ii} \sim v_i / L$. Consequently,
the electron-electron and the electron-ion collision frequencies
are of the order of the electron transit time, $\nu_{ee} \sim
\nu_{ei} \sim v_e / L$. With this assumption, I can treat low
collisionality cases by sub-expanding the final results in the
parameter $\nu_{ii} L/v_i \sim \nu_{ee} L/v_e \sim \nu_{ei} L/v_e
\ll 1$.

The average velocities are assumed to be in the low flow or drift
ordering, where the $E \times B$ drift is of order $\delta_i v_i$.
Therefore, the electric field is of order $\mathbf{E} = - \nabla
\phi \sim T_e/e L$, and the total electrostatic potential drop
across the characteristic length $L$ is of order $T_e/e$. The
spatial gradients of the distribution functions are assumed to be
$\nabla f_i \sim f_{Mi}/L$ and $\nabla f_e \sim f_{Me}/L$, where
$f_{Mi}$ and $f_{Me}$ are the zeroth order distribution functions.
Since I am primarily interested in the core plasma in tokamaks, I
will assume that the zeroth order distribution functions are
stationary Maxwellians, with ion and electron temperatures of the
same order, $T_i \sim T_e$. The Maxwellians are stationary to be
consistent with the drift ordering.

To include the turbulence, I allow wavelengths perpendicular to
the magnetic field that are on the order of the ion gyroradius,
$k_\bot \rho_i \sim 1$. At the same time, due to low flow or drift
ordering, I assume $\nabla \phi \sim T_e/eL$, $\nabla f_i \sim
f_{Mi}/L$ and $\nabla f_e \sim f_{Me}/L$. This ordering requires
that the pieces of the potential and the distribution functions
with short perpendicular wavelengths $\phi_k$, $f_{i,k}$ and
$f_{e,k}$ be small in size, in particular
\begin{equation}
\frac{f_{i,k}}{f_{Mi}} \sim \frac{f_{e,k}}{f_{Me}} \sim
\frac{e\phi_k}{T_e} \sim \frac{1}{k_\bot L} \ssim 1,
\label{order_kbot}
\end{equation}

\noindent with $k_\bot \rho_i \ssim 1$. According to
\eq{order_kbot}, the pieces of the distribution function that have
wavelengths on the order of the ion gyroradius are next order in
the expansion in $\delta_i$. Notice that $k_\bot f_{i,k} \sim
\nabla_\bot f_{i,k} \sim \nabla f_{Mi} \sim f_{Mi}/L$, and since
$\partial/\partial t \sim k_\bot \rho_i v_i/L$, $\partial f_{i,k}
/ \partial t \sim \delta_i f_{Mi} v_i/L$. I could allow
wavelengths on the order of the electron gyroradius following a
similar ordering, but I ignore these small wavelengths to simplify
the presentation. Unlike the perpendicular wavelengths, the
wavelengths along the magnetic field, $k_{||}^{-1}$, are taken to
be on the order of the larger scale $L$. Moreover, except for
initial transients, the variations along the magnetic field of
$f_i$, $f_e$ and $\phi$ are slow, i.e., in general $\bun \cdot
\nabla f_i \sim \delta_i f_{Mi}/L$, $\bun \cdot \nabla f_e \sim
\delta_i f_{Me}/L$ and $\bun \cdot \nabla \phi \sim \delta_i
T_e/eL$.

Both the potential and the distribution function may be viewed as
having a piece with slow spatial variations (representing the
average value in the plasma) plus some rapid oscillations of small
amplitude. The zonal flow, for example, will be included in the
small piece if its characteristic wavelength is comparable to the
gyroradius, but its amplitude may be larger for longer
wavelengths. This ordering implies that the rapid spatial
potential fluctuations seen by a particle in its gyromotion are
small compared to the average energy of the particle. Then, the
gyromotion remains almost circular, and the distribution function
of the gyrocenters is equal, to zeroth order, to the distribution
function of the particles. The difference, coming from the rapidly
oscillating pieces, is small in our ordering. Notice that the
$\delta f$ codes \cite{dorland00, dannert05, candy03, chen03,
dimits96, lin02} explicitly adopt this treatment for the
components of $\phi$, $f_i$ and $f_e$ that satisfy $k_\bot \rho_i
\sim 1$, and, as in this thesis, they order them as $O
(\delta_i)$.

\section{Gyrokinetic variables \label{sect_GKvariables}}

In this section, I derive the gyrokinetic variables for ions, the
only species where the finite gyroradius effects matter, since
$k_\bot \rho_i \sim 1$. It is possible to find a similar
gyrokinetic equation for electrons, but for this thesis the drift
kinetic equation is all that is required. I begin by defining the
Vlasov operator for ions in the usual $\br$, $\bv$ variables for
an electrostatic electric field as the following total derivative
\begin{equation}
\frac{d}{dt} = \frac{\partial}{\partial t} + \bv \cdot \nabla +
\left ( - \frac{Ze}{M} \nabla \phi + \Omega_i \bv \times \bun
\right ) \cdot \nabla_v, \label{tot_ddt}
\end{equation}

\noindent where $\Omega_i = Z e B/M c$ is the gyrofrequency. The
Fokker--Planck equation for ions is then simply
\begin{equation}
\frac{d f_i}{d t} = C \{ f_i \}, \label{FPeq_rv}
\end{equation}

\noindent where $C \{ f_i \}$ is the relevant Fokker-Planck
collision operator for the ions. The ion-electron collision
operator is small in $\sqrt{m/M}$, leaving only the ion-ion
collision operator.

The objective of gyrokinetics is to change the Fokker-Planck
equation to gyrokinetic variables, defined such that the
gyromotion time scale disappears from equation \eq{FPeq_rv}. The
nonlinear gyrokinetic variables to be employed are the guiding
center location $\bR$, the kinetic energy $E$, the magnetic moment
$\mu$, and the gyrophase $\varphi$. These variables will be
defined to higher order than is customary by employing an
extension of the procedure presented in \cite{leecatto83,
bernstein85}. The general idea is to construct the gyrokinetic
variables to higher order by adding in $\delta_i$ corrections such
that the total derivative of a generic gyrokinetic variable $Q$ is
gyrophase independent to the desired order, and we may safely
employ
\begin{equation}
\frac{dQ}{dt} \simeq \left \langle \frac{dQ}{dt} \right \rangle,
\label{general_cond}
\end{equation}

\noindent where the gyroaverage $\langle \ldots \rangle$ is
performed holding $\bR$, $E$, $\mu$ and $t$ fixed. This
gyrokinetic gyroaverage can be understood as a fast time average
where $\tau = - \varphi/\Omega_i$ is the fast gyromotion time and
$t$ is the slow time of the turbulent fluctuations (ion gyromotion
is such that $d\varphi/dt < 0$, hence the sign difference between
$\tau$ and $\varphi$). The gyrokinetic variable $Q$ is expanded in
powers of $\delta_i$,
\begin{equation}
Q = Q_0 + Q_1 + Q_2 + \ldots, \label{expansion_Q}
\end{equation}

\noindent where $Q_0$ is the lowest order gyrokinetic variable
(kinetic energy, magnetic moment, etc.), and $Q_1 = O( \delta_i
Q_0 )$, $Q_2 = O (\delta_i^2 Q_0)$... are the order $\delta_i$,
$\delta_i^2$~... corrections. The first correction $Q_1$ is
constructed so that $dQ/dt = \langle dQ / dt \rangle + O (
\delta_i^2 \Omega_i Q)$, while the second correction $Q_2$ is
evaluated such that $dQ/dt = \langle dQ / dt \rangle + O (
\delta_i^3 \Omega_i Q)$. In principle this process can be
continued indefinitely. Any $Q_k$ can be found once the functions
$Q_1$, $Q_2$ ..., $Q_{k-1}$ are known. The functions $Q_1$, $Q_2$
..., $Q_{k-1}$ are constructed so that
\begin{equation}
\frac{dQ}{dt} \simeq \frac{d}{dt} ( Q_0 + \ldots + Q_{k-1} ) =
\left \langle \frac{d}{dt} ( Q_0 + \ldots + Q_{k-1} ) \right
\rangle + O ( \delta_i^k \Omega_i Q_0 ). \label{dQ_dt_1}
\end{equation}

\noindent Adding $Q_k$ to \eq{expansion_Q} means adding $d Q_k /
dt$ to \eq{dQ_dt_1}. To lowest order, $d Q_k / dt \simeq -
\Omega_i \:
\partial Q_k /
\partial \varphi$, which to the requisite order leads to an equation for
$Q_k$,
\begin{equation}
\frac{dQ}{dt} \simeq \frac{d}{dt} ( Q_0 + \ldots + Q_{k-1} ) -
\Omega_i \frac{\partial Q_k}{\partial \varphi} = \left \langle
\frac{d}{dt} ( Q_0 + \ldots + Q_{k-1} ) \right \rangle,
\label{dQ_dt_2}
\end{equation}

\noindent where $\langle \partial Q_k / \partial \varphi \rangle =
0$ is employed. Notice that the gyrophase derivative is holding
the gyrokinetic variables fixed, and not $\br$, $v_{||}$ and
$v_\bot$ fixed (in some cases these two distinct derivatives with
respect to gyrophase are almost equal). Using \eq{dQ_dt_2}, $Q_k =
O( \delta_i^k Q_0 )$ is found to be periodic in gyrophase and
given by
\begin{equation}
Q_k = \frac{1}{\Omega_i} \int^\varphi d \varphi^\prime \left [
\frac{d}{dt} ( Q_0 + \ldots + Q_{k-1} ) - \left \langle
\frac{d}{dt} ( Q_0 + \ldots + Q_{k-1} ) \right \rangle \right ].
\end{equation}

More explicitly, through the first two orders, $Q_1$ and $Q_2$ are
determined to be
\begin{equation}
Q_1 = \frac{1}{\Omega_i} \int^\varphi d \varphi^\prime \left (
\frac{d Q_0}{dt} - \left \langle \frac{dQ_0}{dt} \right \rangle
\right ) \label{o1_correction}
\end{equation}

\noindent and
\begin{equation}
Q_2 = \frac{1}{\Omega_i} \int^\varphi d \varphi^\prime \left [
\frac{d}{dt} ( Q_0 + Q_1 ) - \left \langle \frac{d}{dt} ( Q_0 +
Q_1 ) \right \rangle \right ] . \label{o2_correction}
\end{equation}

\noindent By adding $Q_1$ and $Q_2$, the total derivative of the
gyrokinetic variable $Q = Q_0 + Q_1 + Q_2$ is
\begin{equation}
\frac{dQ}{dt} = \left \langle \frac{d}{dt} (Q_0 + Q_1) \right
\rangle + O ( \delta_i^3 \Omega_i Q_0 ). \label{final_dQdt}
\end{equation}

In the remainder of this subsection, I present the gyrokinetic
variables that result from this process. I begin with the guiding
center position expanded as
\begin{equation}
\bR = \bR_0 + \bR_1 + \bR_2, \label{RGK_def}
\end{equation}

\noindent where $\bR_0 = \br$, $|\bR_1| = O( \rho_i )$ and
$|\bR_2| = O( \delta_i \rho_i )$. I construct $\bR_1$ and $\bR_2$
such that the gyrocenter position time derivative is gyrophase
independent to order $\delta_i v_i$,
\begin{equation}
\frac{d\bR}{dt} = \left \langle \frac{d\bR}{dt} \right \rangle + O
( \delta_i^2 v_i ).
\end{equation}

\noindent The explicit details of the calculation are presented in
sections~\ref{sectapp_GKvar1} and \ref{sectapp_GKvar2} of
Appendix~\ref{app_GKvar}. To first order, I find the usual result
\cite{catto78}
\begin{equation}
\bR_1 = \frac{1}{\Omega_i} \bv \times \bun. \label{RGK1_def}
\end{equation}

\begin{figure}

\includegraphics[width = \textwidth]{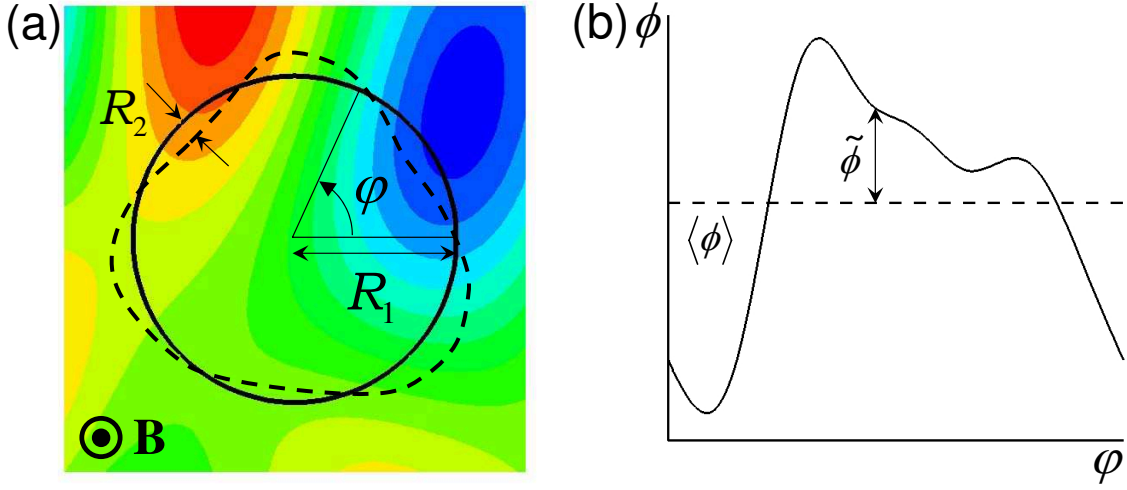}

\vspace{-4cm} \caption[Gyrokinetic gyromotion and functions
$\phiave$ and $\phiwig$]{(a) Gyrokinetic gyromotion $\br = \bR -
\bR_1 - \bR_2$ (dashed line) in a turbulent potential. The solid
circle $\bR - \bR_1$ describes the gyromotion to first order. (b)
Potential $\phi$ along gyromotion as a function of $\varphi$. The
average $\phiave$ is the dashed line, and the fast time variation
$\phiwig$ is the difference $\phi - \phiave$.}
\label{fig_phiavewig}
\end{figure}

\noindent The gyromotion is approximately circular, as sketched in
figure~\ref{fig_phiavewig}(a), even in the presence of
fluctuations with wavelengths of the order of the ion gyroradius.
The ordering in \eq{order_kbot} that bounds the electric field to
be $O(T_e/eL)$ leads to this significant simplification. To next
order, I obtain
\begin{eqnarray}
\bR_2 = \frac{1}{\Omega_i} \left [ \left ( v_{||} \bun +
\frac{1}{4} \bv_\bot \right ) (\bv \times \bun) + (\bv \times
\bun) \left ( v_{||} \bun + \frac{1}{4} \bv_\bot \right ) \right ]
\dotcross \nabla \left ( \frac{\bun}{\Omega_i} \right ) \nonumber
\\ + \frac{v_{||}}{\Omega_i^2} \bv_\bot \cdot \nabla \bun +
\frac{\bun}{\Omega_i^2} \left \{ v_{||} \bun \cdot \nabla \bun
\cdot \bv_\bot + \frac{1}{8} [ \bv_\bot \bv_\bot - ( \bv \times
\bun ) ( \bv \times \bun ) ] : \nabla \bun \right \} \nonumber \\
- \frac{c}{B \Omega_i} \nabla_\bR \Phiwig \times \bun,
\label{R_1_main}
\end{eqnarray}

\noindent which is the same as \cite{leecatto83} except for the
nonlinear term given last. My vector conventions are $\mathbf{x}
\mathbf{y}\hspace{-1.2mm} : \matrixtop{\mathbf{M}} = \mathbf{y}
\cdot \matrixtop{\mathbf{M}} \cdot \mathbf{x}$ and $\mathbf{x}
\mathbf{y} \dotcross \hspace{-1.5mm} \matrixtop{\mathbf{M}} =
\mathbf{x} \times (\mathbf{y} \cdot \matrixtop{\mathbf{M}} )$. One
of the terms in the second order correction $\bR_2$ includes the
function $\Phiwig (\bR, E, \mu, \varphi, t)$. It will appear
several times during derivations, as will the functions $\phiave$
and $\phiwig$. These are functions related to the electrostatic
potential that depend on the new gyrokinetic variables. Their
definitions are
\begin{equation}
\phiave \equiv \phiave (\bR, E, \mu, t) = \frac{1}{2\pi} \oint
d\varphi\: \phi(\br (\bR, E, \mu, \varphi, t), t ), \label{phiave}
\end{equation}
\begin{equation}
\phiwig \equiv \phiwig (\bR, E, \mu, \varphi, t) = \phi(\br (\bR,
E, \mu, \varphi, t), t ) - \phiave (\bR, E, \mu, t) \label{phiwig}
\end{equation}

\noindent and
\begin{equation}
\Phiwig \equiv \Phiwig (\bR, E, \mu, \varphi, t) = \int^\varphi
d\varphi^\prime\: \phiwig (\bR, E, \mu, \varphi^\prime, t),
\label{Phiwig}
\end{equation}

\noindent such that $\langle \Phiwig \rangle = 0$. These are
definitions similar to those used by Dubin \cite{dubin83}. I have
discussed the subtle differences between them in \cite{parra09a},
as summarized in Appendix~\ref{app_equivalence}. From their
definitions \eq{phiave} and \eq{phiwig}, we see that the function
$\phiave$ is the fast time average of the potential in a
gyromotion, and $\phiwig$ is the difference between the potential
at the position of the particle and $\phiave$. Both are
represented in figure~\ref{fig_phiavewig}(b) as a function of the
gyrophase $\varphi$. The function $\Phiwig$ is proportional to the
fast time integral of $\phiwig$, with $\tau = - \varphi/\Omega_i$
the fast time variation. The term $-(c/B\Omega_i) \nabla_\bR
\Phiwig \times \bun$ in \eq{R_1_main} is the correction to the
gyromotion due to the fast time component of the $E\times B$ drift
$- (c/B) \nabla_\bR \phiwig \times \bun$. This fast time
contribution integrated over fast times $\tau = -
\varphi/\Omega_i$ gives the correction to the gyrocenter position.
The other corrections in $\bR_2$ are fast time contributions due
to magnetic geometry.

It is important to comment on the size of functions $\phiave$,
$\phiwig$ and $\Phiwig$. Both $\phi$ and $\phiave$ are of the same
order as the temperature for long wavelengths, but small for short
wavelengths [recall \eq{order_kbot}]. However, $\phiwig$ is always
small as it accounts for the variation in the electrostatic
potential that a particle sees as it moves in its gyromotion. Of
course, since the potential is small for short wavelengths, the
variation observed by the particle is also small. For long
wavelengths, even though the potential is comparable to the
temperature, the particle motion is small compared to the
wavelength, and the variations that it sees in its motion are
small. Therefore, $\phiwig \sim \delta_i T_e / e$ for all
wavelengths in my ordering, making $\Phiwig$ small as well.

The Vlasov operator acting on $\bR$ gives
\begin{equation}
\frac{d \bR}{dt} = \left \langle \frac{d \bR}{dt} \right \rangle +
O ( \delta_i^2 v_i ) = u \bun (\bR) + \bv_d + O ( \delta_i^2 v_i
), \label{dR_dt_main}
\end{equation}

\noindent where $\bv_d$ is the total drift velocity,
\begin{equation}
\bv_d = \bv_E + \bv_M, \label{drift}
\end{equation}

\noindent composed of the $E \times B$ drift
\begin{equation}
\bv_E = - \frac{c}{B(\bR)} \nabla_\bR \phiave \times \bun (\bR)
\label{drift_E}
\end{equation}

\noindent and the magnetic drift
\begin{equation}
\bv_M = \frac{u^2}{\Omega_i(\bR)} \bun (\bR) \times \kappabf (\bR)
+ \frac{\mu}{\Omega_i(\bR)} \bun (\bR) \times \nabla_\bR B (\bR).
\label{drift_M}
\end{equation}

\noindent In the preceding equations, $u$ is the gyrocenter
parallel velocity defined by
\begin{equation}
\frac{u^2}{2} + \mu B (\bR) = E. \label{par_drift}
\end{equation}

\noindent Notice that in \eq{dR_dt_main}, \eq{drift_E},
\eq{drift_M} and \eq{par_drift}, all the terms are given as a
function of the new gyrokinetic variables, $\bR$, $E$ and $\mu$.

\begin{figure}
\begin{center}
\includegraphics[width = 0.65\textwidth]{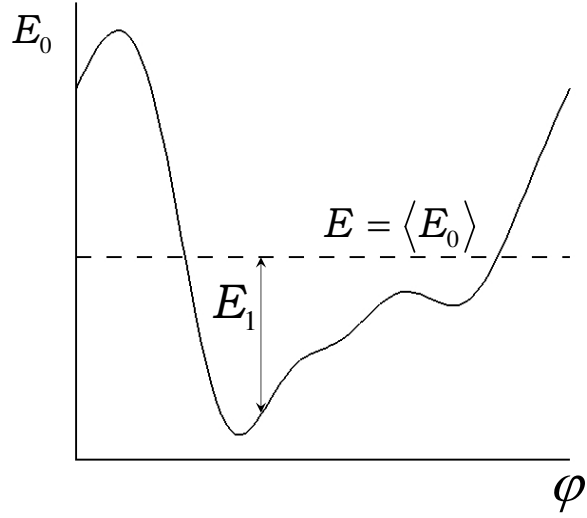}
\end{center}

\vspace{-1cm} \caption[Kinetic energy along the
gyromotion]{Kinetic energy $E_0 = v^2/2$ along the gyromotion. The
dashed line is the average value $E = \langle E_0 \rangle$, and
the difference $E - E_0$ is the correction $E_1$. Notice that the
variation of the kinetic energy and the variation of the potential
have opposite signs because maxima of potential correspond to
minima of kinetic energy.} \label{fig_Ewig}
\end{figure}

Following the same procedure as for the gyrocenter position $\bR$,
the gyrokinetic kinetic energy is defined as
\begin{equation}
E = E_0 + E_1 + E_2, \label{Etotal_def}
\end{equation}

\noindent where $E_0 = v^2/2$, $E_1 = O(\delta_i v_i^2)$ and $E_2
= O(\delta_i^2 v_i^2)$. The details of the calculation are given
in sections~\ref{sectapp_GKvar1} and \ref{sectapp_GKvar2} of
Appendix~\ref{app_GKvar}. I find
\begin{equation}
E_1 = \frac{Z e \phiwig}{M} \label{correc_E_o1}
\end{equation}

\noindent and
\begin{equation}
E_2 = \frac{c}{B} \frac{\partial \Phiwig}{\partial t}.
\label{correc_E_o2}
\end{equation}

\noindent The Vlasov operator acting on $E$ is shown in
section~\ref{sectapp_GKvar2} of Appendix~\ref{app_GKvar} to give
\begin{equation}
\frac{d E}{d t} = \left \langle \frac{dE}{dt} \right \rangle + O
\left ( \delta_i^2 \frac{v_i^3}{L} \right ) = - \frac{Z e}{M}
\left [ u \bun (\bR) + \bv_M \right ] \cdot \nabla_\bR \phiave + O
\left ( \delta_i^2 \frac{v_i^3}{L} \right ). \label{dE_dt_main}
\end{equation}

\noindent The corrections $E_1$ and $E_2$ are necessary because
otherwise the kinetic energy would vary in the fast gyromotion
time scale. Figure~\ref{fig_phiavewig} shows how the potential
changes rapidly along the gyromotion due to the short wavelength
turbulent structures. The kinetic energy variations are
accordingly rapid, as sketched in figure~\ref{fig_Ewig}. The
gyrokinetic energy is the average kinetic energy $E = \langle E_0
\rangle$, where the fast time variation has been extracted.

The gyrokinetic gyrophase is obtained in a similar way as the
energy and the gyrocenter position by defining
\begin{equation}
\varphi = \varphi_0 + \varphi_1, \label{varphitotal_def}
\end{equation}

\noindent with $\varphi_0$ the original gyrophase. The details are
again in Appendix~\ref{app_GKvar}, in
section~\ref{sectapp_GKvar1}. Notice that only the first order
correction $\varphi_1$ is calculated since gyrokinetics will make
the gyrophase dependence weak and hence next order corrections
unnecessary. The first order correction is
\begin{eqnarray}
\varphi_1 = - \frac{Ze}{MB} \frac{\partial \Phiwig}{\partial \mu}
- \frac{1}{\Omega_i} \bv_\bot \cdot \left [ \nabla \ln B +
\frac{v^2_{||}}{v^2_\bot} \bun \cdot \nabla \bun - \bun \times
\nabla \eun_2 \cdot \eun_1 \right ] \nonumber \\ - \frac{v_{||}}{4
\Omega_i v^2_\bot} [ \bv_\bot \bv_\bot - ( \bv \times \bun ) ( \bv
\times \bun )  ] : \nabla \bun, \label{correc_varphi_o1}
\end{eqnarray}

\noindent where $\Phiwig$ is defined in \eq{Phiwig}. With this
correction, $d \varphi / dt$ is gyrophase independent to order
$O(v_i/L)$, that is,
\begin{equation}
\frac{d \varphi}{d t} = \left \langle \frac{d \varphi}{d t} \right
\rangle + O( \delta_i^2 \Omega_i ) = - \overline{\Omega}_i + O(
\delta_i^2 \Omega_i ),
\end{equation}

\noindent where
\begin{equation}
\overline{\Omega}_i = \Omega_i ( \bR ) + \frac{v_{||}}{2} \bun
\cdot \nabla \times \bun - v_{||} \bun \cdot \nabla \eun_2 \cdot
\eun_1 + \frac{Z^2 e^2}{M^2 c} \frac{\partial \phiave}{\partial
\mu}. \label{omeg_mod}
\end{equation}

\begin{figure}
\includegraphics[width = 1.4\textwidth]{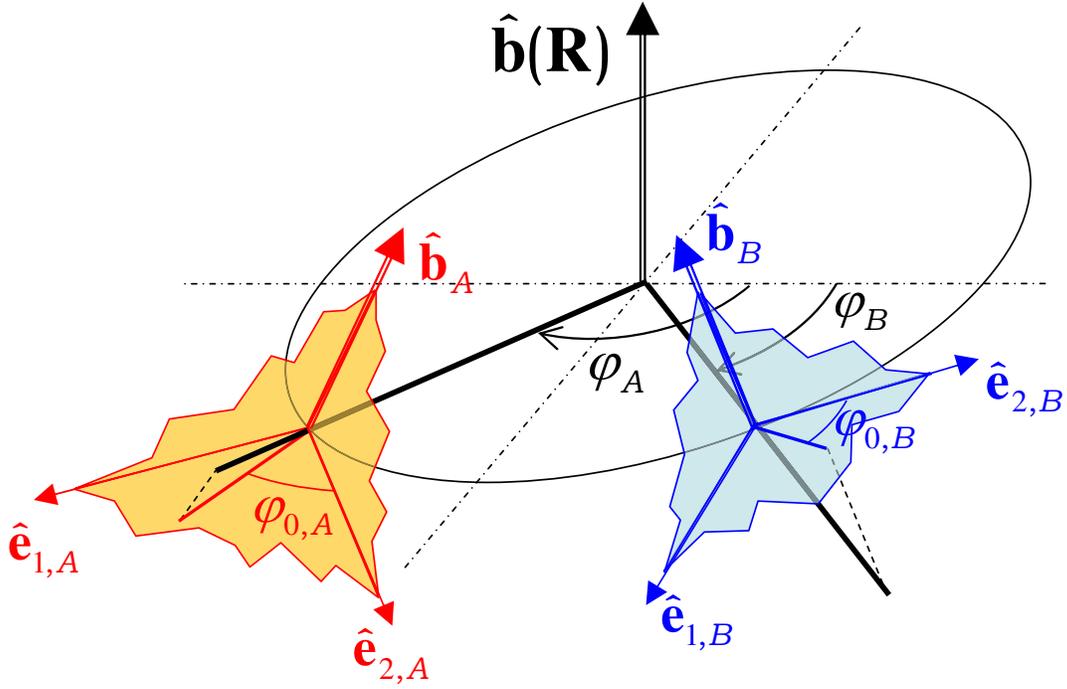}

\vspace{-7cm} \caption[Geometric effects on the gyrokinetic
gyrophase]{Geometric effects on the gyrokinetic gyrophase. The
circle represents the gyromotion of a given particle. The
variations of $\eun_1$, $\eun_2$ and $\bun$ are exaggerated.}
\label{fig_gyrophase}
\end{figure}

\noindent The function $\overline{\Omega}_i$ is equal to
$\Omega_i$ to lowest order. It might be surprising that the
gyrophase $\varphi_0$ needs to be corrected by $\varphi_1$. The
gyrokinetic gyrophase $\varphi$ is in reality the fast time
variation $\tau = - \varphi/\overline{\Omega}_i$. The correction
$\varphi_1$ is necessary because $d\varphi_0/dt$ changes in the
fast gyromotion time scale, making the dependence of $\varphi_0$
on $\tau$ non-trivial. There are two effects that contribute to
the fast variation of $d\varphi_0/dt$, namely, the electromagnetic
fields and geometric corrections. The electromagnetic fields
contribute through the magnetic field strength $B(\br)$ in
$\Omega_i (\br)$, giving the correction $- \Omega_i^{-1} \bv_\bot
\cdot \nabla \ln B$ in \eq{correc_varphi_o1}, and through the
short wavelength turbulent structures of the potential. The
particle feels a rapidly changing potential along its gyromotion.
The perpendicular velocity then has a variation similar to the
kinetic energy [recall figure~\ref{fig_Ewig}], and the gyromotion
accelerates and decelerates accordingly, giving the correction
$-(Ze/MB) (\partial \Phiwig/ \partial \mu)$ in
\eq{correc_varphi_o1}. The other contributions to the fast
variation of $d\varphi_0/dt$ are geometrical. The gyrophase
$\varphi_0$ is defined with respect to the vectors $\eun_1(\br)$,
$\eun_2(\br)$ and $\bun(\br)$ at the position $\br$ of the
particle, not the gyrocenter position $\bR$.
Figure~\ref{fig_gyrophase} shows that to find the gyrophase
$\varphi_{0,A}$ that corresponds to the particle position $A$, the
position vector $\br - \bR$ must be projected onto the plane
defined by the local vectors $\eun_{1,A}$ and $\eun_{2,A}$;
$\varphi_{0,A}$ is the angle of the projection with respect to
$\eun_{2,A}$ [recall from \eq{gyrophase_def} that $\varphi_0$ is
the angle of $\bv_\bot$ with respect to the vector $\eun_1$ and
hence it is the angle of $\br - \bR = - \Omega_{i}^{-1} \bv \times
\bun$ with respect to $\eun_2$]. In figure~\ref{fig_gyrophase},
the geometric construction to determine $\varphi_0$ at point $A$
is sketched in red, and the corresponding construction at another
point $B$ is given in blue, making explicit the distinction
between the gyrokinetic gyrophases $\varphi_A$ and $\varphi_B$,
and the ``local" gyrophases $\varphi_{0,A}$ and $\varphi_{0,B}$.
Notice that both variation in $\bun$, that determines the ``local"
plane of $\eun_1$ and $\eun_2$, and rotation of $\eun_1$ and
$\eun_2$ within that plane change the value of $\varphi_0$. For
this reason, both $\nabla \bun$ and $\nabla \eun_2 \cdot \eun_1$
enter in $\varphi_1$ and $d\varphi/dt$.

Finally, the gyrokinetic magnetic moment variable is dealt with
somewhat differently since we want to construct it to remain an
adiabatic invariant order by order. The condition for the magnetic
moment is not only that its derivative must be gyrophase
independent, but that $\langle d \mu / dt \rangle$ must vanish
order by order. In this thesis, I will define the magnetic moment
$\mu$ such that $d\mu/dt$ has a gyrophase dependent component of
order $\delta_i v_i^3/BL$ but satisfies
\begin{equation}
\left \langle \frac{d \mu}{dt} \right \rangle = O \left (
\delta_i^2 \frac{v_i^3}{BL} \right ) \simeq 0, \label{dmu_dt_main}
\end{equation}

\noindent for $\mu$ to remain an adiabatic invariant. This
variable $\mu$ is
\begin{equation}
\mu = \mu_0 + \mu_1, \label{mutotal_def}
\end{equation}

\noindent where $\mu_0 = v_\bot^2/2B$ is the usual lowest order
result, and $\mu_1 = O(\delta_i \mu_0)$. The correction $\mu_2 =
O(\delta_i^2 \mu_0)$ is not necessary because the distribution
function is assumed to be a stationary Maxwellian to zeroth order,
making the dependence on $\mu$ weak. For $\mu$ to remain an
adiabatic invariant, $\mu_1$ must contain gyrophase independent
contributions such that $\langle d \mu / dt \rangle = 0$ to the
requisite order, given in \eq{dmu_dt_main}. Solving for $\mu_1$ as
outlined in section~\ref{sectapp_GKvar1} of
Appendix~\ref{app_GKvar} gives
\begin{equation}
\mu_1 = \frac{Z e \phiwig}{M B (\bR)} - \frac{1}{B} \bv_\bot \cdot
\bv_M - \frac{v_{||}}{4 B \Omega_i} [ \bv_\bot ( \bv \times \bun )
+ ( \bv \times \bun ) \bv_\bot ]:\nabla \bun - \frac{v_{||}
v_{\bot}^2}{2B\Omega_i} \bun \cdot \nabla \times \bun.
\label{correc_mu_o1}
\end{equation}

\noindent To keep $\mu$ an adiabatic invariant, $\langle \mu_1
\rangle = - (v_{||} v_\bot^2/2 B \Omega_i) (\bun \cdot \nabla
\times \bun) \neq 0$. In subsection~\ref{subapp_mu1ave} of
Appendix~\ref{app_GKvar}, $\langle \mu_1 \rangle$ is proven to
make $\mu$ an adiabatic invariant to next order.

The procedure presented in this section is compared to the Lie
transform techniques \cite{brizard07} in
Appendix~\ref{app_equivalence}. Both methods yield the same
results, although the final equations look somewhat different.
These apparent discrepancies are shown to be due to subtleties in
the definitions of the functions $\phiave$, $\phiwig$ and
$\Phiwig$. Finally, I also compare this derivation with drift
kinetics. In particular, in Appendix~\ref{app_gyrodepend_kL}, I
show that with the higher order corrections $\bR_2$ and $E_2$ it
is possible to recover the drift kinetic gyrophase dependent
portion of $f_i$ up to order $\delta_i^2 f_{Mi}$.

\section{Gyrokinetic Fokker-Planck equation \label{sect_GKFPequation}}

The Fokker-Planck equation \eq{FPeq_rv} becomes
\begin{equation}
\frac{\partial f_i}{\partial t} + \dot{\bR} \cdot \nabla_\bR f_i +
\dot{E}\, \frac{\partial f_i}{\partial E} + \dot{\mu}\,
\frac{\partial f_i}{\partial \mu} + \dot{\varphi}\, \frac{\partial
f_i}{\partial \varphi} = C \{ f_i \} \label{FPeq_full}
\end{equation}

\noindent when written in gyrokinetic variables, where $\dot{Q}
\equiv dQ/dt$, and $Q$ is any of the gyrokinetic variables. The
gyroaverage of this equation is
\begin{equation}
\frac{\partial \langle f_i \rangle}{\partial t} + \dot{\bR} \cdot
\nabla_\bR \langle f_i \rangle + \dot{E}\, \frac{\partial \langle
f_i \rangle}{\partial E} = \langle C \{ f_i \} \rangle,
\label{FPeq_ave_2}
\end{equation}

\noindent where $\langle f_i \rangle \equiv \langle f_i \rangle
(\bR, E, \mu, t)$ is the gyroaveraged ion distribution function.
Here, I have used that $E$ and $\bR$ are defined such that their
time derivatives are gyrophase independent to the orders given by
\eq{dR_dt_main} and \eq{dE_dt_main}. The term $\dot{\mu} (\partial
f_i/\partial \mu)$ is neglected because the magnetic moment $\mu$
is defined such that $d\mu/dt = O(\delta_i v_i^3/BL)$ and the
zeroth order distribution function is assumed to be a stationary
Maxwellian, making $\partial f_i / \partial \mu = O(\delta_i
f_{Mi} B/v_i^2)$. Therefore, in \eq{FPeq_ave_2} I have neglected
pieces that are $O(f_{Mi} \delta_i^2 v_i / L)$. I have also
neglected the term $\langle \dot{ \varphi }\, \partial f_i /
\partial \varphi \rangle = O( \fwig_i \delta_i v_i / L )$, where
$\fwig_i = f_i - \langle f_i \rangle$ is the gyrophase dependent
piece of the distribution function. I will prove in the next
paragraph that $\fwig_i$ is $O( f_{Mi} \delta_i \nu_{ii} /
\Omega_i)$, making all the neglected terms comparable to or
smaller than $f_{Mi} \delta_i^2 v_i / L$, and the distribution
function gyrophase independent to first order, $f_i \simeq \langle
f_i \rangle$. Notice that, due to the missing pieces, I can only
obtain contributions to the distribution function that are $O (
\delta_i f_{Mi} )$, as well as all terms with $k_\bot \rho_i \sim
1$.

The explicit equation for the gyrophase dependent part of the
distribution function is obtained by subtracting from the full
Fokker-Planck equation \eq{FPeq_full} its gyroaverage, giving to
lowest order
\begin{equation}
- \Omega_i \frac{\partial \fwig_i}{\partial \varphi} = C \{ f_i \}
- \langle C \{ f_i \} \rangle.
\end{equation}

\noindent Therefore, the collisional term is the one that sets the
size of $\fwig_i$. Since the distribution function is a Maxwellian
to zeroth order, the collision operator vanishes to zeroth order,
$C \{ f_i \} = O( \delta_i \nu_{ii} f_{Mi} )$, giving $C \{ f_i \}
- \langle C \{ f_i \} \rangle = O ( \delta_i \nu_{ii} f_{Mi} )$.
As a result, $\fwig_i$ is
\begin{equation}
\fwig_i \simeq - \frac{1}{\Omega_i} \int^{\varphi} d
\varphi^\prime \: ( C \{ f_i \} - \langle C \{ f_i \} \rangle ) =
O \left ( \frac{\delta_i \nu_{ii}}{\Omega_i} f_{Mi} \right ),
\label{fwig_sol}
\end{equation}

\noindent where $\nu_{ii} / \Omega_i \ll 1$.

Using the values of $d \bR / dt$ from \eq{dR_dt_main} and $d E /
dt$ from \eq{dE_dt_main}, and using $\langle f_i \rangle \simeq
f_i$, the equation for $f_i$ in gyrokinetic variables is
\begin{equation}
\frac{\partial f_i}{\partial t} + [ u \bun (\bR) + \bv_d ] \cdot
\left ( \nabla_\bR f_i - \frac{Ze}{M} \nabla_\bR \phiave
\frac{\partial f_i}{\partial E} \right ) = \langle C \{ f_i \}
\rangle, \label{FP_final}
\end{equation}

\noindent where $\phiave$ is defined in \eq{phiave}, and $f_i
\equiv f_i (\bR, E, \mu, t)$ is gyrophase independent.

The gyrokinetic equation can be also written in conservative form.
To do so, the Jacobian of the gyrokinetic transformation is
needed. Conservation of particles in phase space requires the
Jacobian of the transformation, $J = \partial ( \br, \bv ) /
\partial ( \bR, E, \mu, \varphi )$, to satisfy
\begin{equation}
\frac{\partial J}{\partial t} + \nabla_\bR \cdot ( \dot{\bR} J ) +
\frac{\partial}{\partial E} ( \dot{E} J ) +
\frac{\partial}{\partial \mu} ( \dot{\mu} J ) +
\frac{\partial}{\partial \varphi} ( \dot{\varphi} J ) = 0.
\label{Jacob_condition}
\end{equation}

\noindent [This is the equality $\nabla \cdot \dot{\br} + \nabla_v
\cdot \dot{\bv} = 0$ written in gyrokinetic variables]. Employing
this property, equation~\eq{FPeq_full} can be written in
conservative form by multiplying it by $J$ to obtain
\begin{equation}
\frac{\partial}{\partial t} ( J f_i ) + \nabla_\bR \cdot \left (
\dot{\bR} J f_i \right ) + \frac{\partial}{\partial E} \left (
\dot{E} J f_i \right ) + \frac{\partial}{\partial \mu} \left (
\dot{\mu} J f_i \right ) + \frac{\partial}{\partial \varphi} \left
( \dot{\varphi} J f_i \right ) = J C \{f_i\}.
\end{equation}

\noindent The gyroaverage of this equation is
\begin{equation}
\frac{\partial}{\partial t} ( J f_i ) + \nabla_\bR \cdot (
\dot{\bR} J f_i ) + \frac{\partial}{\partial E} ( \dot{E} J f_i )
= J \langle C \{f_i\} \rangle. \label{FPeq_cons_ave}
\end{equation}

\noindent Here, I have taken into account that the Jacobian $J$ is
independent of $\varphi$ to the order of interest, as can be seen
by using \eq{Jacob_condition}. The equation for the gyrophase
dependent part of the Jacobian is obtained by subtracting from
\eq{Jacob_condition} its gyroaverage. Notice that $J - \langle J
\rangle$ depends on the differences $\dot{\bR} - \langle \dot{\bR}
\rangle$, $ \dot{E} - \langle \dot{E} \rangle$..., and those
differences are small by definition of the gyrokinetic variables.
The gyrophase-dependent part of the Jacobian is estimated to be $J
- \langle J \rangle = O ( \delta_i^2 B / v_i )$. Finally, I
substitute $d\bR/dt$ and $dE/dt$ in \eq{FPeq_cons_ave} to get
\begin{equation}
\frac{\partial}{\partial t} (Jf_i) + \nabla_\bR \cdot \{ J f_i [ u
\bun (\bR) + \bv_d ] \} - \frac{\partial}{\partial E} \left \{ J
f_i \frac{Ze}{M} [ u \bun (\bR) + \bv_M ] \cdot \nabla_\bR \phiave
\right \} = J \langle C \{ f_i\} \rangle. \label{FP_conserv}
\end{equation}

The calculation of the Jacobian is described in
section~\ref{sectapp_jacob} of Appendix~\ref{app_GKvar}. The final
result is
\begin{equation}
J = \frac{\partial  (\br, \bv)}{\partial ( \bR, E, \mu, \varphi )}
= \frac{B (\bR)}{u} + \frac{Mc}{Ze} \bun(\bR) \cdot \nabla_\bR
\times \bun (\bR). \label{result_Jacob}
\end{equation}

\noindent In section~\ref{sectapp_jacob}, it is also proven that
$J$ satisfies the gyroaverage of \eq{Jacob_condition}.

Similar gyrokinetic equations to \eq{FP_final} and \eq{FP_conserv}
can be found for the gyrokinetic variables $\bR$, $u$ and $\mu$,
where $u$ is defined by \eq{par_drift}. Combining equations
\eq{dR_dt_main}, \eq{par_drift}, \eq{dE_dt_main} and $d\mu / dt
\simeq 0$ to obtain
\begin{equation}
\dot{u} = - \left [ \bun (\bR) + \frac{u}{\Omega_i} \bun (\bR)
\times \kappabf(\bR) \right ] \cdot  \nabla_\bR \left [ \mu B (
\bR ) + \frac{Ze\phiave}{M} \right ]
\end{equation}

\noindent gives the gyrokinetic equation
\begin{equation}
\frac{\partial f_i}{\partial t} + [ u \bun (\bR) + \bv_d ] \cdot
\nabla_\bR f_i + \dot{u} \frac{\partial f_i}{\partial u} = \langle
C\{f_i\} \rangle. \label{FP_u}
\end{equation}

\noindent This gyrokinetic equation can be written in conservative
form by noticing that the new Jacobian is given by
\begin{equation}
J_u = \frac{\partial ( \br, \bv )}{\partial ( \bR, u, \mu, \varphi
)} = \frac{\partial ( \br, \bv )}{\partial ( \bR, E, \mu, \varphi
)} \frac{\partial E}{\partial u} = B (\bR) + \frac{Mcu}{Ze} \bun
(\bR) \cdot \nabla_\bR \times \bun (\bR). \label{Jacob_u}
\end{equation}

\noindent Using the new Jacobian, the gyrokinetic equation may be
written as
\begin{equation}
\frac{\partial}{\partial t} (J_u f_i) + \nabla_\bR \cdot \{ J_u
f_i [ u \bun ( \bR ) + \bv_d ] \} + \frac{\partial}{\partial u} (
J_u f_i \dot{u} ) = J_u \langle C\{f_i\} \rangle.
\label{FP_u_conserv}
\end{equation}

The ion gyrokinetic Fokker-Planck equations \eq{FP_final},
\eq{FP_conserv}, \eq{FP_u} and \eq{FP_u_conserv} have their
counterpart for electrons. However, since in this thesis the
wavelengths shorter than the ion gyroradius are not considered,
the electron gyrokinetic equation reduces to the drift kinetic
equation
\begin{equation}
\frac{\partial \overline{f}_e}{\partial t} + ( v_{||} \bun +
\bv_{de} ) \cdot \left ( \nablaave \, \overline{f}_e + \frac{e}{m}
\nabla \phi \frac{\partial \overline{f}_e}{\partial E_0} \right )
= C \{ \overline{f}_e \}, \label{drift_kinetic_e}
\end{equation}

\noindent with $\overline{f}_e = \overline{f}_e (\br, E_0, \mu_0,
t)$ the gyrophase independent piece of the distribution function,
$\nablaave$ the gradient holding $E_0$, $\mu_0$, $\varphi_0$ and
$t$ fixed, and
\begin{equation}
\bv_{de} = - \frac{\mu_0}{2\Omega_e} \bun \times \nabla B -
\frac{v_{||}^2}{\Omega_e} \bun \times \kappabf - \frac{c}{B}
\nabla \phi \times \bun \label{drift_electrons}
\end{equation}

\noindent the electron drifts. Here, $\Omega_e = eB/mc$ is the
electron gyrofrequency. The total distribution function for
electrons contains the gyrophase independent piece
$\overline{f}_e$ and the gyrophase dependent piece
\begin{equation}
f_e - \overline{f}_e = - \frac{f_{Me}}{\Omega_e} (\bv \times \bun)
\cdot \left [ \frac{\nabla n_e}{n_e} - \frac{e}{T_e} \nabla \phi +
\left ( \frac{mE_0}{T_e} - \frac{3}{2} \right ) \frac{\nabla
T_e}{T_e} \right ]. \label{e_distribution}
\end{equation}

In the rest of this thesis, the gyrokinetics variables to be used
are $\bR$, $E$ and $\mu$ for ions and $\br$, $E_0$ and $\mu_0$ for
electrons. Then, the relevant equations are \eq{FP_final} for
ions, and \eq{drift_kinetic_e} for electrons.

\section{Gyrokinetic quasineutrality equation \label{sect_GKpoisson}}

Modern gyrokinetics employs a low order quasineutrality condition
to calculate the electrostatic potential. The ion and electron
distribution functions $f_i$ and $f_e$ are calculated using the
lower order equations \eq{FP_final} and \eq{drift_kinetic_e}.
These distribution functions are integrated over velocity space to
obtain the densities $n_i$ and $n_e$ that are substituted into the
quasineutrality condition $Zn_i = n_e$, from which the potential
is solved. Since equations \eq{FP_final} and \eq{drift_kinetic_e}
give distribution functions good to order $\delta_i f_{Mi}$ and
$\delta_e f_{Me}$, respectively, the quasineutrality equation is
correct only to order $\delta_i n_e$. In
chapter~\ref{chap_gyrovorticity}, I will show that this is not
enough to solve for the long wavelength axisymmetric radial
electric field, and employing such a low order quasineutrality
equation can lead to unphysical results. In this section, I will
derive the modern gyrokinetic quasineutrality equation to the
order that is usually implemented, so I can demonstrate later, in
chapter~\ref{chap_gyrovorticity}, that it cannot provide the
correct long wavelength axisymmetric radial electric field.

I will begin with Poisson's equation to explicitly show that the
quasineutral approximation is valid for the range of wavelengths
of interest. The distribution function $f_i$ in Poisson's
equation,
\begin{equation}
- \nabla^2 \phi = 4 \pi e \left [ Z \int d^3 v \: f_i (\bR, E,
\mu, t) - n_e (\br, t ) \right ], \label{poisson_real}
\end{equation}

\noindent is obtained from the Fokker-Planck equation
\eq{FP_final}. Therefore, it is known as a function of the
gyrokinetic variables. The distribution function can be rewritten
more conveniently as a function of $\br + \Omega_i^{-1} \bv \times
\bun$, $E_0$ and $\mu_0$ by Taylor expanding. However, it is
important to remember that there are missing pieces of order
$\delta_i^2 f_{Mi}$ in the distribution function since terms of
this order must be neglected to derive \eq{FP_final}. Thus, the
expansion can only be carried out to the order where the
distribution function is totally known, resulting in
\begin{equation}
f_i (\bR, E, \mu, t) = f_i(\bR_g, E_0, \mu_0, t) + E_1
\frac{\partial f_i}{\partial E_0} + \mu_1 \frac{\partial
f_i}{\partial \mu_0} + O ( \delta_i^2 f_{Mi} ).
\end{equation}

\noindent Notice that $f_i (\bR_g, E_0, \mu_0, t)$, with $\bR_g
\equiv \br + \Omega_i^{-1} \bv \times \bun$, cannot be Taylor
expanded around $\br$ because $k_\bot \rho_i \sim 1$. In the
higher order terms proportional to $E_1$ and $\mu_1$, the function
$f_i$ is valid only to lowest order, i.e., $f_i \simeq f_{Mi}$,
$\partial f_i /\partial E_0 \simeq (-M/T_i) f_{Mi}$ and $\partial
f_i /\partial \mu_0 \simeq 0$. Moreover, according to the ordering
in \eq{order_kbot}, the corrections arising from using $\br$
instead of $\bR_g$ are small by $\delta_i$ because, even though
small wavelengths are allowed, the amplitude of the fluctuations
with small wavelengths is assumed to be of the next order.
Therefore, in the higher order integrals, only the long wavelength
distribution function $f_{Mi}$ depending on $\bR_g \simeq \br$
need be retained.

Since the turbulent wavelengths are much larger than the Debye
length, the term in the left side of Poisson's equation
\eq{poisson_real} may be neglected. The resulting quasineutrality
equation reduces to
\begin{equation}
Z n_{ip} \{ \phi \} \simeq n_e(\br, t) - Z \hat{N}_i(\br, t),
\label{qn_mod}
\end{equation}

\noindent with $n_e = \int d^3v\, \overline{f}_e$ the electron
density,
\begin{equation}
n_{ip} \{ \phi \} = - \int d^3v\, \frac{Ze\phiwig}{T_i} f_{Mi}
\label{polar_n}
\end{equation}

\noindent the ion polarization density that depends explicitly in
the potential, and
\begin{eqnarray}
\hat{N}_i (\br, t) = \int d^3v \: f_i(\bR_g, E_0, \mu_0, t)
\label{eq_Nhat}
\end{eqnarray}

\noindent the ion guiding center density. In equation \eq{qn_mod},
terms of order $\delta_i^2 n_e$ and $n_e k_\bot \lambda_D^2 /L \ll
\delta_i k_\bot \rho_i n_e$ have been neglected, with $\lambda_D =
\sqrt{T_e/4\pi e^2 n_e}$ the Debye length.

\begin{figure}
\begin{center}
\includegraphics[width = 0.8\textwidth]{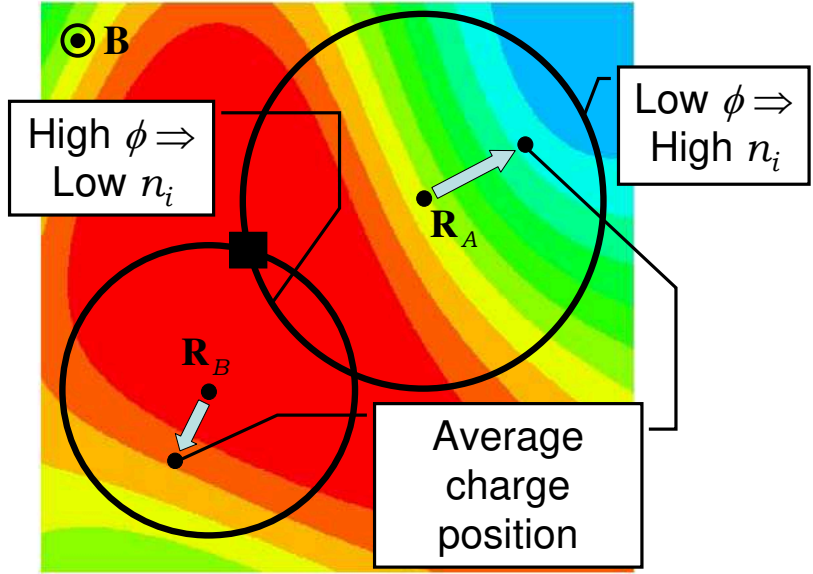}
\end{center}

\vspace{-1cm} \caption[Ion density in gyrokinetics]{Ion density in
the black square calculated using a gyrokinetic distribution
function. The electrostatic potential is the background contour
plot.} \label{fig_polarization}
\end{figure}

In the quasineutrality equation \eq{qn_mod}, the ion density is
composed of two terms, the guiding center density $\hat{N}_i$ and
the polarization density $n_{ip} \{ \phi \}$. Both terms have a
clear physical meaning. In figure~\ref{fig_polarization}, the ion
density calculation is sketched. In gyrokinetics, ions are
substituted by rings of charge with radius the ion gyroradius. The
ion density at a point is then calculated by counting the rings
that pass through that point. In figure~\ref{fig_polarization},
the ion density in the black square is computed by summing all the
gyrocenters whose rings of charge cross the square. In the figure,
two of them, $\bR_A$ and $\bR_B$, are shown. Importantly, the
charge density along a ring is not constant, but depends on the
potential. In figure~\ref{fig_phiavewig} we saw that the potential
changes rapidly along the gyromotion. There is a Maxwell-Boltzmann
response to this variation, $-Ze\phiwig/T_e$, that gives a varying
ion charge density along the ring, as indicated in
figure~\ref{fig_polarization}, where the left region of the ring
$\bR_A$ has lower density than the right side. In the ring,
therefore, the average charge position is displaced towards the
lower potential values, i.e., the charge moves in the same
direction as the local electric field, hence the name
polarization. For this reason, the ion density is separated into
the guiding center density $\hat{N}_i$, due to the average charge
in the ring, and the polarization density $n_{ip}$ that originates
in the non-uniform density along the ring that is induced by the
short wavelength pieces of the potential.

Equation~\eq{qn_mod} is used to calculate $\phi$ for wavelengths
of the order of the gyroradius, including zonal flow, in $\delta
f$ turbulence codes such as GS2 \cite{dorland00}, GENE
\cite{dannert05} or GYRO \cite{candy03}. In most cases, the
distribution function is obtained from the gyrokinetic equation
\eq{FP_final} written for $f_i = f_{Mi} + h_i$, with $|h_i(\bR, E,
\mu, t)| \ll f_{Mi}$ and $f_{Mi}$ only depending on $\psi$. The
resulting equation is
\begin{eqnarray}
\frac{\partial h_i}{\partial t} + [u \bun (\bR) + \bv_d] \cdot
\nabla_\bR h_i - \left \langle C^{(\ell)} \left \{ h_i -
\frac{Ze\phiwig}{T_i} f_{Mi} \right \} \right \rangle = \nonumber
\\ \frac{Ze}{T_i} f_{Mi} \left \{ i \omega_\ast^{n,T} \phiave - [u
\bun (\bR) + \bv_M ] \cdot \nabla_\bR \phiave \right \},
\label{deltaf_FP}
\end{eqnarray}

\noindent with $i = \sqrt{-1}$ and
\begin{equation}
\omega_\ast^{n,T} \equiv - i \frac{cT_i}{ZeB} (\bun \times \nabla
\psi) \cdot \nabla_\bR \ln \phiave \left [ \frac{1}{n_i} \frac{d
n_i}{d \psi} + \left ( \frac{ME}{T_i} - \frac{3}{2} \right )
\frac{1}{T_i} \frac{d T_i}{d \psi} \right ]
\label{general_omegastar}
\end{equation}

\noindent the drift wave frequency. In equation \eq{deltaf_FP},
$\phi$ appears nonlinearly in $\bv_d \cdot \nabla_\bR h_i$ and
linearly on the right side of the equation. The linear terms are
usually solved implicitly. Then, $h_i$ has a linear dependence on
$\phi$ that will appear as a linear dependence in $\hat{N}_i$, and
can be used to solve for $\phi$ in equation \eq{qn_mod}. The ion
polarization density \eq{polar_n} also depends linearly on $\phi$.
However, at long wavelengths $n_{ip}$ becomes too small to be
important. For $\delta_i \ll k_\bot \rho_i \ll 1$, the
polarization density is
\begin{equation}
n_{ip} \simeq \nabla \cdot \left ( \frac{cn_i}{B\Omega_i}
\nabla_\bot \phi \right ) = O(\delta_i k_\bot \rho_i n_e).
\label{nip_krhosmall}
\end{equation}

\noindent The details of this calculation are given in
section~\ref{sectapp_polarn_simple} of
Appendix~\ref{app_poisson_kL}. To estimate the size of $n_{ip}$, I
have used \eq{order_kbot} to order $\nabla_\bot \phi \sim T_e/eL$
and $\nabla_\bot \nabla_\bot \phi \sim k_\bot T_e/eL$. With this
estimate, for $k_\bot L \sim 1$ equation \eq{qn_mod} becomes
\begin{equation}
Z\hat{N}_i(\br, t) = n_e(\br, t), \label{qn_highorder_1}
\end{equation}

\noindent where terms of order $\delta_i^2 n_e$ have been
neglected.

It is possible to obtain a higher order long wavelength
quasineutrality equation for a non-turbulent plasma if the ion
distribution function is assumed to be known to high enough order.
The resulting equation is
\begin{equation}
\nabla \cdot \left ( \frac{Z c n_i}{B \Omega_i} \nabla_\bot \phi
\right ) - \frac{ZM c^2 n_i}{2T_i B^2} | \nabla_\bot \phi |^2 =
n_e - Z \hat{N}_i, \label{qn_highorder}
\end{equation}

\noindent where $\hat{N}_i$ must be defined to higher order,
\begin{equation}
\hat{N}_i (\br, t) = \int d^3 v \, f_i (\br, E_0, \mu_0, t ) \left
( 1 + \frac{v_{||}}{\Omega_i} \bun \cdot \nabla \times \bun \right
) + ( \matrixtop{\mathbf{I}} - \bun \bun ) : \frac{\nabla \nabla
p_i}{2 M \Omega_i^2}. \label{nhat_final_lk}
\end{equation}

\noindent The derivation of \eq{qn_highorder} is shown in
section~\ref{sectapp_poisson_kL} of Appendix~\ref{app_poisson_kL}.
Even though equation \eq{qn_highorder} is correct, it is only
useful if we are able to evaluate the missing $O(\delta_i^2
f_{Mi})$ pieces in $f_i$ that are of the same order as the left
side in \eq{qn_highorder}. Equation~\eq{FP_final} misses these
pieces. Equation \eq{qn_highorder} will only serve to demonstrate
the problems that arise from the use of equation \eq{qn_mod}.

In chapter~\ref{chap_gyrovorticity}, I will prove that neither
equation \eq{qn_highorder_1} or its higher order version
\eq{qn_highorder} are able to provide the self-consistent, long
wavelength radial electric field. I will even give an example in
section~\ref{sect_thetapinch} in which the higher order equation
\eq{qn_highorder} leaves the radial electric field undetermined.

\chapter{Gyrokinetic vorticity equation \label{chap_gyrovorticity}}

In this chapter, I rewrite the vorticity equation \eq{vorticity}
in a convenient form for gyrokinetics. The gyrokinetic change of
variables found in chapter~\ref{chap_gyrokinetics} is especially
well suited for simulation of drift wave turbulence. However, the
gyrokinetic quasineutrality equation, traditionally used to
calculate the electrostatic potential, has problems. By writing
the vorticity equation \eq{vorticity} in gyrokinetic form, I can
study the behavior of the quasineutrality equation at different
time scales and wavelengths. In particular, I am able to prove
that the radial current vanishes to a very high order for any
radial electric field, i.e., the radial drift of ions and
electrons is intrinsically ambipolar.

In section~\ref{sect_assumptions}, I explain the notation employed
in this chapter and I list the assumptions. These assumptions
restrict the treatment to turbulence that has reached a
statistical equilibrium and, therefore, only has small variations
along magnetic flux surfaces. Under the assumptions of
sections~\ref{sect_GKorderings} and \ref{sect_assumptions}, I
evaluate the size of the terms in the vorticity equation
\eq{vorticity} and in the toroidal angular momentum conservation
equation \eq{toro_angmom}. The size of the different contributions
to the vorticity equation depends on the perpendicular wavelength
of interest, and this dependence makes some terms important for
structures of the size of the ion gyroradius and negligible at
wavelengths on the order of the minor radius of the device. In
section~\ref{sect_vorticity_1}, it will become clear that direct
evaluation of the terms in equation \eq{vorticity} is too
difficult to be of interest. Then, in the rest of the chapter, a
different approach is taken. The equations for particle and
momentum conservation are derived from the gyrokinetic equation in
section~\ref{sect_transportGK}. These conservation equations are
then combined in section~\ref{sect_vorticityGK} to derive two
different vorticity equations equivalent to \eq{vorticity} to
lowest order. Importantly, these equations are easier to study and
to evaluate numerically. With them, I show that the dependence of
quasineutrality on the long wavelength radial electric field is
not meaningful for gyrokinetic codes to retain since gyrokinetics
will be shown to be intrinsically ambipolar as already stressed in
section~\ref{sect_fluxsurfaceaverage}. The problems that arise
from quasineutrality are exposed in a simplified example in
section~\ref{sect_thetapinch}. I finish this chapter with a
discussion in section~\ref{sect_discussion_GKvorticity}. All the
details of the calculation are relegated to
Appendices~\ref{app_transportGK}-\ref{app_gyrovisc}.

\section{Notation and assumptions \label{sect_assumptions}}

In this chapter, I work in both the gyrokinetic phase space $\{
\bR, E, \mu, \varphi \}$ and the ``physical" phase space $\{ \br,
\bv \}$. I refer to it as physical phase space because spatial and
velocity coordinates do not get mixed as they do in gyrokinetic
phase space. I use the variables $\br$, $E_0$, $\mu_0$ and
$\varphi_0$ to describe this physical phase space. Whenever I
write $\partial/\partial E_0$, it is implied that $\br$, $\mu_0$,
$\varphi_0$ and $t$ are held fixed, and similarly for
$\partial/\partial \mu_0$ and $\partial/\partial \varphi_0$. The
gradient holding $E_0$, $\mu_0$, $\varphi_0$ and $t$ fixed will be
written as $\nablaave$. In addition, any derivative with respect
to a gyrokinetic variable is performed holding the other
gyrokinetic variables constant. The partial derivative with
respect to the time variable $t$ deserves a special mention since
it is necessary to indicate which variables are kept fixed. In
this formulation, the time derivative holding $\br$ and $\bv$
fixed is equivalent to holding $\br$, $E_0$, $\mu_0$ and
$\varphi_0$ fixed because the magnetic field is constant in time.
Also, a gyroaverage holding $\br$, $E_0$, $\mu_0$ and $t$ fixed is
denoted as $\overline{(\ldots)}$, as opposed to the gyrokinetic
gyroaverage $\langle \ldots \rangle$ performed holding $\bR$, $E$,
$\mu$ and $t$ fixed.

All the assumptions in section~\ref{sect_GKorderings} are
applicable here. Then, the zeroth order ion and electron
distribution functions are assumed to be stationary Maxwellians,
$f_{Mi}$ and $f_{Me}$. In this chapter, the only spatial
dependence allowed for these zeroth order solutions is in the
radial variable $\psi$. Therefore, $\bun \cdot \nabla f_{Mi} =
\bun \cdot \nabla f_{Me} = 0$. I assume that the radial gradients
of $f_{Mi}$ and $f_{Me}$ are $O(1/L)$, with $L$ of the order of
the minor radius of the tokamak. The zeroth order potential $\phi$
works in a similar fashion, depending only on $\psi$ and with a
radial gradient on the longer scale $L$.

As in section~\ref{sect_GKorderings}, I allow wavelengths
perpendicular to the magnetic field that are on the order of the
ion gyroradius, $k_\bot \rho_i \sim 1$. The pieces of the
potential and the distribution function with short perpendicular
wavelengths are small in size, following the ordering in
\eq{order_kbot}. Except for initial transients, I assume that the
variation along the magnetic field of $f_i$, $f_e$ and $\phi$ is
slow, i.e., in general $\bun \cdot \nabla f_i \sim \delta_i
f_{Mi}/L$, $\bun \cdot \nabla f_e \sim \delta_i f_{Me}/L$ and
$\bun \cdot \nabla \phi \sim \delta_i T_e/eL$.

The ion distribution function $f_i (\bR, E, \mu, t)$ is found
employing the gyrokinetic equation \eq{FP_final} [the gyrophase
dependent piece is $O(\delta_i f_{Mi} \nu_{ii}/\Omega_i)$ and
given by \eq{fwig_sol}]. After the initial transient, equation
\eq{FP_final} becomes $u \bun \cdot \nabla_\bR f_i = \langle C
\{f_i\} \rangle$ to zeroth order. At long wavelengths, this
requires that $f_i$ approach a Maxwellian $f_{Mi}$ with $\bun
\cdot \nabla f_{Mi} = 0$, giving $f_{Mi} \equiv f_{Mi} (\psi, E)$.
The assumed long wavelength piece of the distribution function
satisfies this condition. Importantly, $\bun \cdot \nabla f_{Mi} =
0$ does not impose any condition on the radial dependence of
$f_{Mi}$. Consequently, the density and temperature in $f_{Mi}$
may have short wavelength components as long as they satisfy the
orderings in \eq{order_kbot}, i.e., $\nabla_\bot n_i \sim k_\bot
n_{i,k} \sim n_{i, k=0}/L$, $\nabla_\bot T_i \sim k_\bot T_{i,k}
\sim T_{i,k=0}/L$ and $\nabla_\bot \nabla_\bot f_{Mi, k} \sim
k_\bot f_{Mi, k=0}/L$. Solving for the next order correction $f_i
- f_{Mi}$ in equation \eq{FP_final} gives $f_i - f_{Mi} \sim
\delta_i f_{Mi}$. Then, the average velocity $\bV_i = n_i^{-1}
\int d^3v \bv f_i$ is of order $\delta_i v_i$. Furthermore, any
variation of the distribution function within a flux surface is
due to $f_i - f_{Mi}$, and thus small by $\delta_i$ as compared to
the long wavelength piece of $f_{Mi}$. This means that when we
consider average velocities or the gradients $\bun \cdot
\nabla_\bR f_i$ and $\zun \cdot \nabla_\bR f_i$, it will be useful
to think about the distribution function as it is done in $\delta
f$ codes where $f_i = f_{Mi} + h_i$, with $h_i \sim \delta_i
f_{Mi} \ll f_{Mi}$. Comparing the estimate for $f_i - f_{Mi}$ with
the orderings in \eq{order_kbot}, I find that the gradients of
$f_i$ and $\phi$ parallel to the flux surfaces are smaller than
the maximum allowed in gyrokinetics, i.e., $\bun \cdot \nabla_\bR
f_i \sim \delta_i f_{Mi}/L \ssim \zun \cdot \nabla_\bR f_i \sim
k_\bot (f_i - f_{Mi}) \sim k_\bot \rho_i f_{Mi}/L \ssim f_{Mi}/L$
and $\bun \cdot \nabla \phi \sim \delta_i T_e/eL \ssim \zun \cdot
\nabla \phi \sim k_\bot \rho_i T_e/eL \ssim T_e/eL$. These
estimates may fail for the initial transient, but I am interested
in the electric field evolution at long times, when the transient
has died away.

Interestingly, these assumptions imply that the long wavelength
axisymmetric flows are neoclassical. At long wavelengths, the ion
distribution function $f_i (\bR, E, \mu, t)$ can be Taylor
expanded around $\br$, $E_0$ and $\mu_0$. Then, the gyroaveraged
piece $\overline{f}_i$ is approximately $f_i (\br, E_0, \mu_0,
t)$, where $\bR$, $E$ and $\mu$ have been replaced by $\br$, $E_0$
and $\mu_0$. This gyroaveraged piece of the ion distribution
function $\overline{f}_i$ satisfies the ion drift kinetic equation
\begin{equation}
\frac{\partial \overline{f}_i}{\partial t} + ( v_{||} \bun + \bv_M
) \cdot \left ( \nablaave\, \overline{f}_i - \frac{Ze}{M} \nabla
\phi \frac{\partial \overline{f}_i}{\partial E_0} \right ) + \bv_E
\cdot \nabla_\bR f_i = C \{ \overline{f}_i \}.
\label{drift_kinetic_i}
\end{equation}

\noindent This equation is obtained from \eq{FP_final} by
realizing that the functional dependence of $f_i$ on the
gyrokinetic variables $\bR$, $E$ and $\mu$ is the same as the
dependence of $\overline{f}_i$ on $\br$, $E_0$ and $\mu_0$. Then,
equation \eq{drift_kinetic_i} is derived from \eq{FP_final} by
replacing $\bR$, $E$ and $\mu$ by $\br$, $E_0$ and $\mu_0$, and
employing that $\phiave \simeq \phi$ for long wavelengths. The
difference between the long wavelength ion equation
\eq{drift_kinetic_i} and a drift kinetic equation [see, for
example, the electron equation \eq{drift_kinetic_e}] is in the
nonlinear term $\bv_E \cdot \nabla_\bR f_i$. In this term, the
short wavelength components of $f_i$ and $\phi$ beat together to
give a long wavelength contribution. Due to the presence of these
short wavelength pieces, $f_i$ cannot be Taylor expanded and
$\phiave \neq \phi$. Importantly, this term gives a negligible
contribution to the axisymmetric piece of $\overline{f}_i$. For
$\bv_E \cdot \nabla_\bR f_i$ to have an $n = 0$ toroidal mode
number, the beating components $f_{i,n}$ and $\nabla_\bR
\phiave_{-n}$ must have toroidal mode numbers of the same
magnitude and opposite sign. Moreover, these components must have
$n \neq 0$ because otherwise $\nabla_\bR \phiave$ is parallel to
$\nabla_\bR f_i$ and $\bv_E \cdot \nabla_\bR f_i$ vanishes
exactly. Thus, only the non-axisymmetric components of $\phiave$
and $f_i$ contribute to $\bv_E \cdot \nabla_\bR f_i$ and these are
of order $\delta_i T_e/e$ and $\delta_i f_{Mi}$, respectively.
Writing $\bv_E \cdot \nabla_\bR f_i = - (c/B) \bun \cdot
\nabla_\bR \times ( f_i \nabla_\bR \phiave )$, it is easy to see
that $\bv_E \cdot \nabla_\bR f_i \sim (c/B) k_\bot f_{i,n}
|\nabla_\bR \phiave_{-n}|$, with $k_\bot$ the radial wavenumber of
the axisymmetric piece of $\overline{f}_i$. Since $f_{i,n} \sim
\delta_i f_{Mi}$ and $(c/B) |\nabla_\bR \phiave_{-n}| \sim
\delta_i v_i$, the largest possible size for $\bv_E \cdot
\nabla_\bR f_i$ is $\delta_i k_\bot \rho_i f_{Mi} v_i/L$.
Consequently, at long wavelengths, $\bv_E \cdot \nabla_\bR f_i$ is
negligibly small compared with the other terms in
\eq{drift_kinetic_i}, and the equation for the axisymmetric
component of $\overline{f}_i$ is the neoclassical drift kinetic
equation. For this reason, the long wavelength axisymmetric flows
must be neoclassical, i.e., they are given by
\begin{equation}
n_i \bV_i = - \frac{cR}{Ze} \left( \frac{\partial p_i}{\partial
\psi} + Ze n_i \frac{\partial \phi}{\partial \psi} \right ) + U
(\psi) \bB, \label{neo_flows}
\end{equation}

\noindent where $U(\psi)$ is proportional to $\partial
T_i/\partial \psi$ in neoclassical theory \cite{hinton76,
helander02bk}. For $k_\bot \rho_i \sim \delta_i$, the turbulent
term $\bv_E \cdot \nabla_\bR f_i$ is of order $\delta_i^2 f_{Mi}
v_i/L$. By comparing its size with the smaller term in the drift
kinetic equation, usually the collisional operator $C\{f_i\}
\simeq C^{(\ell)} \{ f_i - f_{Mi} \} \sim \delta_i f_{Mi}
\nu_{ii}$, I obtain that the turbulent correction to the
neoclassical flow $U(\psi)$ is of order $\delta_i v_i/L \nu_{ii}$.

To finish this section, I will present the different contributions
to the distribution function. To order $\delta_i f_{Mi}$, the
gyrokinetic distribution function can be written as
\begin{equation}
f_i \equiv f_i(\bR, E, \mu, t) \simeq f_{ig} -
\frac{Ze\phiwig}{T_i} f_{Mi}, \label{fi1}
\end{equation}

\noindent where
\begin{equation}
f_{ig} \equiv f_i (\bR_g, E_0, \mu_0, t) \label{fig}
\end{equation}

\noindent and
\begin{equation}
\bR_g = \br + \frac{1}{\Omega_i} \bv \times \bun.
\end{equation}

\noindent In equation \eq{fi1}, I have Taylor expanded $f_i(\bR,
E, \mu, t)$ around $E_0$ and $\mu_0$, and I have used the zeroth
order Maxwellian $f_{Mi}$ in the higher order terms. In the
function $\phiwig$ in \eq{fi1}, it is enough to use the lowest
order variables $\bR_g$, $\mu_0$ and $\varphi_0$ instead of $\bR$,
$\mu$ and $\varphi$ (the dependence of $\phiwig$ on $E$ is weak).
The piece $f_{ig}$ of the distribution function will be useful in
section~\ref{sect_transportGK} to obtain moment equations from the
gyrokinetic equation \eq{FP_final}. However, the pieces of the
distribution function given in \eq{fi1} are not useful to evaluate
terms in physical phase space since the variable $\bR_g$ still
mixes spatial and velocity space variables. In physical phase
space, it is useful to distinguish between the gyrophase
independent piece of the distribution function $\overline{f}_i$
and the gyrophase dependent piece. At long wavelengths, $f_{ig}
\equiv f_i (\bR_g, E_0, \mu_0, t)$ can be Taylor expanded around
$\br$, and I can employ the lowest order result $\phiwig \simeq
\Omega_i^{-1} (\bv \times \bun) \cdot \nabla \phi$, giving
\begin{equation}
\overline{f}_i = f_{i0} + O(\delta_i k_\bot \rho_i f_{Mi})
\label{fave_krho}
\end{equation}

\noindent and
\begin{equation}
f_i - \overline{f}_i = \frac{1}{\Omega_i} (\bv \times \bun) \cdot
\left [ \frac{\nabla p_i}{p_i} + \frac{Ze \nabla \phi}{T_i} +
\left ( \frac{Mv^2}{2T_i} - \frac{5}{2} \right ) \frac{\nabla
T_i}{T_i} \right ] f_{Mi} + O(\delta_i k_\bot \rho_i f_{Mi}),
\label{fwig_krho}
\end{equation}

\noindent with
\begin{equation}
f_{i0} \equiv f_i (\br, E_0, \mu_0, t). \label{fi0}
\end{equation}

\noindent Notice that $f_{i0}$ differs from $f_{ig}$ in that the
gyrocenter position $\bR$ in $f_i (\bR, E, \mu, t)$ is replaced by
the particle position $\br$ and not the intermediate variable
$\bR_g$.

\section{General vorticity equation in gyrokinetics \label{sect_vorticity_1}}

In this section, I evaluate the size of the different terms in the
vorticity equation \eq{vorticity} and the toroidal momentum
equation \eq{toro_angmom}. For these estimates I will use the
assumptions in sections~\ref{sect_GKorderings} and
\ref{sect_assumptions}. The estimates are summarized in
table~\ref{table_FPvorticity}. The final result will be that the
vorticity equation \eq{vorticity} is not the most convenient to
evaluate the electric field and a new vorticity equation is
needed.

\begin{table}
\begin{center}
\renewcommand{\arraystretch}{1.5}
\begin{tabular}{|c|c|}
\hline Term & Order of magnitude \\
\hline $\varpi$ & $\delta_i k_\bot \rho_i e n_e$ \\
$\nabla \cdot (J_{||} \bun + \bJ_d)$ & $\delta_i e n_e v_i/L$ \\
$\nabla \cdot [ (c/B) \bun \times (\nabla \cdot \matrixtop{\pibf}_i) ]$ & $\delta_i (k_\bot \rho_i)^2 e n_e v_i/L$ \\
\hline
\end{tabular}
\end{center}

\caption[Order of magnitude estimates for vorticity equation
\eq{vorticity}]{Order of magnitude estimates for vorticity
equation \eq{vorticity}.} \label{table_FPvorticity}
\end{table}

In equation \eq{vorticity}, $\varpi \sim \delta_i k_\bot \rho_i e
n_e$ since $\bV_i \sim \delta_i v_i$ and $\nabla_\bot \sim
k_\bot$. Both currents $\bJ_d$ and $J_{||}$ are of order $\delta_i
e n_e v_i$. The divergence of $\bJ_d$ is of order $ \delta_i e n_e
v_i / L$ because, according to the orderings in \eq{order_kbot},
$\nabla p_i \sim k_\bot p_{i, k} \sim p_{i, k = 0} / L$. The
divergence of $J_{||} \bun$ is also of order $\delta_i e n_e v_i /
L$, but in this case it is due to the small parallel gradients.
With these estimates, all the terms $\partial \varpi / \partial
t$, $\nabla \cdot \bJ_d$ and $\nabla \cdot ( J_{||} \bun )$
compete with each other to determine the electric field. The
remaining term, $\nabla \cdot [ (c/B) \bun \times (\nabla \cdot
\matrixtop{\pibf}_i)]$, is more difficult to evaluate.

The ion viscosity $\matrixtop{\pibf}_i$, given by \eq{pi_def}, is
of order $O(\delta_i k_\bot \rho_i p_i)$. It only depends on the
gyrophase dependent part of the distribution function, and at long
wavelengths, the gyrophase dependent piece is given by
\eq{fwig_krho}, making $\matrixtop{\pibf}_i \sim \delta_i k_\bot
\rho_i p_i$ because the lowest order gyrophase dependent piece
vanishes. The estimate is also valid for electrons, for which it
is assumed that the shortest wavelength is of the order of the ion
gyroradius, giving an electron viscosity $m/M$ times smaller than
the ion viscosity, thereby justifying its neglect.

With this estimate for $\matrixtop{\pibf}_i$, the term $\nabla
\cdot [ (c/B) \bun \times (\nabla \cdot \matrixtop{\pibf}_i) ]$ in
\eq{vorticity} is formally of order $(k_\bot \rho_i)^3 e n_e v_i /
L$, while the rest of the terms are of order $\delta_i e n_e v_i /
L$. However, I will prove in section \ref{sect_vorticityGK} that
the formal estimate is too high, and in reality
\begin{equation}
\nabla \cdot \left [ \frac{c}{B} \bun \times (\nabla \cdot
\matrixtop{\pibf}_i) \right ] \sim \delta_i (k_\bot \rho_i)^2 e
n_e v_i/L. \label{proof_intrinsic}
\end{equation}

\noindent This term is the only one that enters in the equation
for the radial electric field, as proven by \eq{toro_angmom}.
Since it becomes small for long wavelengths, the vorticity
equation and hence its time integral, quasineutrality, are almost
independent of the long wavelength radial electric field, i.e.,
the tokamak is intrinsically ambipolar even in the presence of
turbulence. The order of magnitude estimate in equation
\eq{proof_intrinsic} is the proof of intrinsic ambipolarity.

Except for the flux surface averaged vorticity equation, dominated
by the term $\nabla \cdot [ (c/B) \bun \times (\nabla \cdot
\matrixtop{\pibf}_i) ] \sim \delta_i (k_\bot \rho_i)^2 e n_e v_i /
L$, the lower order terms in the vorticity equation are always of
order $\delta_i e n_e v_i / L$. Then, since the vorticity $\varpi$
is $O(\delta_i k_\bot \rho_i e n_e)$, becoming small for the
longer wavelengths, the time evolution of equation \eq{vorticity}
requires a decreasing time step as $k_\bot \rho_i \rightarrow 0$,
or an implicit numerical method that will ensure that the right
side of \eq{vorticity} vanishes for long wavelengths. Solving
implicitly for the potential is routinely done in gyrokinetic
simulations \cite{candy03, kotschen95}.

The flux surface averaged vorticity equation gives toroidal
angular momentum conservation equation \eq{toro_angmom}. According
to the estimates in previous paragraphs, the radial toroidal
viscosity would be $\langle R \zun \cdot \matrixtop{\pibf}_i \cdot
\nabla \psi \rangle_\psi \sim \delta_i k_\bot \rho_i p_i R |\nabla
\psi|$. Then, the characteristic time derivative for the toroidal
velocity is $\partial / \partial t \sim k_\bot \rho_i v_i/L$. This
estimate is in contradiction with neoclassical calculations
\cite{catto05a, catto05b, wong07, rosenbluth71, wong05} and the
random walk estimate at the end of
section~\ref{sect_fluxsurfaceaverage}, where the long wavelength
toroidal-radial component of viscosity is found to be of order
$\delta_i^3 p_i R |\nabla \psi|$. This discrepancy probably only
occurs transiently for short periods of time. It is to be expected
that for longer times, the transport of toroidal angular momentum
$\langle R \zun \cdot \matrixtop{\pibf}_i \cdot \nabla \psi
\rangle_\psi \sim \delta_i k_\bot \rho_i p_i R |\nabla \psi|$
gives a net zero contribution, and the time averaged
toroidal-radial component of viscosity is actually of order
$\delta_i^3 p_i R |\nabla \psi|$. The size of the time averaged
$\langle R \zun \cdot \matrixtop{\pibf}_i \cdot \nabla \psi
\rangle_\psi$ is discussed in chapter~\ref{chap_angularmomentum}.

To summarize, the vorticity equation \eq{vorticity} has the right
physics, and makes explicit the different times scales (from the
fast turbulence times to the slow radial transport time). However,
I show in section \ref{sect_vorticityGK} that the divergence of
$(c/B) \bun \times (\nabla \cdot \matrixtop{\pibf}_i)$ is an order
smaller than its formal estimate suggests for $k_\bot \rho_i \ll
1$, going from $(k_\bot \rho_i)^3 e n_i v_i / L$ to $\delta_i
(k_\bot \rho_i)^2 e n_i v_i / L$. In other words, $(c/B) \bun
\times (\nabla \cdot \matrixtop{\pibf}_i)$ has a large divergence
free piece. This difference between the real size and the formal
ordering makes theoretical studies cumbersome and it may lead to
numerical problems upon implementation. The rest of this chapter
is devoted to finding a more convenient vorticity equation. In
section~\ref{sect_transportGK}, I will derive the particle and
momentum conservation equations from the gyrokinetic equation
\eq{FP_final}. In the same way that particle and momentum
conservation equations were used in
section~\ref{sect_FPvorticity}, I will employ the gyrokinetic
conservation equations to find two gyrokinetic vorticity equations
in section~\ref{sect_vorticityGK}.

\section{Transport in gyrokinetics \label{sect_transportGK}}

It is necessary to understand the transport of particles and
momentum at wavelengths that are of the order of the ion
gyroradius. It is at those wavelengths that the divergence of the
viscosity becomes as important as the gradient of pressure, and we
need to determine which one dominates in the vorticity equation.
The gyrokinetic equation \eq{FP_final} is especially well suited
to this task. In this section, I derive moment equations from the
gyrokinetic Fokker-Planck equation, in particular, conservation
equations for particles and momentum. They will provide powerful
insights, but we have to remember that the gyrokinetic equation is
correct only to $O(\delta_i f_{Mi} v_i/L )$. Then, the
conservation equations for particles and momentum are missing
terms of order $\delta_i^2 n_e v_i/L$ and $\delta_i^2 p_i/L$.

\begin{figure}
\begin{center}
\includegraphics[width = 0.8\textwidth]{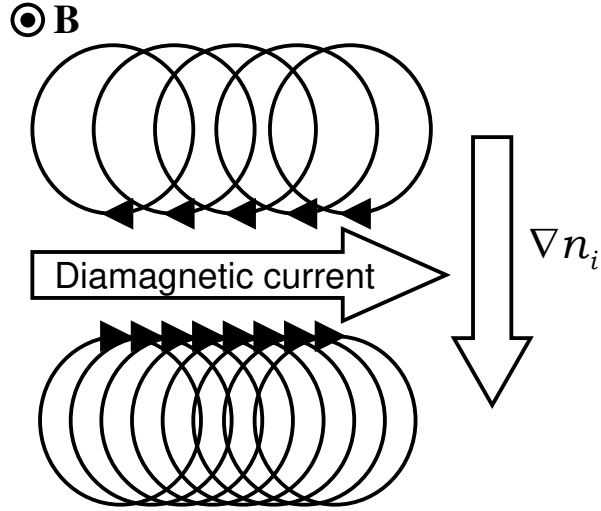}
\end{center}

\vspace{-3cm} \caption[Ion diamagnetic flow]{Ion diamagnetic flow
due to a gradient in ion density.} \label{fig_diamagnetic}
\end{figure}

With the conservation equations for particles and momentum derived
in this section, I will obtain two gyrokinetic vorticity equations
in section~\ref{sect_vorticityGK}. These new vorticity equations
will be equivalent to equation \eq{vorticity} up to, but not
including, $O(\delta_i^2 e n_e v_i/L)$. They will have the
advantage of explicitly cancelling the problematic divergence free
component of the current density $(c/B) \bun \times (\nabla \cdot
\matrixtop{\pibf}_i)$ discussed in section~\ref{sect_vorticity_1}.
The simplification originates in the fact that the ion
distribution function is gyrophase independent in gyrokinetic
variables. Physically, ions are replaced by rings of charge,
thereby eliminating divergence free terms due to the particle
gyromotion. A simple example is the diamagnetic current $- \nabla
\times ( cp_{i\bot} \bun /B )$. The physical mechanism responsible
for the ion diamagnetic flow in the presence of a gradient of
density is sketched in figure~\ref{fig_diamagnetic}. In the
figure, the ion density is higher in the bottom half. Then, due to
the gyration, in the middle of the figure there are more ions
moving towards the right than towards the left, giving a
divergence free ion flow. Its divergence vanishes because the ions
just gyrate around the fixed guiding centers and there is no net
ion motion. In gyrokinetics, the gyromotion velocities are not
considered because the gyromotion is replaced by rings of charge,
removing the divergence free terms automatically and leaving only
the net gyrocenter drifts.

The conservation equations for particles and momentum are of order
$\delta_i n_e v_i/L$ and $\delta_i p_i/L$, respectively, and they
miss terms of order $\delta_i^2 n_e v_i/L$ and $\delta_i^2 p_i/L$.
In these equations, it is possible to study what happens for
wavelengths longer than the ion gyroradius, $k_\bot \rho_i \ll 1$.
Different terms will have different scalings in $k_\bot \rho_i$,
and these scalings will define which terms dominate at longer
wavelengths. For this reason, I will determine the scalings along
with the conservation equations. It is important to keep in mind
that there are missing terms of order $\delta_i^2 n_e v_i/L$ and
$\delta_i^2 p_i/L$, and any terms from a subsidiary expansion in
$k_\bot \rho_i$ are not meaningful in this limit. In particular, I
will show in section~\ref{sect_vorticityGK} that the terms that
determine the axisymmetric radial electric field are too small at
long wavelengths to be determined by this subsidiary expansion.

In this section, I present the general method to obtain
conservation equations from gyrokinetics. In subsection
\ref{sub_GKeq}, I derive the gyrokinetic equation in the physical
phase space variables $\br$, $E_0$, $\mu_0$ and $\varphi_0$, and I
write it in a conservative form that is convenient for deriving
moment equations. The details of the calculation are contained in
Appendix~\ref{app_transportGK}. In subsection \ref{sub_conservGK},
I derive the general moment equation for a quantity $G(\br, \bv,
t)$. I will apply this general equation to obtain particle and
momentum transport in subsections~\ref{sub_n} and \ref{sub_v},
respectively. The details of the calculations are in
Appendix~\ref{app_n_v}. In Appendix \ref{app_collisiontransport},
I show how to treat the effect of the finite gyroradius on
collisions.

\subsection{Gyrokinetic equation in physical phase space \label{sub_GKeq}}

\begin{figure}
\begin{center}
\includegraphics[width = \textwidth]{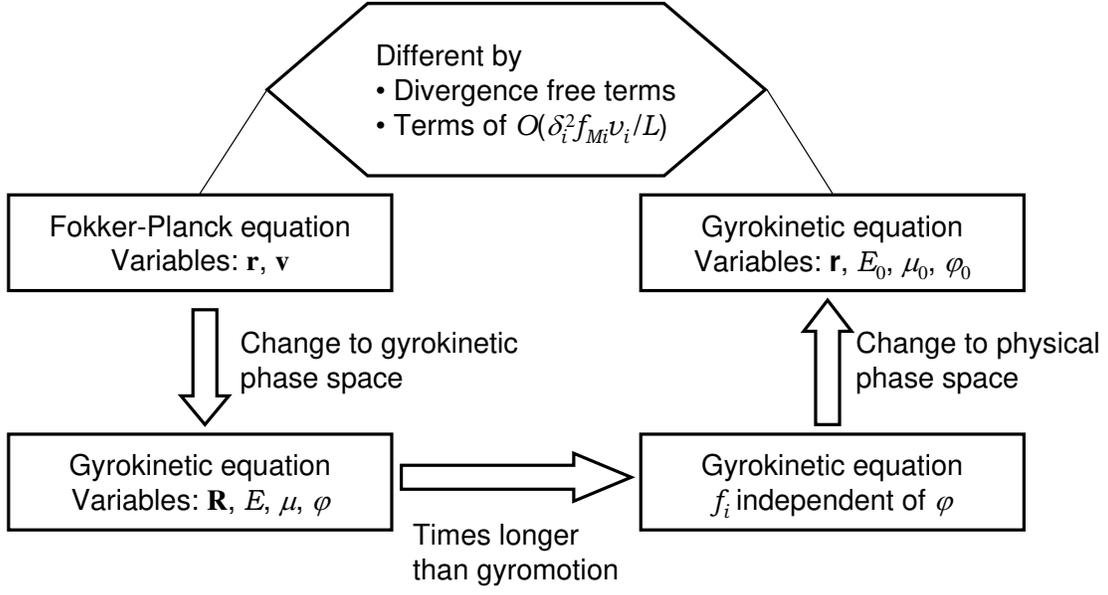}
\end{center}

\vspace{-3.5cm} \caption[Gyrokinetic equation in physical phase
space]{Gyrokinetic equation in physical phase space.}
\label{fig_FokkerPlanckgyro}
\end{figure}

The distribution function shows a simpler structure when written
in gyrokinetic variables, namely, it is independent of the
gyrophase except for the piece in \eq{fwig_sol} responsible for
classical collisional transport. The goal of this subsection is
writing the Fokker-Planck equation, $df_i/dt = C\{f_i\}$, in the
physical phase space variables $\br$, $E_0$, $\mu_0$ and
$\varphi_0$, while preserving the simple form obtained by
employing the gyrokinetic variables. The relation between the full
Fokker-Planck equation and the gyrokinetic equation written in
physical phase space variables is sketched in
figure~\ref{fig_FokkerPlanckgyro}. They differ in two aspects. On
the one hand, the gyrokinetic equation \eq{FP_final} misses terms
of order $\delta_i^2 f_{Mi} v_i/L$. On the other hand,
gyrokinetics is not only a change of variables, but it also
implies a time scale separation between the gyromotion and the
evolution of the slowly varying electrostatic potential. The ion
distribution function is then gyrophase independent in the
gyrokinetic phase space, i.e., the motion of the particles may be
replaced by drifting rings of charge. The advantage of this
distribution function is the lack of divergence free terms in the
moment equations constructed from it, as explained in the
introduction to this section.

I write the Fokker-Planck equation to order $\delta_i f_{Mi} v_i /
L$, the order to which the gyrokinetic equation is deduced, by
starting with
\begin{equation}
 \frac{df_i}{dt} \equiv \left. \frac{\partial f_i}{\partial t}
\right |_{\br, \bv} + \bv \cdot \nabla f_i + \mathbf{a} \cdot
\nabla_v f_i \simeq \left. \frac{\partial f_i}{\partial t} \right
|_{\bR, E, \mu, \varphi} + \dot{\bR} \cdot \nabla_\bR f_i +
\dot{E} \frac{\partial f_i}{\partial E} + \dot{\varphi}
\frac{\partial f_i}{\partial \varphi}, \label{VlasovGK}
\end{equation}

\noindent where $\mathbf{a} = - Ze \nabla \phi / M + \Omega_i (\bv
\times \bun)$ is the acceleration of particles and I have written
the Vlasov operator $d/dt$ in both $\br$, $\bv$ and gyrokinetic
variables. The term $\dot{\mu} (\partial f_i/\partial \mu)$ does
not appear in equation \eq{VlasovGK} because I assume that $f_i$
is a stationary Maxwellian to zeroth order and $\dot{\mu}$ is
small by definition of $\mu$. The derivative respect to the
gyrokinetic gyrophase $\varphi$ is small and related to the
collision operator by \eq{fwig_sol},
\begin{equation}
\dot{\varphi} \frac{\partial f_i}{\partial \varphi} = C\{ f_i \} -
\langle C \{ f_i \} \rangle. \label{varphidot_ddvarphi}
\end{equation}

\noindent The difference between time derivatives of $f_i$ can be
written as
\begin{equation}
\left. \frac{\partial f_i}{\partial t} \right |_{\bR, E, \mu,
\varphi} - \left. \frac{\partial f_i}{\partial t}  \right |_{\br,
\bv} \simeq - \frac{Ze}{M} \left. \frac{\partial \phiwig}{\partial
t} \right |_{\br, \bv} \frac{\partial f_i}{\partial E},
\label{diff_ddt}
\end{equation}

\noindent where I have employed \eq{correc_E_o1} and that
$\mathbf{B}$ is independent of time. Combining equations
\eq{VlasovGK}, \eq{varphidot_ddvarphi} and \eq{diff_ddt}, the
Fokker-Planck equation $df_i/dt = C\{f_i\}$ becomes, to
$O(\delta_i f_{Mi} v_i/L)$,
\begin{equation}
\left. \frac{\partial f_i}{\partial t} \right |_{\br, \bv} +
\left. \frac{\partial}{\partial t} \left ( \frac{Ze\phiwig}{T_i}
f_{Mi} \right )  \right |_{\br, \bv} + \dot{\bR} \cdot \nabla_\bR
f_i + \dot{E} \frac{\partial f_i}{\partial E} = \langle C\{f_i\}
\rangle. \label{eqGK_ini}
\end{equation}

\noindent Here, I have used that the time evolution of $\partial
f_i / \partial E \simeq (-M/T_i) f_{Mi}$ is slow.

It is necessary to rewrite equation \eq{eqGK_ini} in the variables
$\br$, $E_0$, $\mu_0$ and $\varphi_0$. Using equation \eq{fi1} and
considering that both the zeroth order distribution function and
the zeroth order potential are almost constant along magnetic
field lines, I can rewrite part of equation \eq{eqGK_ini} in terms
of the variables $\br$, $E_0$, $\mu_0$ and $\varphi_0$. The
details are in section~\ref{sectapp_drifts} of
Appendix~\ref{app_transportGK}, and the final result is
\begin{equation}
\dot{\bR} \cdot \nabla_\bR f_i + \dot{E} \frac{\partial
f_i}{\partial E} \simeq \dot{\bR} \cdot \nabla_{\bR_g} \br \cdot
\left ( \nablaave f_{ig} - \frac{Ze}{M} \nablaave \phiave
\frac{\partial f_{Mi}}{\partial E_0} \right ), \label{transf_phys}
\end{equation}

\noindent where $f_{ig}$ is missing the piece proportional to
$\phiwig$ [see equation \eq{fi1}]. The gradient $\nabla_{\bR_g}$
is taken with respect to $\bR_g$ holding $E_0$, $\mu_0$,
$\varphi_0$ and $t$ fixed, and $\nablaave$ is the gradient with
respect to $\br$ holding $E_0$, $\mu_0$, $\varphi_0$ and $t$
fixed. The quantity $\dot{\bR} \cdot \nabla_{\bR_g} \br$ is given
by
\begin{equation}
\dot{\bR} \cdot \nabla_{\bR_g} \br = v_{||0} \bun + \bv_{M0} +
\bv_{E0} + \tilde{\bv}_1, \label{Rdot_gradRr}
\end{equation}

\noindent with
\begin{eqnarray}
v_{||0} = v_{||} + \frac{v_\bot^2}{2\Omega_i} \bun \cdot \nabla
\times \bun + \frac{2v_{||}}{\Omega_i} \kappabf \cdot ( \bv \times
\bun ) - \frac{v_{||}}{2B\Omega_i} (\bv \times \bun) \cdot \nabla
B \nonumber \\ + \frac{v_{||}}{\Omega_i} \bv_\bot \cdot \nabla
\eun_2 \cdot \eun_1 + \frac{1}{4 \Omega_i} [ \bv_\bot ( \bv \times
\bun ) + ( \bv \times \bun ) \bv_\bot ]:\nabla \bun, \label{vpar0}
\end{eqnarray}
\begin{equation}
\bv_{M0} = \frac{\mu_0}{\Omega_i} \bun \times \nabla B +
\frac{v_{||}^2}{\Omega_i} \bun \times \kappabf, \label{v_M0}
\end{equation}
\begin{equation}
\bv_{E0} = - \frac{c}{B} \nablaave \phiave \times \bun
\label{v_E0}
\end{equation}

\noindent and using equation \eq{curl_vbot}
\begin{equation}
\tilde{\bv}_1 = \frac{v_{||}}{\Omega_i} \nablaave \times \bv_\bot.
\label{vtilde1}
\end{equation}

\noindent In equation \eq{transf_phys}, it is important to be
aware of higher order terms (like $\bv_{M0} \cdot \nablaave
f_{ig}$), in which the full distribution function, not just
$f_{Mi}$, must be retained. In these terms, the steep
perpendicular gradients make the higher order pieces of the
distribution function important [recall the orderings in
\eq{order_kbot}].

Equation \eq{transf_phys} can be written in conservative form,
more convenient for transport calculations. The details of this
calculation are in section~\ref{sectapp_conserv} of
Appendix~\ref{app_transportGK}, and the result is
\begin{eqnarray}
\dot{\bR} \cdot \nabla_\bR f_i + \dot{E} \frac{\partial
f_i}{\partial E} \simeq \frac{v_{||}}{B} \Bigg [ \nablaave \cdot
\left ( \frac{B}{v_{||}} f_{ig} \dot{\bR} \cdot \nabla_{\bR_g} \br
\right ) - \frac{\partial}{\partial \mu_0} \left ( f_{Mi}
\mathbf{B} \cdot \nablaave \mu_{10} \right )  \nonumber \\ -
\frac{\partial}{\partial \varphi_0} \left (  f_{Mi} \mathbf{B}
\cdot \nablaave \varphi_{10} \right ) - \frac{\partial}{\partial
E_0} \left ( \frac{B}{v_{||}} f_{Mi} \frac{Ze}{M} \dot{\bR} \cdot
\nabla_{\bR_g} \br \cdot \nablaave \phiave \right ) \Bigg ],
\label{cons_Rdot}
\end{eqnarray}

\noindent where $B/v_{||}$ is the Jacobian $\partial
(\bv)/\partial(E_0, \mu, \varphi_0)$, and the quantities
$\mu_{10}$ and $\varphi_{10}$ are the pieces of the first order
corrections $\mu_1$ and $\varphi_1$ that do not depend on the
potential. They are given by
\begin{equation}
\mu_{10} = \mu_1 - \frac{Ze\phiwig}{MB} \label{mu10}
\end{equation}

\noindent and
\begin{equation}
\varphi_{10} = \varphi_1 + \frac{Ze}{MB} \frac{\partial
\Phiwig}{\partial \mu}. \label{varphi10}
\end{equation}

\noindent The definitions of $\varphi_1$ and $\mu_1$ are in
equations \eq{correc_varphi_o1} and \eq{correc_mu_o1},
respectively.

Finally, substituting equation \eq{cons_Rdot} into equation
\eq{eqGK_ini}, I find
\begin{eqnarray}
\left. \frac{\partial f_{ig}}{\partial t} \right |_{\br, \bv} +
\frac{v_{||}}{B} \Bigg [ \nablaave \cdot \left ( \frac{B}{v_{||}}
f_{ig} \dot{\bR} \cdot \nabla_{\bR_g} \br \right ) -
\frac{\partial}{\partial \mu_0} \left ( f_{Mi} \mathbf{B} \cdot
\nablaave \mu_{10} \right ) \nonumber \\ -
\frac{\partial}{\partial \varphi_0} \left (  f_{Mi} \mathbf{B}
\cdot \nablaave \varphi_{10} \right ) - \frac{\partial}{\partial
E_0} \left ( \frac{B}{v_{||}} f_{Mi} \frac{Ze}{M} \dot{\bR} \cdot
\nabla_{\bR_g} \br \cdot \nablaave \phiave \right ) \Bigg ] =
\langle C \{ f_i \} \rangle. \label{eqGK_final}
\end{eqnarray}

\noindent Here, for $\phiave$ and $\phiwig$, it is enough to
consider the dependence on the lowest order variables, i.e.,
$\bR_g$, $\mu_0$ and $\varphi_0$ (the dependence of $\phiave$ and
$\phiwig$ on $E$ is weak).

\subsection{Transport of a general function $G(\br, \bv, t)$ at $k_\bot \rho_i \sim 1$ \label{sub_conservGK}}
Multiplying equation \eq{eqGK_final} by a function $G(\br, \bv,
t)$ and integrating over velocity space, I find the conservation
equation for that function $G$ to be
\begin{eqnarray}
 \frac{\partial}{\partial t} \left ( \int d^3v\, G f_{ig}
\right ) + \nabla \cdot \left [ \int d^3v\, f_{ig} \left (
\dot{\bR} \cdot \nabla_{\bR_g} \br \right ) G \right ] = \int
d^3v\, f_{ig} K \{G\} \nonumber \\ + \int d^3v\, G \langle C \{
f_i \} \rangle, \label{div_GKtransp}
\end{eqnarray}

\noindent with
\begin{eqnarray}
K\{G\} = \left. \frac{\partial G}{\partial t} \right |_{\br, \bv}
+ \dot{\bR} \cdot \nabla_{\bR_g} \br \cdot \left ( \nablaave G -
\frac{Ze}{M} \nablaave \phiave \frac{\partial G}{\partial E_0}
\right ) \nonumber \\ - v_{||} \bun \cdot \left ( \nablaave
\mu_{10} \frac{\partial G}{\partial \mu_0} + \nablaave
\varphi_{10} \frac{\partial G}{\partial \varphi_0} \right ).
\label{op_K}
\end{eqnarray}

\noindent In the next two subsections, I will use this formalism
to study the transport of particles and momentum at short
wavelengths.

\subsection{Transport of particles at $k_\bot \rho_i \sim 1$ \label{sub_n}}

Particle transport for electrons is easy to obtain since I only
need to consider $k_\bot \rho_e \ll 1$. In this limit, drift
kinetics is valid giving
\begin{equation}
\frac{\partial n_e}{\partial t} + \nabla \cdot \left ( n_e V_{e||}
\bun + n_e \bV_{ed} - \frac{cn_e}{B} \nabla \phi \times \bun
\right ) = 0, \label{ntransp_elect}
\end{equation}

\noindent with $n_e \bV_{e||} = \int d^3 v\, f_e v_{||}$ and
\begin{equation}
n_e \bV_{ed} = - \frac{cp_e}{eB} \bun \bun \cdot \nabla \times
\bun - \frac{cp_e}{eB^2} \bun \times \nabla B - \frac{cp_e}{eB}
\bun \times \kappabf.
\end{equation}

\noindent The same result may be deduced from equation
\eq{div_GKtransp} by neglecting electron pressure anisotropy terms
that are small by a factor $\sqrt{m/M}$ compared to the ion
pressure anisotropy. Obviously, the ion particle transport must be
exactly the same as for the electrons due to quasineutrality.
Nonetheless, we still must obtain the particle transport equation
for ions to be able to calculate the electric field by requiring
that both ions and electrons have the same density.

The conservation equation for ion particle number is given by
equation \eq{div_GKtransp} with $G = 1$. Employing
section~\ref{sectapp_ntransp} of Appendix~\ref{app_n_v}, it can be
written as
\begin{equation}
\frac{\partial}{\partial t} \left ( n_i - n_{ip} \right ) + \nabla
\cdot \left ( n_i V_{i||} \bun + n_i \bV_{igd} + n_i \bV_{iE} +
n_i \tilde{\bV}_i + n_i \bV_{iC} \right ) = 0,
\label{ntransp_krho}
\end{equation}

\noindent where $n_{ip}$ is the polarization density, defined in
\eq{polar_n}, the parallel flow is
\begin{equation}
n_i V_{i||} = \int d^3v\, f_i v_{||} = \int d^3v\, f_{ig} v_{||},
\label{n_parV}
\end{equation}

\noindent the term $n_i \tilde{\bV}_i$ is a perpendicular flow
that originates in finite gyroradius effects, given by
\begin{equation}
n_i \tilde{\bV}_i = \int d^3v\, f_{ig} \tilde{\bv}_1 = \int d^3v\,
f_{ig} \frac{v_{||}}{\Omega_i} \nablaave \times \bv_\bot,
\label{Vtilde_i}
\end{equation}

\noindent and the flows due to the $E \times B$ and magnetic
drifts are
\begin{equation}
n_i \bV_{iE} = \int d^3v\, f_{ig} \bv_{E0} = - \frac{c}{B} \int
d^3v\, f_{ig} \nablaave \phiave \times \bun \label{VE_i}
\end{equation}

\noindent and
\begin{equation}
n_i \bV_{igd} =  \frac{cp_{ig\bot}}{ZeB} \bun \bun \cdot \nabla
\times \bun + \frac{cp_{ig\bot}}{ZeB^2} \bun \times \nabla B +
\frac{cp_{ig||}}{ZeB} \bun \times \kappabf, \label{VM_i}
\end{equation}

\noindent with $p_{ig||} = \int d^3v\, f_{ig} Mv_{||}^2$ and
$p_{ig\bot} = \int d^3v\, f_{ig} Mv_\bot^2/2$. The collisional
flow $n_i \bV_{iC}$ is evaluated in
Appendix~\ref{app_collisiontransport} and is caused by ion-ion
collisions due to finite gyroradius effects. It is given by
\begin{equation}
n_i \bV_{iC} = - \frac{\gamma}{\Omega_i} \int d^3v\, \left (
\langle \Gammabf \rangle \times \bun - \frac{1}{v_\bot^2} \langle
\Gammabf \cdot \bv_\bot \rangle \bv \times \bun \right ),
\label{nViC}
\end{equation}

\noindent with $\gamma = 2\pi Z^4 e^4 \ln \Lambda/M^2$ and
\begin{equation}
\Gammabf = \int d^3v^\prime\, f_{Mi} f_{Mi}^\prime \nabla_g
\nabla_g g \cdot \left [ \nabla_v \left ( \frac{f_i}{f_{Mi}}
\right ) - \nabla_{v^\prime} \left (
\frac{f_i^\prime}{f_{Mi}^\prime} \right ) \right ].
\label{Gamma_c}
\end{equation}

\noindent Here, $f = f(\bv)$, $f^\prime = f(\bv^\prime)$,
$\mathbf{g} = \bv - \bv^\prime$, $g = |\mathbf{g}|$ and $\nabla_g
\nabla_g g = (g^2 \matI - \mathbf{g} \mathbf{g})/g^3$.

In the presence of potential structure on the order of the ion
gyroradius, the contributions to $n_i \tilde{\bV}_i$ no longer
average to zero in a gyration since they can add coherently. In
the integration $n_i \tilde{\bV}_i = \int d^3v\, f_{ig}
(v_{||}/\Omega_i) \nablaave \times \bv_\bot$ only two terms
contribute to its divergence so that $\nabla \cdot (n_i
\tilde{\bV}_i) = \nabla \cdot (n_i \tilde{\bV}_{i0})$ with
\begin{equation}
n_i \tilde{\bV}_{i0} = \int d^3v\, f_{ig} \left [
\frac{v_{||}}{\Omega_i} (\bv \times \bun) \cdot \nabla \bun +
\frac{v_{||}\bun \cdot \nabla B}{2B \Omega_i} \bv \times \bun
\right ]. \label{flowtilde_v2}
\end{equation}

\begin{figure}
\begin{center}
\includegraphics[width = \textwidth]{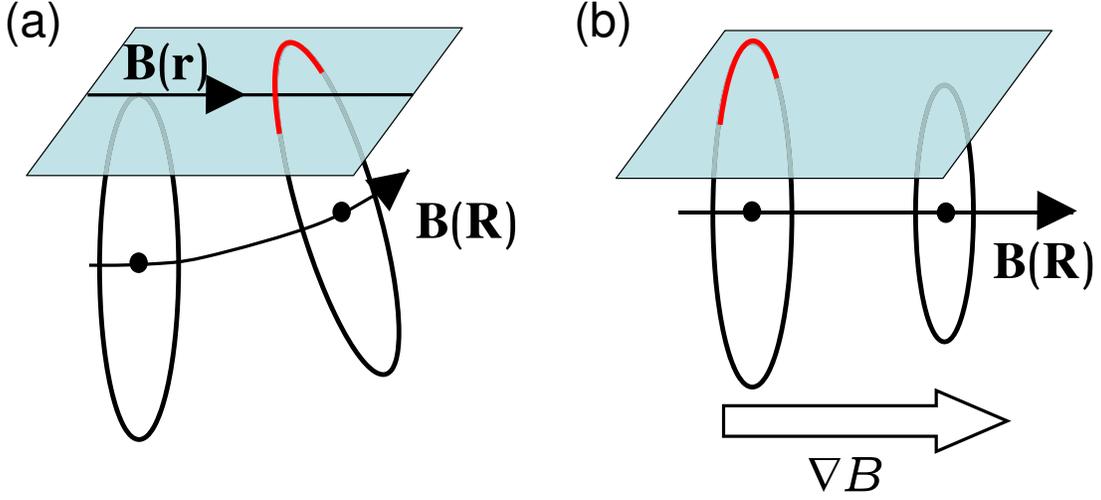}
\end{center}

\vspace{-5cm} \caption[Finite gyroradius effects in $n_i
\tilde{\bV}_i$]{Finite gyroradius effects in $n_i \tilde{\bV}_i$.
(a) The magnetic field line along which the gyrocenter lies is the
line that guides the parallel motion, making it possible for
parallel motion to transport particles across magnetic field
lines. (b) As the size of the gyromotion changes, a particle that
spent time on both sides of the blue plane is now only on one of
its sides, leading to an effective particle transport. }
\label{fig_vtilde1}
\end{figure}

\noindent In section~\ref{sectapp_ntransp} of
Appendix~\ref{app_n_v}, I prove that all the other terms in
$\tilde{\bv}_1$ can be neglected. The physical origin of $n_i
\tilde{\bV}_{i0}$ is sketched in figure~\ref{fig_vtilde1}. The
drift $(v_{||}/\Omega_i) (\bv \times \bun) \cdot \nabla \bun$ is
presented in figure~\ref{fig_vtilde1}(a), where there is
difference between the direction of the magnetic field at the
gyrocenter $\bun (\bR)$, and the direction of the magnetic field
at the real position of the particle $\bun(\br)$. Due to this
difference, the parallel motion of the gyrocenter drives part of
the gyromotion, plotted in red in the figure, across the blue
plane. Notice that the magnetic field line $\bB(\br)$ lies in the
plane, leading to the ``paradoxical" parallel motion across
magnetic field lines. The other term in $n_i \tilde{\bV}_{i0}$,
$(v_{||}\bun \cdot \nabla B/2B \Omega_i) \bv \times \bun$, is
explained in figure~\ref{fig_vtilde1}(b). Here, the change in the
size of the gyroradius due to the change in magnitude of the
magnetic field, $dB/dt = v_{||} \bun \cdot \nabla B$, drives part
of the gyromotion, plotted in red, across the blue plane.

\begin{table}
\begin{center}
\renewcommand{\arraystretch}{1.5}
\begin{tabular}{|c|c|}
\hline Term & Order of magnitude \\
\hline $n_{ip}$ & $\delta_i k_\bot \rho_i n_e$ \\
$\nabla \cdot (n_i V_{i||} \bun + n_i \bV_{igd} + n_i \bV_{iE})$ & $\delta_i n_e v_i/L$ \\
$\nabla \cdot (n_i \tilde{\bV}_i)$ & $\delta_i (k_\bot \rho_i)^2 n_e v_i/L$ \\
$\nabla \cdot (n_i \bV_{iC})$ & $\delta_i (k_\bot \rho_i)^2 n_e \nu_{ii}$ \\
\hline
\end{tabular}
\end{center}

\caption[Order of magnitude estimates for particle conservation
equation \eq{ntransp_krho}]{Order of magnitude estimates for ion
particle conservation equation \eq{ntransp_krho}.}
\label{table_densitycserv}
\end{table}

In section~\ref{sect_vorticityGK}, I will obtain a vorticity
equation for $\phi$ by imposing $Zn_i = n_e$. Equations
\eq{ntransp_elect} and \eq{ntransp_krho} will provide the time
evolution of $Zn_i - n_e$. It will be useful to know the size of
the different terms in equation \eq{ntransp_krho}. I summarize the
estimates of order of magnitude in table~\ref{table_densitycserv}.
These estimates include the scaling with $k_\bot \rho_i$ for long
wavelengths. The wavenumber $\mathbf{k}_\bot$ is the overall
perpendicular wavenumber, i.e., given $\mathbf{k}_\bot$, the
corresponding Fourier component for nonlinear terms like $A(\br)
\times B(\br)$ is $\int d^2 k_\bot^\prime \tilde{A} (
\mathbf{k}_\bot^\prime ) \times \tilde{B} ( \mathbf{k}_\bot -
\mathbf{k}_\bot^\prime )$, with $\tilde{A}$ and $\tilde{B}$ the
Fourier transforms of functions $A$ and $B$. The divergence of the
drift flow, $\nabla \cdot (n_i \bV_{igd} + n_i \bV_{iE})$, is of
order $\delta_i n_e v_i / L$ since $\nablaave f_{Mi} \sim
\nablaave f_{ik}$ [recall \eq{order_kbot}]. The flow $n_i
\tilde{\bV}_i$ is of order $\delta_i^2 k_\bot \rho_i n_e v_i$
because, for $k_\bot \rho_i \ll 1$, the gyrophase dependent piece
of $f_{ig}$, similar to \eq{fwig_krho}, is even in $v_{||}$ to
zeroth order, making the integral vanish. Its divergence, $\nabla
\cdot ( n_i \tilde{\bV}_i )$, is then of order $\delta_i (k_\bot
\rho_i)^2 n_e v_i / L$. The divergence of the collisional flow,
$\nabla \cdot (n_i \bV_{iC})$, is of order $\delta_i (k_\bot
\rho_i)^2 \nu_{ii} n_e$, as proven in
Appendix~\ref{app_collisiontransport}. The polarization density
$n_{ip}$ is of order $\delta_i k_\bot \rho_i n_e$, as shown in
\eq{nip_krhosmall}. This result means that for long wavelengths,
the polarization term becomes unimportant. Therefore, at long
wavelengths only the balance between the time evolution of the
density, the parallel flow, and the magnetic and $E\times B$
drifts matter.

\subsection{Transport of momentum at $k_\bot \rho_i \sim 1$ \label{sub_v}}

From electron momentum conservation, I will only need the parallel
component, given by
\begin{equation}
\bun \cdot \nabla p_{e||} + (p_{e||} - p_{e\bot}) \nabla \cdot
\bun = e n_e \bun \cdot \nabla \phi + F_{ei||},
\label{vtransp_elect_complex}
\end{equation}

\noindent where $F_{ei||} = m \int d^3v\, v_{||} C_{ei} \{f_e\}$
is the collisional parallel momentum exchange. The time derivative
of $n_e m \bV_e$ and the viscosity have been neglected because
they are small by a factor of $m/M$. In this equation, there are
terms of two different orders of magnitude. The dominant terms are
$\bun \cdot \nabla p_{e||} \simeq \bun \cdot \nabla p_e$ and $en_e
\bun \cdot \nabla \phi$, both of order $O(\delta_i p_e/L)$ for
turbulent fluctuations at $k_\bot \rho_i \sim 1$. The friction
force $F_{ei||}$ and the terms that contain the pressure
anisotropy $p_{e||} - p_{e\bot} \sim \delta_e p_e$ are an order
$\sqrt{m/M}$ smaller than the dominant terms. However, these
smaller terms are crucial because they provide the non-adiabatic
behavior and hence allow radial transport of particles. In the
vorticity equation \eq{vorticity}, the non-adiabatic electron
response is kept in the integral $\int d^3v\, f_e v_{||}$ in
$J_{||}$. Thus, for most purposes, equation
\eq{vtransp_elect_complex} can be simplified to
\begin{equation}
\bun \cdot \nabla p_e = e n_e \bun \cdot \nabla \phi.
\label{vtransp_elect}
\end{equation}

\noindent Equation \eq{vtransp_elect_complex} can also be
recovered by using equation \eq{div_GKtransp} and neglecting terms
small by $m/M$.

For the ions, using equation \eq{div_GKtransp} with $G = \bv$ and
employing section~\ref{sectapp_vtransp} of Appendix~\ref{app_n_v},
I find the momentum conservation equation
\begin{eqnarray}
\frac{\partial}{\partial t} \left ( n_i M \bV_{ig} \right ) + \bun
[ \bun \cdot \nabla p_{ig||} + (p_{ig||} - p_{ig\bot}) \nabla
\cdot \bun + \nabla \cdot \pibf_{ig||} ] + \nabla \cdot
\matrixtop{\pibf}_{ig\times} = \nonumber \\ - Zen_i \bun \bun
\cdot \nabla \phi + \tilde{F}_{iE} \bun + \mathbf{F}_{iB} +
\mathbf{F}_{iC}, \label{vtransp_krho}
\end{eqnarray}

\noindent where $n_i \bV_{ig} = \int d^3v\, f_{ig} \bv$ is the
average gyrocenter velocity; the vector $\pibf_{ig||}$ is the
parallel momentum transported by the drifts and is given by
\begin{equation}
\pibf_{ig||} = \int d^3v\, f_{ig} (\bv_{M0} + \bv_{E0} +
\tilde{\bv}_1) M v_{||}; \label{pig_par}
\end{equation}

\noindent the tensor $\matrixtop{\pibf}_{ig\times}$ gives the
transport of perpendicular momentum by the parallel velocity and
the drifts,
\begin{equation}
\matrixtop{\pibf}_{ig\times} = \int d^3v\, f_{ig} ( v_{||} \bun +
\bv_{M0} + \bv_{E0} + \tilde{\bv}_1 ) M \bv_\bot; \label{pig_perp}
\end{equation}

\noindent the vector $\tilde{F}_{iE} \bun$ is a correction due to
the short wavelengths of the electric field with
\begin{equation}
\tilde{F}_{iE} = Ze \int d^3v\, f_{Mi} ( \bun + \Omega_i^{-1}
\nablaave \times \bv_\bot ) \cdot \nablaave \phiwig;
\label{force_E}
\end{equation}

\noindent the vector $\mathbf{F}_{iB}$ contains the effect on the
gyromotion of the variation in the magnetic field and is given by
\begin{equation}
\mathbf{F}_{iB} = \int d^3v\, M f_{ig} v_{||} \bun \cdot \nablaave
\bv_\bot + \int d^3v\, \frac{Mc}{B} f_{Mi} (\nablaave \phiwig
\times \bun) \cdot \nablaave \bv_\bot; \label{force_B}
\end{equation}

\noindent and the finite gyroradius effects on collisions are
included in
\begin{eqnarray}
 \mathbf{F}_{iC} \equiv M \int d^3v \bv \langle C \{ f_i \}
\rangle = - M \gamma \int d^3 v\, \left ( \langle \Gammabf \rangle
\cdot \bun \bun + \frac{1}{v_\bot^2} \langle \Gammabf \cdot
\bv_\bot \rangle \bv_\bot \right ) \nonumber \\ + \nabla \cdot
\left [ \frac{M \gamma}{\Omega_i} \int d^3 v\, \left ( \langle
\Gammabf \rangle \times \bun - \frac{1}{v_\bot^2} \langle \Gammabf
\cdot \bv_\bot \rangle \bv \times \bun \right ) \bv \right ].
\label{FiC}
\end{eqnarray}

\noindent The force $\tilde{F}_{iE}$ originates in the change in
the parallel velocity magnitude due to potential structures on the
size of the ion gyroradius. The force $\mathbf{F}_{iB}$ accounts
for the change in perpendicular velocity due to variation in the
magnetic field that the particle feels during its motion.
Interestingly, the parallel component of equation
\eq{vtransp_krho} is simply
\begin{eqnarray}
\frac{\partial}{\partial t} (n_i M V_{i||}) + \bun \cdot \nabla
p_{ig||} + (p_{ig||} - p_{ig\bot}) \nabla \cdot \bun + \nabla
\cdot \pibf_{ig||} = - Zen_i \bun \cdot \nabla \phi \nonumber \\ +
\tilde{F}_{iE} + \mathbf{F}_{iC} \cdot \bun.
\label{par_vtransp_krho}
\end{eqnarray}

\noindent The parallel components of $\nabla \cdot
\matrixtop{\pibf}_{ig\times}$ and $\mathbf{F}_{iB}$ cancel each
other, as proven in section~\ref{sectapp_vtransp} of
Appendix~\ref{app_n_v}.

\begin{table}
\begin{center}
\renewcommand{\arraystretch}{1.5}
\begin{tabular}{|c|c|}
\hline Term & Order of magnitude \\
\hline $n_i M \bV_{ig}$ & $\delta_i n_e M v_i$ \\
$\bun \cdot \nabla p_{ig||}$, $(p_{ig||} - p_{ig\bot}) \nabla \cdot \bun$, $Zen_i \bun \cdot \nabla \phi$ & $\delta_i p_i/L$ \\
$\nabla \cdot \pibf_{ig||}$, $\nabla \cdot \matrixtop{\pibf}_{ig\times}$, $\tilde{F}_{iE}$, $\mathbf{F}_{iB}$ & $\delta_i k_\bot \rho_i p_i/L$ \\
$\mathbf{F}_{iC}$ & $\delta_i k_\bot \rho_i \nu_{ii} n_e M v_i$ \\
\hline
\end{tabular}
\end{center}

\caption[Order of magnitude estimates for momentum equation
\eq{vtransp_krho}]{Order of magnitude estimates for ion momentum
conservation equation \eq{vtransp_krho}.}
\label{table_momentumcserv}
\end{table}

Equation \eq{vtransp_krho} will be used in
section~\ref{sect_vorticityGK} to get one of the forms of the
vorticity equation. Thus, it is useful to estimate the size of the
different terms in it. The estimates of order of magnitude are
summarized in table~\ref{table_momentumcserv}. The pressure terms,
$\bun \cdot \nabla p_{ig||} + (p_{ig||} - p_{ig\bot}) \nabla \cdot
\bun$, and the electric field term, $Zen_i \bun \cdot \nabla
\phi$, are $O(\delta_i p_i/L)$. The terms $\tilde{F}_{iE}$ and
$\mathbf{F}_{iB}$ are of order $\delta_i k_\bot \rho_i p_i / L$.
These estimates are obvious for the integrals $(Ze/\Omega_i) \int
d^3v\, f_{Mi} (\nablaave \times \bv_\bot) \cdot \nablaave \phiwig
\sim (Mc/B) \int d^3v\, f_{Mi} (\nablaave \phiwig \times \bun)
\cdot \nablaave \bv_\bot$ since $\nablaave \phiwig \sim k_\bot
\rho_i T_e/eL$ due to $e\phiwig/T_e \sim \delta_i$. The integral
$Ze \int d^3v\, f_{Mi} \bun \cdot \nablaave \phiwig$ would seem to
be of order $\delta_i p_i/L$ since $\bun \cdot \nablaave \phiwig
\sim \delta_i T_e/eL$ but it is an order $k_\bot \rho_i$ smaller
because the integral of $\phiwig$ in the gyrophase $\varphi_0$
vanishes to zeroth order. The integral $M \int d^3v f_{ig} v_{||}
\bun \cdot \nablaave \bv_\bot$ is of order $\delta_i k_\bot \rho_i
p_i/L$ because the leading order gyrophase dependent piece of
$f_{ig}$ is even in $v_{||}$ [recall \eq{fwig_krho}]. The
collisional force, $\mathbf{F}_{iC}$ is order $\delta_i k_\bot
\rho_i \nu_{ii} n_e M v_i$, as proven in
Appendix~\ref{app_collisiontransport}. The vector $\pibf_{ig||}$
is $O(\delta_i^2 p_i)$ because $f_{ig}$ is even in $v_{||}$ and
$\bv_\bot$ up to order $\delta_i f_{Mi}$. The matrix
$\matrixtop{\pibf}_{ig\times}$ has three different pieces: the
integral $\int d^3v\, f_{ig} (\bv_{M0} + \tilde{\bv}_1) M
\bv_\bot$ also of order $\delta_i^2 p_i$, the integral $\int
d^3v\, f_{ig} M v_{||} \bun \bv_\bot$ of order $\delta_i k_\bot
\rho_i p_i$, and the integral $\int d^3v\, f_{ig} M \bv_{E0}
\bv_\bot$ of order $\delta_i p_i$. The integral $\int d^3v\,
f_{ig} Mv_{||} \bun \bv_\bot$ is of order $\delta_i k_\bot \rho_i
p_i$ because the leading order gyrophase dependent piece of
$f_{ig}$ is even in $v_{||}$ as in \eq{fwig_krho}. On the other
hand, the size of the integral $\int d^3v\, f_{ig} M \bv_{E0}
\bv_\bot$ is estimated by employing $\bv_{E0} = - (c/B) \nablaave
\phiave \times \bun = - (c/B) \nabla \phi \times \bun + (c/B)
\nablaave \phiwig \times \bun$. Then, I find that $(Mc/B) \int
d^3v\, f_{ig} (\nabla \phi \times \bun) \bv_\bot \sim \delta_i^2
p_i$ since $f_{ig}$ is gyrophase dependent at $O(\delta_i
f_{Mi})$, and $(Mc/B) \int d^3v\, f_{ig} (\nablaave \phiwig \times
\bun) \bv_\bot \ssim \delta_i p_i$. It is difficult to refine this
last estimate because it is a nonlinear term and short wavelength
pieces of $f_{ig}$ and $\phiwig$ can beat to give a long
wavelength result. The divergences of $\pibf_{ig||}$ and
$\matrixtop{\pibf}_{ig\times}$ are both $O(\delta_i k_\bot \rho_i
p_i/L)$. Importantly, the divergence of
$\matrixtop{\pibf}_{ig\times} \ssim \delta_i p_i$ is not of order
$k_\bot \rho_i p_i/L$ but of order $\delta_i k_\bot \rho_i p_i/L$.
The divergence of $\int d^3v\, f_{ig} M v_{||} \bun \bv_\bot$ is
of order $\delta_i k_\bot \rho_i p_i/L$ because it only contains a
parallel gradient, and $\nabla \cdot [ (Mc/B) \int d^3v\, f_{ig}
(\nablaave \phiwig \times \bun) \bv_\bot ] = - \nabla \cdot [ Mc
\int dE_0\,d\mu_0\,d\varphi_0\, \phiwig\, \nablaave \times (\bun
f_{ig} \bv_\bot / v_{||}) ] \sim \delta_i k_\bot \rho_i p_i/L$,
where I have used $d^3v = (B/v_{||}) dE_0\,d\mu_0\,d\varphi_0$ and
$\nablaave \cdot [ \nablaave \times (\ldots) ] = 0$ [since
$(f_{ig}/v_{||}) (\nablaave \phiwig \times \bun) \bv_\bot =
\nablaave \times ( \bun \phiwig f_{ig} \bv_\bot / v_{||} ) -
\phiwig\, \nablaave \times (\bun f_{ig} \bv_\bot / v_{||} )$].

\section{Vorticity equation for gyrokinetics
\label{sect_vorticityGK}}
In section \ref{sect_vorticity_1}, I showed that the term that
contains the ion viscosity in the vorticity equation
\eq{vorticity} seems to dominate at short wavelengths. However, in
reality it is smaller than $\nabla \cdot \bJ_d$, as I demonstrate
in this section. As a result, the viscosity must be evaluated
carefully; otherwise, spurious terms may appear in numerical
simulations. Here, I propose two different vorticity equations
that avoid this numerical problem and are valid for short
wavelengths. Long wavelength, transport time scale phenomena, like
the self-consistent calculation of the radial electric field, can
be included, but this will be the subject of
chapter~\ref{chap_angularmomentum}.

The vorticity equation \eq{vorticity} provides a way to temporally
evolve the electric field perpendicular to the magnetic field.
However, the parallel electric field strongly depends on the
parallel electron dynamics, hidden in the parallel current
$J_{||}$ in equation \eq{vorticity}. Fortunately, it is enough to
use the integral $J_{||} = Ze \int d^3v\, v_{||} f_{ig} - e \int
d^3v\, v_{||} f_e$ for the parallel current since $J_{||}$ does
not alter the higher order calculation of the radial electric
field determined by equation \eq{toro_angmom}. In several codes
\cite{candy03, kotschen95}, the electron and ion distribution
functions are solved implicitly in the potential. These implicit
solutions are then substituted in $J_{||}$ to find the potential
in the next time step from the vorticity equation.

In subsection~\ref{sub_vortquasi}, a vorticity equation is derived
directly from gyrokinetic quasi\-neutrality. The advantage of this
form is its close relation to previous algorithms, but it differs
greatly from the general vorticity equation \eq{vorticity}. In
subsection~\ref{sub_vortvort}, I present a modified vorticity
equation that has more similarities with equation \eq{vorticity}.
I ensure that both forms are equivalent and satisfy the desired
condition at long wavelengths, namely, that they provide a fixed
toroidal velocity.

\subsection{Vorticity from quasineutrality \label{sub_vortquasi}}

The first version of the vorticity equation is obtained by taking
the time derivative of the gyrokinetic quasineutrality ($Zn_i =
n_e$). In other words, I find the time evolution of ion and
electron density and ensure that its difference is constant in
time. This is equivalent to subtracting equation
\eq{ntransp_elect} from $Z$ times \eq{ntransp_krho} to obtain
\begin{equation}
\frac{\partial}{\partial t} (Ze n_{ip}) = \nabla \cdot \left (
J_{||} \bun + \bJ_{gd} + \tilde{\bJ}_i + Zen_i\tilde{\bV}_i +
Zen_i \bV_{iC} \right ), \label{vort_type1}
\end{equation}

\noindent where $\tilde{\bJ}_i$ is the polarization current
\begin{eqnarray}
 \tilde{\bJ}_i \equiv  \frac{Zec}{B} \left ( \nabla \phi \times
\bun \int d^3v\, f_i - \int d^3v\, f_{ig} \nablaave \phiave \times
\bun \right ) = \nonumber \\ \frac{Zec}{B} \left ( \int d^3v\,
f_{ig} \nablaave \phiwig \times \bun - \nabla \phi \times \bun
\int d^3v\, \frac{Ze\phiwig}{T_i} f_{Mi} \right )
\label{Jtilde_i1}
\end{eqnarray}

\noindent and $\bJ_{gd}$ is the drift current
\begin{equation}
 \bJ_{gd} \equiv Zen_i \bV_{igd} - en_e \bV_{ed} =
\frac{cp_{g||}}{B} \bun \bun \cdot \nabla \times \bun +
\frac{cp_{g\bot}}{B^2} \bun \times \nabla B + \frac{cp_{g||}}{B}
\bun \times \kappabf, \label{Jg_drift_M}
\end{equation}

\noindent with $p_{g||} = p_{ig||} + p_e$ and $p_{g\bot} =
p_{ig\bot} + p_e$. Here, to write the second form of
$\tilde{\bJ}_i$, I use $f_i - f_{ig} = - (Ze\phiwig/T_i) f_{Mi}$,
given in \eq{fi1}. In equation \eq{vort_type1}, the ion
polarization density, $n_{ip} = - \int d^3v\, (Ze\phiwig/T_i)
f_{Mi}$ is advanced in time, and the electric field is solved from
$n_{ip}$.

It is necessary to check if equation \eq{vort_type1} satisfies the
right conditions at long wavelengths. In the present form, though,
it is a tedious task. To perform this check, I will use the much
more convenient form in subsection~\ref{sub_vortvort}, that I will
prove is equivalent.

\begin{table}
\begin{center}
\renewcommand{\arraystretch}{1.5}
\begin{tabular}{|c|c|}
\hline Term & Order of magnitude \\
\hline $Zen_{ip}$ & $\delta_i k_\bot \rho_i e n_e$ \\
$\nabla \cdot ( J_{||} \bun + \bJ_{gd} )$ & $\delta_i e n_e v_i/L$ \\
$\nabla \cdot (\tilde{\bJ}_i + Zen_i \tilde{\bV}_i)$ & $\delta_i (k_\bot \rho_i)^2 e n_e v_i/L$ \\
$\nabla \cdot (Ze n_i \bV_{iC})$ & $\delta_i (k_\bot \rho_i)^2 e n_e \nu_{ii}$ \\
\hline
\end{tabular}
\end{center}

\caption[Order of magnitude estimates for vorticity equation
\eq{vort_type1}]{Order of magnitude estimates for vorticity
equation \eq{vort_type1}.} \label{table_vort_type1}
\end{table}

Finally, I give estimates for the size of all the terms in
\eq{vort_type1} in table~\ref{table_vort_type1}. These estimates
will be useful in subsection~\ref{sub_vortvort} to study the
behavior of the toroidal angular momentum for $k_\bot \rho_i \ll
1$. The size of most of the terms in equation \eq{vort_type1} can
be obtained from the estimates given in
table~\ref{table_densitycserv}, giving $\nabla \cdot \bJ_{gd} \sim
\delta_i e n_e v_i/L$, $\nabla \cdot (Zen_i \tilde{\bV}_i) \sim
\delta_i (k_\bot \rho_i)^2 e n_e v_i/L$ and $\nabla \cdot (Zen_i
\bV_{iC}) \sim \delta_i (k_\bot \rho_i)^2 \nu_{ii} e n_e$. The
size of $\nabla \cdot \tilde{\bJ}_i$ requires more work. Even
though a cancellation between the drift kinetic $E\times B$ flow,
$(cn_i/B) \nabla \phi \times \bun$, and the corresponding
gyrokinetic flow, $(c/B) \int d^3v\, f_{ig} \nablaave \phiave
\times \bun$, is expected, due to the nonlinear character of these
terms, where the short wavelength components of $f_{ig}$ and
$\phi$ can beat to give a long wavelength term, I can only give a
bound for the size of $\tilde{\bJ}_i$. Given the definition of
$\tilde{\bJ}_i$ in equation \eq{Jtilde_i1}, its size is bounded by
$|\tilde{\bJ}_i| \ssim \delta_i e n_e v_i$. Then, it would seem
that its divergence must be $|\nabla \cdot \tilde{\bJ}_i| \ssim
k_\bot \rho_i e n_e v_i / L$, but the lowest order terms contain
$\nabla \times \nabla \phi = 0$ and $\nablaave \times \nablaave
\phiave = 0$, leading to $|\nabla \cdot \tilde{\bJ}_i| \ssim
\delta_i e n_e v_i / L$. To refine this bound, I use that $\nabla
\cdot \tilde{\bJ}_i = \int dE_0\,d\mu_0\,d\varphi_0\, \nablaave
\cdot \tilde{\mathbf{j}}_i$, where $\nablaave \cdot
\tilde{\mathbf{j}}_i$ is found from equation \eq{Jtilde_i1} to be
\begin{eqnarray}
\nablaave \cdot \tilde{\mathbf{j}}_i \equiv \nablaave \cdot \left
[ \frac{Zec}{v_{||}} \left ( f_i \nabla \phi \times \bun - f_{ig}
\nablaave \phiave \times \bun \right ) \right ] = \nonumber
\\ - \nablaave \cdot \left [ \phiwig \, \nablaave \times \left (
\bun \frac{Zec}{v_{||}} f_{ig} \right ) + \frac{Zec}{v_{||}}
\frac{Ze\phiwig}{T_i} f_{Mi} \nabla \phi \times \bun \right ] \sim
\delta_i k_\bot \rho_i e f_{Mi} B/L. \label{small_Jtilde_i}
\end{eqnarray}

\noindent I have used $d^3v\, = (B/v_{||})
dE_0\,d\mu_0\,d\varphi_0$ and $\nablaave \cdot ( v_{||}^{-1}
f_{ig} \nablaave \phiwig \times \bun ) = - \nablaave \cdot
[\phiwig \, \nablaave \times ( \bun  v_{||}^{-1} f_{ig} ) ]$ to
find this result. The second form of \eq{small_Jtilde_i} is useful
to estimate the size of $\nablaave \cdot \tilde{\mathbf{j}}_i$
because I can use $e\phiwig/T_e \sim \delta_i$ and $\nablaave
f_{ig} \sim f_{Mi} / L$ to find $\nablaave \cdot
\tilde{\mathbf{j}}_i \sim \delta_i k_\bot \rho_i e f_{Mi} B/L$. In
$\nablaave \cdot \tilde{\mathbf{j}}_i$, given in
\eq{small_Jtilde_i}, there are short wavelength components of
$f_{ig}$ and $\phi$ that beat together to give a long wavelength
component. These short wavelength components of $f_{ig} \equiv f_i
(\bR_g, E_0, \mu_0, t)$ and $\phiwig (\bR_g, \mu_0, \varphi_0, t)$
cannot be expanded around $\br$, but the long wavelength component
of $\nablaave \cdot \tilde{\mathbf{j}}_i$ can be expanded as a
whole to find
\begin{equation}
\nablaave \cdot \tilde{\mathbf{j}}_i (\bR_g, E_0, \mu_0,
\varphi_0) \simeq \nablaave \cdot \tilde{\mathbf{j}}_i (\br, E_0,
\mu_0, \varphi_0) + \frac{1}{\Omega_i} (\bv \times \bun) \cdot
\nablaave (\nablaave \cdot \tilde{\mathbf{j}}_i).
\label{beating_lw}
\end{equation}

\noindent The difference between $\nablaave \cdot
\tilde{\mathbf{j}}_i(\br, E_0, \mu_0, \varphi_0)$ and $\nablaave
\cdot \tilde{\mathbf{j}}_i(\bR_g, E_0, \mu_0, \varphi_0)$ is
negligible in the higher order term. The interesting property of
equation \eq{beating_lw} is that the velocity integral of the
zeroth order term $\nablaave \cdot \tilde{\mathbf{j}}_i (\br, E_0,
\mu_0, \varphi_0)$ can be done because the gyrophase dependence in
$\bR_g$ has disappeared. Employing the second form of equation
\eq{small_Jtilde_i} for $\nablaave \cdot \tilde{\mathbf{j}}_i
(\br, E_0, \mu_0, \varphi_0)$, I find $\int
dE_0\,d\mu_0\,d\varphi_0\, \nablaave \cdot \tilde{\mathbf{j}}_i
(\br, E_0, \mu_0, \varphi_0) = - \nabla \cdot [ (Zec/B) \int
d^3v\, (Ze\phiwig/T_i) f_{Mi} \nablaave \phiwig \times \bun ] \sim
\delta_i^2 k_\bot \rho_i e n_e v_i/L$ since $\phiwig \nablaave
\phiwig = \nablaave (\phiwig^2/2)$. This result is negligible
compared with the term $\Omega_i^{-1} (\bv \times \bun) \cdot
\nablaave (\nablaave \cdot \tilde{\mathbf{j}}_i)$ in
\eq{beating_lw}, which gives $\nabla \cdot \tilde{\bJ}_i \sim
\delta_i (k_\bot \rho_i)^2 e n_e v_i/L$ since $\nablaave \cdot
\tilde{\mathbf{j}}_i \sim \delta_i k_\bot \rho_i e f_{Mi} B/L$.
Using this result, I find that the $k_\bot \rho_i \ll 1$ limit of
$\nabla \cdot \tilde{\bJ}_i$ is
\begin{equation}
\nabla \cdot \tilde{\bJ}_i \simeq  \nabla \cdot \left [ \int
dE_0\,d\mu_0\,d\varphi_0 \frac{1}{\Omega_i} (\bv \times \bun)
\nablaave \cdot \tilde{\mathbf{j}}_i \right ],
\label{div_Jtilde_i}
\end{equation}

\noindent where I use that $\Omega_i^{-1} (\bv \times \bun) \cdot
\nablaave (\nablaave \cdot \tilde{\mathbf{j}}_i) \simeq \nablaave
\cdot [ \Omega_i^{-1} (\bv \times \bun) (\nablaave \cdot
\tilde{\mathbf{j}}_i) ]$ because the gradient of $\Omega_i^{-1}
\bv \times \bun$ is of order $1/L$ and the gradient of $\nablaave
\cdot \tilde{\mathbf{j}}_i$ is of order $k_\bot$.

\subsection{Vorticity from moment description \label{sub_vortvort}}
Equation \eq{vort_type1} has the advantage of having a direct
relation with the gyrokinetic quasineutrality. However, its
relation with the full vorticity equation \eq{vorticity} and the
evolution of toroidal angular momentum is not explicit. For those
reasons, I next derive an alternative vorticity equation.

I define a new useful function
\begin{equation}
\varpi_G = \nabla \cdot \left ( \frac{Ze}{\Omega_i} \int d^3v\,
f_{ig} \bv \times \bun \right ) - \int d^3v\,
\frac{Z^2e^2\phiwig}{T_i} f_{Mi} \label{vortGKdef}
\end{equation}

\noindent that I will call gyrokinetic ``vorticity" because, for
$k_\bot \rho_i \ll 1$, it tends to $\varpi = \nabla \cdot [
(Ze/\Omega_i) \bV_i \times \bun ]$. To see this, I use $- \int
d^3v\, (Ze\phiwig/T_i) f_{Mi} \rightarrow \nabla \cdot [
(cn_i/B\Omega_i) \nabla_\bot \phi ]$ and $\int d^3v\, f_{ig} \bv
\times \bun \rightarrow (c/ZeB) \nabla_\bot p_i$ to obtain
\begin{equation}
\varpi_G \simeq \nabla \cdot \left ( \frac{Zecn_i}{B\Omega_i}
\nabla_\bot \phi + \frac{c}{B\Omega_i} \nabla_\bot p_i \right ),
\label{vortGK_lw}
\end{equation}

\noindent as given by $\varpi$ in equation \eq{vort_1o} to first
order. The new version of the vorticity equation will evolve in
time the gyrokinetic ``vorticity" \eq{vortGKdef}. The advantage of
this new equation is that it will tend to a form similar to the
moment vorticity equation \eq{vorticity} for $k_\bot \rho_i \ll
1$. It is important to point out that the new version of the
vorticity equation, to be given in \eq{vort_type2}, and equations
\eq{vorticity} and \eq{vort_type1} are totally equivalent up to
$O(\delta_i^2 e n_e v_i/L)$. The advantage of the new version is a
similarity with the full vorticity equation \eq{vorticity}. This
similarity helps to study the size of the term $\nabla \cdot
[(c/B) \bun \times (\nabla \cdot \matrixtop{\pibf}_i)]$ and the
behavior of the transport of toroidal angular momentum for $k_\bot
\rho_i \ll 1$. I will prove that the toroidal velocity varies
slowly, as expected. Additionally, since the new vorticity
equation is derived from quasineutrality and the gyrokinetic
equation, it is equivalent to the gyrokinetic quasineutrality
equation and provides a way to study its limitations.

The new version of the vorticity equation is obtained by adding
equations \eq{vort_type1} and $\nabla \cdot \{ (c/B) [
\mathrm{equation}\; \eq{vtransp_krho} ] \times \bun \}$ to obtain
\begin{equation}
 \frac{\partial \varpi_G}{\partial t} = \nabla \cdot \Bigg [
J_{||} \bun + \bJ_{gd} + \tilde{\bJ}_{i\phi} + \frac{c}{B} \bun
\times (\nabla \cdot \matrixtop{\pibf}_{iG}) + Zen_i\bV_{iC} -
\frac{c}{B} \bun \times \mathbf{F}_{iC} \Bigg ].
\label{vort_type2}
\end{equation}

\noindent The terms $\tilde{\bJ}_i$, $Zen_i \tilde{\bV}_i$, $(c/B)
\bun \times (\nabla \cdot \matrixtop{\pibf}_{ig\times})$ and
$(c/B) \bun \times \mathbf{F}_{iB}$ are recombined to give
$\tilde{\bJ}_{i\phi}$, $(c/B) \bun \times (\nabla \cdot
\matrixtop{\pibf}_{iG})$ and some other terms that have vanished
because they are divergence free. The details on how to obtain
equation \eq{vort_type2} are in Appendix~\ref{app_vortGK}. Here I
have defined the new viscosity tensor
\begin{eqnarray}
\matrixtop{\pibf}_{iG} \equiv M \int d^3v\, f_{ig} \bv_\bot v_{||}
\bun + \matrixtop{\pibf}_{ig\times} = \nonumber \\ \int d^3v\,
f_{ig} M \left [ v_{||} ( \bv_\bot \bun + \bun \bv_\bot ) + \left
( \bv_{M0} + \bv_{E0} + \tilde{\bv}_1 \right ) \bv_\bot \right ]
\label{piGfinal}
\end{eqnarray}

\noindent and the new polarization current
\begin{equation}
\tilde{\bJ}_{i\phi} = \tilde{\bJ}_i - \frac{Zec}{B\Omega_i} \bun
\times \int d^3v\, f_{Mi} (\nablaave \phiwig \times \bun) \cdot
\nablaave \bv_\bot, \label{Jtilde_iphi}
\end{equation}

\noindent with $\tilde{\bJ}_i$ given in \eq{Jtilde_i1}. The
derivation of equation \eq{vort_type2} implies that both equations
\eq{vort_type1} and \eq{vort_type2} are equivalent as long as the
perpendicular component of equation \eq{vtransp_krho} is
satisfied, and any property proved for one of them is valid for
the other.

Equation \eq{vort_type2} gives the evolution of $\varpi_G$ and the
potential is then found by solving equation \eq{vortGKdef}.
Equation \eq{vort_type2} does not contain terms that are almost
divergence free, as was the case of $(c/B) \bun \times (\nabla
\cdot \matrixtop{\pibf}_i)$ in equation \eq{vorticity}. This will
ease implementation in existing simulations.

\begin{table}
\begin{center}
\renewcommand{\arraystretch}{1.5}
\begin{tabular}{|c|c|}
\hline Term & Order of magnitude \\
\hline $\varpi_G$ & $\delta_i k_\bot \rho_i e n_e$ \\
$\nabla \cdot ( J_{||} \bun + \bJ_{gd} )$ & $\delta_i e n_e v_i/L$ \\
$\nabla \cdot [ \tilde{\bJ}_{i\phi} + (c/B) \bun \times (\nabla \cdot \matrixtop{\pibf}_{iG})]$ & $\delta_i (k_\bot \rho_i)^2 e n_e v_i/L$ \\
$\nabla \cdot [Ze n_i \bV_{iC} - (c/B) \bun \times \mathbf{F}_{iC}]$ & $\delta_i (k_\bot \rho_i)^2 e n_e \nu_{ii}$ \\
\hline
\end{tabular}
\end{center}

\caption[Order of magnitude estimates for vorticity equation
\eq{vort_type2}]{Order of magnitude estimates for vorticity
equation \eq{vort_type2}.} \label{table_vort_type2}
\end{table}

It is important to know the size of the different terms in
\eq{vort_type2} for implementation purposes. The order of
magnitude of the different terms is summarized in
table~\ref{table_vort_type2}. In subsection~\ref{sub_vortquasi},
in table~\ref{table_vort_type1}, I showed $\nabla \cdot \bJ_{gd}
\sim \delta_i e n_e v_i/L$ and $\nabla \cdot (Zen_i \bV_{iC}) \sim
\delta_i (k_\bot \rho_i)^2 \nu_{ii} e n_e$. The term $\nabla \cdot
[ (c/B) \bun \times \mathbf{F}_{iC} ]$ is of order $\delta_i
(k_\bot \rho_i)^2 \nu_{ii} e n_e$ according to the results in
Appendix~\ref{app_collisiontransport}. For the flow $(c/B) \bun
\times (\nabla \cdot \matrixtop{\pibf}_{iG})$, there are two
different pieces in $\matrixtop{\pibf}_{iG}$, given in
\eq{piGfinal}, namely, $M \int d^3v\, f_{ig} \bv_\bot v_{||} \bun$
and $\matrixtop{\pibf}_{ig\times}$ as defined in \eq{pig_perp}.
The first component gives $(Mc/B) \bun \times [\nabla \cdot (\int
d^3v f_{ig} \bv_\bot v_{||} \bun) ] = (Ze/\Omega_i) \bun \times (
\int d^3v\, f_{ig} v_{||} \bv_\bot \cdot \nabla \bun) \sim
\delta_i^2 k_\bot \rho_i e n_e v_i$, where I have used that the
lowest order gyrophase dependent piece of $f_{ig}$ is even in
$v_{||}$ [recall \eq{fwig_krho}]. The divergence $\nabla \cdot
\matrixtop{\pibf}_{ig\times}$ is of order $\delta_i k_\bot \rho_i
p_i/L$ as proven in subsection~\ref{sub_v}, giving $\nabla \cdot [
(c/B) \bun \times (\nabla \cdot \matrixtop{\pibf}_{iG})] \sim
\delta_i (k_\bot \rho_i)^2 e n_e v_i / L$. Finally,
$\tilde{\bJ}_{i\phi}$ is also composed of two pieces shown in
\eq{Jtilde_iphi}. The divergence of $\nabla \cdot \tilde{\bJ}_i$
was already found in \eq{div_Jtilde_i}, and it is of order
$\delta_i (k_\bot \rho_i)^2 e n_e v_i/L$. The second term in
\eq{Jtilde_iphi} is of order $\delta_i^2 k_\bot \rho_i e n_e v_i$,
giving $\nabla \cdot \tilde{\bJ}_{i\phi} \sim \delta_i (k_\bot
\rho_i)^2 e n_e v_i/L$. Interestingly, employing equations
\eq{small_Jtilde_i} and \eq{div_Jtilde_i}, and the definition of
$\tilde{\bJ}_{i\phi}$ in \eq{Jtilde_iphi}, I find that for $k_\bot
\rho_i \ll 1$, $\nabla \cdot \tilde{\bJ}_{i\phi}$ tends to
\begin{equation}
\nabla \cdot \tilde{\bJ}_{i\phi} = \nabla \cdot \left \{
\frac{c}{B} \bun \times \left [ \nabla \cdot \left \{ \int d^3v\,
\left ( f_{ig} \frac{c}{B} \nablaave \phiave \times \bun - f_i
\frac{c}{B} \nabla \phi \times \bun \right ) M \bv_\bot \right \}
\right ] \right \}. \label{div_Jtilde_iphi}
\end{equation}

\noindent To obtain this expression I neglect
\begin{equation}
\nabla \cdot \left \{ \frac{Ze}{\Omega_i} \bun \times \left [ \int
d^3v\, ( f_i - f_{ig} ) \frac{c}{B} (\nabla \phi \times \bun)
\cdot \nablaave \bv_\bot \right ] \right \} \sim \delta_i^2 k_\bot
\rho_i e n_e v_i/L.
\end{equation}

\noindent Equation \eq{div_Jtilde_iphi} will be useful in the
$k_\bot \rho_i \ll 1$ limit worked out in
Appendix~\ref{app_tormom_long}.

The estimates in table~\ref{table_vort_type2} are useful to
determine the size of the problematic term $\nabla \cdot [ (c/B)
\bun \times (\nabla \cdot \matrixtop{\pibf}_i)]$. Subtracting
equation \eq{vorticity} from \eq{vort_type2} gives
\begin{eqnarray}
\nabla \cdot \left [ \frac{c}{B} \bun \times ( \nabla \cdot
\matrixtop{\pibf}_i ) \right ] = \frac{\partial}{\partial t} (
\varpi - \varpi_G) + \nabla \cdot \Bigg [ \bJ_{gd} - \bJ_d +
\tilde{\bJ}_{i\phi} + \frac{c}{B} \bun \times (\nabla \cdot
\matrixtop{\pibf}_{iG}) \nonumber \\ + Zen_i\bV_{iC} - \frac{c}{B}
\bun \times \mathbf{F}_{iC} \Bigg ] \sim \delta_i (k_\bot
\rho_i)^2 e n_e v_i/L. \label{estim_realvisc}
\end{eqnarray}

\noindent This result was anticipated in equation
\eq{proof_intrinsic} and proves that turbulent tokamaks are
intrinsically ambipolar!

Most of the estimates for the terms in equation
\eq{estim_realvisc} are obtained from
table~\ref{table_vort_type2}. Only $\varpi - \varpi_G$ and $\nabla
\cdot ( \bJ_{gd} - \bJ_d)$ need clarification. The long wavelength
limit of $\varpi_G \sim \delta_i k_\bot \rho_i e n_e$, given in
\eq{vortGK_lw}, is the same as the long wavelength limit of
$\varpi$ in \eq{vort_1o}. Thus, they can only differ in the next
order in $k_\bot \rho_i$, giving $\varpi - \varpi_G \sim \delta_i
(k_\bot \rho_i)^2 e n_e$. The difference $\bJ_{gd} - \bJ_d$ can be
rewritten using $f_i - f_{ig} = - (Ze\phiwig/T_i) f_{Mi}$ to
obtain
\begin{equation}
\bJ_{gd} - \bJ_d = \int d^3v\, \frac{Z^2e^2\phiwig}{T_i} f_{Mi}
\left ( \frac{v_\bot^2}{2\Omega_i} \bun \cdot \nabla \times \bun +
\frac{v_\bot^2}{2B\Omega_i} \bun \times \nabla B +
\frac{v_{||}^2}{\Omega_i} \bun \times \kappabf \right ).
\end{equation}

\noindent The size of $\bJ_{gd} - \bJ_d$ is $\delta_i^2 k_\bot
\rho_i e n_e v_i$ since the integral of $\phiwig$ in the gyrophase
vanishes to zeroth order. Then, $\nabla \cdot ( \bJ_{gd} - \bJ_d )
\sim \delta_i (k_\bot \rho_i)^2 e n_e v_i/L$. Employing these
estimates, I find that $\nabla \cdot [ (c/B) \bun \times (\nabla
\cdot \matrixtop{\pibf}_i)]$ is of order $\delta_i (k_\bot
\rho_i)^2 e n_e v_i/L$, as given in \eq{estim_realvisc} and
asserted in equation \eq{proof_intrinsic} in section
\ref{sect_vorticity_1}. Therefore, there is a piece of the
viscosity of order $\delta_i k_\bot \rho_i p_i$ that vanishes to
zeroth order in $\nabla \cdot [ (c/B) \bun \times (\nabla \cdot
\matrixtop{\pibf}_i)]$. In equations \eq{vort_type1} and
\eq{vort_type2} this piece has already been cancelled.

Finally, I study the evolution of the toroidal velocity implicit
in equation \eq{vort_type2}. To do so, I flux surface average
equation \eq{vort_type2} as I did in
section~\ref{sect_fluxsurfaceaverage} for equation \eq{vorticity}.
The result is
\begin{eqnarray}
\frac{\partial}{\partial t} \langle  \varpi_G \rangle_\psi =
\frac{1}{V^\prime} \frac{\partial}{\partial \psi} V^\prime \Bigg
\langle \bJ_{gd} \cdot \nabla \psi + \tilde{\bJ}_{i\phi} \cdot
\nabla \psi + Zen_i \bV_{iC} \cdot \nabla \psi \nonumber \\ -
\frac{c}{B} (\nabla \cdot \matrixtop{\pibf}_{iG} -
\mathbf{F}_{iC}) \cdot (\bun \times \nabla \psi) \Bigg
\rangle_\psi.
\end{eqnarray}

\noindent The term $\bJ_{gd} \cdot \nabla \psi$ can be manipulated
in the same way as the term $\bJ_d$ in equation \eq{vort_surfave}
to give $\langle \bJ_{gd} \cdot \nabla \psi \rangle_\psi = -
\langle (cI/B) [ \bun \cdot \nabla p_{g||} + (p_{g||} - p_{g\bot})
\nabla \cdot \bun ] \rangle_\psi$. Employing the parallel momentum
equation for ions, given by \eq{par_vtransp_krho}, and electrons,
given by \eq{vtransp_elect}, to write
\begin{equation}
\bun \cdot \nabla p_{g||} + (p_{g||} - p_{g\bot}) \nabla \cdot
\bun = - \frac{\partial}{\partial t} (n_i M V_{i||}) - \nabla
\cdot \pibf_{ig||} + \tilde{F}_{iE} + \mathbf{F}_{iC} \cdot \bun,
\end{equation}

\noindent I find
\begin{eqnarray}
\frac{\partial}{\partial t} \left ( \langle  \varpi_G \rangle_\psi
- \frac{1}{V^\prime} \frac{\partial}{\partial \psi} V^\prime \left
\langle \frac{ZeI}{\Omega_i} n_i V_{i||} \right \rangle_\psi
\right ) = \frac{1}{V^\prime} \frac{\partial}{\partial \psi}
V^\prime \Bigg \langle \frac{cI}{B} (\nabla \cdot \pibf_{ig||} -
\tilde{F}_{iE}) \nonumber \\ + \tilde{\bJ}_{i\phi} \cdot \nabla
\psi - \frac{c}{B} (\nabla \cdot \matrixtop{\pibf}_{iG}) \cdot
(\bun \times \nabla \psi) + Zen_i \bV_{iC} \cdot \nabla \psi - cR
\mathbf{F}_{iC} \cdot \zun \Bigg \rangle_\psi, \label{tormomGK_1}
\end{eqnarray}

\noindent where I have employed $(I/B) \mathbf{F}_{iC} \cdot \bun
- B^{-1} \mathbf{F}_{iC} \cdot ( \bun \times \nabla \psi ) = R
\mathbf{F}_{iC} \cdot \zun$ [recall \eq{bxgradpsi}]. Taking the
limit $k_\bot \rho_i \ll 1$, for which $\varpi_G \rightarrow
\nabla \cdot [ (Zen_i/\Omega_i) \bV_i \times \bun ]$, equation
\eq{tormomGK_1} can be shown to give
\begin{equation}
\frac{\partial}{\partial t} \langle R n_i M \bV_i \cdot \zun
\rangle_\psi = - \frac{1}{V^\prime} \frac{\partial}{\partial \psi}
V^\prime \langle R \zun \cdot \matrixtop{\pibf}_i^{(0)} \cdot
\nabla \psi \rangle_\psi, \label{tormom_long}
\end{equation}

\noindent where I have integrated once in $\psi$. The details of
the calculation are in Appendix~\ref{app_tormom_long}. The zeroth
order off-diagonal viscosity is given by
\begin{equation}
\langle R \zun \cdot \matrixtop{\pibf}_i^{(0)} \cdot \nabla \psi
\rangle_\psi = \left \langle \int d^3v\, f_i R M(\bv \cdot \zun)
\left (\bv_{M0} + \tilde{\bv}_1 - \frac{c}{B} \nabla \phi \times
\bun \right ) \cdot \nabla \psi \right \rangle_\psi.
\label{viscosity_long}
\end{equation}

\noindent The distribution function $f_i = f_{ig} -
(Ze\phiwig/T_i) f_{Mi}$ has both the adiabatic and the
non-adiabatic pieces. The viscosity in \eq{viscosity_long}
includes the nonlinear Reynolds stress, describing the $E\times B$
transport of toroidal angular momentum, and the transport due to
the magnetic drifts $\bv_{M0}$ and finite gyroradius effects
$\tilde{\bv}_1$. In the absence of collisions, only the Reynolds
stress gives a non-vanishing contribution as the other terms
correspond to the gyroviscosity. In section~\ref{sectapp_gyrovisc}
of Appendix~\ref{app_gyrovisc}, I prove that
\begin{eqnarray}
\left \langle \int d^3v\, f_i R M(\bv \cdot \zun) (\bv_{M0} +
\tilde{\bv}_1 ) \cdot \nabla \psi \right \rangle_\psi =
\frac{\partial}{\partial t} \left \langle \frac{1}{2B\Omega_i} (
|\nabla \psi|^2 p_{i\bot} + I^2 p_{i||} ) \right \rangle_\psi
\nonumber \\ - \left \langle \frac{M}{2B\Omega_i} \int d^3v\, C \{
f_i \} \left ( |\nabla \psi|^2 \frac{v_\bot^2}{2} + I^2 v_{||}^2
\right ) \right \rangle_\psi. \label{gyrovisc}
\end{eqnarray}

\noindent It becomes apparent that when statistical equilibrium is
reached and the net radial transport of energy is slow so that
$\partial/\partial t \simeq 0$, the magnetic drifts only provide
momentum transport proportional to the collision frequency. Since
collisions are usually weak, this term will tend to be small.
Moreover, in section~\ref{sectapp_collgyrovisc} of
Appendix~\ref{app_gyrovisc}, I show that this collisional piece
vanishes exactly in up-down symmetric tokamaks, leaving only the
Reynolds stress,
\begin{equation}
\langle R \zun \cdot \matrixtop{\pibf}_i^{(0)} \cdot \nabla \psi
\rangle_\psi \simeq - \left \langle \int d^3v\, f_i R M(\bv \cdot
\zun) \frac{c}{B} (\nabla \phi \times \bun ) \cdot \nabla \psi
\right \rangle_\psi. \label{Reynolds stress}
\end{equation}

\noindent In any case, the zeroth order viscosity is of order
$\delta_i^2 p_i$, and the corresponding piece in the vorticity
equation \eq{vort_type2} is of order $\delta_i (k_\bot \rho_i)^2 e
n_e v_i/L$, which becomes of order $\delta_i^3 e n_e v_i/L$ as
$k_\bot \rightarrow 1/L$.

An important conclusion that can be derived from equation
\eq{tormom_long} is that the rate of change of the toroidal
velocity is $\partial/\partial t \sim k_\bot \rho_i v_i/L$,
becoming slower and slower as we reach longer wavelengths. This
behavior must be reproduced by any equation used to calculate the
radial electric field. Equations \eq{vort_type1} and
\eq{vort_type2} satisfy this condition, but in addition they have
the advantage of showing this property explicitly. The terms that
determine the radial electric field turn out to be $\tilde{\bJ}_i
\cdot \nabla \psi$, $Zen_i \tilde{\bV}_i \cdot \nabla \psi$ and
$Zen_i \bV_{iC} \cdot \nabla \psi$ in equation \eq{vort_type1},
and $\tilde{\bJ}_{i\phi} \cdot \nabla \psi$, $(c/B) [ \bun \times
(\nabla \cdot \matrixtop{\pibf}_{iG}) ] \cdot \nabla \psi$, $Zen_i
\bV_{iC} \cdot \nabla \psi$ and $(c/B) (\bun \times
\mathbf{F}_{iC}) \cdot \nabla \psi$ in equation \eq{vort_type2}.
Additionally, it is necessary to keep the terms $\nabla \cdot
\pibf_{ig||}$, $\tilde{F}_{iE}$ and $\mathbf{F}_{iC} \cdot \bun$
in the parallel momentum equation \eq{par_vtransp_krho}. Any
simulation must make sure that these terms have the correct
behavior at long wavelengths and give equation \eq{tormom_long}.
In the traditional gyrokinetic approach, the terms $\tilde{\bJ}_i
\cdot \nabla \psi$, $Zen_i \tilde{\bV}_i \cdot \nabla \psi$ and
$Zen_i \bV_{iC} \cdot \nabla \psi$ of equation \eq{vort_type1} can
be tracked back to terms in the gyrokinetic Fokker-Planck
equation. They correspond to the difference between the ion and
electron gyroaveraged $E\times B$ flows, $(c/B) ( f_i \nabla \phi
\times \bun - f_{ig} \nabla_\bR \phiave \times \bun )$, and the
finite gyroradius effects that make $\bun ( \bR_g ) \neq \bun
(\br)$, $\nabla_{\bR_g} \neq \nablaave$ and $\langle C \{f_i\}
\rangle \neq C\{ f_i \}$. This identification is the advantage of
equation \eq{vort_type1} since it allows easier analysis of
existing simulations. Equation \eq{vort_type1} can be used to
check if the simulations reproduce the correct transport of
toroidal angular momentum.

Since the vorticity equations \eq{vort_type1} and \eq{vort_type2}
give equation \eq{tormom_long} for $k_\bot \rho_i \ll 1$, it may
seem that they provide the correct radial electric field at long
wavelengths. Moreover, I have deduced these vorticity equations
employing only the gyrokinetic Fokker-Planck equation and the
corresponding quasineutrality, making it tempting to argue that
the traditional gyrokinetic method is good enough to find the
radial electric field. This argument is flawed because there are
missing terms of order $\delta_i^2 e n_e v_i/L$ in equations
\eq{vort_type1} and \eq{vort_type2}. Then, the transport of
toroidal angular momentum \eq{tormom_long}, that corresponds to a
term of order $\delta_i (k_\bot \rho_i)^2 e n_e v_i/L$, will
remain correct only if $(k_\bot \rho_i)^2 \gg \delta_i$.
Consequently, the gyrokinetic quasineutrality should provide the
correct radial electric field up to wavelengths of order
$\sqrt{\rho_i L}$. For longer wavelengths, there will be missing
terms. This estimate only considers the terms that the gyrokinetic
equation is missing and neglects possible numerical inaccuracies.

In the next section, I will show with a simplified example that
gyrokinetic indeed has problems determining the radial electric
field in axisymmetric configurations. This example is intended to
illustrate the difficulties that arise from the use of gyrokinetic
Fokker-Planck equation together with the gyrokinetic
quasineutrality equation \eq{qn_mod}.

\section{Example: quasineutrality in a $\theta$-pinch \label{sect_thetapinch}}

In this section, I try to find the solution to the non-turbulent,
axisymmetric $\theta$-pinch. Without turbulence, the perpendicular
wavelengths are of the order of the characteristic size of the
$\theta$-pinch, $k_\bot L \sim 1$. This section is intended only
as an example, and neglecting the turbulence greatly simplifies
the calculation without fundamentally changing the properties of
quasineutrality. In this simplified problem, I find that current
gyrokinetic treatments, even if extended to a higher order in
$\delta_i = \rho_i/L \ll 1$ than in
chapter~\ref{chap_gyrokinetics}, do not yield a solution for the
long wavelength radial electric field, leaving it as a free
parameter. The gyrokinetic Fokker-Planck equation and the
quasineutrality equation are intrinsically ambipolar and cannot
determine the radial electric field. In
chapter~\ref{chap_angularmomentum}, I will show that the radial
electric field is recovered if a different approach is employed.

\begin{figure}
\begin{center}
\includegraphics[width = \textwidth]{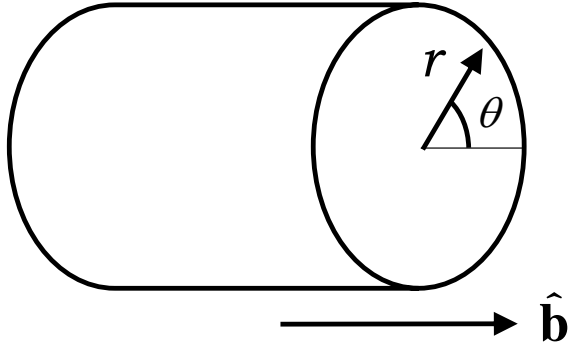}
\end{center}

\vspace{-7cm} \caption[Geometry of the $\theta$-pinch]{Geometry of
the $\theta$-pinch.} \label{fig_thetapinch}
\end{figure}

In the $\theta$-pinch, the magnetic field is given by $\mathbf{B}
= B ( r ) \bun$, where here $\bun$ is a constant unit vector in
the axial direction, and $r$ is the radial coordinate in
cylindrical geometry. The geometry is sketched in
figure~\ref{fig_thetapinch}. For long wavelengths, the gyrokinetic
equation can be found to order $\delta_i^2 f_{Mi} v_i/L$. The
simplified geometry of the magnetic field yields more manageable
expressions for the gyrokinetic variables, i.e., $\mu_1$ and
$\bR_2$ become
\begin{equation}
\mu_1 = \frac{Ze\phiwig}{MB} - \frac{v_\bot^2}{2B^2 \Omega_i} (
\bv \times \bun ) \cdot \nabla B
\end{equation}

\noindent and
\begin{equation}
\bR_2 = \frac{1}{4 B \Omega_i^2} [ \bv_\bot \bv_\bot - ( \bv
\times \bun ) ( \bv \times \bun ) ] \cdot \nabla B,
\end{equation}

\noindent where the term $(c/B \Omega_i) \nabla_\bR \Phiwig \times
\bun \sim \delta_i^2 k_\bot \rho_i L$ has been neglected because I
assume that $k_\bot L \sim 1$. Using $\bR_2$, the gyroaverage of
$\dot{\bR}$ is calculated to be
\begin{equation}
\langle \dot{\bR} \rangle \simeq \langle v_{||} \rangle \bun +
\left \langle \frac{\mu_0}{\Omega_i} \bun \times \nabla B -
\frac{c}{B} \nabla \phi \times \bun + \bv_\bot \cdot \nabla \bR_2
- \frac{Ze}{M} \nabla \phi \cdot \nabla_v \bR_2 \right \rangle,
\end{equation}

\noindent where I have used that in a $\theta$-pinch $\bun \cdot
\nabla B = 0$ to write $\bv \cdot \nabla \bR_2 = \bv_\bot \cdot
\nabla \bR_2$. The gyroaverages are performed by employing the
long wavelength approximation $\nabla \phi \simeq \nabla_\bR \phi
- \Omega_i^{-1} ( \bv \times \bun ) \cdot \nabla_\bR \phi$ and the
relation $\langle \bv_\bot \bv_\bot \bv_\bot \rangle = 0$ to get
\begin{equation}
\langle \dot{\bR} \rangle \simeq \langle v_{||} \rangle \bun +
\frac{\mu}{\Omega_i( \bR )} \bun \times \nabla_\bR B -
\frac{c}{B(\bR)} \nabla_\bR \phi \times \bun.
\end{equation}

\noindent The gyroaverage of $\dot{E}$ is found by using
\eq{vgradphi} to write
\begin{equation}
\dot{E} = - \frac{Ze}{M} \left (\bv \cdot \nabla \phi - \frac{d
\phiwig}{dt} \right ) \simeq - \frac{Ze}{M} \left ( \dot{\bR}
\cdot \nabla_\bR \phiave + \dot{\mu} \frac{\partial
\phiave}{\partial \mu} - \frac{\partial \phiwig}{\partial t}
\right ), \label{dEdt_theta}
\end{equation}

\noindent where I employ that $\partial /\partial t \sim \delta_i
v_i / L$ for long wavelengths and $\partial \phi /
\partial E = O ( \delta_i^3 M / e )$ in the $\theta$-pinch.
Considering that $\langle \dot{\mu} \rangle = O ( \delta_i^2
v_i^3/BL )$ and $\partial \phiave/\partial \mu = O ( \delta_i M B
/ e )$, the gyroaverage of \eq{dEdt_theta} is calculated to be
\begin{equation}
\langle \dot{E} \rangle \simeq - \frac{Ze}{M} \langle \dot{\bR}
\rangle \cdot \nabla_\bR \phiave.
\end{equation}

\noindent Thus, the gyrokinetic equation to order $O(\delta_i^2
f_{Mi} v_i/L)$ is
\begin{equation}
\frac{\partial f_i}{\partial t} + \langle \dot{\bR} \rangle \cdot
\left ( \nabla_\bR f_i - \frac{Ze}{M} \nabla_\bR \phiave
\frac{\partial f_i}{\partial E} \right ) = \langle C \{ f_i \}
\rangle. \label{FP_lk_tpinch}
\end{equation}

\noindent I have neglected the derivative $\partial f_i /\partial
\mu$ because the distribution function is Maxwellian to zeroth
order and $\langle \dot{\mu} \rangle$ is already small by
definition of $\mu$. For an axisymmetric steady state solution,
the terms on the left side of \eq{FP_lk_tpinch} vanish, the second
term because the gyrocenter parallel and perpendicular drifts,
$\langle \dot{\bR} \rangle$, remain in surfaces of constant $f_i$
and $\phi$ (for this reason, $\langle v_{||} \rangle$ need not be
evaluated to second order). Thus, equation~\eq{FP_lk_tpinch}
becomes $\langle C\{f_i\} \rangle = 0$. Such an equation can be
solved for a simplified collision operator. I use a Krook
operator, $C\{f_i\} = - \nu ( f_i - f_M )$, with constant
collision frequency $\nu$ and a shifted Maxwellian,
\begin{equation}
f_M = n_i \left ( \frac{M}{2 \pi T_i} \right )^{3/2} \exp \left [
- \frac{M ( \bv - \mathbf{V}_i )^2}{2 T_i} \right ],
\end{equation}

\noindent where $n_i$, $T_i$ and $\mathbf{V}_i$ are functions of
the position $\br$. I assume that the parallel average velocity,
$V_{i||} = \bun \cdot \mathbf{V}_i$, is zero and I order
$\mathbf{V}_i$ as $O(\delta_i v_i)$ to obtain
\begin{equation}
f_M \simeq f_{M0} \left [ 1 + \frac{M \bv_\bot \cdot
\mathbf{V}_i}{T_i} + \frac{M^2 ( \bv_\bot \cdot \mathbf{V}_i
)^2}{2T_i^2} - \frac{MV_i^2}{2T_i} \right ], \label{f_M_tpinch}
\end{equation}

\noindent with
\begin{equation}
f_{M0} = n_i \left ( \frac{M}{2 \pi T_i} \right )^{3/2} \exp \left
( - \frac{M v^2}{2 T_i} \right ). \label{f_M0_tpinch}
\end{equation}

\noindent With the Krook operator, the gyrokinetic solution is
\begin{equation}
f_i = \langle f_M \rangle = \langle f_{M0} \rangle + \left \langle
\frac{M\bv_\bot \cdot \bV_i}{T_i} f_{M0} \right \rangle +
\frac{M^2 v_\bot^2 V_i^2}{4T_i^2} F_M - \frac{MV_i^2}{2T_i} F_M,
\label{fave_tpinch_1}
\end{equation}

\noindent where I have used that in the higher order terms,
$f_{M0} \simeq F_M$, with
\begin{equation}
F_M = n_i (\bR) \left [ \frac{M}{2 \pi T_i (\bR)} \right ]^{3/2}
\exp \left [ - \frac{ME}{T_i(\bR)} \right ]. \label{low_Maxw_GK}
\end{equation}

\noindent For the first two terms in equation \eq{fave_tpinch_1},
it is necessary to Taylor expand $f_{M0} ( \br, E_0 )$, $\bV_i
(\br)$ and $\bv_\bot = \sqrt{2\mu_0 B(\br)} [ \eun_1 (\br) \cos
\varphi_0 + \eun_2 (\br) \sin \varphi_0 ]$ around $\bR$, $E$,
$\mu$ and $\varphi$. The final result is
\begin{eqnarray}
f_i = \langle f_M \rangle = F_M \Bigg [ 1 - \frac{x_\bot^2}{n_i}
\nabla \cdot \left ( \frac{cn_i}{B \Omega_i} \nabla_\bot \phi
\right ) + \frac{Mc^2}{2T_iB^2} (2 - x_\bot^2) | \nabla_\bot \phi
|^2 + \frac{MV_i^2}{2T} ( x_\bot^2 - 1 ) \nonumber \\ +
\frac{x_\bot^2}{2M\Omega_i^2} \left ( x^2 - \frac{5}{2} \right )
\nabla_\bot^2 T_i - \frac{x_\bot^2} {2n_iM\Omega_i^2}
\nabla_\bot^2 p_i + \frac{x_\bot^2}{2 T_i M \Omega_i^2} \left (
\frac{35}{4} - 7 x^2 + x^4 \right ) | \nabla_\bot T_i |^2
\nonumber \\ + \frac{2 x_\bot^2}{M B \Omega_i^2} \left (
\frac{5}{2} - x^2 \right ) \nabla_\bot B \cdot \nabla_\bot T_i  +
\frac{c}{T_i B\Omega_i} \nabla_\bot \phi \cdot \nabla_\bot T_i
\left ( \frac{5}{2} - x_\bot^2 - x^2 \right ) \Bigg
], \nonumber \\
\label{solution_o2_theta}
\end{eqnarray}

\noindent where $x^2 = Mv^2/2T_i \simeq ME/T_i$, $x_\bot^2 =
Mv_\bot^2/2T_i \simeq M\mu B/T_i$ and I have employed
\begin{equation}
\mathbf{V}_i = \frac{1}{n_i M \Omega_i} \bun \times \nabla p_i -
\frac{c}{B} \nabla \phi \times \bun. \label{Vi_drift_thetapinch}
\end{equation}

\noindent The distribution function in \eq{solution_o2_theta} has
been calculated by using a gyrokinetic equation that is correct to
order $\delta_i^2 f_{Mi} v_i/L$ for both the Vlasov operator and
the gyroaveraged collision operator. Using the definitions $\bR =
\br + \bR_1 + \bR_2$, $E = E_0 + E_1 + E_2$ and $\mu = \mu_0 +
\mu_1$, and the gyrophase dependent collisional piece $\fwig_i$
given in \eq{fwig_sol}, I can find the distribution function $f_i$
in $\br$, $\bv$ variables to order $O(\delta_i^2 f_{Mi})$. As a
check, the same solution has been also obtained without resorting
to gyrokinetics to order $O(\delta_i^2 f_{Mi})$. This check is
omitted here.

If I had gyroaveraged $C\{f_i\}$ only to order $\delta_i$, as most
gyrokinetic models do, the solution would have been simply
\begin{equation}
f_i \simeq F_M. \label{solution_o1_theta}
\end{equation}

\noindent Substituting this solution into the higher order
gyrokinetic quasineutrality equation \eq{qn_highorder}, I find the
inconsistent result
\begin{equation}
\nabla \cdot \left ( \frac{Z c n_i}{B \Omega_i} \nabla_\bot \phi
\right ) - \frac{ZM c^2 n_i}{2T_i B^2} | \nabla_\bot \phi |^2 =
n_e - Z n_i - \frac{Z}{2M\Omega_i^2} \nabla_\bot^2 p_i.
\label{fake_qn_theta}
\end{equation}

\noindent However, this quasineutrality equation is very different
from the one we obtain by using the full $O(\delta_i^2 f_{Mi})$
solution in \eq{solution_o2_theta}, which simply gives
\begin{equation}
Z n_i = n_e. \label{real_qn_theta}
\end{equation}

\noindent Therefore, the gyrokinetic quasineutrality equation
reduces to the quasineutrality condition when the exact
$O(\delta_i^2 f_{Mi})$ distribution function of
\eq{solution_o2_theta} is employed. Equation~\eq{fake_qn_theta} is
wrong because the $O(\delta_i f_{Mi})$ result of
\eq{solution_o1_theta} is either inducing an $O(\delta_i^2 e n_e)$
charge difference or imposing the non-physical condition
\begin{equation}
\nabla \cdot \left ( \frac{Z c n_i}{B \Omega_i} \nabla_\bot \phi
\right ) - \frac{ZM c^2 n_i}{2T_i B^2} | \nabla_\bot \phi |^2 = -
\frac{Z}{2M\Omega_i^2} \nabla_\bot^2 p_i. \label{fake_qn_theta_2}
\end{equation}

\noindent The difference between \eq{fake_qn_theta} and
\eq{real_qn_theta}, given by \eq{fake_qn_theta_2}, originates in
$O(\delta_i^2 n_e)$ terms that should have been cancelled by
pieces of the distribution function of the same order.

The $\theta$-pinch example illustrates the problem of using a
lower order gyrokinetic equation than needed, but it also
highlights another issue. The potential does not appear in the
quasineutrality equation \eq{real_qn_theta}, and, therefore, it
cannot be found using it. In a computer simulation, the potential
is obtained from the gyrokinetic quasineutrality equation
\eq{qn_highorder}, and the distribution function is evolved
employing the gyrokinetic equation \eq{FP_lk_tpinch}. A possible
initial condition $f_{i, t = 0} (\bR, E, \mu)$ is the stationary
solution \eq{solution_o2_theta}, where the potential $\phi(r)$ is
a free function that is set to be $\phi_{t = 0}(r)$. Equation
\eq{real_qn_theta} proves that the solution to the gyrokinetic
quasineutrality equation \eq{qn_highorder} at $t = 0$ must be
$\phi (r, t=0) = \phi_{t = 0}(r)$, where $\phi_{t = 0}(r)$ is the
free function that I chose for the initial condition. Since the
initial condition $f_{i, t = 0} (\bR, E, \mu)$ is a stationary
solution, I find that the solution for all times is $f_i = f_{i, t
= 0}$ and $\phi = \phi_{t = 0}$, and the radial electric field is
solely determined by the initial condition. If there were any
numerical errors that made the solution invalid to order
$\delta_i^2 f_{Mi}$, the radial electric field would suffer a
non-physical evolution. In modern gyrokinetics, the radial
electric field is then determined by the initial condition, in
particular, by a piece of order $\delta_i^2 f_{Mi}$ of the initial
condition. This result is not surprising since in this chapter I
have proved that, in axisymmetric configurations, the axisymmetric
piece of the vorticity equation, or time derivative of $Zn_i -
n_e$, is of order $\delta_i^3 n_e v_i/L$ at the most, as given by
\eq{proof_intrinsic}. The gyrokinetic equation is only calculated
to order $\delta_i^2 f_{Mi} v_i/L$ in this section (and only to
order $\delta_i f_{Mi} v_i/L$ in codes), leading to an effectively
constant $Zn_i - n_e$ and hence a radial electric field dependent
only on the initial condition. In section~\ref{sect_thetapinch_2},
I will show that in reality the time derivative of $Zn_i - n_e$ is
even smaller than $\delta_i^3 n_e v_i/L$ by a factor of
$\nu/\Omega_i$.

\section{Discussion \label{sect_discussion_GKvorticity}}

In this chapter, I have shown how the vorticity equation recovers
the physics of quasineutrality and at the same time retains the
effect of the transport of toroidal angular momentum in the radial
electric field. I have proposed two possible vorticity equations,
\eq{vort_type1} and \eq{vort_type2}. With these two equations, I
estimate the size of the term that determines the radial electric
field, given by equation \eq{proof_intrinsic}. In this manner, I
prove that, with the usual gyrokinetic equation, setting the
radial current to zero, $\langle \bJ \cdot \nabla \psi
\rangle_\psi = 0$, cannot determine the long wavelength
axisymmetric radial electric field. Therefore, modern gyrokinetic
formulations are intrinsically ambipolar, and thereby unable to
determine the long wavelength axisymmetric radial electric field.

I illustrated the problems that arise from a failure to satisfy
intrinsic ambipolarity with a simplified problem in
section~\ref{sect_thetapinch}. In the example, the long
wavelength, axisymmetric radial electric field was left
undetermined by the gyrokinetic quasineutrality equation even if
the distribution was calculated to an order higher than current
codes can achieve. More importantly, if there is an error in the
density as small as $\delta_i^2 n_e$, the gyrokinetic
quasineutrality equation yields an erroneous long wavelength
radial electric field. This feature places a strong requirement on
the accuracy of any code that calculates the radial electric
field. The vorticity equation, on the other hand, makes the
dependence of the radial electric field on the toroidal transport
of angular momentum explicit. Both vorticity equations
\eq{vort_type1} and \eq{vort_type2} would yield a long wavelength
radial electric field constant for the short turbulence saturation
time.

Between the two vorticity equations, equation \eq{vort_type1} is
closer to the gyrokinetic quasineutrality and is probably the best
candidate to implement and compare with existing results. In fact,
a similar, but less complete, vorticity equation has already been
implemented in the PIC gyrokinetic code GEM \cite{chen01}. On the
other hand, equation \eq{vort_type2} is similar to the traditional
vorticity equation \eq{vorticity}, making the study of
conservation of toroidal angular momentum straightforward.

These vorticity equations are valid for short wavelengths on the
order of the ion gyroradius. They must be supplemented with long
wavelength physics to be extended to wavelengths longer than
$\sqrt{\rho_i L}$. Only then will the transport of toroidal
angular momentum be correctly described. The extension to longer
wavelengths is treated in chapter~\ref{chap_angularmomentum}.

Finally, any numerical implementation of either of the vorticity
equations needs to make sure that the properties derived and
discussed are satisfied, namely, the scaling of the different
terms with $k_\bot \rho_i$ should be ensured, and the
cancellations that take place due to the flux surface average
should also be maintained in codes. It is for this reason that I
give all the details of the analytical calculations including
detailed appendices.

\chapter{Solving for the radial electric field
\label{chap_angularmomentum}}

In this chapter, I describe the method that I propose to calculate
the radial electric field. As I showed in
section~\ref{sect_fluxsurfaceaverage}, the flux surface averaged
vorticity equation reduces to the transport of toroidal angular
momentum. Then, to obtain the radial electric field, it is
necessary to solve the conservation equation for the transport of
toroidal angular momentum.

In section~\ref{sect_ionvisc}, I obtain an equation for the
toroidal-radial component of the ion viscosity to order
$\delta_i^3 p_i$ that only requires the ion distribution function
to order $\delta_i^2 f_{Mi}$. I already argued in
section~\ref{sect_fluxsurfaceaverage} that $\delta_i^3 p_i$ is the
order to which the ion viscosity should be found to recover
gyroBohm transport of angular momentum. In this chapter, I give
some arguments that suggest that the radial transport of angular
momentum is indeed of order $\delta_i^3 p_i$. Interestingly, the
transport of toroidal angular momentum found in
\eq{viscosity_long}, of order $\delta_i k_\bot \rho_i p_i$, should
then vanish for long wavelengths. In reality, the transport
probably is of order $\delta_i^2 p_i$ at each time step yet its
time average vanish to that order. In other words, there might be
fast local exchange of toroidal angular momentum, leading to zonal
flow structure, but the irreversible transport of angular momentum
from the edge to the core is much slower. In
section~\ref{sect_thetapinch_2}, I apply the equation for the ion
viscosity obtained in section~\ref{sect_ionvisc} to solve for the
radial electric field in the example presented in
section~\ref{sect_thetapinch}.

Even with the convenient equation that gives the radial transport
of toroidal angular momentum to order $\delta_i^3 p_i$ with only
an $O(\delta_i^2 f_{Mi})$ distribution function, this still makes
it necessary to find the ion distribution function and the
potential to an order higher than the order to which they are
usually calculated. At the end of this chapter, in
section~\ref{sect_iondistribution}, I prove that under certain
assumptions, the lowest order full $f$ gyrokinetic equation
\eq{FP_final} is valid even to calculate the second order
distribution function, and the vorticity equations \eq{vort_type1}
and \eq{vort_type2} are easily extended to yield a higher order
potential. Finally, I discuss all the results of this chapter in
section~\ref{sect_discussion_toroidal}.

\section{Ion viscosity and the axisymmetric potential} \label{sect_ionvisc}

The evolution of the toroidal velocity, given in \eq{toro_angmom},
is determined by the flux surface averaged toroidal-radial
component of the ion viscosity $\langle R\zun \cdot
\matrixtop{\pibf}_i \cdot \nabla \psi \rangle_\psi$. According to
the gyroBohm estimates at the end of
section~\ref{sect_fluxsurfaceaverage}, the ion viscosity has to be
obtained to order $\delta_i^3 p_i$. If the ion viscosity is to be
determined directly from its definition in \eq{pi_def}, the ion
distribution function must be calculated to order $\delta_i^3
f_{Mi}$, too high of an order to be practical or implementable.

To avoid direct evaluation of the ion viscosity, I propose using
moments of the Fokker-Planck equation. This is the approach
followed in drift kinetics \cite{simakov07} and to formulate a
hybrid gyrokinetic-fluid description \cite{catto08}. The ion
viscosity can be solved from the $M \bv \bv$ moment of the
Fokker-Planck equation, given by
\begin{equation}
\Omega_i ( \matrixtop{\pibf}_i \times \bun - \bun \times
\matrixtop{\pibf}_i ) = \matrixtop{\mathbf{K}}, \label{moment_pi}
\end{equation}

\noindent with
\begin{equation}
\matrixtop{\mathbf{K}} = \frac{\partial
\matrixtop{\mathbf{P}}_i}{\partial t} + \nabla \cdot \left ( M
\int d^3v\, f_i \bv \bv \bv \right ) + Zen_i (\nabla \phi \bV_i +
\bV_i \nabla \phi) - M \int d^3v\, C\{ f_i \} \bv \bv.
\end{equation}

\noindent Here, $\matrixtop{\mathbf{P}}_i = M \int d^3v\, f_i \bv
\bv$. From the moment equation \eq{moment_pi}, the off diagonal
elements of $\matrixtop{\pibf}_i$ can be solved for as a function
of $\matrixtop{\mathbf{K}}$. Additionally, equation \eq{moment_pi}
contains the energy conservation equation, $\mathrm{Trace}
(\matrixtop{\mathbf{K}}) = 0$, and the parallel pressure equation,
$\bun \cdot \matrixtop{\mathbf{K}} \cdot \bun = 0$.

To solve for the toroidal-radial component $R \zun \cdot
\matrixtop{\pibf}_i \cdot \nabla \psi$, I pre-multiply and
post-multiply equation \eq{moment_pi} by $R\zun$, giving
\begin{eqnarray}
R \zun \cdot \matrixtop{\pibf}_i \cdot \nabla \psi =
\frac{Mc}{2Ze} \frac{\partial}{\partial t} (R^2 \zun \cdot
\matrixtop{\mathbf{P}}_i \cdot \zun) + \frac{M^2c}{2Ze} \nabla
\cdot \left [ \int d^3v\, \bv f_i R^2 (\bv \cdot \zun)^2 \right ]
\nonumber \\ + M c n_i R^2 (\zun \cdot \nabla \phi) (\bV_i \cdot
\zun) - \frac{M^2c}{2Ze} \int d^3v\, C\{ f_i \} R^2 (\bv \cdot
\zun)^2,
\end{eqnarray}

\noindent where I use $R (\bun \times \zun) = \nabla \psi/B$ and
$\nabla (R\zun) = (\nabla R) \zun - \zun (\nabla R)$. Flux surface
averaging this expression gives
\begin{eqnarray}
\langle R \zun \cdot \matrixtop{\pibf}_i \cdot \nabla \psi
\rangle_\psi = \frac{M^2c}{2Ze} \frac{1}{V^\prime}
\frac{\partial}{\partial \psi} V^\prime \left \langle \int d^3v\,
(f_i - \overline{f}_i) (\bv \cdot \nabla \psi) R^2 (\bv \cdot
\zun)^2 \right \rangle_\psi \nonumber \\ + \langle Mc n_i R^2
(\zun \cdot \nabla \phi) (\bV_i \cdot \zun) \rangle_\psi +
\frac{Mc}{2Ze} \frac{\partial}{\partial t} \langle R^2 \zun \cdot
\matrixtop{\mathbf{P}}_i \cdot \zun \rangle_\psi \nonumber \\ -
\frac{M^2 c}{2Ze} \left \langle \int d^3v\, C\{ f_i \} R^2 (\bv
\cdot \zun)^2 \right \rangle_\psi. \label{pi_rzeta_ave}
\end{eqnarray}

\noindent In the first term of the right side, I use that
$\overline{(\bv \cdot \nabla \psi) (\bv \cdot \zun)^2} = 0$ to
write the integral only as a function of the gyrophase dependent
piece of the distribution function. Equation \eq{pi_rzeta_ave} has
the advantage that a distribution function correct to order
$\delta_i^2 f_{Mi}$ gives a viscosity good to order $\delta_i^3
p_i$! The method by which we have gained an order in $\delta_i$ is
similar to the calculation of the perpendicular ion flow employing
the momentum equation. To evaluate the perpendicular ion flow $n_i
\bV_{i\bot} = \int d^3v\, f_i \bv_\bot$ to order $\delta_i n_e
v_i$ by direct integration over velocity space, the distribution
function $f_i$ must be correct to order $\delta_i f_{Mi}$.
Instead, it is possible to use the ion perpendicular momentum
equation to order $p_i/L$, where the Lorentz force $(Ze/c) n_i
\bV_i \times \bB$ balances the perpendicular pressure gradient
$\nabla_\bot p_i$ and the perpendicular electric field $Zen_i
\nabla_\bot \phi$, giving the ion flow $n_i \bV_{i\bot} = (c/ZeB)
\bun \times \nabla p_i - (cn_i/B) \nabla \phi \times \bun$. Notice
that only the lowest order Maxwellian $f_{Mi}$ has been used to
find $p_i$. We have gained an order in $\delta_i$.

If the gyroBohm estimates done at the end of
section~\ref{sect_fluxsurfaceaverage} are correct, the ion
viscosity must identically vanish to order $\delta_i^2 p_i$
without determining the evolution of the long wavelength
axisymmetric radial electric field on transport time scales. To
obtain the toroidal-radial component of $\matrixtop{\pibf}_i$ to
this order, it is enough to use a distribution function good to
order $\delta_i f_{Mi}$ in \eq{pi_rzeta_ave}. For long wavelengths
and to the order of interest, the gyrophase dependent piece of the
distribution function is given by \eq{fwig_krho}, i.e., it is
proportional to $(\bv \times \bun) \cdot \nabla \psi$. Then, the
first integral in \eq{pi_rzeta_ave} vanishes. Additionally, I find
that for this gyrophase dependence, $\matrixtop{\mathbf{P}}_i
\simeq p_{i\bot} (\matI - \bun \bun) + p_{i||} \bun \bun$ and
$\int d^3v\, \bv \bv C \{ f_i \} \simeq \int d^3v\, [ (v_\bot^2/2)
(\matI - \bun \bun) + v_{||}^2 \bun \bun ] C \{ f_i \}$. With all
these simplifications, $\langle R\zun \cdot \matrixtop{\pibf}_i
\cdot \nabla \psi \rangle_\psi$ becomes to $O(\delta_i^2 p_i
R|\nabla \psi|)$
\begin{eqnarray}
\langle R\zun \cdot \matrixtop{\pibf}_i \cdot \nabla \psi
\rangle_\psi \simeq - \left \langle n_i R M (\bV_i \cdot \zun)
\frac{c}{B} (\nabla \phi \times \bun) \cdot \nabla \psi \right
\rangle_\psi \nonumber \\ + \frac{\partial}{\partial t} \left
\langle \frac{1}{2B \Omega_i} ( p_{i\bot} |\nabla \psi|^2 +
p_{i||} I^2 ) \right \rangle_\psi - \left \langle
\frac{M}{2B\Omega_i} \int d^3v\, C\{ f_i \} \left (
\frac{v_\bot^2}{2} |\nabla \psi|^2 + v_{||}^2 I^2 \right ) \right
\rangle_\psi, \label{pi_rzeta_ave_delta2}
\end{eqnarray}

\noindent where I have used $\bun \cdot \nabla \phi \simeq 0$ and
$R\zun = I\bun/B - (\bun \times \nabla \psi)/B$ to write
$cR\zun\cdot \nabla \phi \simeq - (c/B) (\nabla \phi \times \bun)
\cdot \nabla \phi$. Equation \eq{pi_rzeta_ave_delta2} is exactly
the transport of momentum found from the gyrokinetic vorticity
equation, given in \eq{viscosity_long} and \eq{gyrovisc}. For
turbulence in statistical equilibrium, the time derivative term
can be neglected. If in addition the tokamak is up-down symmetric,
the collisional term vanishes as proven in
section~\ref{sectapp_collgyrovisc} of Appendix~\ref{app_gyrovisc},
leaving only the Reynolds stress
\begin{eqnarray}
\langle R\zun \cdot \matrixtop{\pibf}_i \cdot \nabla \psi
\rangle_\psi = - \left \langle n_i R M (\bV_i \cdot \zun)
\frac{c}{B} (\nabla \phi \times \bun) \cdot \nabla \psi \right
\rangle_\psi = \nonumber \\ - \left \langle \int d^3v\, f_i R M
(\bv \cdot \zun) \frac{c}{B} (\nabla \phi \times \bun) \cdot
\nabla \psi \right \rangle_\psi. \label{Reynolds_stress_2}
\end{eqnarray}

\noindent This Reynolds stress is formally of order $\delta_i^2
p_i R|\nabla \psi|$. If the Reynolds stress were this big, the
transport of toroidal angular momentum would be much larger than
the gyroBohm estimate. It is more plausible that the Reynolds
stress averaged over time is almost zero. Therefore, the Reynolds
stress to order $\delta_i^2 p_i$ does not determine the evolution
of the long wavelength axisymmetric radial electric field on
transport time scales. This possibility does not conflict with
possible fast growth and evolution of zonal flow structure, that
happens in relatively short times, but does not transport angular
momentum through large distances.

It is difficult to prove unarguably that the Reynolds stress
\eq{Reynolds_stress_2} must vanish to order $\delta_i^2 p_i$. In
$\delta f$ flux tube codes like GS2 \cite{dorland00} and GENE
\cite{dannert05}, only the gradients of density and temperature
enter the equation for the correction to the Maxwellian [recall
\eq{deltaf_FP}]. The gradient of the velocity and hence of the
long wavelength axisymmetric radial electric field is ordered out
because the average velocity in the plasma is assumed to be small
by $\delta_i$. Then, the system does not have a preferred
direction and it is unlikely that there is any transport of
angular momentum. Quasilinear calculations suggest that in up-down
symmetric tokamaks, $\delta f$ flux tube formulations must give
zero transport \cite{peeters05}. If the average velocity is
ordered as large as the thermal velocity, the symmetry in the flux
tube is broken and there is a net radial momentum transport
\cite{waltz07}, but such a description is not relevant in many
tokamaks.

It seems more reasonable to assume that, at least in a time
averaged sense, the Reynolds stress \eq{Reynolds_stress_2} becomes
of order $\delta_i^3 p_i$. Therefore, from now on, I consider the
fast time average of equation \eq{pi_rzeta_ave} to filter the
fluctuations in the transport of toroidal angular momentum. Fast
time here is an intermediate time between the transit time of the
particle motion around the tokamak, $L/v_i$, and the much slower
transport time scale, $\delta_i^{-2} L/v_i$. Since this time
average should make the Reynolds stress of order $\delta_i^3 p_i$,
the rest of the terms in equation \eq{pi_rzeta_ave} must be
evaluated to order $\delta_i^3 p_i$. To that end, the ion
distribution function and the potential must be known to order
$\delta_i^2 f_{Mi}$ and order $\delta_i^2 T_e/e$; an order higher
than solved for in gyrokinetic codes. In
section~\ref{sect_iondistribution}, I will prove that this problem
can be circumvented under some simplifying assumptions. The rest
of this section is on how to evaluate the first term in
\eq{pi_rzeta_ave} in a convenient way.

The first term in equation \eq{pi_rzeta_ave} only depends on the
gyrophase dependent piece of the ion distribution function. For
this reason, it can be solved by employing the moment $\bv \bv
\bv$ of the Fokker-Planck equation, given by
\begin{eqnarray}
\Omega_i \int d^3v\, f_i M [ (\bv \times \bun) \bv \bv + \bv (\bv
\times \bun) \bv + \bv \bv (\bv \times \bun) ] =
\frac{\partial}{\partial t} \left ( \int d^3v\, f_i M \bv \bv \bv
\right ) \nonumber \\ + \nabla \cdot \left ( \int d^3v\, f_i M \bv
\bv \bv \bv \right ) + Ze \int d^3v\, f_i ( \nabla \phi \bv \bv +
\bv \nabla \phi \bv + \bv \bv \nabla \phi ) \nonumber \\ - \int
d^3v\, C\{ f_i \} M \bv \bv \bv.
\end{eqnarray}

\noindent Multiplying every index in this tensor by $R\zun$,
employing $R(\bun \times \zun) = \nabla \psi/B$ and flux surface
averaging gives
\begin{eqnarray}
\left \langle M \int d^3v\, (f_i - \overline{f}_i) (\bv \cdot
\nabla \psi) R^2 (\bv \cdot \zun)^2 \right \rangle_\psi =
\frac{M^2c}{3Ze} \frac{\partial}{\partial t} \left \langle \int
d^3v\, f_i R^3 (\bv \cdot \zun)^3 \right \rangle_\psi \nonumber \\
+ \frac{M^2c}{3Ze} \frac{1}{V^\prime} \frac{\partial}{\partial
\psi} V^\prime \left \langle \int d^3v\, f_i (\bv \cdot \nabla
\psi) R^3 (\bv \cdot \zun)^3 \right \rangle_\psi + c \langle R^3
(\zun \cdot \nabla \phi) (\zun \cdot \matrixtop{\mathbf{P}}_i
\cdot \zun) \rangle_\psi \nonumber \\ - \frac{M^2c}{3Ze} \left
\langle \int d^3v\, C\{ f_i \} R^3 (\bv \cdot \zun)^3 \right
\rangle_\psi. \label{gyrovisc_gyrophase_1}
\end{eqnarray}

\noindent This equation has to be evaluated to order $\delta_i^2
p_i v_i R^2 |\nabla \psi|$ to give terms of order $\delta_i^3 p_i
R | \nabla \psi |$ in \eq{pi_rzeta_ave}. The first term in the
right side is then negligible because it has a time derivative.
With turbulence that has reached statistical equilibrium and after
fast time averaging, the time derivative becomes of the order of
the transport time scale at long wavelengths, i.e., $\partial
/\partial t \sim D_{gB}/L^2 \sim \delta_i^2 v_i/L$. The
contribution of such a time derivative is negligible since it
gives a term of order $\delta_i^3 p_i v_i R^2 |\nabla \psi|$. The
second term in equation \eq{gyrovisc_gyrophase_1} is also
negligible since $\overline{(\bv \cdot \nabla \psi) (\bv \cdot
\zun)^3} = 0$ means that only the gyrophase dependent piece of the
distribution function contributes. To the order of interest, the
gyrophase dependent piece is given by \eq{fwig_krho}, and its
contribution vanishes. Then, the only terms left are
\begin{eqnarray}
\left \langle M \int d^3v\, (f_i - \overline{f}_i) (\bv \cdot
\nabla \psi) R^2 (\bv \cdot \zun)^2 \right \rangle_\psi = c
\langle R^3 (\zun \cdot \nabla \phi) (\zun \cdot
\matrixtop{\mathbf{P}}_i \cdot \zun) \rangle_\psi \nonumber \\ -
\frac{M^2c}{3Ze} \left \langle \int d^3v\, C\{ f_i \} R^3 (\bv
\cdot \zun)^3 \right \rangle_\psi. \label{gyrovisc_gyrophase_2}
\end{eqnarray}

\noindent Substituting this relation into equation
\eq{pi_rzeta_ave} gives
\begin{eqnarray}
\langle R \zun \cdot \matrixtop{\pibf}_i \cdot \nabla \psi
\rangle_\psi = \frac{Mc}{2Ze} \langle R^2 \rangle_\psi
\frac{\partial p_i}{\partial t} + \langle Mc n_i R^2 (\zun \cdot
\nabla \phi) (\bV_i \cdot \zun) \rangle_\psi \nonumber \\ -
\frac{M^2 c}{2Ze} \left \langle \int d^3v\, C\{ f_i \} R^2 (\bv
\cdot \zun)^2 \right \rangle_\psi + \frac{Mc^2}{2Ze}
\frac{1}{V^\prime} \frac{\partial}{\partial \psi} V^\prime \langle
R^3 (\zun \cdot \nabla \phi) (\zun \cdot \matrixtop{\mathbf{P}}_i
\cdot \zun) \rangle_\psi \nonumber \\ - \frac{M^3c^2}{6Z^2e^2}
\frac{1}{V^\prime} \frac{\partial}{\partial \psi} V^\prime \left
\langle \int d^3v\, C\{ f_i \} R^3 (\bv \cdot \zun)^3 \right
\rangle_\psi. \label{pi_rzeta_ave_2}
\end{eqnarray}

\noindent In the time derivative term, I used that for statistical
equilibrium and after fast time averaging, only the slow transport
time scales are left. Then, the dominant term in $\partial (\zun
\cdot \matrixtop{\mathbf{P}}_i \cdot \zun)/\partial t$ is
$\partial p_i / \partial t$. This term is determined by the
turbulent heat transport and heating in the plasma. Of the rest of
the terms in equation \eq{pi_rzeta_ave_2}, the second and the
third term require a distribution function good to order
$\delta_i^2 f_{Mi}$, and a potential good to $\delta_i^2 T_e/e$.
The last two terms only need the distribution function to order
$\delta_i f_{Mi}$ and the potential to order $\delta_i T_e/e$.

In the next section, I apply the methodology suggested in this
section to solve for the radial electric field in the
$\theta$-pinch problem presented in section~\ref{sect_thetapinch}.
In section~\ref{sect_iondistribution}, I will explain how the
distribution function can be found to second order in tokamak
geometry. Notice that this piece of the distribution function is
only needed for the irreversible transport of angular momentum
since the transport of angular momentum to order $\delta_i^2 p_i$
is enough to capture the fast evolution of zonal flow.

\section{Example: the solution of a $\theta$-pinch \label{sect_thetapinch_2}}

In section~\ref{sect_thetapinch}, I showed that, in the
$\theta$-pinch, the quasineutrality condition applied to the
second order solution \eq{solution_o2_theta}, valid to
$O(\delta_i^2 f_{Mi})$, does not determine the radial electric
field. However, I will show in this section that the electrostatic
potential can be obtained from the conservation of azimuthal
angular momentum, equivalent to the conservation of toroidal
angular momentum in tokamaks since both momentums are in the
direction of symmetry. The momentum equation has the advantage of
showing how quasineutrality depends on the long wavelength
axisymmetric potential without having to calculate the
distribution function to higher order than $O(\delta_i^2 f_{Mi})$.
The methodology I use here is presented for screw pinches and
dipolar configurations in \cite{simakov06}.

For a steady state solution, the azimuthal angular momentum must
be conserved, giving
\begin{equation}
\frac{1}{r} \frac{\partial}{\partial r} ( r^2 \hat{\br} \cdot
\matrixtop{\pibf}_i \cdot \hat{\thetabf} ) = 0,
\label{cons_theta_ang}
\end{equation}

\noindent where $\hat{\br}$ and $\hat{\thetabf}$ are the unit
vectors in the radial and azimuthal directions, with $\hat{\br}
\times \hat{\thetabf} = \bun$ [recall
figure~\ref{fig_thetapinch}], and $\matrixtop{\pibf}_i$ is the ion
viscosity, given by \eq{pi_def}. In a case without sources or
sinks of momentum, the final equation for the potential is $r^2
\hat{\br} \cdot \matrixtop{\pibf}_i \cdot \hat{\thetabf} = 0$.
Finding $r^2 \hat{\br} \cdot \matrixtop{\pibf}_i \cdot
\hat{\thetabf}$ directly from the distribution function requires a
higher order solution than the one provided by the $O(\delta_i^2
f_{Mi} v_i/L)$ gyrokinetic equation \eq{FP_lk_tpinch} used so far.
However, this problem can be circumvented by using the equivalent
to equation \eq{pi_rzeta_ave_2} for $\theta$-pinches, given by
\begin{eqnarray}
r^2 \hat{\br} \cdot \matrixtop{\pibf}_i \cdot \hat{\thetabf} =
\frac{Mr^2}{2\Omega_i} \int d^3v\, C\{ f_i \} (\bv \cdot
\hat{\thetabf})^2 - \frac{M}{6r \Omega_i} \frac{\partial}{\partial
r} \left [ \frac{r^3}{\Omega_i} \int d^3v\, C\{ f_i \} (\bv \cdot
\hat{\thetabf})^3 \right ]. \label{pi_thetapinch_1}
\end{eqnarray}

\noindent To obtain this equation, I have used that
$\hat{\thetabf} \cdot \nabla \phi = 0$ due to axisymmetry. In this
particular case, this expression can be reduced to integrals of
the gyrophase dependent piece of the distribution function, terms
much simpler to obtain to order $\delta_i^2 f_{Mi}$. To see this,
I use $\overline{(\bv \cdot \hat{\thetabf})^2} = v_\bot^2/2$,
$(\bv \cdot \hat{\thetabf})^2 - \overline{(\bv \cdot
\hat{\thetabf})^2} = (1/2)\hat{\thetabf} \cdot [ \bv_\bot \bv_\bot
- (\bv \times \bun)(\bv \times \bun) ] \cdot \hat{\thetabf} =
(1/2)[ (\bv \cdot \hat{\thetabf})^2 - (\bv \cdot \hat{\br})^2 ]$
and $\overline{(\bv \cdot \hat{\thetabf})^3} = 0$ to write
\begin{eqnarray}
r^2 \hat{\br} \cdot \matrixtop{\pibf}_i \cdot \hat{\thetabf} = -
\frac{Mr^2}{4\Omega_i} \int d^3v\, \left ( C\{ f_i \} -
\overline{C \{ f_i \}} \right ) [(\bv \cdot \hat{\br})^2 - (\bv
\cdot \hat{\thetabf})^2] \nonumber \\ + \frac{Mr^2}{2\Omega_i}
\int d^3v\, C\{ f_i \} \frac{v_\bot^2}{2} - \frac{M}{6r \Omega_i}
\frac{\partial}{\partial r} \left [ \frac{r^3}{\Omega_i} \int
d^3v\, \left ( C\{ f_i \} - \overline{C \{f_i \}} \right ) (\bv
\cdot \hat{\thetabf})^3 \right ]. \label{pi_thetapinch_2}
\end{eqnarray}

\noindent For the Krook operator $C \{ f_i \} = - \nu (f_i -
f_M)$, with $f_M$ given in \eq{f_M_tpinch},
\begin{eqnarray}
C \{ f_i \} - \overline{ C \{ f_i \} } = - \nu \Bigg \{ f_i -
\overline{f}_i - \frac{M \bv_\bot \cdot \bV_i}{T_i} f_{M0}
\nonumber \\ - \frac{M^2}{2T_i^2}  ( \bv_\bot \bv_\bot -
\overline{\bv_\bot \bv_\bot} ) : (\bV_i \bV_i) f_{M0} \Bigg \},
\label{Cwig_tpinch}
\end{eqnarray}

\noindent where, according to \eq{Vi_drift_thetapinch},
\begin{equation}
\bV_i = \frac{c}{B} \hat{\thetabf} \left ( \frac{1}{Zen_i}
\frac{\partial p_i}{\partial r} + \frac{\partial \phi}{\partial r}
\right ). \label{Vi_drift_thetapinch_2}
\end{equation}

\noindent The term $\int d^3v\, C\{ f_i \} (Mv_\bot^2/2)$ in
\eq{pi_thetapinch_2} can be found from the equation for the
perpendicular pressure, given by
\begin{equation}
\frac{\partial p_{i\bot}}{\partial t} + \frac{1}{r}
\frac{\partial}{\partial r} \left [ r M \int d^3v\, f_i (\bv \cdot
\hat{\br}) \frac{v_\bot^2}{2} \right ] = - Ze n_i \bV_{i\bot}
\cdot \nabla \phi + M \int d^3v\, C\{f_i\} \frac{v_\bot^2}{2}.
\end{equation}

\noindent Since in this case, $\partial/\partial t = 0$ and
$\bV_{i\bot} \cdot \nabla \phi = (\bV_{i\bot} \cdot \hat{\br})
(\partial \phi/\partial r) = 0$ [recall
\eq{Vi_drift_thetapinch_2}], equation \eq{pi_thetapinch_2} finally
becomes
\begin{eqnarray}
r^2 \hat{\br} \cdot \matrixtop{\pibf}_i \cdot \hat{\thetabf} = -
\frac{Mr^2}{4\Omega_i} \int d^3v\, \left ( C\{ f_i \} -
\overline{C\{f_i\}} \right ) [(\bv \cdot \hat{\br})^2 - (\bv \cdot
\hat{\thetabf})^2] \nonumber \\ + \frac{Mr}{2\Omega_i}
\frac{\partial}{\partial r} \left [r \int d^3v\, (f_i -
\overline{f}_i) (\bv \cdot \hat{\br}) \frac{v_\bot^2}{2} \right ]
\nonumber \\ - \frac{M}{6r \Omega_i} \frac{\partial}{\partial r}
\left [ \frac{r^3}{\Omega_i} \int d^3v\, \left ( C\{ f_i \} -
\overline{C\{f_i\}} \right ) (\bv \cdot \hat{\thetabf})^3 \right
]. \label{pi_thetapinch_3}
\end{eqnarray}

\noindent In this equation, only the gyrophase dependent piece of
the distribution function enters in the integrals. The corrections
$\bR_1$, $\bR_2$, $E_1$, $E_2$ and $\mu_1$ depend on the
gyrophase. Then $f_i (\bR, E, \mu, t)$ must be Taylor expanded
around $\br$, $E_0$ and $\mu_0$ to get the second order gyrophase
dependent piece. The calculation is done in
section~\ref{sectapp_gyro_tpinch} of
Appendix~\ref{app_gyrodepend_kL}, and the final result is
\begin{eqnarray}
(f_i - \overline{f}_i)_g = \frac{1}{\Omega_i} (\bv \cdot
\hat{\thetabf}) \left ( \frac{\partial f_{M0}}{\partial r} +
\frac{Ze}{T_i} \frac{\partial \phi}{\partial r} f_{M0} \right )
\nonumber \\ - \frac{r}{4\Omega_i} [ (\bv \cdot \hat{\br})^2 -
(\bv \cdot \hat{\thetabf})^2] \frac{\partial}{\partial r} \left [
\frac{1}{r\Omega_i} \left ( \frac{\partial f_{M0}}{\partial r} +
\frac{Ze}{T_i} \frac{\partial \phi}{\partial r} f_{M0} \right )
\right ] \nonumber \\ - \frac{Mc}{4B\Omega_i} [ (\bv \cdot
\hat{\br})^2 - (\bv \cdot \hat{\thetabf})^2] \frac{\partial
\phi}{\partial r} \left [ \frac{\partial}{\partial r} \left (
\frac{f_{M0}}{T_i} \right ) + \frac{Ze}{T_i^2} \frac{\partial
\phi}{\partial r} f_{M0} \right ], \label{gp_tpinch}
\end{eqnarray}

\noindent where $f_{M0}$ is given in \eq{f_M0_tpinch}, and the
subindex $g$ indicates the non-collisional origin of this
gyrophase dependence. The gyrophase dependent piece given by
\eq{fwig_sol} is also necessary. For the Krook operator it becomes
\begin{equation}
(f_i - \overline{f}_i)_c = - \frac{\nu}{\Omega_i^2} f_{M0} \left (
\frac{Mv^2}{2T_i} - \frac{5}{2} \right ) \bv_\bot \cdot \nabla \ln
T_i. \label{gp_coll_tpinch}
\end{equation}

\noindent Employing equations \eq{Cwig_tpinch},
\eq{Vi_drift_thetapinch_2}, \eq{gp_tpinch} and
\eq{gp_coll_tpinch}, I find
\begin{eqnarray}
\frac{Mr^2}{4\Omega_i} \int d^3v\, (C\{ f_i \} -
\overline{C\{f_i\}} ) [(\bv \cdot \hat{\br})^2 - (\bv \cdot
\hat{\thetabf})^2] = \nonumber \\ \frac{\nu r^3 p_i}{4\Omega_i^2}
\frac{\partial}{\partial r} \left [ \frac{c}{rB} \left (
\frac{\partial \phi}{\partial r} + \frac{1}{Zen_i} \frac{\partial
p_i}{\partial r} \right ) \right ] + \frac{\nu r^3}{4\Omega_i^2}
\frac{\partial}{\partial r} \left ( \frac{1}{r} \frac{p_i}{M
\Omega_i} \frac{\partial T_i}{\partial r} \right ),
\end{eqnarray}
\begin{equation}
\frac{Mr}{2\Omega_i} \frac{\partial}{\partial r} \left [r \int
d^3v\, (f_i - \overline{f}_i) (\bv \cdot \hat{\br})
\frac{v_\bot^2}{2} \right ] = - \frac{r}{\Omega_i}
\frac{\partial}{\partial r} \left ( \frac{\nu p_i r}{M \Omega_i^2}
\frac{\partial T_i}{\partial r} \right )
\end{equation}

\noindent and
\begin{equation}
\frac{M}{6r \Omega_i} \frac{\partial}{\partial r} \left [
\frac{r^3}{\Omega_i} \int d^3v\, (C\{ f_i \} - \overline{C\{ f_i
\}}) (\bv \cdot \hat{\thetabf})^3 \right ] = - \frac{1}{2r
\Omega_i} \frac{\partial}{\partial r} \left ( \frac{\nu p_i r^3}{M
\Omega_i^2} \frac{\partial T_i}{\partial r} \right ).
\end{equation}

\noindent Substituting these results into \eq{pi_thetapinch_3},
$r^2 \hat{\br} \cdot \matrixtop{\pibf}_i \cdot \hat{\thetabf} = 0$
gives
\begin{equation}
c \left ( \frac{\partial \phi}{\partial r} + \frac{1}{Zen_i}
\frac{\partial p_i}{\partial r} \right ) = r B(r) \int_0^r
dr^\prime\; \frac{U ( r^\prime )}{r^\prime} \left [
\frac{\partial}{\partial r^\prime} \ln B(r^\prime) - \frac{3}{2}
\frac{\partial}{\partial r^\prime} \ln \left ( \frac{p_i
(r^\prime) U (r^\prime)}{r^\prime} \right ) \right ],
\label{MaxBoltz_tpinch}
\end{equation}

\noindent where $U = (2/M\Omega_i) (\partial T_i/\partial r)$.
Notice the difference between this equation and
\eq{fake_qn_theta_2}. In particular, notice that for an isothermal
$f_{M0}$, $\partial T_i / \partial r = 0$, a radial
Maxwell-Boltzmann response is recovered from \eq{MaxBoltz_tpinch}
as expected, but this is not a feature of the non-physical forms
\eq{fake_qn_theta} and \eq{fake_qn_theta_2}.

Finally, I remark that equation \eq{pi_thetapinch_3} gives a
radial transport of toroidal angular momentum $\hat{\br} \cdot
\matrixtop{\pibf}_i \cdot \hat{\thetabf} \sim \delta_i^2
(\nu/\Omega_i) p_i$, corresponding to the term in the vorticity
equation
\begin{equation}
\nabla \cdot \left [ \frac{c}{B} \bun \times (\nabla \cdot
\matrixtop{\pibf}_i ) \right ] = - \frac{1}{r}
\frac{\partial}{\partial r} \left [ \frac{c}{rB}
\frac{\partial}{\partial r} (r^2 \hat{\br} \cdot
\matrixtop{\pibf}_i \cdot \hat{\thetabf} ) \right ] \sim \frac{\nu
\delta_i^3}{\Omega_i} e n_e v_i/L.
\end{equation}

\noindent The radial current density represented by this term is
too small to be recovered with a gyrokinetic equation good only to
order $\delta_i^2 f_{Mi} v_i/L$, as already shown in
section~\ref{sect_thetapinch}.

\section{Distribution function and potential to second order} \label{sect_iondistribution}

To evaluate \eq{pi_rzeta_ave_2}, the ion distribution function and
the potential have to be found to order $\delta_i^2 f_{Mi}$ and
$\delta_i^2 T_e/e$, respectively. In this section, I show how both
the distribution function and the potential can be calculated to
higher order without the full second order Fokker-Planck and
vorticity equations. I take advantage of the usually small ratio
$1/q \sim B_p/B \ll 1$, where $q(\psi) = (2\pi)^{-1} \oint d\theta
(\bB \cdot \nabla \zeta/\bB \cdot \nabla \theta)$ is the safety
factor, and $B_p = |\nabla \psi|/R$ is the poloidal component of
the magnetic field. Expanding in $q \gg 1$, I will find the ion
distribution function to order $q \delta_i^2 f_{Mi}$, neglecting
terms of order $\delta_i^2 f_{Mi}$. The potential is calculated
consistently with this higher order solution for $f_i$ by
employing a higher order vorticity equation. Importantly, the
vorticity equations obtained in this section only give to higher
order the short wavelength, non-axisymmetric part of the
potential. The axisymmetric piece of the potential must be
calculated employing another equation. The axisymmetric component
is composed of the flux surface averaged piece, given by the
conservation of toroidal angular momentum equation
\eq{toro_angmom} and the higher order viscosity
\eq{pi_rzeta_ave_2}, and a poloidally varying modification. The
poloidal variation is the Geodesic Acoustic Mode (GAM) response
\cite{winsor68, gao08}; the initial transient of an axisymmetric
perturbation in the potential. The perturbation initially induces
poloidal density variations that rapidly Landau damp towards a
constant zonal flow known as the Hinton-Rosenbluth residual
\cite{rosenbluth98, hinton99}. This initial decay or GAM is
axisymmetric and thus does not drive radial transport. It can,
however, shear the turbulence. The lower order vorticity equations
\eq{vort_type1} and \eq{vort_type2} reproduce the lower order GAM
response. The new higher order vorticity equations derived in this
subsection will not have, however, higher order corrections to
GAMs. It is relatively straightforward to calculate the higher
order corrections analytically, but they complicate the vorticity
equations unnecessarily since in reality the GAM response is
believed to be less important than the Hinton-Rosenbluth residual,
that is adequately kept by the gyrokinetic ion Fokker-Planck
equation \eq{FP_final} and the toroidal angular momentum
conservation equation \eq{toro_angmom}. For this reason, I will
drop the higher order axisymmetric corrections to the gyrokinetic
vorticity equations \eq{vort_type1} and \eq{vort_type2}.

In subsection~\ref{sub_iondistribution}, I show that equation
\eq{FP_final} is enough to calculate the distribution function to
order $q \delta_i^2 f_{Mi}$. I also argue that the second order
corrections $\bR_2$ and $E_2$ are not needed since they only
provide corrections of order $\delta_i^2 f_{Mi}$. Employing these
two results in subsection~\ref{sub_potential}, I extend the
gyrokinetic vorticity equations \eq{vort_type1} and
\eq{vort_type2} to give the electrostatic potential consistent
with the higher order $f_i$. To do so, I develop an extended
gyrokinetic equation in the physical phase space, as I did in
subsection~\ref{sub_GKeq}, but now to order $q \delta_i^2 f_{Mi}
v_i/L$. Taking moments of this equation, I obtain the new extended
vorticity equations that retain the short wavelength,
non-axisymmetric pieces of the potential to higher order.

\subsection{Higher order ion distribution function} \label{sub_iondistribution}

To find $f_i (\bR, E, \mu, t)$ to order $\delta_i^2 f_{Mi}$, it is
necessary to solve a higher order gyrokinetic Fokker-Planck
equation. Similarly, for the higher order potential, it is
necessary to find a higher order gyrokinetic vorticity equation.
In this section, I show that under certain assumptions, the second
order gyrokinetic Fokker-Planck and vorticity equations can be
easily deduced from their first order versions.

\begin{figure}
\begin{center}
\includegraphics[width = \textwidth]{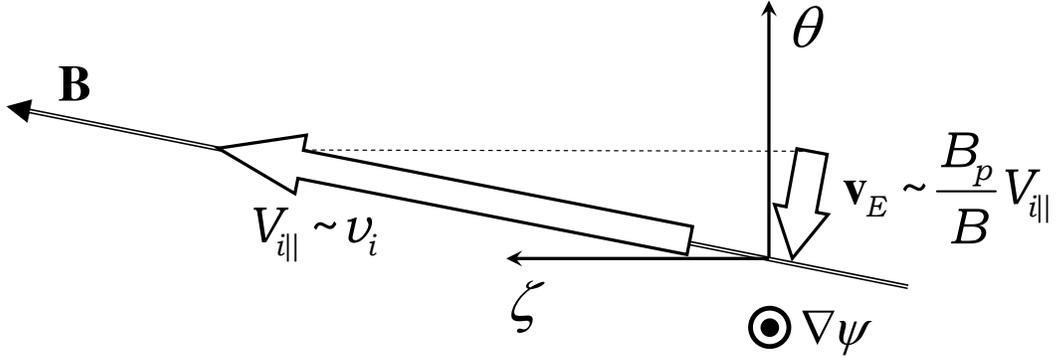}
\end{center}

\vspace{-7cm} \caption[High flow ordering for the ion parallel
velocity with $B_p/B \ll 1$]{High flow ordering for the ion
parallel velocity with $B_p/B \ll 1$. Notice that the poloidal
projection of the $E\times B$ drift must be comparable to the
poloidal projection of the ion parallel velocity, giving $\bv_E
\sim (B_p/B) V_{i||} \sim (B_p/B) v_i \ll v_i$.} \label{fig_lowBp}
\end{figure}

There has already been some work in transport of toroidal angular
momentum in gyrokinetics. For these studies, it was necessary to
realize that the Reynolds stress tends to vanish to order
$\delta_i^2 p_i$ in the low flow limit, becoming of order
$\delta_i^3 p_i$. In references \cite{peeters05, waltz07,
peeters07} the revised approach is ordering the parallel velocity
as comparable to the ion thermal speed. Since for sonic
velocities, the plasma can only rotate toroidally \cite{hinton85,
catto87}, a sonic parallel velocity requires, in general, a sonic
$E\times B$ drift to cancel its poloidal component. However, sonic
$E\times B$ drifts invalidate the gyrokinetic derivation of
chapter~\ref{chap_gyrokinetics}. To avoid this problem, references
\cite{peeters05, waltz07, peeters07} do not reach sonic $E\times
B$ velocities because they take advantage of the expansion
parameter $B_p/B \ll 1$. The perpendicular $E\times B$ drift is
small compared to the thermal speed by $B_p/B$, making the
traditional gyrokinetic formulation based on subsonic $E\times B$
motion still valid [see figure~\ref{fig_lowBp}]. Under these
assumptions, the toroidal velocity is given by $- cR (\partial
\phi/\partial \psi)$, and the radial transport of toroidal angular
momentum is of order $\delta_i^2 p_i$, as can be found from
equation \eq{pi_rzeta_ave_delta2}. The term with the time
derivative in equation \eq{pi_rzeta_ave_delta2} is still
negligible if the turbulence reaches its statistical equilibrium,
but the collisional term, proven to vanish for up-down symmetric
tokamaks in section~\ref{sectapp_collgyrovisc} of
Appendix~\ref{app_gyrovisc}, contributes to order $\delta_i^2 p_i$
because the sonic parallel velocity breaks the up-down symmetry by
introducing a preferred direction in the Maxwellian. Similarly,
the Reynolds stress \eq{Reynolds_stress_2} in this case is of
order $\delta_i^2 p_i$. In reference \cite{waltz07}, the Reynolds
stress is calculated employing the $\delta f$ code GYRO with sonic
parallel velocities, and it does not vanish to order $\delta_i^2
p_i$, as expected.

This approach has the disadvantage of making the toroidal velocity
only depend on the radial electric field $\partial \phi / \partial
\psi$. Density and temperature gradients cannot compete with the
radial electric field, and therefore it is not possible to recover
naturally the isothermal radial Maxwell-Boltzmann solution, or the
dependence of the velocity on the temperature gradient. I propose
an alternative approach with subsonic velocities that at the same
time avoids solving a full second order gyrokinetic equation. It
exploits the extra expansion parameter $B_p/B \ll 1$ in a
different manner.

In the new ordering with $B_p/B \ll 1$, the parallel gradient is
of order $1/qR$, with $R$ the major radius and $q \sim B/B_p \gg
1$ the safety factor. As in section~\ref{sect_assumptions}, the
ion and electron zeroth order distribution functions are assumed
to be stationary Maxwellians with only radial dependence, i.e.,
$f_i \simeq f_{Mi} (\psi)$ and $f_e \simeq f_{Me}(\psi)$. With the
new orderings, the size of the first order correction to the
Maxwellian $h_{i1}$ changes depending on the nature of the
correction, i.e., depending on whether it is turbulent, due to
non-axisymmetric potential fluctuations, or neoclassical, given by
the long wavelength, axisymmetric pieces. For the neoclassical
banana regime pieces, the term $\bv_E \cdot \nabla_\bR f_i$ is
negligible [recall the discussion in
section~\ref{sect_assumptions}], and the neoclassical correction
$h_{i1}^\mathrm{nc}$ is determined by a balance between the
parallel streaming term $u \bun \cdot \nabla_\bR
h_{i1}^\mathrm{nc} \sim (v_i/qR) h_{i1}^\mathrm{nc}$ and the
magnetic drift term $\bv_M \cdot \nabla_\bR f_{Mi} \sim (\rho_i/R)
v_i f_{Mi}/a$, giving a neoclassical piece of order
$h_{i1}^\mathrm{nc} \sim q \delta_i f_{Mi}$ when the transit
average collisional constraint is satisfied. On the other hand, in
tokamaks, turbulence is driven by toroidal drift wave modes in
which the parallel streaming term $u \bun \cdot \nabla_\bR f_i$ is
of secondary importance. The turbulent contributions
$h_{i1}^\mathrm{tb}$ are determined by the competition between the
magnetic drift term $\bv_M \cdot \nabla_\bR h_{i1}^\mathrm{tb}$
and the $E\times B$ drift term $\bv_E^\mathrm{tb} \cdot \nabla_\bR
h_{i1}^\mathrm{tb}$. The orderings of \eq{order_kbot} still hold,
giving that for $k_\bot \rho_i \sim 1$, $h_{i1}^\mathrm{tb}/f_{Mi}
\sim e\phi^\mathrm{tb}/T_e \sim \delta_i$. The turbulent
contribution $h_{i1}^\mathrm{tb}$ is then smaller than the
neoclassical piece $h_{i1}^\mathrm{nc}$ by a factor of $1/q$.

To calculate the axisymmetric radial electric field in a
completely general manner, the gyrokinetic treatment needs to be
extended to provide the pieces $h_{i2} \sim \delta_i^2 f_{Mi}$ of
the ion distribution function. This would require calculating the
gyrokinetic Fokker-Planck equation \eq{FP_final} to higher order,
i.e., obtaining the time derivatives of the gyrokinetic variables
$\bR$, $E$ and $\mu$ to an order higher in $\delta_i$. However, as
just noted, there are terms that are larger by $q \gg 1$. Instead
of calculating the complete $O(\delta_i^2 f_{Mi} v_i/L)$
gyrokinetic Fokker-Planck equation, I will only keep the terms
that are larger by $q$. To identify these terms, I let $f_i =
f_{Mi} + h_{i1} + h_{i2} + \ldots$ and then write the gyrokinetic
equation for the second order perturbation as
\begin{eqnarray}
\frac{\partial h_{i2}}{\partial t} + [ u \bun (\bR) + \bv_d ]
\cdot \nabla_\bR h_{i2} - \langle C \{f_i\} \rangle^{(2)} = -
\bv_d \cdot \nabla_\bR h_{i1}^\mathrm{nc} \nonumber \\ -
\dot{\bR}^{(2)} \cdot \nabla_\bR (f_{Mi} + h_{i1}^\mathrm{tb}) +
\frac{Ze}{M} [ u \bun (\bR) + \bv_M ] \cdot \nabla_\bR \phiave
\frac{\partial h_{i1}}{\partial E} + \dot{E}^{(2)} \frac{M
f_{Mi}}{T_i}, \label{delta2f_FP}
\end{eqnarray}

\noindent with $\langle C \{ f_i \} \rangle^{(2)} = \langle C \{
f_i \} \rangle - \langle C^{(\ell)} \{ f_{Mi} + h_{i1} \}
\rangle$, $\dot{\bR}^{(2)} = \langle \dot{\bR} \rangle -
[u\bun(\bR) + \bv_d]$ and $\dot{E}^{(2)} = \langle \dot{E} \rangle
+ (Ze/M) [u\bun(\bR) + \bv_M] \cdot \nabla_\bR \phiave$. Here,
$\langle C \{ f_i \} \rangle$, $\langle \dot{E} \rangle$ and
$\langle \dot{\bR} \rangle$ are calculated to order $\delta_i^2
\nu_{ii} f_{Mi}$, $\delta_i^2 v_i^3/L$ and $\delta_i^2 v_i$,
respectively; an order higher than in equation \eq{FP_final}.
Notice that the first order correction $h_{i1}$ enters differently
depending on its nature. The turbulent short wavelength piece
$h_{i1}^\mathrm{tb}$ has large gradients and it is multiplied by
the small quantity $\dot{\bR}^{(2)}$, while the gradient of the
neoclassical piece $h_{i1}^\mathrm{nc}$ is small but is multiplied
by the lowest order term $\bv_d \gg \dot{\bR}^{(2)}$.

On the right side of equation \eq{delta2f_FP}, the dominant terms
are $- \bv_d \cdot \nabla_\bR h_{i1}^\mathrm{nc}$ and $(Ze/M) [ u
\bun (\bR) + \bv_M ] \cdot \nabla_\bR \phiave (\partial
h_{i1}^\mathrm{nc}/\partial E)$ because $h_{i1}^\mathrm{nc}$ is
larger than all other terms by a factor of $q$. The higher order
corrections $\dot{\bR}^{(2)}$ and $\dot{E}^{(2)}$ are finite
gyroradius correction that do not contain any $q$ factors. Since
$h_{i1}^\mathrm{nc}$ determines the parallel velocity and the
parallel heat flow, the term $\bv_d \cdot \nabla_\bR
h_{i1}^\mathrm{nc}$ represents the effect of the gradient of the
parallel velocity and parallel heat flow on turbulence. This term
is not kept in $\delta f$ flux tube codes because only the short
wavelength pieces of the first order correction to the
distribution function are calculated. We see from \eq{delta2f_FP}
that it is possible to simply add this neoclassical term to the
$\delta f$ gyrokinetic equation along with $(Ze/M) [ u \bun(\bR) +
\bv_M ] \cdot \nabla_\bR \phiave (\partial
h_{i1}^\mathrm{nc}/\partial E)$. In full $f$ codes, the
distribution function is solved from the lowest order equation
\eq{FP_final}. In this equation, the terms $-\bv_d \cdot
\nabla_\bR h_{i1}^\mathrm{nc}$ and $(Ze/M) [ u \bun (\bR) + \bv_M
] \cdot \nabla_\bR \phiave (\partial h_{i1}^\mathrm{nc}/\partial
E)$ are naturally included, so it is not necessary to write the
gyrokinetic equation to higher order than \eq{FP_final}.

As for $h_{i1}$, the function $h_{i2}$ has a turbulent piece
$h_{i2}^\mathrm{tb}$, and a neoclassical piece
$h_{i2}^\mathrm{nc}$. The turbulent piece is given by the balance
between the drifts $\bv_d \cdot \nabla_\bR h_{i2}^\mathrm{tb} \sim
\delta_i v_i k_\bot h_{i2}^\mathrm{tb}$ and the driving term
$\bv_d \cdot \nabla_\bR h_{i1}^\mathrm{nc} \sim q \delta_i^2 v_i
f_{Mi}/a$, giving $h_{i2}^\mathrm{tb} \sim q \delta_i^2 f_{Mi}$
for $k_\bot \rho_i \sim 1$. The neoclassical piece is a result of
a balance between the parallel streaming term $u \bun \cdot
\nabla_\bR h_{i2}^\mathrm{nc} \sim (v_i/qR) h_{i2}^\mathrm{nc}$
and the magnetic drift term $\bv_M \cdot \nabla_\bR
h_{i1}^\mathrm{nc} \sim (\rho_i/R) v_i q \delta_i f_{Mi}/a$,
leading to $h_{i2}^\mathrm{nc} \sim q^2 \delta_i^2 f_{Mi}$.

Since $h_{i2}^\mathrm{tb}$ is larger than $\delta_i^2 f_{Mi}$ by a
factor of $q$, the second order corrections $\bR_2$ and $E_2$ give
negligible contributions to the second order piece of the
distribution function. To see this, Taylor expand $f_i(\bR, E,
\mu, t)$ around $\bR_g = \br + \Omega_i^{-1} \bv \times \bun$,
$E_0$ and $\mu_0$. Then, the terms $\bR_2 \cdot \nabla_{\bR_g}
f_{ig}$ and $E_2 (\partial f_{Mi}/\partial E_0)$, of order
$\delta_i^2 f_{Mi}$, are negligible. This fact simplifies the
integration in velocity space in \eq{pi_rzeta_ave_2}, since it is
enough to keep only the first order corrections $\bR_1$, $E_1$ and
$\mu_1$. Finally, the gyrophase dependent piece $\fwig_i$, defined
in \eq{fwig_sol}, vanishes [$\fwig_i \simeq - \Omega_i^{-1}
\int^\varphi d\varphi^\prime ( C^{(\ell)} \{ h_{i1}^\mathrm{nc} \}
- \langle C^{(\ell)} \{ h_{i1}^\mathrm{nc} \} \rangle ) = 0$ since
$\partial h_{i1}^\mathrm{nc} / \partial \varphi_0 = 0$].

\subsection{Higher order electrostatic potential} \label{sub_potential}

In this subsection, I find the gyrokinetic vorticity equations
\eq{vort_type1} and \eq{vort_type2} to higher order. To simplify
the derivation, I limit myself to the short wavelength,
non-axisymmetric contributions to the vorticity equation -- the
ones responsible for the turbulence. The flux surface averaged
component of the potential will be given by the conservation of
toroidal angular momentum. For the usually unimportant GAM
response \cite{winsor68, gao08}, it is enough to retain the first
order terms, already in the lower order gyrokinetic vorticity
equations \eq{vort_type1} and \eq{vort_type2}.

Finding the higher order vorticity equation becomes a simple task
because the higher order corrections to the gyrokinetic variables
$\bR_2$ and $E_2$ are negligible. Quasineutrality must be enforced
for the gyrokinetic Fokker-Planck equation with the new terms
$\bv_d \cdot \nabla_\bR h_{i1}^\mathrm{nc}$ and $(Ze/M) [ u \bun
(\bR) + \bv_M ] \cdot \nabla_\bR \phiave (\partial
h_{i1}^\mathrm{nc}/\partial E)$, and a $O(q\delta_i^2 f_{Mi})$
distribution function given by
\begin{equation}
f_i (\bR, E, \mu, t) \simeq f_{iG} + \frac{Ze\phiwig}{M} \left [
\frac{\partial}{\partial E_0} (f_{Mi} + h_{i1}^\mathrm{nc}) +
\frac{1}{B} \frac{\partial h_{i1}^\mathrm{nc}}{\partial \mu_0}
\right ], \label{fig_delta2}
\end{equation}

\noindent with $f_{iG} \equiv f_i (\bR_g, E_0, \mu_g, t)$, $\bR_g
= \br + \Omega_i^{-1} \bv \times \bun$ and $\mu_g = \mu -
Ze\phiwig/MB \neq \mu_0$. In expression \eq{fig_delta2},
$h_{i1}^\mathrm{nc} (\bR, E, \mu, t) \simeq h_{i1}^\mathrm{nc}
(\br, E_0, \mu_0, t)$. It is convenient to extract the
$Ze\phiwig/MB$ portion of $\mu$ by introducing $\mu_g$ so we can
take advantage of previous results.

As I did in section~\ref{sect_transportGK}, I will find the
vorticity equations by taking moments of the gyrokinetic
Fokker-Planck equation \eq{FP_final}; in this case to order $q
\delta_i^2 f_{Mi} v_i/L$. To write \eq{FP_final} in physical phase
space, I will first find the evolution equation for $f_{iG}$.
Equation \eq{FP_final} gives the evolution of $f_{iG}$ if $\bR$ is
replaced by $\bR_g$, $E$ by $E_0$ and $\mu$ by $\mu_g$, giving to
order $q \delta_i^2 f_{Mi} v_i/L$
\begin{eqnarray}
\left. \frac{\partial f_{iG}}{\partial t} \right |_{\br, \bv} + [
u_g \bun(\bR_g) + \bv_{dg} ] \cdot \left [ \nabla_{\bR_g} f_{iG} -
\frac{Ze}{M} \nabla_{\bR_g} \phiave \frac{\partial}{\partial E_0}
( f_{Mi} + h_{i1}^\mathrm{nc}) \right ] = \nonumber \\ \langle C
\{ f_i \} \rangle |_{\bR \rightarrow \bR_g, E \rightarrow E_0, \mu
\rightarrow \mu_g}, \label{FP_delta2_phys_1}
\end{eqnarray}

\noindent with
\begin{eqnarray}
\bv_{dg} \equiv \frac{\mu_g}{\Omega_i(\bR_g)} \bun(\bR_g) \times
\nabla_{\bR_g} B (\bR_g) + \frac{u_g^2}{\Omega_i(\bR_g)}
\bun(\bR_g) \times \kappabf (\bR_g) \nonumber \\ -
\frac{c}{B(\bR_g)} \nabla_{\bR_g} \phiave ( \bR_g, \mu_g, t )
\times \bun (\bR_g) \label{vdg_o2}
\end{eqnarray}

\noindent and
\begin{equation}
u_g \equiv \sqrt{ 2[E_0 - \mu_g B (\bR_g)]} = \sqrt{ 2[E - \mu B
(\bR)] } = u, \label{ug_o2}
\end{equation}

\noindent where $E_1 = Ze\phiwig/M$ and $\mu - \mu_g = Ze
\phiwig/MB(\bR_g)$ cancel exactly to give the second equality. In
equation \eq{FP_delta2_phys_1}, I have neglected terms of order
$\delta_i^2 f_{Mi}$ by taking the approximation $\partial f_{iG}
/\partial E_0 \simeq \partial ( f_{Mi} + h_{i1}^\mathrm{nc})/
\partial E_0$.

Equation \eq{FP_delta2_phys_1} needs to be written in the physical
phase space variables. To the order of interest, $\nabla_{\bR_g}
f_{iG} = \nabla_{\bR_g} \br \cdot \nablaave f_{iG} +
\nabla_{\bR_g} \mu_0 (\partial h_{i1}^\mathrm{nc}/\partial \mu_0)$
and $\nabla_{\bR_g} \phiave \simeq \nabla_{\bR_g} \br \cdot
\nablaave \phiave$, with $\nabla_{\bR_g} \mu_0 = - \nabla_{\bR_g}
\mu_{10} \simeq - \nablaave \mu_{10}$ and $\mu_g - \mu_0 =
\mu_{10} = \mu_1 - Ze\phiwig/MB$. Then, equation
\eq{FP_delta2_phys_1} becomes
\begin{eqnarray}
\left. \frac{\partial f_{iG}}{\partial t} \right |_{\br, \bv} + [
u \bun(\bR_g) + \bv_{M0} + \bv_{E0} ] \cdot \nabla_{\bR_g} \br
\cdot \left [ \nablaave f_{iG} - \frac{Ze}{M} \nablaave \phiave
\frac{\partial}{\partial E_0} ( f_{Mi} + h_{i1}^\mathrm{nc})
\right ] \nonumber \\ - v_{||} \bun \cdot \nablaave \mu_{10}
\frac{\partial h_{i1}^\mathrm{nc}}{\partial \mu_0} = \langle C \{
f_i \} \rangle. \label{FP_delta2_phys_2a}
\end{eqnarray}

\noindent Here, the term $(\bv_{dg} - \bv_{M0} - \bv_{E0}) \cdot
\nabla_\bR f_{iG} \sim \delta_i^2 f_{Mi} v_i/L$ has been
neglected. In the collisional term $\langle C \{ f_i \} \rangle$
in \eq{FP_delta2_phys_2a}, it is necessary to consider pieces of
order $q \delta_i^2 \nu_{ii} f_{Mi}$ that come from the
gyroaverage of $C^{(\ell)} \{ h_{i1}^\mathrm{nc} \}$ performed
holding the higher order gyrokinetic variables $\bR_g$, $E$ and
$\mu$ fixed, with $C^{(\ell)}$ the linearized collision operator.
These terms are not considered in
Appendix~\ref{app_collisiontransport}, where the gyrokinetic
variables are approximated by $\bR_g$, $E_0$ and $\mu_0$.
Fortunately, the collision frequency is usually small, making
these terms negligible. Ignoring these terms and assuming that the
collision operator can be treated as in
Appendix~\ref{app_collisiontransport} is reasonable and simplifies
the rest of the derivation. Finally, employing the results in
section~\ref{sectapp_conserv} of Appendix~\ref{app_transportGK},
equation \eq{FP_delta2_phys_2a} gives
\begin{eqnarray}
\left. \frac{\partial f_{iG}}{\partial t} \right |_{\br, \bv} +
\frac{v_{||}}{B} \Bigg \{ \nablaave \cdot \left [ \frac{B}{v_{||}}
f_{iG} \left ( \dot{\bR} \cdot \nabla_{\bR_g} \br \right ) \right
] - \frac{\partial}{\partial \mu_0} \left [ ( f_{Mi} +
h_{i1}^\mathrm{nc} ) \bB \cdot \nablaave \mu_{10} \right ]
\nonumber \\ - \frac{\partial}{\partial \varphi_0} \left [ (
f_{Mi} + h_{i1}^\mathrm{nc} ) \bB \cdot \nablaave \varphi_{10}
\right ] - \frac{\partial}{\partial E_0} \left [ \frac{B}{v_{||}}
(f_{Mi} + h_{i1}^\mathrm{nc}) \frac{Ze}{M} \left ( \dot{\bR} \cdot
\nabla_{\bR_g} \br \right ) \cdot \nablaave \phiave \right ] \Bigg
\} \nonumber \\ = \langle C \{ f_i \} \rangle,
\label{FP_delta2_phys_2}
\end{eqnarray}

\noindent with $\dot{\bR} \cdot \nabla_{\bR_g} \br$ from
\eq{Rdot_gradRr}. Notice that equation \eq{FP_delta2_phys_2} is
equivalent to equation \eq{eqGK_final} except for the changes
$f_{Mi} \rightarrow f_{Mi} + h_{i1}^\mathrm{nc}$ and $f_{ig}
\rightarrow f_{iG}$.

The same moment equations that were obtained with \eq{eqGK_final}
can be found for equation \eq{FP_delta2_phys_2}, but now with
$f_{iG}$ instead of $f_{ig}$ and $f_{Mi} + h_{i1}^\mathrm{nc}$
instead of $f_{Mi}$. The differences between equations
\eq{eqGK_final} and \eq{FP_delta2_phys_2} are enough to invalidate
some of the cancellations that were found in
Appendix~\ref{app_n_v}. For example, in the momentum conservation
equation for ions \eq{vtransp_krho}, the piece of the viscosity $M
\int d^3v\, f_{ig} v_{||0} \bv_\bot$, with $v_{||0}$ given in
\eq{vpar0}, reduced to $M \int d^3v\, f_{ig} v_{||} \bv_\bot$
because the small correction $M (v_{||0} - v_{||}) \bv_\bot$,
composed of terms that are either odd in $v_{||}$ or in
$\bv_\bot$, vanished when integrated over the lowest order
Maxwellian. In the new vorticity equation, terms of order $q
\delta_i^2 p_i/L$ must be retained in the momentum conservation
equation, leading to a new non-vanishing term $M \int d^3v\,
h_{i1}^\mathrm{nc} (v_{||0} - v_{||}) \bv_\bot$. Fortunately, most
terms like this one are axisymmetric and only enter in the
calculation of the flux surface averaged radial electric field and
the higher order GAM response. To calculate the flux surface
averaged radial electric field, the toroidal angular momentum
conservation equation is to be used, and to simplify the equations
I ignore the higher order corrections to the GAM response that is
expected to be unimportant, as already discussed. Therefore, the
vorticity equation is only employed to solve for the
non-axisymmetric, turbulent fluctuations in the potential, and
many of the cancellations employed in Appendix~\ref{app_n_v} are
recovered. Additionally, the difference $f_{iG} - f_{ig} \simeq
(\mu_g - \mu_0) (\partial h_{i1}^\mathrm{nc}/\partial \mu_0) \sim
q \delta_i^2 f_{Mi}$ is also a long wavelength, axisymmetric
piece, and it will not enter in the equations for the
non-axisymmetric electric field. Then, $f_{iG}$ can be
approximated by the simpler distribution function $f_{ig} \equiv
f_i (\bR_g, E_0, \mu_0, t)$. The generalized particle conservation
equation can be found following section~\ref{sectapp_ntransp} of
Appendix~\ref{app_n_v} by ignoring the purely axisymmetric
contributions of $h_{i1}^\mathrm{nc}$. Then, the non-axisymmetric
component of particle conservation is
\begin{equation}
\frac{\partial}{\partial t} \left ( n_i - n_{ip}^{(2)} \right ) +
\nabla \cdot \left ( n_i V_{ig||}^{(2)} \bun + n_i \bV_{igd} + n_i
\bV_{iE} + n_i \tilde{\bV}_i + n_i \bV_{iC} \right ) = 0,
\label{ntransp_krho_delta2}
\end{equation}

\noindent with
\begin{equation}
n_{ip}^{(2)} = - \int d^3v\, \frac{Ze\phiwig}{T_i} f_{Mi} + \int
d^3v\, \frac{Ze\phiwig}{M} \left ( \frac{\partial
h_{i1}^\mathrm{nc}}{\partial E_0} + \frac{1}{B} \frac{\partial
h_{i1}^\mathrm{nc}}{\partial \mu_0} \right ) \label{nip_o2}
\end{equation}

\noindent and
\begin{equation}
n_i V_{ig||}^{(2)} = \int d^3v\, f_{ig} v_{||} \neq n_i V_{i||}.
\label{Vigpar_o2}
\end{equation}

\noindent The rest of the terms in \eq{ntransp_krho_delta2} are as
defined in subsection~\ref{sub_n}, although now the turbulent
second order contribution $h_{i2}^\mathrm{tb}$ implicitly enters
in the integrals via the solution $f_i$ to the full gyrokinetic
equation \eq{FP_final}. The non-axisymmetric piece of the momentum
conservation equation can be obtained following
section~\ref{sectapp_vtransp} of Appendix~\ref{app_n_v}, finally
giving
\begin{eqnarray}
\frac{\partial}{\partial t} \left ( n_i M \bV_{ig} \right ) + \bun
[ \bun \cdot \nabla p_{ig||} + (p_{ig||} - p_{ig\bot}) \nabla
\cdot \bun + \nabla \cdot \pibf_{ig||} ] + \nabla \cdot
\matrixtop{\pibf}_{ig\times} = \nonumber \\ - Zen_i \bun \left (
\bun + \frac{V_{i||}^\mathrm{nc}}{\Omega_i} \bun \times \kappabf
\right ) \cdot \nabla \phi + \tilde{F}_{iE}^{(2)} \bun +
\mathbf{F}_{iB}^{(2)} + \mathbf{F}_{iC},
\label{vtransp_krho_delta2}
\end{eqnarray}

\noindent with $n_i V_{i||}^\mathrm{nc} = \int d^3v\, v_{||}
h_{i1}^\mathrm{nc}$,
\begin{equation}
\tilde{F}_{iE}^{(2)} = Ze \int d^3v\, (f_{Mi} +
h_{i1}^\mathrm{nc}) \left ( \bun + \frac{v_{||}}{\Omega_i} \bun
\times \kappabf + \frac{1}{\Omega_i} \nablaave \times \bv_\bot
\right ) \cdot \nablaave \phiwig \label{force_E_o2}
\end{equation}

\noindent and
\begin{equation}
\mathbf{F}_{iB}^{(2)} = \int d^3v\, M f_{ig} v_{||} \bun \cdot
\nablaave \bv_\bot + \int d^3v\, \frac{Mc}{B} (f_{Mi} +
h_{i1}^\mathrm{nc}) (\nablaave \phiwig \times \bun) \cdot
\nablaave \bv_\bot. \label{force_B_o2}
\end{equation}

\noindent Again, the rest of the terms are as defined in
subsection~\ref{sub_v}, but with the higher order piece
$h_{i2}^\mathrm{tb}$ implicitly included.

The moment equations \eq{ntransp_krho_delta2} and
\eq{vtransp_krho_delta2} can be used to extend the gyrokinetic
vorticity equations \eq{vort_type1} and \eq{vort_type2} to order
$q \delta_i^2 e n_e v_i/L$. Combining equation
\eq{ntransp_krho_delta2} with the electron number conservation
equation \eq{ntransp_elect} gives the vorticity equation
\begin{equation}
\frac{\partial}{\partial t} (Ze n_{ip}^{(2)}) = \nabla \cdot \left
( J_{g||}^{(2)} \bun + \bJ_{gd} + \tilde{\bJ}_i +
Zen_i\tilde{\bV}_i + Zen_i \bV_{iC} \right ),
\label{vort_type1_delta2}
\end{equation}

\noindent with
\begin{equation}
J_{g||}^{(2)} = Zen_i V_{ig||}^{(2)} - en_e V_{e||},
\label{Jgpar_o2}
\end{equation}

\noindent and the rest of the terms as defined in
subsection~\ref{sub_vortquasi} with $h_{i2}^\mathrm{tb}$ implicit.
Finally, combining \eq{vort_type1_delta2} with
\eq{vtransp_krho_delta2} gives
\begin{equation}
\frac{\partial \varpi_G^{(2)}}{\partial t} = \nabla \cdot \Bigg [
J_{g||}^{(2)} \bun + \bJ_{gd} + \tilde{\bJ}_{i\phi}^{(2)} +
\frac{c}{B} \bun \times (\nabla \cdot \matrixtop{\pibf}_{iG}) +
Zen_i\bV_{iC} - \frac{c}{B} \bun \times \mathbf{F}_{iC} \Bigg ],
\label{vort_type2_delta2}
\end{equation}

\noindent with
\begin{equation}
\varpi_G^{(2)} = \nabla \cdot \left (\frac{Ze}{\Omega_i} \int
d^3v\ f_{ig} \bv \times \bun \right ) - \int d^3v\,
\frac{Z^2e^2\phiwig}{T_i} f_{Mi} + \int d^3v\,
\frac{Z^2e^2\phiwig}{M} \left ( \frac{\partial
h_{i1}^\mathrm{nc}}{\partial E_0} + \frac{1}{B} \frac{\partial
h_{i1}^\mathrm{nc}}{\partial \mu_0} \right ), \label{vortGKdef_o2}
\end{equation}
\begin{equation}
\tilde{\bJ}_{i\phi}^{(2)} = \tilde{\bJ}_i - \frac{Zec}{B\Omega_i}
\bun \times \int d^3v\, (f_{Mi} + h_{i1}^\mathrm{nc}) (\nablaave
\phiwig \times \bun) \cdot \nablaave \bv_\bot,
\end{equation}

\noindent and the rest of the terms as defined in
subsection~\ref{sub_vortvort}. Both equations
\eq{vort_type1_delta2} and \eq{vort_type2_delta2} can be used to
find the short wavelength non-axisymmetric pieces of the potential
consistent with the higher order $f_i$. The flux surface averaged
component of the potential is given by the conservation equation
of toroidal angular momentum. Finally, it is important to remember
that dropping the higher order axisymmetric terms in equations
\eq{vort_type1_delta2} and \eq{vort_type2_delta2} implies dropping
the higher order corrections to the GAM response \cite{winsor68,
gao08}. I expect this response to be unimportant for core
turbulence based on previous experience with tokamak core
simulations.

\section{Discussion} \label{sect_discussion_toroidal}

The fast time average of equation \eq{pi_rzeta_ave_2} provides the
irreversible transport of toroidal angular momentum across the
tokamak. This irreversible transport determines the toroidal
rotation profile and hence the self-consistent radial electric
field.

It is expected that the fast time average of equation
\eq{pi_rzeta_ave_2} is of order $\delta_i^3 p_i$, requiring then a
distribution function good to order $\delta_i^2 f_{Mi}$ and a
potential consistent with this higher order distribution function.
It is in principle possible (if not in practice) to obtain a
gyrokinetic equation able to provide such accurate results, but in
this thesis I propose an alternative approach. To simplify the
problem, I take advantage of the expansion in $B_p/B \ll 1$ to
prove that the gyrokinetic Fokker-Planck equation \eq{FP_final} is
enough to obtain the distribution function up to order $q
\delta_i^2 f_{Mi}$. In $\delta f$ flux tube codes, the
distribution function cannot be calculated to higher order than
$\delta_i f_{Mi}$ because of the present implementation technique,
but these codes can be adapted by adding terms that contain the
first order neoclassical correction $h_{i1}^\mathrm{nc}$. In full
$f$ codes, the gyrokinetic equation \eq{FP_final} is fully
implemented. Once the turbulence has reached statistical
equilibrium, the long wavelength axisymmetric flows must be close
to the neoclassical solution, and the orderings described in this
chapter hold, leading to solutions valid up to $q \delta_i^2
f_{Mi}$.

Finally, I have extended the gyrokinetic vorticity equations
\eq{vort_type1} and \eq{vort_type2}, and written them in
\eq{vort_type1_delta2} and \eq{vort_type2_delta2}. These vorticity
equations have only been found for non-axisymmetric pieces of the
potential, giving then the non-axisymmetric turbulent fluctuations
consistent with the higher order $f_i$. The flux surface averaged
component of the potential cannot be calculated from these
vorticity equations, but it can be solved from the conservation of
toroidal angular momentum \eq{toro_angmom}.

\begin{figure}
\begin{center}
\includegraphics[width = \textwidth]{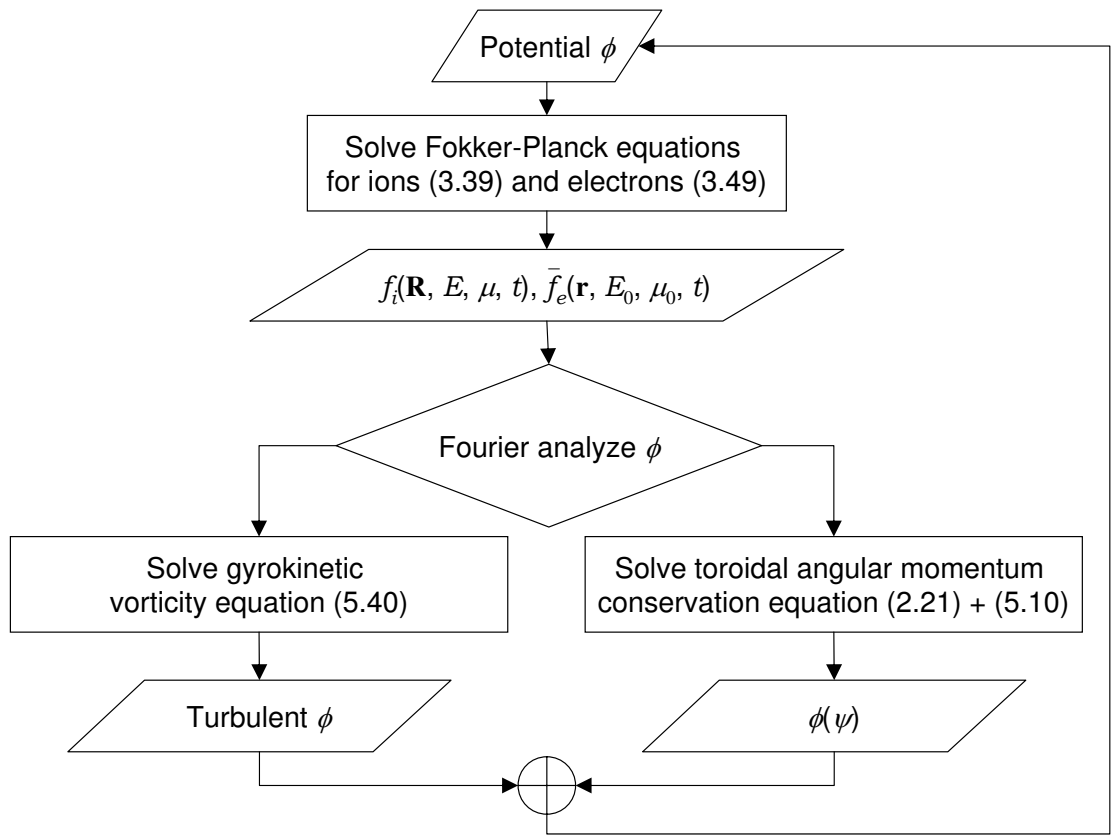}
\end{center}

\caption[Procedure to obtain the radial electric field]{Procedure
proposed to obtain the long wavelength axisymmetric radial
electric field.} \label{fig_solution}
\end{figure}

The method proposed to self-consistently solve for the long
wavelength axisymmetric radial electric field in the presence of
drift wave turbulence is summarized in figure~\ref{fig_solution}.
To be specific I employ the higher order gyrokinetic vorticity
equation \eq{vort_type2_delta2}. The solution must be reached by
time evolution of $\phi$, $f_i$ and $\overline{f}_e$. Employing
the gyrokinetic Fokker-Planck equation \eq{FP_final} and the
electron drift kinetic equation \eq{drift_kinetic_e}, the ion and
electron distribution functions $f_i (\bR, E, \mu, t)$ and
$\overline{f}_e (\br, E_0, \mu_0, t)$ are evolved in time. To find
the corresponding electrostatic potential, we first split it into
the flux averaged component and the turbulent piece by, for
example, Fourier analyzing it into toroidal and poloidal modes.
For the turbulent pieces, we must employ the higher order
gyrokinetic vorticity equation \eq{vort_type2_delta2}. The flux
surface averaged component of the potential is obtained by
evolving the axisymmetric toroidal rotation $\bV_i \cdot \zun$
with equation \eq{toro_angmom} and the viscosity given in
\eq{pi_rzeta_ave_2}. Once the toroidal rotation is found, the
axisymmetric radial electric field is obtained using the lower
order result \eq{fig} to write
\begin{eqnarray}
R n_i \bV_i \cdot \zun = \int d^3v\, f_{ig} R (\bv \cdot \zun) -
\int d^3v\, \frac{Ze\phiwig}{T_i} f_{Mi} R (\bv_\bot \cdot \zun)
\nonumber \\ _{\overrightarrow{\; \; \; \; k_\bot \rho_i
\rightarrow 0 \;\; \; \;}} \; IU(\psi) - \frac{cR^2}{Ze} \left (
\frac{\partial p_i}{\partial \psi} + Zen_i \frac{\partial
\phi}{\partial \psi} \right ),
\end{eqnarray}

\noindent where to write this last expression I have used the
neoclassical relation \eq{neo_flows}. Here, $f_{ig}$ depends
implicitly on $\phi(\psi)$. Finally, the turbulent and flux
surface averaged components of the electrostatic potential are
added to obtain the total electric field, and the distribution
functions $f_i$ and $\overline{f}_e$ may be evolved in time again.
Notice that equation \eq{pi_rzeta_ave_2}, employed here to find
the radial electric field, requires that the radial wavelengths be
longer than the ion gyroradius. This limitation is probably
unimportant since the zonal flow is characterized by $k_\bot
\rho_i \sim 0.1$ \cite{diamond05}.

\chapter{Conclusions \label{chap_conclusion}}

In this thesis, I have proven that the current gyrokinetic
treatments, composed of a gyrokinetic Fokker-Planck equation and a
gyrokinetic quasineutrality equation, cannot provide the long
wavelength, axisymmetric radial electric field. Employing the
vorticity equation \eq{vorticity}, I first showed that setting the
radial current to zero to obtain the axisymmetric radial electric
field is equivalent to solving the toroidal angular momentum
conservation equation, given in \eq{toro_angmom}.

In chapter~\ref{chap_gyrokinetics}, I present a new derivation of
electrostatic gyrokinetics that generalizes the linear treatment
of \cite{leecatto83, bernstein85}. This derivation is useful in
chapter~\ref{chap_gyrovorticity} to study the current conservation
or vorticity equation in steady state turbulence. To simplify the
problem, I assume that, in statistical equilibrium, the turbulent
fluctuations within a flux surface must be small by $\delta_i =
\rho_i/L$ because of the fast transport along magnetic field
lines. Then, the long wavelength axisymmetric flows must remain
neoclassical, and the tokamak is intrinsically ambipolar even in
the presence of turbulence, i.e., $\langle \bJ \cdot \nabla \psi
\rangle_\psi \simeq 0$ for any long wavelength axisymmetric radial
electric field. According to the estimate in \eq{proof_intrinsic},
the radial current density associated with transport of toroidal
angular momentum is so small that modern gyrokinetic treatments
are unable to self-consistently calculate the long wavelength
radial electric field. For most codes, long wavelengths are those
above $\sqrt{\rho_i L}$.

To solve this issue, I propose to solve a vorticity equation
instead of the gyrokinetic quasineutrality equation. The vorticity
equation has the advantage of showing explicitly the dependence on
the transport of toroidal angular momentum. I have derived two
approximate gyrokinetic vorticity equations. Vorticity equation
\eq{vort_type1} is similar to the gyrokinetic quasineutrality
equation, and vorticity equation \eq{vort_type2}, on the other
hand, is constructed to resemble the full vorticity equation
\eq{vorticity}. The two gyrokinetic vorticity equations
\eq{vort_type1} and \eq{vort_type2} are equivalent to the full
vorticity equation \eq{vorticity} within terms of order
$\delta_i$. The long wavelength radial electric field cannot be
found from these gyrokinetic vorticity equations because they are
missing crucial terms. However, they satisfy a very desirable
property explicitly, namely, the long wavelength toroidal velocity
tends to be constant for the short turbulence saturation time
scales.

To complement the gyrokinetic vorticity equation, I propose using
the conservation equation for the toroidal angular momentum
\eq{toro_angmom}, where the toroidal-radial component of the ion
viscosity $\langle R\zun \cdot \matrixtop{\pibf}_i \cdot \nabla
\psi \rangle_\psi$ is given to order $\delta_i^3 p_i$ in
\eq{pi_rzeta_ave_2}. Unfortunately, expression \eq{pi_rzeta_ave_2}
requires a distribution function and a potential of order higher
than calculated in gyrokinetic codes. In
section~\ref{sect_iondistribution}, I show that, for $q \gg 1$,
the ion distribution function can be found to high enough order by
employing the full $f$ gyrokinetic equation \eq{FP_final}. The
gyrokinetic vorticity equations \eq{vort_type1} and
\eq{vort_type2}, however, have to be extended to determine the
higher order potential. Equation \eq{vort_type1_delta2} and
\eq{vort_type2_delta2} are the higher order versions of
\eq{vort_type1} and \eq{vort_type2}.

To summarize, to obtain the self-consistent electric field, it is
necessary to refrain from using the lower order gyrokinetic
quasineutrality equation. Instead, the electric field has to be
found by employing a higher order formulation like the proposed
vorticity equation \eq{vorticity}. Moreover, since the
axisymmetric contributions to the vorticity equation are
equivalent to the conservation of toroidal angular momentum, only
the non-axisymmetric pieces of the potential must be found by
employing higher order gyrokinetic vorticity equations like
equation \eq{vort_type1_delta2} or equation
\eq{vort_type2_delta2}. The axisymmetric electric field should be
found from \eq{toro_angmom} employing the toroidal-radial
component of the ion viscosity in \eq{pi_rzeta_ave_2}. At the same
time, the ion distribution function evolves according to
\eq{FP_final}. In order to implement this methodology, I foresee
several steps. First, the lowest order gyrokinetic vorticity
equations \eq{vort_type1} and \eq{vort_type2} should be
implemented in $\delta f$ flux tube codes. These codes are well
understood and easy to study. Equation \eq{vort_type1} is
appealing since it is very similar to the gyrokinetic
quasineutrality equation. If the vorticity equations show good
numerical behavior in $\delta f$ flux tube codes, they should be
then implemented in full $f$ codes. For runs that stay below
transport time scales, these vorticity equations are still valid.
Finally, the transport of toroidal angular momentum given in
\eq{pi_rzeta_ave_2} must be studied. To do so, it is necessary to
calculate the potential and distribution function to higher order,
requiring then the higher order versions of the gyrokinetic
vorticity equations \eq{vort_type1_delta2} and
\eq{vort_type2_delta2}. Again, it will probably be easier to study
these equations and the transport of toroidal angular momentum in
$\delta f$ flux tube codes first, and then, finally, implement
this method in full $f$ models.

\appendix
\chapter{Derivation of the gyrokinetic variables \label{app_GKvar}}

In this Appendix the detailed calculation of the gyrokinetic
variables is carried out. In section~\ref{sectapp_GKvar1}, the
gyrokinetic variables are computed to first order in $\delta_i$.
In section~\ref{sectapp_GKvar2}, the gyrokinetic variables $\bR$
and $E$ are extended to second order, and the gyrokinetic magnetic
moment $\mu$ is proven to be an adiabatic invariant to higher
order. Finally, in section~\ref{sectapp_jacob}, the Jacobian of
the transformation from the variables $\br$, $\bv$ to the
gyrokinetic variables is calculated. The Jacobian is employed to
write the gyrokinetic equation in conservative form.

\section{First order gyrokinetic variables \label{sectapp_GKvar1}}
It is convenient to express any term that contains the
electrostatic potential $\phi$ in gyrokinetic variables, mainly
because the electrostatic potential components with $k_\bot \rho_i
\sim 1$ cannot be Taylor expanded. In order to do so, I will
develop some useful relations involving the potential $\phi$ in
subsection~\ref{subapp_relationphi}. With these relations, the
first order corrections, $\bR_1$, $E_1$, $\varphi_1$ and $\mu_1$,
are derived.

\subsection{Useful relations for $\phi$ \label{subapp_relationphi}}

I first derive all possible gyrokinetic partial derivatives of
$\phi$ and their relation to one another. To do so, only $\bR =
\br + \Omega_i^{-1} \bv \times \bun + O (\delta_i^2 L)$ is needed.

The derivative respect to the gyrocenter position is
\begin{equation}
\nabla_\bR \phi (\br) = \nabla \phi + \nabla_\bR ( \br - \bR )
\cdot \nabla \phi = \nabla \phi + O ( \delta_i T_e/eL ) \simeq
\nabla \phi.  \label{dphidR}
\end{equation}

The derivative respect to the energy is
\begin{equation}
\frac{\partial \phi}{\partial E} = \frac{\partial}{\partial E}
(\br - \bR) \cdot \nabla \phi = O ( \delta_i^2 M/e ) \simeq 0,
\label{dphidE}
\end{equation}

\noindent since $\br - \bR$ only depends on $E$ at $O(\delta_i^2
L)$.

Using $\br - \bR \propto \sqrt{\mu} ( \eun_1 \sin \varphi - \eun_2
\cos \varphi)$, the derivatives with respect to $\mu$ and
$\varphi$ are calculated to be
\begin{equation}
\frac{\partial \phi}{\partial \mu} = \frac{\partial}{\partial \mu}
( \br - \bR ) \cdot \nabla \phi \simeq - \frac{Mc}{Ze v^2_\bot}
(\bv \times \bun) \cdot \nabla \phi \label{dphidmu}
\end{equation}

\noindent and
\begin{equation}
\frac{\partial \phi}{\partial \varphi} = \frac{\partial}{\partial
\varphi} ( \br - \bR ) \cdot \nabla \phi \simeq -
\frac{1}{\Omega_i} \bv_\bot \cdot \nabla \phi. \label{dphidvphi}
\end{equation}

I will need more accurate relationship than \eq{dphidR} and
\eq{dphidvphi} for the second order corrections. They will be
developed in subsection~\ref{subapp_morerelationphi}.

\subsection{Calculation of $\bR_1$}
The first order correction $\bR_1$ is given by \eq{o1_correction},
where in this case, $Q_0 = \bR_0 = \br$. The total derivative of
$\bR_0$ is $d\bR_0/dt = \bv = v_{||} \bun + \bv_\bot$, and its
gyroaverage gives $\langle d \bR_0/dt \rangle = v_{||} \bun + O
(\delta_i v_i)$. By employing $\bv_\bot = \partial( \bv \times
\bun)/\partial \varphi_0$, equation~\eq{o1_correction} gives
\eq{RGK1_def}.

\subsection{Calculation of $E_1$}
The first order correction $E_1$ is given by \eq{o1_correction},
where $Q_0 = E_0 = v^2/2$ and $dQ_0/dt = dE_0/dt = - (Ze/M) \bv
\cdot \nabla \phi$. It is convenient to write $E_1$ as a function
of $\bR$, $E$, $\mu$ and $\varphi$. To do so, I use \eq{dphidR}
and \eq{dphidvphi} to find
\begin{equation}
- \bv \cdot \nabla \phi = - v_{||} \bun \cdot \nabla \phi -
\bv_\bot \cdot \nabla \phi \simeq - v_{||} \bun \cdot \nabla_\bR
\phi + \Omega_i \frac{\partial \phi}{\partial \varphi}.
\label{vgradphi_1o}
\end{equation}

\noindent Notice that $\bun \cdot \nabla_\bR \phiwig \ll \bun
\cdot \nabla_\bR \phiave$ because $\phiwig$ is smaller than
$\phiave$. As a result, $dE_0/dt \simeq - (Ze/M) v_{||} \bun \cdot
\nabla_\bR \phiave + (Ze\Omega_i/M) \partial \phi/\partial
\varphi$ and $\langle d E_0/d t \rangle = - (Ze/M) v_{||} \bun
\cdot \nabla_\bR \phiave + O(\delta_i v_i^3/L)$. Then, equation
\eq{o1_correction} gives \eq{correc_E_o1}.

\subsection{Calculation of $\varphi_1$}

The first order correction $\varphi_1$ is given by
\eq{o1_correction}, where $Q_0 = \varphi_0$. The zeroth order
gyrophase $\varphi_0$ is defined by equation \eq{gyrophase_def}.
According to this definition, upon using $\nabla_v \varphi_0 = -
v_\bot^{-2} \bv \times \bun$ and $\nabla \varphi_0 =
(v_{||}/v_\bot^2) \nabla \bun \cdot ( \bv \times \bun ) + \nabla
\eun_2 \cdot \eun_1$, the total derivative of $\varphi_0$ is
\begin{eqnarray}
\frac{d \varphi_0}{dt} = - \overline{\Omega}_i -
\frac{Z^2e^2}{M^2c} \frac{\partial \phiwig}{\partial \mu} + (\bv
\times \bun) \cdot \left [ \nabla \ln \Omega_i +
\frac{v^2_{||}}{v^2_\bot} \bun \cdot \nabla \bun - \bun \times
\nabla \eun_2 \cdot \eun_1 \right ] \nonumber \\  +
\frac{v_{||}}{2 v^2_\bot} [ \bv_\bot ( \bv \times \bun ) + ( \bv
\times \bun ) \bv_\bot ] : \nabla \bun, \label{dvphi0dt_3}
\end{eqnarray}

\noindent where the potential $\phi (\br, t)$ and the
gyrofrequency $\Omega_i( \br )$ have been written as functions of
the gyrokinetic variables by using \eq{dphidmu} and $\Omega_i
(\br) \simeq \Omega_i (\bR) + (\br - \bR) \cdot \nabla \Omega_i$,
respectively, and I have used the relations $\langle \bv_\bot
\bv_\bot \rangle \simeq \overline{\bv_\bot \bv_\bot} =
(v_\bot^2/2) ( \matrixtop{\mathbf{I}} - \bun \bun )$ and
\begin{equation}
\bv_\bot \bv_\bot - \langle \bv_\bot \bv_\bot \rangle =
\frac{1}{2} [ \bv_\bot \bv_\bot - ( \bv \times \bun ) ( \bv \times
\bun ) ]. \label{v2wig}
\end{equation}

\noindent Here, $\overline{(\ldots)}$  is the gyroaverage holding
$\br$, $v_{||}$, $v_\bot$ and $t$ fixed, and $\langle \ldots
\rangle$ is the gyroaverage holding $\bR$, $E$, $\mu$ and $t$
fixed. These two gyroaverages are equivalent in this case because
the functions involved do not have short wavelengths. A detailed
derivation of \eq{v2wig} and other velocity relations can be found
in Appendix~\ref{app_vrelations}.

In equation \eq{dvphi0dt_3}, the function $\overline{\Omega}_i$ is
given by equation~\eq{omeg_mod}. Upon gyroaveraging
\eq{dvphi0dt_3}, I obtain $\langle d \varphi_0/d t \rangle = -
\overline{\Omega}_i + O ( \delta_i^2 \Omega_i )$. Finally,
$\varphi_1$ is obtained from \eq{o1_correction} by employing $\bv
\times \bun = - \partial \bv_\bot/\partial \varphi_0$ and
\begin{equation}
\bv_\bot \bv_\bot - (\bv \times \bun) (\bv \times \bun) =
\frac{1}{2} \frac{\partial}{\partial \varphi_0} [ \bv_\bot ( \bv
\times \bun ) + ( \bv \times \bun ) \bv_\bot ], \label{v2int}
\end{equation}

\noindent giving equation~\eq{correc_varphi_o1}. Relation
\eq{v2int} is proven in Appendix~\ref{app_vrelations}.

\subsection{Calculation of $\mu_1$}
Calculating $\mu_1$ requires more work than calculating any of the
other first order corrections since we want $\mu$ to be an
adiabatic invariant to all orders of interest. This requirement
imposes two conditions to $\mu_1$. One of them is similar to the
requirements already imposed to $\bR_1$, $E_1$ and $\varphi_1$,
$d\mu_0/dt - \Omega_i (\partial \mu_1/\partial \varphi) = \langle
d\mu_0/dt \rangle = 0$, but there is an additional condition
making $\mu_0 + \mu_1$ an adiabatic invariant to first order,
\begin{equation}
\left \langle \frac{d}{dt} (\mu_0 + \mu_1) \right \rangle = O
\left ( \delta_i^2 \frac{v_i^3}{BL} \right ). \label{mu1_cond2}
\end{equation}

\noindent The solution to both conditions is given by
\begin{equation}
\mu_1 = \frac{1}{\Omega_i} \int^\varphi d\varphi^\prime \left (
\frac{d \mu_0}{dt} - \left \langle \frac{d\mu_0}{dt} \right
\rangle \right ) + \langle \mu_1 \rangle. \label{muGK1_1def}
\end{equation}

\noindent Notice that the only difference with the result in
\eq{o1_correction} is that the gyrophase independent term,
$\langle \mu_1 \rangle$, must be retained, making it possible to
satisfy condition~\eq{mu1_cond2}.

Employing $\nabla_v \mu_0 = \bv_\bot/B$ and $\nabla \mu_0 = -
(v_\bot^2/2B^2) \nabla B - (v_{||}/B) \nabla \bun \cdot \bv$, I
find that the total derivative for $\mu_0 = v_\bot^2 / 2B$ is
\begin{eqnarray}
\frac{d \mu_0}{d t}  = - \frac{Ze}{MB} \bv_\bot \cdot \nabla \phi
- \frac{v_\bot^2}{2B^2} \bv_\bot \cdot \nabla B -
\frac{v^2_{||}}{B} \bun \cdot \nabla \bun \cdot \bv_\bot \nonumber
\\ - \frac{v_{||}}{2B} [ \bv_\bot \bv_\bot - ( \bv \times \bun ) ( \bv
\times \bun ) ] : \nabla \bun, \label{dmu0dt_app}
\end{eqnarray}

\noindent where I have used the relations $\langle \bv_\bot
\bv_\bot \rangle \simeq \overline{\bv_\bot \bv_\bot} =
(v_\bot^2/2) ( \matrixtop{\mathbf{I}} - \bun \bun )$ and
\eq{v2wig}.

Notice that the gyrophase independent terms in \eq{dmu0dt_app}
cancel exactly due to $\bun \cdot \nabla \ln B + \nabla \cdot \bun
= 0$, making $\mu_0$ an adiabatic invariant to zeroth order. The
term that contains $\phi$ in \eq{dmu0dt_app} is rewritten as a
function of the gyrokinetic variables by using \eq{dphidvphi}, to
give $ - (Ze/MB) \bv_\bot \cdot \nabla \phi = (Ze\Omega_i/MB)
\partial \phi/\partial \varphi$.

Applying \eq{muGK1_1def}, $\mu_1$ is found to be given by
\eq{correc_mu_o1}. To get this result, I have employed $\bv_\bot =
\partial( \bv \times \bun )/\partial \varphi_0$ and \eq{v2int}.
The average value $\langle \mu_1 \rangle = - (v_{||} v_\bot^2/2 B
\Omega_i) (\bun \cdot \nabla \times \bun)$ was chosen to ensure
that condition \eq{mu1_cond2} is satisfied. In previous works
\cite{catto81, northrop78}, it has been noticed that solving
\eq{mu1_cond2} may be avoided and replaced by imposing the
relation $E = [ d\bR/dt \cdot \bun(\bR) ]^2/2 + \mu B ( \bR )$ on
the gyrokinetic variables. This procedure works in this case, and
allows me to find $\langle \mu_1 \rangle$. I will prove that the
chosen $\langle \mu_1 \rangle$ satisfies condition \eq{mu1_cond2}
in subsection~\ref{subapp_mu1ave}.

\section{Second order gyrokinetic variables \label{sectapp_GKvar2}}
To construct the gyrokinetic variables to second order, higher
order relations than the ones developed in
subsection~\ref{subapp_relationphi} are needed to express $\phi$
as a function of the gyrokinetic variables. These extended
relations are deduced in subsection~\ref{subapp_morerelationphi}.
Using them, the second order corrections $\bR_2$ and $E_2$ and the
gyrophase independent piece of the first order correction $\langle
\mu_1 \rangle$ are calculated. The magnetic moment and the
gyrophase are not required to higher order.

\subsection{More useful relations for $\phi$
\label{subapp_morerelationphi}}

To calculate the second order correction $E_2$ and the gyrophase
independent piece $\langle \mu_1 \rangle$, the expressions $\bun
\cdot (\nabla \phi - \nabla_\bR \phi)$ and $\bv \cdot \nabla \phi$
must be given in gyrokinetic variables to order $\delta_i T_e/eL$
and $\delta_i T_e v_i/eL $, respectively.

For $\bun \cdot (\nabla \phi - \nabla_\bR \phi)$, I use $\nabla
\phi = \nabla \bR \cdot \nabla_\bR \phi + \nabla E (\partial
\phi/\partial E) + \nabla \mu (\partial \phi/\partial \mu) +
\nabla \varphi (\partial \phi/\partial \varphi)$ to write
\begin{equation}
\bun \cdot ( \nabla \phi - \nabla_\bR \phi ) \simeq \bun \cdot
\nabla \bR_1 \cdot \nabla_\bR \phi + \bun \cdot \nabla \mu_0
\frac{\partial \phi}{\partial \mu} + \bun \cdot \nabla \varphi_0
\frac{\partial \phi}{\partial \varphi},
\end{equation}

\noindent where I have neglected higher order terms. Here, it is
important that the gradient is parallel to the magnetic field,
since $\bR_2$, $\mu_1$ and $\varphi_1$ have pieces with short
perpendicular wavelengths that are important for the perpendicular
component of the gradient. Employing $\nabla \bR_1 = -
\Omega_i^{-1} [ (\nabla \ln B) (\bv \times \bun) + \nabla \bun
\times \bv ]$, $\nabla \mu_0 = - (v_\bot^2/2B^2) \nabla B -
(v_{||}/B) \nabla \bun \cdot \bv$, $\nabla \varphi_0 =
(v_{||}/v_\bot^2) \nabla \bun \cdot (\bv \times \bun) + \nabla
\eun_2 \cdot \eun_1$ and the lowest order relations \eq{dphidmu}
and \eq{dphidvphi} for $\partial \phi/\partial \mu$ and $\partial
\phi/\partial \varphi$, I obtain
\begin{eqnarray}
\bun \cdot (\nabla \phi - \nabla_\bR \phi) = -
\frac{1}{2B\Omega_i} (\bun \cdot \nabla B) (\bv \times \bun) \cdot
\nabla \phi \nonumber \\ - \frac{1}{\Omega_i} \bun \cdot \nabla
\bun \cdot (\bv \times \bun) (\bun \cdot \nabla \phi) + \bun \cdot
\nabla \eun_2 \cdot \eun_1 \frac{\partial \phi}{\partial \varphi}.
\label{par_grad_phi}
\end{eqnarray}

\noindent To find this result, I have used
\begin{equation}
- \frac{v_{||}}{B} \bun \cdot \nabla \bun \cdot \bv_\bot
\frac{\partial \phi}{\partial \mu} + \frac{v_{||}}{v_\bot^2} \bun
\cdot \nabla \bun \cdot (\bv \times \bun) \frac{\partial
\phi}{\partial \varphi} = \frac{v_{||}}{\Omega_i} \bun \cdot
\nabla \bun \cdot (\bun \times \nabla \phi),
\end{equation}

\noindent where I employ \eq{dphidmu} and \eq{dphidvphi} for
$\partial \phi/\partial \mu$ and $\partial \phi/\partial \varphi$,
and the relation $\bv_\bot (\bv \times \bun) - (\bv \times \bun)
\bv_\bot = v_\bot^2 (\matI \times \bun)$. This relation is
obtained from the fact that $\bv_\bot$ and $\bv \times \bun$
expand the vector space perpendicular to the magnetic field,
giving $\bv_\bot \bv_\bot + (\bv \times \bun) (\bv \times \bun) =
v_\bot^2 ( \matI - \bun\bun)$.

To calculate $\bv \cdot \nabla \phi$, I use that the total time
derivative for $\phi$ in $\br$, $\bv$ variables is
\begin{equation}
\frac{d \phi}{dt} = \left. \frac{\partial \phi}{\partial t} \right
|_{\br} + \bv\cdot \nabla \phi, \label{dphidt_r}
\end{equation}

\noindent while as a function of the new gyrokinetic variables it
becomes
\begin{equation}
\frac{d \phi}{dt} = \left. \frac{\partial \phi}{\partial t} \right
|_{\bR, \; E, \; \mu,\; \varphi} + \dot{\bR} \cdot \nabla_\bR \phi
+ \dot{E} \frac{\partial \phi}{\partial E} + \dot{\varphi}
\frac{\partial \phi}{\partial \varphi}. \label{dphidt_R}
\end{equation}

\noindent Combining these equations gives an equation for $\bv
\cdot \nabla \phi$,
\begin{equation}
- \bv \cdot \nabla \phi  = \left ( \left. \frac{\partial
\phi}{\partial t} \right |_{\br} - \left. \frac{\partial
\phi}{\partial t} \right |_{\bR, \; E, \; \mu,\; \varphi} \right )
- \dot{E} \frac{\partial \phi}{\partial E} - \dot{\bR} \cdot
\nabla_\bR \phi - \dot{\varphi} \frac{\partial \phi}{
\partial \varphi}, \label{vgradphi}
\end{equation}

\noindent where the left side of the equation is of order $O ( T_e
v_i/eL )$. I analyze the right side term by term, keeping terms up
to order $\delta_i T_e v_i/eL$. Noticing that $\phi ( \br, t) =
\phi ( \bR + (\br - \bR), t )$, the partial derivatives with
respect to time give the negligible contribution $( \partial
\phi/\partial t|_{\br} -
\partial \phi/\partial t |_{\bR, \; E, \; \mu,\; \varphi} ) = -
\partial(\br - \bR)/\partial t \cdot \nabla \phi = O ( \delta_i^2
T_ev_i/eL )$, since the time derivative of $\br - \bR$ can only be
of order $\delta_i^2 v_i$ for a static magnetic field. The partial
derivative with respect to $E$ is estimated in \eq{dphidE}, giving
that $\dot{E} \, \partial \phi/\partial E = O ( \delta_i^2 T_e
v_i/e L )$ is negligible. The total derivative $\dot{\bR}$ has two
different components, which I will calculate in detail in
subsection~\ref{sub_R2}. These components are the parallel
velocity of the gyrocenter, $u \bun ( \bR )$, of order $v_i$, and
the drift velocity, $\bv_d$, of order $\delta_i v_i$. Using this
information, I find $u \bun ( \bR ) \cdot \nabla_\bR \phi = O (
T_ev_i/eL )$ and $\bv_d \cdot \nabla_\bR \phi = O ( \delta_i
T_ev_i/eL )$. Finally, the last term in the right side of
\eq{vgradphi} is $\dot{\varphi} (\partial \phi/\partial \varphi) =
O ( T_ev_i/eL )$, since $\dot{\varphi} \sim \Omega_i$ and
$\partial \phi / \partial \varphi = \partial \phiwig /
\partial \varphi \sim \delta_i T_e/e$ according to \eq{phiwig}.
Neglecting all the terms smaller than $\delta_i T_e v_i/eL$,
equation \eq{vgradphi} becomes
\begin{equation}
- \bv \cdot \nabla \phi = - u \bun (\bR) \cdot \nabla_\bR \phi -
\bv_d \cdot \nabla_\bR \phi + \overline{\Omega}_i \frac{\partial
\phiwig}{\partial \varphi}. \label{vgradphi2}
\end{equation}

\subsection{Calculation of $\bR_2$} \label{sub_R2}

The second order correction $\bR_2$ is given by
\eq{o2_correction}, where $Q_0 = \bR_0 = \br$ and $Q_1 = \bR_1 =
\Omega_i^{-1} \bv \times \bun$. The total time derivative of
$\bR_0 + \bR_1$ is
\begin{equation}
\frac{d}{dt} (\bR_0 + \bR_1) = v_{||} \bun - \bv \cdot \nabla
\left ( \frac{\bun}{\Omega_i} \right ) \times \bv - \frac{c}{B}
\nabla \phi \times \bun, \label{dR01dt}
\end{equation}

\noindent and its gyroaverage may be written as $\langle d( \bR_0
+ \bR_1 )/dt \rangle = u \bun (\bR) + \bv_d$, where $u = \langle
v_{||} \rangle + (v^2_\bot/2 \Omega_i) (\bun \cdot \nabla \times
\bun)$, and $\bv_d$ has been already defined in \eq{drift}. The
function $u$ can be written as a function of the gyrokinetic
variables. I express $v_{||}$ as a function of $\br$, $E_0$ and
$\mu_0$, expand around $\bR$, $E$ and $\mu$, and insert $\bR_1$,
$\mu_1$ and $E_1$ to obtain
\begin{eqnarray}
v_{||} = \sqrt{2 (E_0 - \mu_0 B(\br)) } \simeq \sqrt{2 (E - \mu
B(\bR)) } - \frac{v^2_\bot}{2 \Omega_i} \bun \cdot \nabla \times
\bun \nonumber \\ - \frac{v_{||}}{\Omega_i} \bun \cdot \nabla \bun
\cdot ( \bv \times \bun ) - \frac{1}{4 \Omega_i} [ \bv_\bot ( \bv
\times \bun ) + ( \bv \times \bun ) \bv_\bot ]:\nabla \bun
\label{vparfGK}.
\end{eqnarray}

\noindent Finally, gyroaveraging and using $\langle \bv_\bot ( \bv
\times \bun ) + ( \bv \times \bun ) \bv_\bot \rangle = 0$ [a
result that is deduced from \eq{v2wig}] give $u = \sqrt{2 [E - \mu
B(\bR)]}$, which can be rewritten as \eq{par_drift}.

Using \eq{dR01dt} and \eq{vparfGK}, Taylor expanding $\bun ( \br
)$ about $\bR$ and inserting the result into \eq{o2_correction}
gives \eq{R_1_main} and \eq{dR_dt_main}. To integrate over
gyrophase, $\bv \times \bun = - \partial \bv_\bot/\partial
\varphi_0$ and \eq{v2int} have been used.

\subsection{Calculation of $E_2$}

Equation \eq{o2_correction} gives $E_2$, where $Q_0 = E_0 = v^2/2$
and $Q_1 = E_1 = Ze\phiwig / M$. The total derivative of $E_0 =
v^2/2$ can be expressed as a function of the new gyrokinetic
variables to the requisite order by using \eq{vgradphi2} to obtain
\begin{equation}
\frac{d E_0}{d t} = - \frac{Ze}{M} \bv \cdot \nabla \phi \simeq
\frac{Z e}{M} \left \{ \overline{\Omega}_i \frac{\partial
\phiwig}{\partial \varphi} - [ u \bun (\bR) + \bv_d ] \cdot
\nabla_\bR \phi \right \}.
\end{equation}

\noindent From the definition of $E_1 = Ze \phiwig /M$, use of
gyrokinetic variables yields
\begin{equation}
\frac{d E_1}{d t} = \frac{Z e}{M} \left \{ \frac{\partial
\phiwig}{\partial t} + [ u \bun(\bR) + \bv_d ] \cdot \nabla_\bR
\phiwig - \overline{\Omega}_i \frac{\partial \phiwig}{\partial
\varphi} \right \}.
\end{equation}

\noindent Adding both contributions together leaves
\begin{equation}
\frac{d}{dt}(E_0 + E_1) = - \frac{Z e}{M} [ u \bun(\bR) + \bv_d ]
\cdot \nabla_\bR \phiave + \frac{Z e}{M} \frac{\partial
\phiwig}{\partial t}.
\end{equation}

\noindent As a result, $E_2$ is as shown in \eq{correc_E_o2}, and
to this order, $dE/dt$ is given by \eq{dE_dt_main}.

\subsection{Calculation of $\langle \mu_1 \rangle$ \label{subapp_mu1ave}}

In this subsection, I will check that $\langle \mu_1 \rangle = -
(v_{||} v_\bot^2/2B\Omega_i) \bun \cdot \nabla \times \bun$
satisfies the condition in \eq{mu1_cond2}. To do so, it is going
to be useful to distinguish between the part of $d ( \mu_0 + \mu_1
)/dt$ that depends on $\phi$ and the part that does not depend on
$\phi$ at all since these pieces will vanish independently of each
other. I will write the piece of $d (\mu_0 + \mu_1)/dt$ that
depends on $\phi$ as a function of the gyrokinetic variables,
finding
\begin{equation}
\left. \frac{d}{dt} ( \mu_0 + \mu_1 ) \right |_\phi = - \frac{Z
e}{M} \nabla \phi \cdot \nabla_v \mu_0 - \frac{Z e}{M} \nabla \phi
\cdot \nabla_v \mu_1|_{\br, \bv} + \frac{d}{d t} \mu_1 |_\phi,
\label{dmudt_phi}
\end{equation}

\noindent with $\mu_1|_\phi = Ze\phiwig/MB$ and $\mu_1|_{\br, \bv}
= \mu_1 - Ze\phiwig/MB$. The piece of $d (\mu_0 + \mu_1)/dt$ that
does not depend on $\phi$ is
\begin{equation}
\left. \frac{d}{dt} (\mu_0 + \mu_1) \right |_{\br, \bv} = \bv
\cdot \nabla \mu_0 - \Omega_i \frac{\partial}{\partial \varphi_0}
\mu_1 |_{\br, \bv} + \bv \cdot \nabla \mu_1 |_{\br, \bv}.
\end{equation}

\noindent In this equation, the two first terms cancel by
definition of $\mu_1 |_{\br, \bv}$, leaving
\begin{equation}
\left. \frac{d}{dt} (\mu_0 + \mu_1) \right |_{\br, \bv} = \bv
\cdot \nabla \mu_1 |_{\br, \bv}. \label{dmudt_rv}
\end{equation}

\noindent I will first prove that the gyroaverage of
\eq{dmudt_rv}, $\langle \bv \cdot \nabla \mu_1|_{\br, \bv}
\rangle$, vanishes to $O(\delta_i v_i^3/BL)$ due to the choice of
$\langle \mu_1 \rangle$. Afterwards, I will prove that the
gyroaverage of \eq{dmudt_phi} vanishes to the same order,
demonstrating then that $\mu_0 + \mu_1$ satisfies condition
\eq{mu1_cond2}.

To prove that $\langle \bv \cdot \nabla \mu_1|_{\br, \bv} \rangle
= 0$, the function $\mu_1|_{\br, \bv}$, is conveniently rewritten
as
\begin{eqnarray}
\mu_1 |_{\br, \bv} = - \frac{v_\bot^2}{2B^2\Omega_i} ( \bv \times
\bun ) \cdot \nabla B - \frac{v_{||}^2}{B\Omega_i} \bv_\bot \cdot
\nabla \times \bun \nonumber \\ - \frac{v_{||}}{2B\Omega_i} (\bv
\times \bun) \cdot \nabla \bun \cdot \bv - \frac{v_{||}
v_\bot^2}{4 B\Omega_i} \bun \cdot \nabla \times \bun,
\label{mu1_rv_conv}
\end{eqnarray}

\noindent where I use $\bun \cdot \nabla \bun = \kappabf$ and
\eq{id_kappa} to write $\bun \cdot \nabla \bun \cdot (\bv \times
\bun) = \bv_\bot \cdot \nabla \times \bun$, and I employ equation
\eq{v2wig} and $\langle \bv_\bot \bv_\bot \rangle = (v_\bot^2/2)
(\matI - \bun\bun)$ to find
\begin{equation}
\frac{1}{2} [ \bv_\bot (\bv \times \bun) + (\bv \times \bun)
\bv_\bot ]:\nabla \bun = (\bv \times \bun) \cdot \nabla \bun \cdot
\bv_\bot - \frac{v_\bot^2}{2} \bun \cdot \nabla \times \bun.
\label{reverse_v4wig}
\end{equation}

\noindent I will examine term by term the gyroaverage of $\bv
\cdot \nabla [ \mathrm{expression\; \eq{mu1_rv_conv}} ]$.
Employing $\bv \cdot \nabla (v_\bot^2/2) = - \bv \cdot \nabla
(v_{||}^2/2) = - v_{||} \bv \cdot \nabla v_{||}$, $\bv \cdot
\nabla v_{||} = \bv \cdot \nabla \bun \cdot \bv$ and $\langle \bv
\bv \rangle = (v_\bot^2/2) \matI + [v_{||}^2 - (v_\bot^2/2)] \bun
\bun$, the first, second and fourth terms in \eq{mu1_rv_conv} give
\begin{eqnarray}
\left \langle \bv \cdot \nabla \left [ \frac{v_\bot^2}{2B^2
\Omega_i} ( \bv \times \bun ) \cdot \nabla B \right ] \right
\rangle = - \frac{v_{||}^2 v_\bot^2}{2B^2\Omega_i} \bun \cdot
\nabla \bun \cdot (\bun \times \nabla B) \nonumber \\ +
\frac{v_\bot^4}{4} \nabla \cdot \left ( \frac{\bun \times \nabla
B}{B^2\Omega_i} \right ) + \frac{v_\bot^2}{2} \left ( v_{||}^2 -
\frac{v_\bot^2}{2} \right ) \bun \cdot \nabla \left ( \frac{\bun
\times \nabla B}{B^2\Omega_i} \right ) \cdot \bun,
\label{dmudt_term1_1}
\end{eqnarray}
\begin{eqnarray}
\left \langle \bv \cdot \nabla \left ( \frac{v_{||}^2}{B\Omega_i}
\bv_\bot \cdot \nabla \times \bun \right ) \right \rangle =
\frac{v_{||}^2 v_\bot^2}{B\Omega_i} \bun \cdot \nabla \bun \cdot (
\nabla \times \bun) \nonumber \\ + \frac{v_{||}^2 v_\bot^2}{2}
\nabla \cdot \left [ \frac{ (\nabla \times \bun)_\bot}{B\Omega_i}
\right ] + v_{||}^2 \left ( v_{||}^2 - \frac{v_\bot^2}{2} \right )
\bun \cdot \nabla \left [ \frac{ (\nabla \times
\bun)_\bot}{B\Omega_i} \right ] \cdot \bun \label{dmudt_term2_1}
\end{eqnarray}

\noindent and
\begin{eqnarray}
\left \langle \bv \cdot \nabla \left ( \frac{v_{||}
v_\bot^2}{4B\Omega_i} \bun \cdot \nabla \times \bun \right )
\right \rangle = \frac{v_\bot^4}{8B\Omega_i} (\nabla \cdot \bun)
\bun \cdot \nabla \times \bun \nonumber \\ - \frac{v_{||}^2
v_\bot^2}{4B\Omega_i} (\nabla \cdot \bun) \bun \cdot \nabla \times
\bun + \frac{v_{||}^2 v_\bot^2}{4} \bun \cdot \nabla \left (
\frac{ \bun \cdot \nabla \times \bun}{B\Omega_i} \right ).
\label{dmudt_term4_1}
\end{eqnarray}

\noindent The contribution to $\langle \bv \cdot \nabla \mu_1
|_{\br, \bv} \rangle$ of the third term in \eq{mu1_rv_conv} is
calculated by using $\bv \cdot \nabla v_{||} = \bv \cdot \nabla
\bun \cdot \bv$, $\langle \bv \bv \rangle = (v_\bot^2/2) \matI +
[v_{||}^2 - (v_\bot^2/2)] \bun \bun$ and
\begin{eqnarray}
\langle v_{\bot,\, j} v_{\bot,\, k} v_{\bot,\, l} v_{\bot,\, m}
\rangle = \frac{v_\bot^4}{8} \big [ (\delta_{jk} - \hat{b}_j
\hat{b}_k) (\delta_{lm} - \hat{b}_l \hat{b}_m) + (\delta_{jl} -
\hat{b}_j \hat{b}_l) (\delta_{km} - \hat{b}_k \hat{b}_m) \nonumber
\\ + (\delta_{jm} - \hat{b}_j \hat{b}_m) (\delta_{kl} - \hat{b}_k
\hat{b}_l) \big ]. \label{v4ave}
\end{eqnarray}

\noindent This result is proven in Appendix~\ref{app_vrelations}.
With these relations, the third term in \eq{mu1_rv_conv} gives
\begin{eqnarray}
\left \langle \bv \cdot \nabla \left [ \frac{v_{||}}{2B\Omega_i}
(\bv \times \bun) \cdot \nabla \bun \cdot \bv \right ] \right
\rangle = \frac{v_\bot^4}{16B\Omega_i} (\nabla \cdot \bun) \bun
\cdot \nabla \times \bun \nonumber \\ +
\frac{v_\bot^4}{16B\Omega_i} \matI : [ (\bun \times \nabla \bun)
\cdot \nabla \bun ] + \frac{v_\bot^4}{16B\Omega_i} \matI : [ (\bun
\times \nabla \bun) \cdot (\nabla \bun)^T ] \nonumber \\ +
\frac{v_{||}^2}{2} \left ( v_{||}^2 - \frac{3}{2} v_\bot^2 \right
) \bun \cdot \left [ \bun \cdot \nabla \left ( \frac{\bun \times
\nabla \bun}{B\Omega_i} \right ) \cdot \bun \right ] +
\frac{v_{||}^2 v_\bot^2}{4} \bun \cdot \nabla \left ( \frac{\bun
\cdot \nabla \times \bun}{B\Omega_i} \right ) \nonumber \\ +
\frac{v_{||}^2 v_\bot^2}{4} \nabla \cdot \left ( \frac{\bun \times
\nabla \bun}{B\Omega_i} \right ) \cdot \bun + \frac{v_{||}^2
v_\bot^2}{4} \nabla \cdot \left [ \frac{(\bun \times \nabla
\bun)^T}{B\Omega_i} \right ] \cdot \bun, \label{dmudt_term3_1}
\end{eqnarray}

\noindent with $(\nabla \bun)^T$ and $(\bun \times \nabla \bun)^T$
the transposes of $\nabla \bun$ and $\bun \times \nabla \bun$.

Several terms in equations \eq{dmudt_term1_1}, \eq{dmudt_term2_1},
\eq{dmudt_term4_1} and \eq{dmudt_term3_1} simplify because they
vanish. In particular, equation \eq{id_kappa}, with $\bun \cdot
\nabla \bun = \kappabf$, leads to $\bun \cdot \nabla \bun \cdot
(\nabla \times \bun) = 0$ and $\bun \cdot \nabla [ (\nabla \times
\bun)_\bot /B\Omega_i] \cdot \bun = - (B\Omega_i)^{-1} \bun \cdot
\nabla \bun \cdot (\nabla \times \bun) = 0$. Also, $\bun \cdot \{
\bun \cdot \nabla [ (\bun \times \nabla \bun)/B\Omega_i] \cdot
\bun \} = 0$, $\matI : [ (\bun \times \nabla \bun) \cdot (\nabla
\bun)^T ] = - \nabla \cdot ( \bun \times \nabla \bun ) \cdot \bun
= 0$ and $\nabla \cdot [ ( \bun \times \nabla \bun )/B\Omega_i ]
\cdot \bun = 0$. With these cancellations, I find that equations
\eq{dmudt_term2_1} and \eq{dmudt_term3_1} reduce to
\begin{eqnarray}
\left \langle \bv \cdot \nabla \left ( \frac{v_{||}^2}{B\Omega_i}
\bv_\bot \cdot \nabla \times \bun \right ) \right \rangle = -
\frac{v_{||}^2 v_\bot^2}{2B\Omega_i} \nabla \cdot ( \bun \bun
\cdot \nabla \times \bun ) \nonumber \\ - \frac{v_{||}^2
v_\bot^2}{B^2\Omega_i} (\nabla \times \bun)_\bot \cdot \nabla B,
\label{dmudt_term2_2}
\end{eqnarray}

\noindent where I have employed $\nabla \cdot [ (\nabla \times
\bun)_\bot ] = - \nabla \cdot ( \bun \bun \cdot \nabla \times \bun
)$, and
\begin{eqnarray}
\left \langle \bv \cdot \nabla \left [ \frac{v_{||}}{2B\Omega_i}
(\bv \times \bun) \cdot \nabla \bun \cdot \bv \right ] \right
\rangle = \frac{v_\bot^4}{8B\Omega_i} (\nabla \cdot \bun) \bun
\cdot \nabla \times \bun \nonumber \\ + \frac{v_{||}^2
v_\bot^2}{4B\Omega_i} \nabla \cdot ( \bun \bun \cdot \nabla \times
\bun ), \label{dmudt_term3_2}
\end{eqnarray}

\noindent where I have used $B \Omega_i \bun \cdot \nabla [ (\bun
\cdot \nabla \times \bun) / B\Omega_i ] = \nabla \cdot ( \bun \bun
\cdot \nabla \times \bun ) - (\nabla \cdot \bun + 2 \bun \cdot
\nabla \ln B) \bun \cdot \nabla \times \bun = \nabla \cdot ( \bun
\bun \cdot \nabla \times \bun ) + (\nabla \cdot \bun) \bun \cdot
\nabla \times \bun$, $B \Omega_i \nabla \cdot [ (\bun \times
\nabla \bun)^T/B\Omega_i ] \cdot \bun = - \matI : [ (\bun \times
\nabla \bun) \cdot \nabla \bun ] = - (\nabla \cdot \bun) \bun
\cdot \nabla \times \bun$ and
\begin{equation}
\matI : [ (\bun \times \nabla \bun) \cdot \nabla \bun ] = (\nabla
\cdot \bun) \bun \cdot \nabla \times \bun. \label{dirty_trick1}
\end{equation}

\noindent To prove this last expression, I write $\matI : [ ( \bun
\times \nabla \bun) \cdot \nabla \bun ]$ as a divergence, giving
\begin{equation}
\matI : [ ( \bun \times \nabla \bun) \cdot \nabla \bun ] = \nabla
\cdot [ (\bun \cdot \nabla \bun) \times \bun ] - \matI:(\bun \cdot
\nabla \nabla \bun \times \bun) + \bun \cdot \nabla \bun \cdot (
\nabla \times \bun ).
\end{equation}

\noindent Then, relation \eq{dirty_trick1} is recovered by using
$\matI:(\bun \cdot \nabla \nabla \bun \times \bun) = \bun \cdot
\nabla (\nabla \times \bun) \cdot \bun = \nabla \cdot (\bun \bun
\cdot \nabla \times \bun) - (\nabla \cdot \bun) \bun \cdot \nabla
\times \bun - \bun \cdot \nabla \bun \cdot (\nabla \times \bun)$
and equation \eq{id_kappa} to write $(\bun \cdot \nabla \bun)
\times \bun = - (\nabla \times \bun)_\bot$ and $\bun \cdot \nabla
\bun \cdot (\nabla \times \bun) = 0$.

In addition to equations \eq{dmudt_term2_2} and
\eq{dmudt_term3_2}, equation \eq{dmudt_term1_1} can be written as
\begin{eqnarray}
\left \langle \bv \cdot \nabla \left [ \frac{v_\bot^2}{2B^2
\Omega_i} ( \bv \times \bun ) \cdot \nabla B \right ] \right
\rangle = \frac{v_{||}^2 v_\bot^2}{B^2\Omega_i} (\nabla \times
\bun)_\bot \cdot \nabla B \nonumber \\ -
\frac{v_\bot^4}{4B\Omega_i} (\nabla \cdot \bun) \bun \cdot \nabla
\times \bun, \label{dmudt_term1_2}
\end{eqnarray}

\noindent where I use equation \eq{id_kappa} to write $\bun \cdot
\nabla \bun \cdot (\bun \times \nabla B) = - (\nabla \times
\bun)_\bot \cdot \nabla B$, and I employ $B^2 \Omega_i \nabla
\cdot [ (\bun \times \nabla B)/B^2 \Omega_i ] = (\nabla \times
\bun) \cdot \nabla B$, $B^2 \Omega_i \bun \cdot \nabla [ (\bun
\times \nabla B)/B^2 \Omega_i ] \cdot \bun = - \bun \cdot \nabla
\bun \cdot (\bun \times \nabla B) = (\nabla \times \bun)_\bot
\cdot \nabla B$ and $\bun \cdot \nabla B = - B (\nabla \cdot
\bun)$. Finally, I add equations \eq{dmudt_term4_1},
\eq{dmudt_term2_2}, \eq{dmudt_term3_2} and \eq{dmudt_term1_2} to
obtain
\begin{eqnarray}
\langle \bv \cdot \nabla \mu_1|_{\br, \bv} \rangle =
\frac{v_{||}^2 v_\bot^2}{4B\Omega_i} \Bigg [ \nabla \cdot (\bun
\bun \cdot \nabla \times \bun) + (\nabla \cdot \bun) \bun \cdot
\nabla \times \bun \nonumber \\ - B\Omega_i \bun \cdot \nabla
\left ( \frac{\bun \cdot \nabla \times \bun}{B\Omega_i} \right )
\Bigg ] = 0,
\end{eqnarray}

\noindent the property I was trying to prove. Notice that the time
derivative of $\langle \mu_1 \rangle = - (v_{||}
v_\bot^2/2B\Omega_i) \bun \cdot \nabla \times \bun$ was necessary
to obtain this result. The derivative $d\langle \mu_1 \rangle/dt$
is in part responsible for the term \eq{dmudt_term4_1}.

There is still the piece of $d(\mu_0 + \mu_1)/dt$ that depends
explicitly on the potential, given in \eq{dmudt_phi}. I will next
prove that it also gyroaverages to zero, the desired result. The
terms in \eq{dmudt_phi} must be written as a function of the
gyrokinetic variables in order to make the gyroaverage easier. I
will do so for each term to the required order. The first term in
\eq{dmudt_phi} is $- (Z e/M) \nabla \phi \cdot \nabla_v \mu_0 = -
(Z e/MB) \bv_\bot \cdot \nabla \phi$. Using $\bv_\bot \cdot \nabla
\phi = \bv \cdot \nabla \phi - v_{||} \bun \cdot \nabla \phi$ and
relation \eq{vgradphi2}, I find to order $\delta_i v_i^3/BL$
\begin{eqnarray}
- \frac{Z e}{M} \nabla \phi \cdot \nabla_v \mu_0 = \frac{Z e}{M B}
[v_{||} \bun (\br) \cdot \nabla \phi - u \bun (\bR) \cdot
\nabla_\bR \phi] \nonumber \\ - \frac{Z e}{M B} \bv_d \cdot
\nabla_\bR \phi + \frac{Z e \overline{\Omega}_i}{M B (\br)}
\frac{\partial \phi}{\partial \varphi}. \label{piecemu0fmu1ave}
\end{eqnarray}

\noindent In the lower order term $[Z e \overline{\Omega}_i/M
B(\br)] (\partial \phi/\partial \varphi)$, the difference $B(\br)-
B(\bR) \simeq - \Omega_i^{-1} (\bv \times \bun) \cdot \nabla B$ is
important, giving
\begin{equation}
\frac{Z e \overline{\Omega}_i}{M B (\br)} \frac{\partial
\phi}{\partial \varphi} \simeq \frac{Z e \overline{\Omega}_i}{M B
(\bR)} \frac{\partial \phi}{\partial \varphi} - \frac{c}{B^3}
(\bv_\bot \cdot \nabla \phi) [(\bv \times \bun) \cdot \nabla B],
\label{piecemu0fmu1ave_term2}
\end{equation}

\noindent where I employ the lowest order result $\partial
\phi/\partial \varphi \simeq - \Omega_i^{-1} \bv_\bot \cdot \nabla
\phi$. The term $v_{||} \bun (\br) \cdot \nabla \phi - u \bun
(\bR) \cdot \nabla_\bR \phi$ in \eq{piecemu0fmu1ave} is to the
order of interest
\begin{eqnarray}
v_{||} \bun (\br) \cdot \nabla \phi - u \bun (\bR) \cdot
\nabla_\bR \phi \simeq (v_{||} - u) \bun (\bR) \cdot \nabla_\bR
\phi \nonumber \\ + u [ \bun(\br) - \bun(\bR)] \cdot \nabla_\bR
\phi + u \bun(\bR) \cdot ( \nabla \phi - \nabla_\bR \phi ),
\label{piecemu0fmu1ave_term1}
\end{eqnarray}

\noindent where $\bun(\br) - \bun(\bR) \simeq - \Omega_i^{-1} (\bv
\times \bun) \cdot \nabla \bun$ and $\bun(\bR) \cdot ( \nabla \phi
- \nabla_\bR \phi)$ is given in \eq{par_grad_phi}. The difference
$v_{||} - u$ can be obtained from \eq{vparfGK}, giving
\begin{eqnarray}
v_{||} - u = - \frac{v_{||}}{\Omega_i} \bun \cdot \nabla \bun
\cdot ( \bv \times \bun ) - \frac{1}{4 \Omega_i} [ \bv_\bot ( \bv
\times \bun ) + ( \bv \times \bun ) \bv_\bot ]:\nabla \bun
\nonumber \\ - \frac{v^2_\bot}{2 \Omega_i} \bun \cdot \nabla
\times \bun = - \frac{v_{||}}{\Omega_i} \bun \cdot \nabla \bun
\cdot ( \bv \times \bun ) - \frac{1}{2\Omega_i} (\bv \times \bun)
\cdot \nabla \bun \cdot \bv_\bot \nonumber \\ -
\frac{v^2_\bot}{4\Omega_i} \bun \cdot \nabla \times \bun
\label{vpar_u_1},
\end{eqnarray}

\noindent where I use \eq{reverse_v4wig} to obtain the second
equality. Employing \eq{par_grad_phi}, \eq{piecemu0fmu1ave_term2},
\eq{piecemu0fmu1ave_term1} and the definition of
$\overline{\Omega}_i$ in \eq{omeg_mod}, equation
\eq{piecemu0fmu1ave} becomes
\begin{eqnarray}
- \frac{Z e}{M} \nabla \phi \cdot \nabla_v \mu_0 = \frac{Z e}{M B}
(\bun \cdot \nabla \phi) \left [ (v_{||} - u) -
\frac{v_{||}}{\Omega_i} \bun \cdot \nabla \bun \cdot (\bv \times
\bun) \right ] \nonumber \\ - \frac{cv_{||}}{B^2} (\bv \times
\bun) \cdot \nabla \bun \cdot \nabla \phi - \frac{cv_{||}}{2B^3}
[(\bv \times \bun) \cdot \nabla \phi] (\bun \cdot \nabla B)
\nonumber \\ - \frac{c}{B^3} (\bv_\bot \cdot \nabla \phi) [(\bv
\times \bun) \cdot \nabla B] - \frac{cv_{||}}{2B^2} (\bv_\bot
\cdot \nabla \phi) (\bun \cdot \nabla \times \bun) \nonumber \\ -
\frac{Ze}{MB} \bv_d \cdot \nabla_\bR \phi + \left ( \frac{Z^2
e^2}{M^2 c} + \frac{Z^3 e^3}{M^3 c B} \frac{\partial
\phiave}{\partial \mu} \right ) \frac{\partial \phi}{\partial
\varphi}. \label{piecemu0fmu1ave_1}
\end{eqnarray}

\noindent To obtain this form, I used the lowest order result
$\partial \phi / \partial \varphi \simeq - \Omega_i^{-1} \bv_\bot
\cdot \nabla \phi$ in the term $(v_{||}/2) (\bun \cdot \nabla
\times \bun) (\partial \phi/\partial \varphi)$.

The second term in \eq{dmudt_phi} is calculated employing
\eq{mu1_rv_conv}, giving
\begin{eqnarray}
- \frac{Z e}{M} \nabla \phi \cdot \nabla_v \mu_1 |_{\br, \bv} =
\frac{Ze}{MB} (\bun \cdot \nabla \phi) \Bigg [
\frac{v_{||}}{\Omega_i} \bun \cdot \nabla \bun \cdot ( \bv \times
\bun ) - (v_{||} - u) \Bigg ] \nonumber \\ + \frac{c}{B^3}
(\bv_\bot \cdot \nabla \phi) [(\bv \times \bun) \cdot \nabla B] +
\frac{Ze}{MB} \bv_M \cdot \nabla \phi \nonumber
\\ + \frac{cv_{||}}{2 B^2} (\nabla \phi \times \bun) \cdot \nabla
\bun \cdot \bv_\bot + \frac{cv_{||}}{2B^2} (\bv \times \bun) \cdot
\nabla \bun \cdot \nabla \phi \nonumber \\ + \frac{cv_{||}}{2B^2}
(\bv_\bot \cdot \nabla \phi) (\bun \cdot \nabla \times \bun),
\label{piecemu1fmu1ave}
\end{eqnarray}

\noindent with $v_{||} - u$ from \eq{vpar_u_1}. Substituting
equations~\eq{piecemu0fmu1ave_1} and \eq{piecemu1fmu1ave} in
equation~\eq{dmudt_phi}, the piece of $d(\mu_0 + \mu_1) / dt$ that
depends on $\phi$ is written as
\begin{eqnarray}
\left. \frac{d}{dt} ( \mu_0 + \mu_1 ) \right |_\phi = \left (
\frac{Z^2 e^2}{M^2c} + \frac{Z^3e^3}{M^3cB} \frac{\partial
\phiave}{\partial \mu} \right ) \frac{\partial \phiwig}{\partial
\varphi} \nonumber \\ + \frac{Zec}{MB^2} (\nabla_\bR \phiave
\times \bun) \cdot \nabla_\bR \phiwig + \frac{d}{dt} \mu_1|_\phi.
\label{dmudt_phi_2}
\end{eqnarray}

\noindent Here, I have used
\begin{equation}
(\nabla \phi \times \bun) \cdot \nabla \bun \cdot \bv_\bot - (\bv
\times \bun) \cdot \nabla \bun \cdot \nabla \phi = - (\nabla \cdot
\bun) (\bv \times \bun) \cdot \nabla \phi. \label{dirty_trick2}
\end{equation}

\noindent To prove this last expression, I employ $\nabla_\bot
\phi = v_\bot^{-2} [ \bv_\bot \bv_\bot \cdot \nabla \phi + (\bv
\times \bun) (\bv \times \bun) \cdot \nabla \phi ]$ to find
\begin{eqnarray}
(\nabla \phi \times \bun) \cdot \nabla \bun \cdot \bv_\bot - (\bv
\times \bun) \cdot \nabla \bun \cdot \nabla \phi = \nonumber \\ -
\frac{1}{v_\bot^2} [(\bv \times \bun) \cdot \nabla \phi] [
\bv_\bot \bv_\bot + (\bv \times \bun) (\bv \times \bun) ] : \nabla
\bun.
\end{eqnarray}

\noindent Upon using $\matI - \bun \bun = v_\bot^{-2} [ \bv_\bot
\bv_\bot + (\bv \times \bun) (\bv \times \bun) ]$, equation
\eq{dirty_trick2} is recovered.

Finally, the gyroaverage of equation \eq{dmudt_phi_2} is zero. The
term
\begin{equation}
\frac{d}{dt} \mu_1|_\phi = \frac{\partial}{\partial t} \mu_1|_\phi
+ \dot{\bR} \cdot \nabla_\bR \mu_1|_\phi + \dot{\varphi}
\frac{\partial}{\partial \varphi} \mu_1|_\phi
\end{equation}

\noindent vanishes when gyroaveraged because $\mu_1|_\phi =
Ze\phiwig/MB(\bR)$ gyroaverages to zero and the gyrokinetic
variables are defined such that $\dot{\bR}$ and $\dot{\varphi}$
are gyrophase independent. Notice that here it is important that
$B(\bR)$ in $\mu_1|_\phi = Ze\phiwig/MB(\bR)$ depends on $\bR$ and
not on $\br$.

\section{Jacobian of the gyrokinetic transformation \label{sectapp_jacob}}

In this section the Jacobian of the transformation from variables
$\br$, $\bv$ to variables $\bR$, $E$, $\mu$, $\varphi$ is
calculated, and the gyroaverage of condition \eq{Jacob_condition}
is checked.

The inverse of the Jacobian is
\begin{equation}
 \frac{1}{J} = \left |
\begin{array}{ccc:c:c:c}
\ddots& & &\vdots&\vdots&\vdots\\
 &\nabla \bR& &\nabla E&\nabla \mu&\nabla \varphi\\
 & &\ddots&\vdots&\vdots&\vdots\\
\hdashline
\ddots& & &\vdots&\vdots&\vdots\\
 &\nabla_v \bR& &\nabla_v E&\nabla_v \mu&\nabla_v \varphi\\
 & &\ddots&\vdots&\vdots&\vdots
\end{array}
\right | = \left |
\begin{array}{ccc:c:c:c}
\ddots & & &\vdots &\vdots &\vdots \\
 &\nabla \bR& &\nabla E&\nabla \mu&\nabla \varphi \\
 & &\ddots &\vdots &\vdots &\vdots \\
\hdashline
\ddots & & &\vdots &\vdots &\vdots \\
 &\mathbf{0} & &\partial E &\partial \mu &\partial \varphi \\
 & &\ddots &\vdots &\vdots &\vdots \label{jac_2}
\end{array}
\right |. \label{jac_1}
\end{equation}

\noindent Employing that the terms in the left columns of the
first form are to first approximation $\nabla \bR \simeq
\matrixtop{\mathbf{I}}$ and $\nabla_v \bR \simeq \Omega_i^{-1}
\matrixtop{\mathbf{I}} \times \bun$, the determinant is simplified
by combining linearly the rows in the matrix to determine the
second form, where
\begin{equation}
\partial (\ldots) = \nabla_v (\ldots) - \frac{1}{\Omega_i} \bun \times \nabla
(\ldots).
\end{equation}

\noindent The second form of \eq{jac_1} can be simplified by
noticing that the lower left piece of the matrix is zero. Thus,
the determinant may be written as
\begin{equation}
J^{-1} = \mathrm{det} ( \nabla \bR ) [ \partial E \cdot (
\partial \mu \times \partial \varphi ) ]. \label{jac_3}
\end{equation}

I analyze the two determinants on the right side independently.
The matrix $\nabla \bR$ is $\matrixtop{\mathbf{I}} + \nabla (
\Omega_i^{-1} \bv \times \bun + \bR_2 )$. Hence, $\mathrm{det} (
\nabla \bR ) \simeq 1 + \nabla \cdot ( \Omega_i^{-1} \bv \times
\bun + \bR_2 )$. The Jacobian must be obtained to first order
only. The only important term to that order in $\bR_2$ is the term
that contains the potential $\phi$, since its gradient may be
large, but $\nabla \cdot \bR_2 \simeq - \nabla \cdot [
(c/B\Omega_i) \nabla_\bR \Phiwig \times \bun ] \simeq 0$.
Therefore, the determinant of $\nabla \bR$ becomes
\begin{equation}
\mathrm{det} ( \nabla \bR ) = 1 - \bv \cdot \nabla \times \left (
\frac{\bun}{\Omega_i} \right ). \label{jac_p1}
\end{equation}

For the second determinant in \eq{jac_3}, I evaluate the columns
of the matrix $\partial E$, $\partial \mu$ and $\partial \varphi$
to the order of interest, using
\begin{equation}
\partial E = \bv + \nabla_v E_1 - \frac{1}{\Omega_i} \bun \times \nabla E,
\end{equation}
\begin{equation}
\partial \mu =\frac{\bv_\bot}{B} + \nabla_v \mu_1 -
\frac{1}{\Omega_i} \bun \times \nabla \mu
\end{equation}

\noindent and
\begin{equation}
\partial \varphi = - \frac{1}{v_\bot^2} \bv \times \bun + \nabla_v \varphi_1 -
\frac{1}{\Omega_i} \bun \times \nabla \varphi.
\end{equation}

\noindent The determinant becomes
\begin{eqnarray}
\partial E \cdot ( \partial \mu \times \partial \varphi )
\simeq \frac{v_{||}}{B(\br)} + \frac{\bun}{B} \cdot \nabla_v E_1 +
\left ( \frac{v_{||}}{v_\bot^2} \bv_\bot - \bun \right ) \cdot
\left ( \nabla_v \mu_1 - \frac{1}{\Omega_i} \bun \times \nabla \mu
\right ) \nonumber \\ - \frac{v_{||}}{B} (\bv \times \bun) \cdot
\left ( \nabla_v \varphi_1 - \frac{1}{\Omega_i} \bun \times \nabla
\varphi \right ). \label{jac_p2_1}
\end{eqnarray}

\noindent In the lower order term $v_{||}/B(\br)$, the difference
$B(\br) - B(\bR) \simeq - \Omega_i^{-1} (\bv \times \bun) \cdot
\nabla B \sim \delta_i B$ is important. From the definitions of
$E_1$, $\mu_1$ and $\varphi_1$, I find their gradients in velocity
space. I need the gradients in velocity space of $\phiwig$ and
$\partial \Phiwig /
\partial \mu$. The gradient $\nabla_v \phiwig$ is given by
\begin{eqnarray}
\nabla_v \phiwig = \nabla_v E \frac{\partial \phiwig}{\partial E}
+ \nabla_v \mu \frac{\partial \phiwig}{\partial \mu} + \nabla_v
\varphi \frac{\partial \phiwig}{\partial \varphi} + \nabla_v \bR
\cdot \nabla_\bR \phiwig \nonumber \\ = \frac{\bv_\bot}{B}
\frac{\partial \phiwig}{\partial \mu} - \frac{1}{v_\bot^2} \bv
\times \bun \frac{\partial \phiwig}{\partial \varphi} +
\frac{1}{\Omega_i} \bun \times \nabla_\bR \phiwig.
\end{eqnarray}

\noindent The gradient $\nabla_v ( \partial \Phiwig / \partial \mu
)$ is found in a similar way. The gradients in real space are only
to be obtained to zeroth order. However, some terms of the first
order quantities that contain $\phi$ are important because they
have steep gradients. Considering this, I find
\begin{equation}
\nabla E = \frac{Ze}{M} \nabla \phiwig,
\end{equation}
\begin{equation}
\nabla \mu = - \frac{v_\bot^2}{2 B^2} \nabla B - \frac{v_{||}}{B}
\nabla \bun \cdot \bv_\bot + \frac{Ze}{MB} \nabla \phiwig
\end{equation}

\noindent and
\begin{equation}
\nabla \varphi = \frac{v_{||}}{v_\bot^2} \nabla \bun \cdot ( \bv
\times \bun ) + \nabla \eun_2 \cdot \eun_1 - \frac{Ze}{MB} \nabla
\left ( \frac{\partial \Phiwig}{\partial \mu} \right ).
\end{equation}

\noindent Due to the preceding considerations, equation
\eq{jac_p2_1} becomes
\begin{equation}
\partial E \cdot ( \partial \mu \times \partial \varphi ) =
\frac{u}{B ( \mathbf{r} )} \left [ 1 + \frac{1}{\Omega_i} \bv
\cdot (\bun \times \kappabf) \right ], \label{jac_p2_3}
\end{equation}

\noindent where I have employed \eq{vpar_u_1} to express $v_{||}$
as a function of the gyrokinetic variables, and I have used $(\bv
\times \bun) \cdot \nabla \bun \cdot \bv_\bot - \bv_\bot \cdot
\nabla \bun \cdot (\bv \times \bun) = v_\bot^2 \bun \cdot \nabla
\times \bun$. This last result is deduced from $v_\bot^2
(\matrixtop{\mathbf{I}} - \bun \bun) = \bv_\bot \bv_\bot + (\bv
\times \bun) ( \bv \times \bun )$. In equation \eq{jac_p2_3}, in
the lower order term $u/B(\br)$, the difference $B(\br) - B(\bR) =
- \Omega_i^{-1} (\bv \times \bun) \cdot \nabla B$ is important.
Combining \eq{jac_p1} and \eq{jac_p2_3}, and using \eq{id_kappa}
to write $\bv \cdot ( \bun \times \kappabf) = \bv_\bot \cdot
\nabla \times \bun$, the Jacobian of the transformation is found
to be as given by \eq{result_Jacob}. Notice that to this order $J
= \langle J \rangle$ as required by \eq{Jacob_condition}.

Finally, I prove that $J$ satisfies the gyroaverage of
\eq{Jacob_condition} to the required order, namely,
\begin{equation}
\frac{\partial J}{\partial t} + \nabla_\bR \cdot \{ J [u \bun (
\bR ) + \bv_d ] \} - \frac{Ze}{M} \frac{\partial}{\partial E} \{ J
[u \bun ( \bR ) + \bv_d ] \cdot \nabla_\bR \phiave \} = 0,
\label{Jcond_1o}
\end{equation}

\noindent where $\bv_d$ is given by \eq{drift}. To first order, I
obtain
\begin{equation}
J [u \bun ( \bR ) + \bv_d ] \simeq \mathbf{B} ( \bR ) + \frac{Mc
\mu}{Ze u} \bun \times \nabla_\bR B + \frac{Mc u}{Ze} \nabla_\bR
\times \bun - \frac{c}{u} \nabla_\bR \phiave \times \bun,
\label{JvGK}
\end{equation}

\noindent where I have employed \eq{id_kappa}. Inserting \eq{JvGK}
into \eq{Jcond_1o} and recalling that $u = \sqrt{2[E - \mu B ( \bR
)]}$ is enough to prove that \eq{Jcond_1o}, and thus
\eq{Jacob_condition} gyroaveraged, are satisfied by the Jacobian
to first order.

\chapter{Useful gyroaverages and gyrophase derivatives
\label{app_vrelations}}

The various derivations require various averages and integrals
respect to the gyrophase $\varphi_0$. In particular, the integrals
for terms that contain several $\bv_\bot$ are recurrent. In this
Appendix, I show how to work out these type of terms.

According to \eq{gyrophase_def}, the perpendicular velocity,
$\bv_\bot$, is $\bv_\bot = v_\bot (\eun_1 \cos \varphi_0 + \eun_2
\sin \varphi_0)$. In order to find gyroaverages and integrals, it
is useful to express the perpendicular velocity as
\begin{equation}
\bv_\bot = \mathrm{Re} [ v_\bot \exp (i \varphi_0) \mathbf{u} ] =
\frac{v_\bot}{2} [ \exp(i \varphi_0) \mathbf{u} + \exp (- i
\varphi_0) \mathbf{u}^\ast ], \label{vp_complex}
\end{equation}

\noindent where $i = \sqrt{-1}$, $\mathbf{u} = \eun_1 - i \eun_2$,
$\mathbf{u}^\ast$ is the conjugate of the complex vector
$\mathbf{u}$ and $\mathrm{Re} ( \mathbf{a} )$ is the real part of
the complex tensor $\mathbf{a}$. Notice also that
\begin{equation}
v_\bot \exp(i \varphi_0) \mathbf{u} = \bv_\bot + i \bv \times \bun
. \label{vp_complex_2}
\end{equation}

It is very common to find tensors composed of tensor products of
$\bv_\bot$,
\begin{equation}
(\bv_\bot)^n \equiv \ubrace{\bv_\bot \bv_\bot \ldots \bv_\bot}{n\;
\mathrm{times} }.
\end{equation}

\noindent It is possible to find a more convenient form for this
tensor product by using equation~\eq{vp_complex}. It is useful to
distinguish between odd and even $n$. For even $n$, $n = 2m$,
\begin{eqnarray}
(\bv_\bot)^{2m} = \frac{v_\bot^{2m}}{2^{2m-1}} \mathrm{Re} \Big \{
\exp(i2m\varphi_0) \mathbf{W}^{(2m, 0)} + \exp[i2(m-1)\varphi_0]
\mathbf{W}^{(2m, 1)} \nonumber \\ + \ldots + \exp(i2\varphi_0)
\mathbf{W}^{(2m, m-1)} \Big \} + \frac{v_\bot^{2m}}{2^{2m}}
\mathbf{W}^{(2m, m)}, \label{even_product}
\end{eqnarray}

\noindent and for odd $n$, $n = 2m + 1$,
\begin{eqnarray}
(\bv_\bot)^{2m+1} = \frac{v_\bot^{2m+1}}{2^{2m}} \mathrm{Re} \Big
\{ \exp[i(2m+1)\varphi_0] \mathbf{W}^{(2m+1, 0)} \nonumber \\
+ \exp[i(2m-1)\varphi_0] \mathbf{W}^{(2m+1, 1)} + \ldots +
\exp(i\varphi_0) \mathbf{W}^{(2m+1, m)} \Big \},
\label{odd_product}
\end{eqnarray}

\noindent where the tensor $\mathbf{W}^{(n, p)}$ is the tensor
formed by the addition of all the possible different tensor
products between $(n-p)$ $\mathbf{u}$ vectors and $p$
$\mathbf{u}^\ast$ vectors, i.e.,
\begin{equation}
\mathbf{W}^{(n,p)} \equiv \ubrace{\mathbf{u} \ldots \mathbf{u}
\mathbf{u}}{n - p} \ubrace{\mathbf{u}^\ast \mathbf{u}^\ast \ldots
\mathbf{u}^\ast}{p} + \ubrace{\mathbf{u} \ldots \mathbf{u}}{n-p-1}
\mathbf{u}^\ast \mathbf{u} \ubrace{\mathbf{u}^\ast \ldots
\mathbf{u}^\ast}{p-1} + \ldots + \ubrace{\mathbf{u}^\ast \ldots
\mathbf{u}^\ast}{p} \ubrace{\mathbf{u} \ldots \mathbf{u}}{n-p}.
\label{tensor_W}
\end{equation}

\noindent There are $n!/[p!(n-p)!]$ different terms in the
summation. For example, $\mathbf{W}^{(5,2)}$ has 10 summands,
given by
\begin{eqnarray}
\mathbf{W}^{(5,2)} \equiv \mathbf{u} \mathbf{u} \mathbf{u}
\mathbf{u}^\ast \mathbf{u}^\ast + \mathbf{u} \mathbf{u}
\mathbf{u}^\ast \mathbf{u} \mathbf{u}^\ast + \mathbf{u}
\mathbf{u}^\ast \mathbf{u} \mathbf{u} \mathbf{u}^\ast +
\mathbf{u}^\ast \mathbf{u} \mathbf{u} \mathbf{u} \mathbf{u}^\ast +
\mathbf{u} \mathbf{u} \mathbf{u}^\ast \mathbf{u}^\ast \mathbf{u}
\nonumber \\ + \mathbf{u} \mathbf{u}^\ast \mathbf{u}
\mathbf{u}^\ast \mathbf{u} + \mathbf{u}^\ast \mathbf{u} \mathbf{u}
\mathbf{u}^\ast \mathbf{u} + \mathbf{u} \mathbf{u}^\ast
\mathbf{u}^\ast \mathbf{u} \mathbf{u} + \mathbf{u}^\ast \mathbf{u}
\mathbf{u}^\ast \mathbf{u} \mathbf{u} + \mathbf{u}^\ast
\mathbf{u}^\ast \mathbf{u} \mathbf{u} \mathbf{u}. \label{ex_W_52}
\end{eqnarray}

The tensor $\mathbf{W}^{(n,p)}$ can be written in a form in which
only $\mathbf{W}^{(n-2p,0)}$ and the matrix $\matI - \bun \bun$
appear. The tensor of the form $\mathbf{W}^{(m,0)}$ is more
convenient because it is part of $\mathrm{Re} [ v_\bot^m \exp (i m
\varphi_0) \mathbf{W}^{(m,0)} ]$ and easy to write in a
recognizable manner. For example, for $m = 2$, employing
\eq{vp_complex_2}, I find
\begin{equation}
\mathrm{Re} [ v_\bot^2 \exp (i 2 \varphi_0) \mathbf{W}^{(2,0)} ] =
\bv_\bot \bv_\bot - (\bv \times \bun) (\bv \times \bun).
\label{v2wig_1}
\end{equation}

\noindent For $m = 3$,
\begin{eqnarray}
\mathrm{Re} [ v_\bot^3 \exp (i 3 \varphi_0) \mathbf{W}^{(3,0)} ] =
\bv_\bot \bv_\bot \bv_\bot - \bv_\bot (\bv \times \bun) (\bv
\times \bun) - (\bv \times \bun) \bv_\bot (\bv \times \bun)
\nonumber \\ - (\bv \times \bun) (\bv \times \bun) \bv_\bot.
\label{v3wig}
\end{eqnarray}

\noindent These tensors are not only easy to write but also easy
to integrate in $\varphi_0$ because
\begin{equation}
\frac{\partial}{\partial \varphi_0} \mathrm{Im} [ v_\bot^m \exp (i
m \varphi_0) \mathbf{W}^{(m,0)} ] = m \mathrm{Re} [ v_\bot^m \exp
(i m \varphi_0) \mathbf{W}^{(m,0)} ], \label{deriv_vtensor}
\end{equation}

\noindent with $\mathrm{Im} (\mathbf{a})$ the imaginary part of
the complex tensor $\mathbf{a}$. Equation \eq{deriv_vtensor} for
$m = 2$ is
\begin{equation}
\frac{\partial}{\partial \varphi_0} [ \bv_\bot ( \bv \times \bun )
+ (\bv \times \bun) \bv_\bot ] = 2 [ \bv_\bot \bv_\bot - (\bv
\times \bun) (\bv \times \bun) ], \label{v2int_1}
\end{equation}

\noindent and for $m = 3$ is
\begin{eqnarray}
\frac{\partial}{\partial \varphi_0} [ \bv_\bot \bv_\bot ( \bv
\times \bun ) + \bv_\bot ( \bv \times \bun ) \bv_\bot + ( \bv
\times \bun ) \bv_\bot \bv_\bot - (\bv \times \bun) (\bv \times
\bun) (\bv \times \bun) ] \nonumber \\ = 3 [ \bv_\bot \bv_\bot
\bv_\bot - \bv_\bot (\bv \times \bun) (\bv \times \bun) - (\bv
\times \bun) \bv_\bot (\bv \times \bun) - (\bv \times \bun) (\bv
\times \bun) \bv_\bot ].
\end{eqnarray}

\noindent Equation \eq{v2int_1} is used in the derivation of the
gyrokinetic variables.

The decomposition of $\mathbf{W}^{(n,p)}$ into
$\mathbf{W}^{(n-2p,0)}$ and matrices $\matI - \bun \bun$ is also
interesting for gyroaverages. According to equation
\eq{even_product}, the tensor products of an even number of
vectors $\bv_\bot$ has the gyroaverage
\begin{equation}
\overline{(\bv_\bot)^{2m}} = \frac{v_\bot^{2m}}{2^{2m}}
\mathbf{W}^{(2m, m)}, \label{gyroave_1}
\end{equation}

\noindent where $\mathbf{W}^{(2m,m)}$ is a summation of tensor
products of $p$ $(\matI - \bun \bun)$ matrices. At the end of this
appendix, I will give the result as a function of $\matI - \bun
\bun$. The gyroaverage of an odd number of $\bv_\bot$ is zero, as
shown in \eq{odd_product}.

I will prove now that $\mathbf{W}^{(n,p)}$ can be written such
that it only contains $\mathbf{W}^{(n-2p,0)}$ and the matrix
$\matI - \bun \bun$. To do so, I will first prove that
$\mathbf{W}^{(n,p)}$ can be decomposed into a summation of tensors
formed by the product of $\mathbf{W}^{(n-2p,0)}$ and
$\mathbf{W}^{(2p,p)}$. After that, I will show that
$\mathbf{W}^{(2p,p)}$ is formed by a summation of tensor products
of $p$ $(\matI - \bun\bun)$ matrices.

By multiplying $\mathbf{W}^{(n,p)}$ by $(n-p)!/[p!(n-2p)!]$, $(n -
2p)$ $\mathbf{u}$ vectors can be distinguished from the rest of
$\mathbf{u}$ vectors. I will denote them as $\mathbf{u}^d$ (this
is a trick to make the derivation easier). Then, a new tensor
$\mathbf{T}^{(n,p,p)}$ is defined as the tensor formed by $p$
$\mathbf{u}$ vectors, $p$ $\mathbf{u}^\ast$ vectors and $(n - 2p)$
$\mathbf{u}^d$ vectors, giving
\begin{eqnarray}
\frac{(n-p)!}{p! (n-2p)!} \mathbf{W}^{(n,p)} =
\mathbf{T}^{(n,p,p)} \equiv \ubrace{\mathbf{u}^d \ldots
\mathbf{u}^d}{n - 2p} \ubrace{\mathbf{u} \ldots \mathbf{u}
\mathbf{u}}{p} \ubrace{\mathbf{u}^\ast \mathbf{u}^\ast \ldots
\mathbf{u}^\ast}{p} \nonumber \\ + \ubrace{\mathbf{u}^d \ldots
\mathbf{u}^d}{n - 2p - 1} \mathbf{u} \mathbf{u}^d
\ubrace{\mathbf{u} \ldots \mathbf{u} \mathbf{u}}{p}
\ubrace{\mathbf{u}^\ast \mathbf{u}^\ast \ldots \mathbf{u}^\ast}{p
- 1} + \ldots + \ubrace{\mathbf{u}^\ast \ldots \mathbf{u}^\ast}{p}
\ubrace{\mathbf{u} \ldots \mathbf{u}}{p} \ubrace{\mathbf{u}^d
\ldots \mathbf{u}^d}{n-2p}. \label{tensor_T}
\end{eqnarray}

\noindent For instance, $3 \mathbf{W}^{(5,2)} =
\mathbf{T}^{(5,2,2)}$ can be deduced from \eq{ex_W_52} to give
\begin{eqnarray}
\mathbf{T}^{(5,2,2)} \equiv \mathbf{u}^d \mathbf{u} \mathbf{u}
\mathbf{u}^\ast \mathbf{u}^\ast + \mathbf{u}^d \mathbf{u}
\mathbf{u}^\ast \mathbf{u} \mathbf{u}^\ast + \mathbf{u}^d
\mathbf{u}^\ast \mathbf{u} \mathbf{u} \mathbf{u}^\ast +
\mathbf{u}^\ast \mathbf{u}^d \mathbf{u} \mathbf{u} \mathbf{u}^\ast
+ \mathbf{u}^d \mathbf{u} \mathbf{u}^\ast \mathbf{u}^\ast
\mathbf{u} \nonumber \\ + \mathbf{u}^d \mathbf{u}^\ast \mathbf{u}
\mathbf{u}^\ast \mathbf{u} + \mathbf{u}^\ast \mathbf{u}^d
\mathbf{u} \mathbf{u}^\ast \mathbf{u} + \mathbf{u}^d
\mathbf{u}^\ast \mathbf{u}^\ast \mathbf{u} \mathbf{u} +
\mathbf{u}^\ast \mathbf{u}^d \mathbf{u}^\ast \mathbf{u} \mathbf{u}
+ \mathbf{u}^\ast \mathbf{u}^\ast \mathbf{u}^d \mathbf{u}
\mathbf{u} \nonumber \\ + \mathbf{u} \mathbf{u}^d \mathbf{u}
\mathbf{u}^\ast \mathbf{u}^\ast + \mathbf{u} \mathbf{u}^d
\mathbf{u}^\ast \mathbf{u} \mathbf{u}^\ast + \mathbf{u}
\mathbf{u}^\ast \mathbf{u}^d \mathbf{u} \mathbf{u}^\ast +
\mathbf{u}^\ast \mathbf{u} \mathbf{u}^d \mathbf{u} \mathbf{u}^\ast
+ \mathbf{u} \mathbf{u}^d \mathbf{u}^\ast \mathbf{u}^\ast
\mathbf{u} \nonumber \\ + \mathbf{u} \mathbf{u}^\ast \mathbf{u}^d
\mathbf{u}^\ast \mathbf{u} + \mathbf{u}^\ast \mathbf{u}
\mathbf{u}^d \mathbf{u}^\ast \mathbf{u} + \mathbf{u}
\mathbf{u}^\ast \mathbf{u}^\ast \mathbf{u}^d \mathbf{u} +
\mathbf{u}^\ast \mathbf{u} \mathbf{u}^\ast \mathbf{u}^d \mathbf{u}
+ \mathbf{u}^\ast \mathbf{u}^\ast \mathbf{u} \mathbf{u}^d
\mathbf{u} \nonumber \\ + \mathbf{u} \mathbf{u} \mathbf{u}^d
\mathbf{u}^\ast \mathbf{u}^\ast + \mathbf{u} \mathbf{u}
\mathbf{u}^\ast \mathbf{u}^d \mathbf{u}^\ast + \mathbf{u}
\mathbf{u}^\ast \mathbf{u} \mathbf{u}^d \mathbf{u}^\ast +
\mathbf{u}^\ast \mathbf{u} \mathbf{u} \mathbf{u}^d \mathbf{u}^\ast
+ \mathbf{u} \mathbf{u} \mathbf{u}^\ast \mathbf{u}^\ast
\mathbf{u}^d \nonumber \\ + \mathbf{u} \mathbf{u}^\ast \mathbf{u}
\mathbf{u}^\ast \mathbf{u}^d + \mathbf{u}^\ast \mathbf{u}
\mathbf{u} \mathbf{u}^\ast \mathbf{u}^d + \mathbf{u}
\mathbf{u}^\ast \mathbf{u}^\ast \mathbf{u} \mathbf{u}^d +
\mathbf{u}^\ast \mathbf{u} \mathbf{u}^\ast \mathbf{u} \mathbf{u}^d
+ \mathbf{u}^\ast \mathbf{u}^\ast \mathbf{u} \mathbf{u}
\mathbf{u}^d. \label{ex_T_522}
\end{eqnarray}

\noindent The tensor $\mathbf{T}^{(n,p,p)}$ can be written as a
combination of a tensor $\mathbf{W}^{(n-2p,0)}$ formed by $(n-2p)$
$\mathbf{u}^d$ vectors and a tensor $\mathbf{W}^{(2p,p)}$ formed
by $p$ $\mathbf{u}$ vectors and $p$ $\mathbf{u}^\ast$ vectors,
leading to
\begin{eqnarray}
T^{(n,p,p)}_{j_1 j_2 \ldots j_n} \equiv W^{(2p,p)}_{j_1 \ldots
j_{2p}} W^{(n-2p,0)}_{j_{2p+1} \ldots j_n} + W^{(2p,p)}_{j_1
\ldots j_{2p-1} j_{2p+1}} W^{(n-2p,0)}_{j_{2p} j_{2p+2} \ldots
j_n} + \ldots \nonumber \\ + W^{(2p,p)}_{j_{n-2p+1} \ldots j_n}
W^{(n-2p,0)}_{j_1 \ldots j_{n-2p}} = \frac{n!}{(n-2p)!(2p)!}
W^{(2p,p)}_{(j_1 \ldots j_{2p}} W^{(n-2p,0)}_{j_{2p+1} \ldots
j_n)}. \label{T_W_tensors}
\end{eqnarray}

\noindent Here, the number of summands is $n!/[(n-2p)!(2p)!]$
because the tensors $\mathbf{W}^{(m,0)}$ and $\mathbf{W}^{(2p,p)}$
are symmetric respect to all of their indexes. Additionally, the
tensor $\mathbf{W}^{(2p,p)}$ is always real, since
$[\mathbf{W}^{(2p,p)}]^\ast = \mathbf{W}^{(2p,p)}$ by definition
of $\mathbf{W}^{(2p,p)}$. The last expression in \eq{T_W_tensors}
is based on the typical tensor notation, where the parenthesis
around the indexes indicate symmetrization of the tensor. As an
example of the equivalence in \eq{T_W_tensors}, the tensor
$\mathbf{T}^{(5,2,2)}$ in \eq{ex_T_522} may be written as
\begin{eqnarray}
T^{(5,2,2)}_{jklmn} = W^{(4,2)}_{jklm} W^{(1,0)}_n +
W^{(4,2)}_{jkln} W^{(1,0)}_m + W^{(4,2)}_{jkmn} W^{(1,0)}_l +
W^{(4,2)}_{jlmn} W^{(1,0)}_k \nonumber \\ + W^{(4,2)}_{klmn}
W^{(1,0)}_j, \label{ex_T522_W10_W42}
\end{eqnarray}

\noindent with $\mathbf{W}^{(1,0)} \equiv \mathbf{u} \equiv
\mathbf{u}^d$, and
\begin{equation}
\mathbf{W}^{(4,2)} \equiv \mathbf{u} \mathbf{u} \mathbf{u}^\ast
\mathbf{u}^\ast + \mathbf{u} \mathbf{u}^\ast \mathbf{u}
\mathbf{u}^\ast + \mathbf{u}^\ast \mathbf{u} \mathbf{u}
\mathbf{u}^\ast + \mathbf{u} \mathbf{u}^\ast \mathbf{u}^\ast
\mathbf{u} + \mathbf{u}^\ast \mathbf{u} \mathbf{u}^\ast \mathbf{u}
+ \mathbf{u}^\ast \mathbf{u}^\ast \mathbf{u} \mathbf{u}.
\label{ex_W_42}
\end{equation}

\noindent The tensor $\mathbf{W}^{(2p,p)}$ can be rewritten
employing yet a third tensor, $\mathbf{I}^{(2p)}$. This tensor is
formed by adding all the possible different tensor products that
are formed by $p$ matrices $(\matI - \bun \bun)_{jk} = \delta_{jk}
- \hat{b}_j \hat{b}_k = \delta^\bot_{jk}$, i.e.,
\begin{eqnarray}
I^{(2p)}_{j_1 j_2 \ldots j_{2p}} = \delta^\bot_{j_1 j_2}
\delta^\bot_{j_3 j_4} \ldots \delta^\bot_{j_{2p-1} j_{2p}} +
\delta^\bot_{j_1 j_3} \delta^\bot_{j_2 j_4} \ldots
\delta^\bot_{j_{2p-1} j_{2p}} + \ldots = \nonumber \\
\frac{(2p)!}{2^p p!} \delta^\bot_{(j_1 j_2} \delta^\bot_{j_3 j_4}
\ldots \delta^\bot_{j_{2p-1} j_{2p} )}. \label{tensor_I}
\end{eqnarray}

\noindent This tensor is formed by $(2p)!/(2^p p!)$ summands. For
example, $\mathbf{I}^{(4)}$ is
\begin{equation}
I^{(4)}_{jklm} = \delta^\bot_{jk} \delta^\bot_{lm} +
\delta^\bot_{jl} \delta^\bot_{km} + \delta^\bot_{jm}
\delta^\bot_{kl}. \label{ex_I_4}
\end{equation}

\noindent The tensor $\mathbf{I}^{(2p)}$ can be written in terms
of $\mathbf{W}^{(2p,p)}$ because
\begin{equation}
2(\matI - \bun \bun) = 2( \eun_1 \eun_1 + \eun_2 \eun_2) =
\mathbf{u} \mathbf{u}^\ast + \mathbf{u}^\ast \mathbf{u}.
\label{u_ustar}
\end{equation}

\noindent Using this relation, the tensor $2^p \mathbf{I}^{(2p)}$,
formed by all possible tensor products of $p$ $2(\matI - \bun
\bun)$ matrices, is written as a summation of tensor products of
$p$ $\mathbf{u}$ vectors and $p$ $\mathbf{u}^\ast$ vectors by
substituting equation \eq{u_ustar} in it. Each summand $2^p
\delta^\bot_{j_1 j_2} \delta^\bot_{j_3 j_4} \ldots
\delta^\bot_{j_{2p-1} j_{2p}}$ of $2^p \mathbf{I}^{(2p)}$ [recall
\eq{tensor_I}] gives $2^p$ different tensors formed by $p$
$\mathbf{u}$ vectors and $p$ $\mathbf{u}^\ast$ vectors. Adding all
the terms in $\mathbf{I}^{(2p)}$, there is a total of $(2p)!/p!$
terms formed by tensor products of $p$ $\mathbf{u}$ vectors and
$p$ $\mathbf{u}^\ast$ vectors. In this summation, each summand of
$\mathbf{W}^{(2p,p)}$ is present $p!$ times. According to
\eq{tensor_W}, a summand of $\mathbf{W}^{(2p,p)}$ is uniquely
determined by the positions that the $p$ $\mathbf{u}$ vectors
occupy, or alternatively, which $p$ indexes of $j_1 j_2 \ldots
j_{2p}$ correspond to the $p$ $\mathbf{u}$ vectors. Without loss
of generality, these indexes can be chosen to be $j_1 j_2 \ldots
j_p$. A tensor of the form $u_{j_1} u_{j_2} \ldots u_{j_p}
u^\ast_{j_{p+1}} \ldots u^\ast_{j_{2p}}$ is obtained from every
term of $2^p \mathbf{I}^{(2p)}$ of the form $2^p \delta^\bot_{j_1
k_1} \delta^\bot_{j_2 k_2} \ldots \delta^\bot_{j_p k_p}$, with
$k_1 k_2 \ldots k_p$ being any permutation of $j_{p+1} j_{p+2}
\ldots j_{2p}$. There are $p!$ of these terms, proving that
\begin{equation}
2^p \mathbf{I}^{(2p)} = p! \mathbf{W}^{(2p,p)}.
\label{I_W_tensors}
\end{equation}

\noindent Considering the examples \eq{ex_W_42} and \eq{ex_I_4},
equation \eq{I_W_tensors} implies that their relation is $4
\mathbf{I}^{(4)} = 2 \mathbf{W}^{(4,2)}$. Combining equations
\eq{tensor_T}, \eq{T_W_tensors} and \eq{I_W_tensors} gives
\begin{eqnarray}
W^{(n,p)}_{j_1 j_2 \ldots j_n} = \frac{2^p (n-2p)!}{(n-p)!} \Big [
I^{(2p)}_{j_1 \ldots j_{2p}} W^{(n-2p,0)}_{j_{2p+1} \ldots j_n} +
I^{(2p)}_{j_1 \ldots j_{2p-1} j_{2p+1}} W^{(n-2p,0)}_{j_{2p}
j_{2p+2} \ldots j_n} + \ldots \nonumber \\ + I^{(2p)}_{j_{n-2p+1}
\ldots j_n} W^{(n-2p,0)}_{j_1 \ldots j_{n-2p}} \Big ] =
\frac{n!}{p! (n-p)!} I^{(2p)}_{(j_1 \ldots j_{2p}}
W^{(n-2p,0)}_{j_{2p+1} \ldots j_n)}.
\end{eqnarray}

\noindent As promised, $\mathbf{W}^{(n,p)}$ is written in terms of
$\mathbf{W}^{(n-2p,0)}$ and $\matI - \bun \bun$ because
$\mathbf{I}^{(2p)}$ is formed by addition of tensor products of
$\matI - \bun \bun$. Continuing with the example in \eq{ex_W_52},
$\mathbf{W}^{(5,2)}$ gives
\begin{equation}
W^{(5,2)}_{jklmn} = \frac{2}{3} \Big [ I^{(4)}_{jklm} u_n +
I^{(4)}_{jkln} u_m + I^{(4)}_{jkmn} u_l + I^{(4)}_{jlmn} u_k +
I^{(4)}_{klmn} u_j \Big ],
\end{equation}

\noindent with $\mathbf{I}^{(4)}$ given in \eq{ex_I_4}. This
tensor appears in $(\bv_\bot)^5 \equiv \bv_\bot \bv_\bot \bv_\bot
\bv_\bot \bv_\bot$ as [recall \eq{odd_product}]
\begin{eqnarray}
\frac{1}{2^4}\mathrm{Re} [ v_\bot^5 \exp(i\varphi_0)
W^{(5,2)}_{jklmn} ] = \frac{v_\bot^4}{24} \Big [ I^{(4)}_{jklm}
v_{\bot, n} + I^{(4)}_{jkln} v_{\bot, m} + I^{(4)}_{jkmn} v_{\bot,
l} \nonumber \\ + I^{(4)}_{jlmn} v_{\bot, k} + I^{(4)}_{klmn}
v_{\bot, j} \Big ].
\end{eqnarray}

To finish, the gyroaverage of an even number of vectors $\bv_\bot$
is calculated. Combining equations \eq{gyroave_1} and
\eq{I_W_tensors} gives
\begin{equation}
\overline{(\bv_\bot)^{2m}} = \frac{v_\bot^{2m}}{2^m m!}
\mathbf{I}^{(2m)}, \label{gyroave_final}
\end{equation}

\noindent with $\mathbf{I}^{(2m)}$ given in \eq{tensor_I}. For $m
= 1$, the familiar result
\begin{equation}
\overline{\bv_\bot \bv_\bot} = \frac{v_\bot^2}{2} (\matI - \bun
\bun)
\end{equation}

\noindent is found. This result, combined with \eq{even_product}
for $m = 2$ and \eq{v2wig_1} gives \eq{v2wig}. The gyroaverage for
$m = 4$ gives \eq{v4ave}.

\chapter{Gyrokinetic equivalence} \label{app_equivalence}

In this Appendix, a summary of \cite{parra09a}, I prove that the
gyrokinetic results in chapter~\ref{chap_gyrokinetics} are
completely consistent with the pioneering results by Dubin
\emph{et al} \cite{dubin83}, obtained using a Hamiltonian
formalism and Lie transforms \cite{brizard07}. This comparison
confirms that both approaches, the recursive method in
chapter~\ref{chap_gyrokinetics} and the Lie transforms, yield the
same final gyrokinetic formalism.

In chapter~\ref{chap_gyrokinetics}, the recursive approach
developed in \cite{leecatto83, bernstein85} was generalized for
nonlinear electrostatic gyrokinetics in a general magnetic field.
In reference \cite{dubin83}, the nonlinear electrostatic
gyrokinetic equation was derived for a constant magnetic field and
a collisionless plasma using a Hamiltonian formalism. The
asymptotic expansion was carried out to higher order in
\cite{dubin83} because the calculation is easier in a constant
magnetic field. When the method proposed in
chapter~\ref{chap_gyrokinetics} is extended to next order, the
results are different in appearance, but I will prove that these
differences are due to subtleties in some definitions.

Both methods are asymptotic expansions in the small parameter
$\delta = \rho/L \ll 1$. Here $L$ is a characteristic macroscopic
length in the problem and $\rho = v_\mathrm{th}/\Omega$ is the
gyroradius, with $\Omega = ZeB/Mc$ the gyrofrequency,
$v_\mathrm{th} = \sqrt{2T/M}$ the thermal velocity and $T$ the
temperature. In both methods, the phase space $\{ \br, \bv \}$,
with $\br$ and $\bv$ the position and velocity of the particles,
is expressed in gyrokinetic variables, defined order by order in
$\delta$. In reference \cite{dubin83}, the gyrokinetic variables
are obtained by Lie transform and the gyrokinetic equation is
found to second order in $\delta$. In
chapter~\ref{chap_gyrokinetics}, the gyrokinetic variables are
found by imposing that their time derivative is gyrophase
independent. Here $d/dt \equiv \partial /
\partial t + \bv \cdot \nabla + ( - Ze \nabla \phi /M + \Omega \bv
\times \bun ) \cdot \nabla_v$ is the Vlasov operator. In
chapter~\ref{chap_gyrokinetics}, the gyrokinetic equation was only
found to first order. In this Appendix, I will calculate the
gyrokinetic equation and the gyrokinetic variables to higher order
for a constant magnetic field, and I will compare the results with
those in \cite{dubin83}. The orderings and assumptions are more
general than in section~\ref{sect_GKorderings}. The pieces of the
distribution function and the potential with short wavelengths
also scale as
\begin{equation}
\frac{f_k}{f_s} \sim \frac{e\phi_k}{T} \sim \frac{1}{k_\bot L}
\ssim 1, \label{order}
\end{equation}

\noindent with $k_\bot \rho \ssim 1$. Here $f_s$ is the lowest
order distribution function with a slow variation in both $\br$
and $\bv$. In chapter~\ref{chap_gyrokinetics}, the lowest order
distribution functions was assumed to be a Maxwellian. In this
Appendix, to ease the comparison with reference \cite{dubin83}, I
relax that assumption. Finally, I order the time derivatives as
$\partial/\partial t \sim v_\mathrm{th} / L$.

\section{Constant magnetic field results}

The general gyrokinetic variables obtained in
section~\ref{sect_GKvariables} are $\bR = \br + \bR_1 + \bR_2$, $E
= E_0 + E_1 + E_2$, $\mu = \mu_0 + \mu_1 + \mu_2$ and $\varphi =
\varphi_0 + \varphi_1 + \varphi_2$. Since the unit vector $\bun$
is assumed constant in space and time, I can define $\eun_1$ and
$\eun_2$ in \eq{gyrophase_def} so that they are also constant, and
I do so to simplify the comparison with \cite{dubin83}. The
corrections found in section~\ref{sect_GKvariables} specialized to
constant magnetic field are, for the gyrocenter position $\bR$,
$\bR_1 = \Omega^{-1} \bv \times \bun$ and $\bR_2 = - (c/B\Omega)
\nabla_\bR \Phiwig \times \bun$; for the kinetic energy $E$, $E_1
= Ze\phiwig/M$ and $E_2 = (c/B) (\partial \Phiwig/\partial t)$;
and for the magnetic moment $\mu$ and the gyrophase $\varphi$,
$\mu_1 = Ze\phiwig/MB$ and $\varphi_1 = - (Ze/MB) (\partial
\Phiwig/\partial \mu)$. The corrections $\mu_2$ and $\varphi_2$
were not calculated because they were not needed to obtain the
gyrokinetic equation to first order in $\delta$ under the
assumptions in chapter~\ref{chap_gyrokinetics}.

I require the gyroaverage of $d\bR/dt$ and $dE/dt$ to higher order
than in section~\ref{sect_GKvariables}, and I need the second
order correction $\mu_2$. For constant magnetic fields, $\langle
d\bR/dt \rangle $, $\langle dE/dt \rangle$ and the correction
$\mu_2$ can be easily calculated by employing the methodology in
chapter~\ref{chap_gyrokinetics}. I will define $\mu_2$ so that the
gyroaverage of $d\mu/dt$ is zero to order $\delta^2
v_\mathrm{th}^3/BL$. The second order correction $\varphi_2$ is
needed to obtain $d\varphi/dt$ to higher order. However, $f$ is
gyroaveraged, making the dependence on $\varphi$ weak. Thus,
$\varphi_2$ never enters in the final results and will not be
necessary for our purposes. Once I have $\bR$, $E$, $\mu$ and
their derivatives to higher order, I will compare these results to
both the gyrokinetic Vlasov equation and the gyrokinetic Poisson's
equation in \cite{dubin83}.

\section{Time derivative of $\bR$}

Employing the definitions of $\bR_1$ and $\bR_2$, I find
\begin{equation}
\frac{d\bR}{dt} = v_{||} \bun - \frac{c}{B} \nabla \phi \times
\bun + \frac{d\bR_2}{dt}.
\end{equation}

\noindent The gyroaverage of this expression is performed holding
the gyrokinetic variables $\bR$, $E$, $\mu$ and $t$ fixed to
obtain
\begin{equation}
\left \langle \frac{d\bR}{dt} \right \rangle = u \bun -
\frac{c}{B} \langle \nabla \phi \rangle \times \bun,
\label{ave_dRdt_1}
\end{equation}

\noindent where $u = \langle v_{||} \rangle$. My gyrokinetic
variables are defined so that when the Vlasov operator is applied
to a function with a vanishing gyroaverage, like $\bR_2 = \bR_2
(\bR, E, \mu, \varphi, t)$, the result also has a zero
gyroaverage; namely $\langle d\bR_2/dt \rangle = 0$.

The gradient $\nabla \phi$ is written in the gyrokinetic variables
by using
\begin{equation}
\nabla \phi = \nabla \bR \cdot \nabla_\bR \phi + \frac{\partial
\phi}{\partial \mu} \nabla \mu + \frac{\partial \phi}{\partial
\varphi} \nabla \varphi \simeq \nabla_\bR \phi + \nabla \bR_2
\cdot \nabla_\bR \phi + \frac{\partial \phi}{\partial \mu} \nabla
\mu_1 + \frac{\partial \phi}{\partial \varphi} \nabla \varphi_1.
\label{gradphi_1}
\end{equation}

\noindent Here, I neglect $\partial \phi / \partial E \simeq -
(\partial \bR_1/\partial E) \cdot \nabla \phi$ because the
function $\bR_1$ does not depend on $E$ to order $\delta L$. To
obtain the second equality, I use that $\nabla \bR_1 = 0 = \nabla
\mu_0 = \nabla \varphi_0$. The gyroaverage of equation
\eq{gradphi_1}, obtained employing the definitions of $\bR_2$,
$\mu_1$ and $\varphi_1$, gives
\begin{equation}
\langle \nabla \phi \rangle \simeq \nabla_\bR \phiave -
\frac{c}{B\Omega} \langle \nabla_\bR \nabla_\bR \Phiwig \cdot
(\bun \times \nabla_\bR \phi) \rangle + \frac{Ze}{MB} \left
\langle \frac{\partial \phi}{\partial \mu} \nabla_\bR \phiwig -
\frac{\partial \phi}{\partial \varphi} \nabla_\bR \left (
\frac{\partial \Phiwig}{\partial \mu} \right ) \right \rangle.
\label{ave_gradphi}
\end{equation}

\noindent This equation can be simplified by integrating by parts
in $\varphi$ to obtain
\begin{equation}
\left \langle \frac{\partial \phi}{\partial \mu} \nabla_\bR
\phiwig - \frac{\partial \phi}{\partial \varphi} \nabla_\bR \left
( \frac{\partial \Phiwig}{\partial \mu} \right ) \right \rangle =
\left \langle  \frac{\partial \phiwig}{\partial \mu} \nabla_\bR
\phiwig + \phiwig \nabla_\bR \left ( \frac{\partial
\phiwig}{\partial \mu} \right ) \right \rangle = \frac{1}{2}
\nabla_\bR \left ( \frac{\partial}{\partial \mu} \langle \phiwig^2
\rangle \right ). \label{int_mu1vphi1dphi}
\end{equation}

\noindent I next demonstrate that
\begin{equation}
\langle \nabla_\bR \nabla_\bR \Phiwig \cdot ( \bun \times
\nabla_\bR \phi) \rangle = \frac{1}{2} \nabla_\bR \langle
\nabla_\bR \Phiwig \cdot ( \bun \times \nabla_\bR \phiwig )
\rangle \label{int_R2gradphi}
\end{equation}

\noindent by first noticing that
\begin{equation}
\langle \nabla_\bR \nabla_\bR \Phiwig \cdot (\bun \times
\nabla_\bR \phi) \rangle = \nabla_\bR \langle \nabla_\bR \Phiwig
\cdot (\bun \times \nabla_\bR \phiwig) \rangle + \langle
\nabla_\bR \nabla_\bR \phiwig \cdot (\bun \times \nabla_\bR
\Phiwig) \rangle.
\end{equation}

\noindent Integrating by parts in $\varphi$ in the second term, I
find $\langle \nabla_\bR \nabla_\bR \phiwig \cdot ( \bun \times
\nabla_\bR \Phiwig ) \rangle = - \langle \nabla_\bR \nabla_\bR
\Phiwig \cdot ( \bun \times \nabla_\bR \phiwig ) \rangle$, giving
the result in \eq{int_R2gradphi}.

Finally, substituting equations \eq{int_mu1vphi1dphi} and
\eq{int_R2gradphi} into equation \eq{ave_gradphi} and using the
result in \eq{ave_dRdt_1} gives
\begin{equation}
\left \langle \frac{d\bR}{dt} \right \rangle = u\bun - \frac{c}{B}
\nabla_\bR \Psi \times \bun, \label{ave_dRdt}
\end{equation}

\noindent with
\begin{equation}
\Psi = \phiave + \frac{Ze}{2MB} \frac{\partial}{\partial \mu}
\langle \phiwig^2 \rangle + \frac{c}{2B\Omega} \langle \nabla_\bR
\phiwig \cdot ( \bun \times \nabla_\bR \Phiwig) \rangle.
\label{Psi}
\end{equation}

\noindent To find $u$, I need $v_{||}$ as a function of the
gyrokinetic variables. To do so, I use
\begin{equation}
\frac{v_{||}^2}{2} = E_0 - \mu_0 B = E - \mu B - (E_2 - \mu_2B),
\label{vpar_energ}
\end{equation}

\noindent where I employ $E_1 - \mu_1 B = 0$. According to this
result, the difference between $u = \langle v_{||} \rangle$ and
$v_{||}$ is necessarily of order $\delta^2 v_\mathrm{th}$. Once I
calculate $\mu_2$, I will be able to find $u$.

\section{Time derivative of $E$}

Employing the definitions of $E_1$ and $E_2$, and gyroaveraging, I
find
\begin{equation}
\left \langle \frac{dE}{dt} \right \rangle = - \frac{Ze}{M}
\langle \bv \cdot \nabla \phi \rangle. \label{ave_dEdt_1}
\end{equation}

\noindent Here, I have used that $\langle dE_1/dt \rangle = 0 =
\langle dE_2/dt \rangle$.

The term $\bv \cdot \nabla \phi$ can be conveniently rewritten by
employing
\begin{equation}
\frac{d\phi}{dt} = \left. \frac{\partial \phi}{\partial t} \right
|_\br + \bv \cdot \nabla \phi = \left. \frac{\partial
\phi}{\partial t} \right |_{\bR, E, \mu, \varphi} +
\frac{d\bR}{dt} \cdot \nabla_\bR \phi + \frac{d\mu}{dt}
\frac{\partial \phi}{\partial \mu} + \frac{d\varphi}{dt}
\frac{\partial \phi}{\partial \varphi}. \label{vgradphi_equiv}
\end{equation}

\noindent Here, I neglect $\partial \phi/\partial E$ again.
Solving for $\bv \cdot \nabla \phi$ and gyroaveraging, I find
\begin{equation}
\langle \bv \cdot \nabla \phi \rangle = \left \langle
\frac{d\bR}{dt} \cdot \nabla_\bR \phi \right \rangle + \left
\langle \frac{d\mu}{dt} \frac{\partial \phi}{\partial \mu} \right
\rangle + \left \langle \frac{d\varphi}{dt} \frac{\partial
\phi}{\partial \varphi} \right \rangle - \left \langle \left.
\frac{\partial \phi}{\partial t} \right |_\br - \left.
\frac{\partial \phi}{\partial t} \right |_{\bR, E, \mu, \varphi}
\right \rangle. \label{ave_vgradphi_1}
\end{equation}

\noindent To simplify the calculation, I will assume that I know
the corrections $\bR_3$, $\mu_3 - \langle \mu_3 \rangle$,
$\varphi_2$ and $\varphi_3$ (obtaining these corrections is
straightforward following the procedure in
section~\ref{sect_GKvariables} but will be unnecessary). With
these corrections, I find that to the order needed, $d\bR/dt =
\langle d\bR/dt \rangle$, given in \eq{ave_dRdt}, $d\mu/dt =
\langle d\mu/dt \rangle \simeq 0$ and $d\varphi/dt = \langle
d\varphi/dt \rangle$. Then, equation \eq{ave_vgradphi_1}
simplifies to
\begin{equation}
\langle \bv \cdot \nabla \phi \rangle = \left ( u \bun -
\frac{c}{B} \nabla_\bR \Psi \times \bun \right ) \cdot \nabla_\bR
\phiave - \left \langle \left. \frac{\partial \phi}{\partial t}
\right |_\br - \left. \frac{\partial \phi}{\partial t} \right
|_{\bR, E, \mu, \varphi} \right \rangle, \label{ave_vgradphi}
\end{equation}

\noindent where $\Psi$ is given in equation \eq{Psi} and $\langle
\partial \phi / \partial \varphi \rangle = 0$. Notice that
assuming that I already have $\bR_3$, $\mu_3 - \langle \mu_3
\rangle$, $\varphi_2$ and $\varphi_3$ is only a shortcut to find
the result in \eq{ave_vgradphi}. To obtain $\phiave$ to the order
required, these higher order corrections are not needed, neither
are they necessary for the difference between time derivatives, as
I will prove next. The difference between time derivatives is
\begin{equation}
\left. \frac{\partial \phi}{\partial t} \right |_\br - \left.
\frac{\partial \phi}{\partial t} \right |_{\bR, E, \mu, \varphi} =
\left. \frac{\partial \bR}{\partial t} \right |_{\br, \bv} \cdot
\nabla_\bR \phi + \left. \frac{\partial \mu}{\partial t} \right
|_{\br, \bv} \frac{\partial \phi}{\partial \mu} + \left.
\frac{\partial \varphi}{\partial t} \right |_{\br, \bv}
\frac{\partial \phi}{\partial \varphi}. \label{diff_ddt_1}
\end{equation}

\noindent The procedure for rewriting \eq{diff_ddt_1} is analogous
to that used on \eq{gradphi_1}. Using the definitions of $\bR_1$,
$\bR_2$, $\mu_1$ and $\varphi_1$, I find that $\partial
\bR/\partial t |_{\br, \bv} \simeq \partial \bR_2/\partial
t|_{\br, \bv}$, $\partial \mu/\partial t |_{\br, \bv} \simeq
\partial \mu_1/\partial t|_{\br, \bv}$ and  $\partial
\varphi / \partial t |_{\br, \bv} \simeq \partial \varphi_1
/\partial t|_{\br, \bv}$, giving
\begin{equation}
\left \langle \left. \frac{\partial \phi}{\partial t} \right |_\br
- \left. \frac{\partial \phi}{\partial t} \right |_{\bR, E, \mu,
\varphi} \right \rangle = \frac{\partial}{\partial t} ( \Psi -
\phiave ), \label{diff_ddt_equiv}
\end{equation}

\noindent where I use the equivalent to equations
\eq{int_mu1vphi1dphi} and \eq{int_R2gradphi} with $\partial /
\partial t$ replacing $\nabla_\bR$. The final result, obtained by
combining equations \eq{ave_dEdt_1}, \eq{ave_vgradphi} and
\eq{diff_ddt_equiv}, is
\begin{equation}
\left \langle \frac{dE}{dt} \right \rangle = \frac{Ze}{M} \left [
\frac{\partial}{\partial t} ( \Psi - \phiave ) - \left ( u \bun -
\frac{c}{B} \nabla_\bR \Psi \times \bun \right ) \cdot \nabla_\bR
\phiave \right ]. \label{ave_dEdt}
\end{equation}

\section{Second order correction $\mu_2$}

The correction $\mu_2$, according to
section~\ref{sect_GKvariables}, is given by
\begin{equation}
\mu_2 = \frac{1}{\Omega} \int^\varphi d\varphi^\prime\, \left [
\frac{d}{dt} (\mu_0 + \mu_1) - \left \langle \frac{d}{dt} (\mu_0 +
\mu_1) \right \rangle \right ] + \langle \mu_2 \rangle,
\label{mu2GK_1}
\end{equation}

\noindent where $\langle \mu_2 \rangle$ is found by requiring that
$\langle d\mu/dt \rangle = 0$ to order $\delta^2
v_\mathrm{th}^3/BL$.

The time derivative of $\mu_0 + \mu_1$ is given by
\begin{equation}
\frac{d}{dt}(\mu_0 + \mu_1) = \frac{Ze}{MB} \left ( - \bv_\bot
\cdot \nabla \phi + \frac{d\phiwig}{dt} \right ).
\label{dmudt_o2_1}
\end{equation}

\noindent To rewrite $\bv_\bot \cdot \nabla \phi$ as a function of
the gyrokinetic variables, I employ $\bv_\bot \cdot \nabla \phi =
\bv \cdot \nabla \phi - v_{||} \bun \cdot \nabla \phi$ and
equation \eq{vgradphi_equiv} to find
\begin{equation}
-\bv_\bot \cdot \nabla \phi + \frac{d\phiwig}{dt} = -
\frac{d\phiave}{dt} + \left. \frac{\partial \phi}{\partial t}
\right |_\br + v_{||} \bun \cdot \nabla \phi.
\end{equation}

\noindent To the order I am interested in, $d\bR/dt \simeq u\bun -
(c/B) \nabla_\bR \phiave \times \bun$, giving
\begin{equation}
- \bv_\bot \cdot \nabla \phi + \frac{d\phiwig}{dt} = - \left.
\frac{\partial \phiave}{\partial t} \right |_{\bR, E, \mu,
\varphi} - u \bun \cdot \nabla_\bR \phiave + \left. \frac{\partial
\phi}{\partial t} \right |_\br + v_{||} \bun \cdot \nabla \phi.
\label{vperp_gradphi}
\end{equation}

\noindent According to equation \eq{vpar_energ}, the difference
between $u = \langle v_{||} \rangle$ and $v_{||}$ is higher order,
and according to equation \eq{diff_ddt_1}, the difference between
$\partial \phi / \partial t|_\br$ and $\partial \phi/
\partial t|_{\bR, E, \mu, \varphi}$ is negligible. Therefore,
equations \eq{dmudt_o2_1} and \eq{vperp_gradphi} give
\begin{equation}
\frac{d}{dt}(\mu_0 + \mu_1) = \frac{Ze}{MB} \left ( \frac{\partial
\phiwig}{\partial t} + u \bun \cdot \nabla_\bR \phiwig \right ),
\label{dmudt_o2}
\end{equation}

\noindent which in turn, using equation \eq{mu2GK_1}, yields
\begin{equation}
\mu_2 = \frac{c}{B^2} \left ( \frac{\partial \Phiwig}{\partial t}
+ u \bun \cdot \nabla_\bR \Phiwig \right ) + \langle \mu_2
\rangle. \label{mu2GK}
\end{equation}

To find $\langle \mu_2 \rangle$ I require that $\langle d\mu/dt
\rangle = 0$ to order $\delta^2 v_\mathrm{th}^3/BL$. The
gyroaverage of $d\mu/dt$ is given by
\begin{equation}
\left \langle \frac{d\mu}{dt} \right \rangle = \left \langle
\frac{d}{dt} (\mu_0 + \mu_1 + \mu_2) \right \rangle = -
\frac{Ze}{MB} \langle \bv_\bot \cdot \nabla \phi \rangle +
\frac{d\langle \mu_2 \rangle}{dt}, \label{ave_dmudt_o3}
\end{equation}

\noindent where the gyroaverages of $d\mu_1/dt$ and $d(\mu_2 -
\langle \mu_2 \rangle)/dt$ vanish. The term $\langle \bv_\bot
\cdot \nabla \phi \rangle$ can be conveniently rewritten to higher
order than in \eq{vperp_gradphi} by employing equation
\eq{ave_vgradphi} to find
\begin{equation}
\langle \bv_\bot \cdot \nabla \phi \rangle = \left ( u \bun -
\frac{c}{B} \nabla_\bR \Psi \times \bun \right ) \cdot \nabla_\bR
\phiave - \frac{\partial}{\partial t} ( \Psi - \phiave ) - \langle
v_{||} \bun \cdot \nabla \phi \rangle, \label{vperp_gradphi_o3}
\end{equation}

\noindent where I used equation \eq{diff_ddt_equiv}. Employing
equation \eq{gradphi_1} and the fact that the difference between
$u = \langle v_{||} \rangle$ and $v_{||}$ is order $\delta^2
v_\mathrm{th}$ \eq{vpar_energ}, I find
\begin{equation}
\langle v_{||} \bun \cdot \nabla \phi \rangle \simeq \langle
v_{||} \bun \cdot \nabla_\bR \phi \rangle + u \bun \cdot
\nabla_\bR ( \Psi - \phiave ) \simeq u \bun \cdot \nabla_\bR \Psi.
\end{equation}

\noindent To obtain the second equality, I employ $\bun \cdot
\nabla_\bR \phiave \gg \bun \cdot \nabla_\bR \phiwig$, which means
that $\langle v_{||} \bun \cdot \nabla_\bR \phi \rangle \simeq
\langle v_{||} \bun \cdot \nabla_\bR \phiave + u \bun \cdot
\nabla_\bR \phiwig \rangle = u \bun \cdot \nabla_\bR \phiave$ to
order $\delta^2 T v_\mathrm{th}/eL$. Then, equation
\eq{vperp_gradphi_o3} becomes $\langle \bv_\bot \cdot \nabla \phi
\rangle = - d(\Psi - \phiave)/dt$, where to this order $d/dt =
\partial/\partial t + [u\bun - (c/B) \nabla_\bR \phiave \times
\bun] \cdot [\nabla_\bR - (Ze/M) \nabla_\bR \phiave (\partial /
\partial E)]$ and $\partial ( \Psi - \phiave ) /
\partial E = 0$. Finally, imposing $\langle d\mu/dt \rangle = 0$ on equation
\eq{ave_dmudt_o3}, I find
\begin{equation}
\langle \mu_2 \rangle = - \frac{Ze}{MB} \left ( \Psi - \phiave
\right ). \label{ave_mu2GK}
\end{equation}

\section{Comparisons with Dubin \emph{et al}}

To compare with reference \cite{dubin83}, I first need to write
the gyrokinetic equation in the same variables that are used in
that reference, i.e., I need to employ $u$ instead of $E$. The
change is easy to carry out. I substitute $E_2$ and \eq{mu2GK}
into \eq{vpar_energ} to write
\begin{equation}
v_{||} = \sqrt{2[E - (\mu - \langle \mu_2 \rangle)B]} +
\frac{c}{B} \bun \cdot \nabla_\bR \Phiwig, \label{vpar_GK}
\end{equation}

\noindent where I Taylor expand $E_2 - (\mu_2 - \langle \mu_2
\rangle)B = - (c/B) u \bun \cdot \nabla_\bR \Phiwig$. Then,
gyroaveraging this equation I find
\begin{equation}
\frac{u^2}{2} = E - ( \mu - \langle \mu_2 \rangle ) B.
\label{u_energ}
\end{equation}

\noindent Applying the Vlasov operator to this expression and
gyroaveraging, I find
\begin{equation}
\left \langle \frac{du}{dt} \right \rangle = - \frac{Ze}{M} \bun
\cdot \nabla_\bR \Psi,
\end{equation}

\noindent where I used equations \eq{ave_dEdt}, \eq{ave_mu2GK} and
$\langle d\mu/dt \rangle = 0$. With this equation, equation
\eq{ave_dRdt} and the fact that $\langle d\mu/dt \rangle = 0$, I
find the same gyrokinetic Vlasov equation as in reference
\cite{dubin83}, namely
\begin{equation}
\frac{\partial f}{\partial t} + \left ( u\bun - \frac{c}{B}
\nabla_\bR \Psi \times \bun \right ) \cdot \nabla_\bR f -
\frac{Ze}{M} \bun \cdot \nabla_\bR \Psi \frac{\partial f}{\partial
u} = 0, \label{GKeq}
\end{equation}

\noindent with $f(\bR, u, \mu, t)$. The differences between my
function $\Psi$ of \eq{Psi} and the function $\psi$ in reference
\cite{dubin83}, given in their equation (19b), come from their
introduction of the potential function $\phi ( \bR + \rhobf, t)
\neq \phi(\br, t)$, leading to subtle differences in the
definitions of $\phiave$, $\phiwig$ and $\Phiwig$. Here, the
vector $\rhobf (\mu, \theta)$ is
\begin{equation}
\rhobf = \frac{\sqrt{2\mu B}}{\Omega} ( \eun_1 \cos \theta -
\eun_2 \sin \theta ), \label{rho_D}
\end{equation}

\noindent with $\theta$ the gyrokinetic gyrophase as defined in
\cite{dubin83}. The relation between the gyrophase $\theta$ and my
gyrophase $\varphi$ is $\theta = - \pi/2 - \varphi$. From now on,
I will denote the functions $\phiave$, $\phiwig$ and $\Phiwig$ as
they are defined in \cite{dubin83} with the subindex $D$. The
definitions in \cite{dubin83} are then
\begin{equation}
\overline{\phi}_D \equiv \overline{\phi}_D(\bR, \mu, t) =
\frac{1}{2\pi} \oint d\theta\, \phi(\bR + \rhobf, t),
\label{phiave_D}
\end{equation}
\begin{equation}
\phiwig_D \equiv \phiwig_D (\bR, \mu, \theta, t) = \phi(\bR +
\rhobf, t) - \overline{\phi}_D \label{phiwig_D}
\end{equation}

\noindent and
\begin{equation}
\Phiwig_D \equiv \Phiwig_D (\bR, \mu, \theta, t) = \int^\theta
d\theta^\prime \, \phiwig_D(\bR, \mu, \theta^\prime, t)
\label{Phiwig_D}
\end{equation}

\noindent such that $\langle \Phiwig_D \rangle = 0$. Notice that
these definitions coincide with mine to order $\delta T/e$, except
for $\Phiwig_D$, for which $\Phiwig_D \simeq - \Phiwig$. The sign
is due to the definition of the gyrophase $\theta$. To second
order, however, Taylor expanding $\phi (\br, t) = \phi ( \bR +
\rhobf - \rhobf - \bR_1 - \bR_2, t)$ gives
\begin{equation}
\phi \simeq \phi(\bR + \rhobf, t) - (\rhobf + \bR_1 + \bR_2) \cdot
\nabla_\bR \phi, \label{phi_phiD_1}
\end{equation}

\noindent where
\begin{equation}
\rhobf + \bR_1 \simeq - \frac{Mc\mu_1}{Zev_\bot^2} \bv \times \bun
- \frac{\varphi_1}{\Omega} \bv_\bot = O(\delta \rho).
\label{correc_rho}
\end{equation}

\noindent To obtain equation \eq{correc_rho}, I Taylor expand
$\rhobf (\mu, \varphi)$ around $\mu_0$ and $\varphi_0$ in equation
\eq{rho_D}. Employing the lowest order results $\partial \phi /
\partial \varphi \simeq - \Omega^{-1} \bv_\bot \cdot \nabla \phi$
and $\partial \phi / \partial \mu \simeq - (Mc/Zev_\bot^2) (\bv
\times \bun) \cdot \nabla \phi$, I write equation \eq{phi_phiD_1}
as
\begin{equation}
\phi \simeq \phi(\bR + \rhobf, t) - \frac{Ze}{MB} \left ( \phiwig
\frac{\partial \phi}{\partial \mu} - \frac{\partial
\Phiwig}{\partial \mu} \frac{\partial \phi}{\partial \varphi}
\right ) + \frac{c}{B\Omega} (\nabla_\bR \Phiwig \times \bun)
\cdot \nabla_\bR \phi, \label{phi_phiD}
\end{equation}

\noindent where I used the definitions of $\bR_2$, $\mu_1$ and
$\varphi_1$. Then, gyroaveraging gives
\begin{equation}
\phiave \simeq \overline{\phi}_D - \frac{Ze}{MB}
\frac{\partial}{\partial \mu} \langle \phiwig^2 \rangle +
\frac{c}{B\Omega} \langle (\nabla_\bR \Phiwig \times \bun) \cdot
\nabla_\bR \phiwig \rangle. \label{ave_phi_phiD}
\end{equation}

\noindent Substituting this equation into the definition \eq{Psi}
of $\Psi$ and employing that to lowest order $\phiwig \simeq
\phiwig_D$ and $\Phiwig \simeq - \Phiwig_D$, I find
\begin{equation}
\Psi = \overline{\phi}_D - \frac{Ze}{2MB} \frac{\partial}{\partial
\mu} \langle \phiwig_D^2 \rangle - \frac{c}{2B\Omega} \langle
(\nabla_\bR \Phiwig_D \times \bun) \cdot \nabla_\bR \phiwig_D
\rangle, \label{Psi_D}
\end{equation}

\noindent exactly as in equation (19b) of reference
\cite{dubin83}.

Finally, I will compare the quasineutrality equations in both
methods. Taylor expanding the ion distribution function around
$\bR_g = \br + \Omega^{-1} \bv \times \bun$, $v_{||}$, $\mu_0$ and
$\varphi_0$, I find
\begin{equation}
f_i (\bR, u, \mu, t) \simeq f_{ig} + \bR_2 \cdot \nabla_{\bR_g}
f_{ig} - \frac{c}{B} \bun \cdot \nabla_\bR \Phiwig \frac{\partial
f_{ig}}{\partial v_{||}} + (\mu_1 + \mu_2) \frac{\partial
f_{ig}}{\partial \mu_0} + \frac{\mu_1^2}{2} \frac{\partial^2
f_{ig}}{\partial \mu_0^2},
\end{equation}

\noindent where $f_{ig} \equiv f_i(\bR_g, v_{||}, \mu_0, t)$. Here
I have used equations \eq{vpar_GK} and \eq{u_energ} to obtain that
$u \simeq v_{||} - (c/B) \bun \cdot \nabla_\bR \Phiwig$. The ion
density is given by
\begin{equation}
n_i = \int d^3v\, f_i \simeq \int d^3v\, \left [ f_{ig} + \bR_2
\cdot \nabla_{\bR_g} f_{ig} + (\mu_1 + \langle \mu_2 \rangle)
\frac{\partial f_{ig}}{\partial \mu_0} + \frac{\mu_1^2}{2}
\frac{\partial^2 f_{ig}}{\partial \mu_0^2} \right ].
\label{dens_1}
\end{equation}

\noindent Here, the integrals of $(c/B) (\bun \cdot \nabla_\bR
\Phiwig) ( \partial f_{ig} / \partial v_{||} )$ and $(\mu_2 -
\langle \mu_2 \rangle) (\partial f_{ig} / \partial \mu_0)$ vanish
because $\oint d\varphi_0 \Phiwig = 0$ and the only gyrophase
dependence is in $\Phiwig$ since $f_{ig}$ is assumed to be a
smooth function of $\br$ and $\bv$ to lowest order, giving $f_{ig}
\equiv f_i (\bR_g, v_{||}, \mu_0, t) \simeq f_i (\br, v_{||},
\mu_0, t)$. The integral $\oint d\varphi_0 \Phiwig$ is performed
holding $\br$, $v_{||}$, $\mu_0$ and $t$ fixed, and it vanishes to
lowest order as proven at the end of this Appendix. On the other
hand, the integral of $\bR_2 \cdot \nabla_{\bR_g} f_{ig}$ does not
vanish. Here the gyrophase dependence of $f_{ig}$ contained in its
short wavelength contributions becomes important due to the steep
gradient [recall the ordering in \eq{order}].

In equation \eq{dens_1}, I can employ $\phiwig \simeq \phiwig_D$
and $\Phiwig \simeq - \Phiwig_D$ in the higher order terms.
However, for $\mu_1$ I need the difference between $\phiwig$ and
$\phiwig_D$. Subtracting \eq{ave_phi_phiD} from \eq{phi_phiD}, I
find
\begin{eqnarray}
\phiwig \simeq \phiwig_D - \frac{Ze}{MB} \left ( \phiwig
\frac{\partial \phi}{\partial \mu} - \frac{\partial
\Phiwig}{\partial \mu} \frac{\partial \phiwig}{\partial \varphi}
\right ) + \frac{c}{B\Omega} ( \nabla_\bR \Phiwig \times \bun )
\cdot \nabla_\bR \phi + \frac{Ze}{MB} \frac{\partial}{\partial
\mu} \langle \phiwig^2 \rangle \nonumber \\ - \frac{c}{B\Omega}
\langle ( \nabla_\bR \Phiwig \times \bun ) \cdot \nabla_\bR
\phiwig \rangle. \label{wig_phi_phiD_1}
\end{eqnarray}

\noindent In this equation, $\phiwig_D = \phiwig_D (\bR, \mu,
\varphi, t)$, but for equation \eq{dens_1}, it is better to use
$\phiwig_{Dg} = \phiwig_D ( \bR_g, \mu_0, \varphi_0, t)$. By
Taylor expanding, I find that
\begin{equation}
\phiwig_D \simeq \phiwig_{Dg} + \bR_2 \cdot \nabla_{\bR_g}
\phiwig_{Dg} + \mu_1 \frac{\partial \phiwig_{Dg}}{\partial \mu_0}
+ \varphi_1 \frac{\partial \phiwig_{Dg}}{\partial \varphi_0}.
\label{g_transf}
\end{equation}

\noindent This equation, combined with equation
\eq{wig_phi_phiD_1} and the definitions of $\bR_2$, $\mu_1$ and
$\varphi_1$, leads to
\begin{eqnarray}
\phiwig \simeq \phiwig_{Dg} - \frac{Ze\phiwig_{Dg}}{MB}
\frac{\partial \overline{\phi}_{Dg}}{\partial \mu} -
\frac{c}{B\Omega} ( \nabla_{\bR_g} \Phiwig_{Dg} \times \bun )
\cdot \nabla_{\bR_g} \overline{\phi}_{Dg} + \frac{Ze}{MB}
\frac{\partial}{\partial \mu} \langle \phiwig_{Dg}^2 \rangle
\nonumber \\ + \frac{c}{B\Omega} \langle ( \nabla_{\bR_g}
\Phiwig_{Dg} \times \bun ) \cdot \nabla_{\bR_g} \phiwig_{Dg}
\rangle, \label{wig_phi_phiD}
\end{eqnarray}

\noindent where I have used that, to lowest order, $\phiwig_D
\simeq \phiwig_{Dg}$, $\overline{\phi}_D \simeq
\overline{\phi}_{Dg} \equiv \overline{\phi}_D (\bR_g, \mu_0, t)$
and $\Phiwig_D \simeq \Phiwig_{Dg} \equiv \Phiwig_D (\bR_g, \mu_0,
\varphi_0, t)$. Substituting equation \eq{wig_phi_phiD} into
\eq{dens_1} yields
\begin{eqnarray}
n_i \simeq \int d^3v\, \Bigg \{ f_{ig} + \frac{Ze\phiwig_{Dg}}{MB}
\frac{\partial f_{ig}}{\partial \mu_0} + \frac{c}{B\Omega}
(\nabla_{\bR_g} \Phiwig_{Dg} \times \bun) \cdot \nabla_{\bR_g}
f_{ig} \nonumber \\ + \frac{Z^2 e^2 \phiwig_{Dg}^2}{2 M^2 B^2}
\frac{\partial^2 f_{ig}}{\partial \mu_0^2} + \frac{Ze}{MB} \Bigg [
- \frac{Ze\phiwig_{Dg}}{MB} \frac{\partial
\overline{\phi}_{Dg}}{\partial \mu_0} - \frac{c}{B\Omega}
(\nabla_{\bR_g} \Phiwig_{Dg} \times \bun) \cdot \nabla_{\bR_g}
\overline{\phi}_{Dg} \nonumber \\ + \frac{Ze}{2MB}
\frac{\partial}{\partial \mu_0} \langle \phiwig_{Dg}^2 \rangle +
\frac{c}{2B\Omega} \langle (\nabla_{\bR_g} \Phiwig_{Dg} \times
\bun) \cdot \nabla_{\bR_g} \phiwig_{Dg} \rangle \Bigg ]
\frac{\partial f_{ig}}{\partial \mu_0} \Bigg \}, \label{dens}
\end{eqnarray}

\noindent where I have used the definitions of $\bR_2$ and
$\langle \mu_2 \rangle$. This result is exactly the same as in
equation (20) in reference \cite{dubin83}. For comparison, I give
$n_i$ to order $\delta^2 n_i$ with the definitions of $\phiave$,
$\phiwig$ and $\Phiwig$ in equations \eq{phiave}, \eq{phiwig} and
\eq{Phiwig},
\begin{eqnarray}
n_i \simeq \int d^3v\, \Bigg \{ f_{ig} + \frac{Ze\phiwig_g}{MB}
\frac{\partial f_{ig}}{\partial \mu_0} - \frac{c}{B\Omega}
(\nabla_{\bR_g} \Phiwig_g \times \bun) \cdot \nabla_{\bR_g} f_{ig}
+ \frac{Z^2 e^2 \phiwig_g^2}{2 M^2 B^2} \frac{\partial^2
f_{ig}}{\partial \mu_0^2} \nonumber \\ + \frac{Ze}{MB} \Bigg [
\frac{Ze\phiwig_g}{MB} \frac{\partial \phiwig_g}{\partial \mu_0} -
\frac{Ze\Phiwig_g}{MB} \frac{\partial \phiwig_g}{\partial
\varphi_0} - \frac{c}{B\Omega} (\nabla_{\bR_g}
\Phiwig_g \times \bun) \cdot \nabla_{\bR_g} \phiwig_g \nonumber \\
- \frac{Ze}{2MB} \frac{\partial}{\partial \mu_0} \langle
\phiwig_g^2 \rangle + \frac{c}{2B\Omega} \langle (\nabla_{\bR_g}
\Phiwig_g \times \bun) \cdot \nabla_{\bR_g} \phiwig_g \rangle
\Bigg ] \frac{\partial f_{ig}}{\partial \mu_0} \Bigg \}.
\end{eqnarray}

\noindent I have found this equation substituting $\mu_1$, $\bR_2$
and $\langle \mu_2 \rangle$ into \eq{dens_1}. From the functions
$\phiwig (\bR, \mu, \varphi, t)$ and $\Phiwig (\bR, \mu, \varphi,
t)$, I have defined $\phiwig_g = \phiwig ( \bR_g, \mu_0,
\varphi_0, t )$ and $\Phiwig_g = \Phiwig ( \bR_g, \mu_0,
\varphi_0, t )$. The relationships between $\phiwig$ and
$\phiwig_g$ and between $\Phiwig$ and $\Phiwig_g$ are similar to
the one given in \eq{g_transf}.

The methodology and results of chapter~\ref{chap_gyrokinetics} are
completely consistent with the results of \cite{dubin83} since
they give the same gyrokinetic equation \eq{GKeq}, generalized
potential $\Psi$ \eq{Psi_D} and quasineutrality condition
\eq{dens}.

\section*{Integral $\oint d\varphi_0 \Phiwig$}

To prove that $\oint d\varphi_0 \Phiwig = 0$ vanishes, I Fourier
analyze $\phi = (2\pi)^{-3} \int d^3k\, \phi_\mathbf{k} \exp (
i\mathbf{k} \cdot \br )$, giving to lowest order
\begin{equation}
\phi(\br, t) \simeq \phi(\bR - \Omega^{-1} \bv \times \bun, t) =
\frac{1}{(2\pi)^3} \int d^3k\, \phi_\mathbf{k} \exp [ i\mathbf{k}
\cdot \bR - i z \sin (\varphi_0 - \varphi_\mathbf{k})],
\end{equation}

\noindent where $z = k_\bot v_\bot/\Omega$. Here, I employ $\br
\simeq \bR - \Omega^{-1} \bv \times \bun$ and I define
$\varphi_\mathbf{k}$ such that $\mathbf{k}_\bot = k_\bot ( \eun_1
\cos \varphi_\mathbf{k} + \eun_2 \sin \varphi_\mathbf{k} )$ to
write $\mathbf{k} \cdot \br \simeq \mathbf{k} \cdot \bR - z \sin
(\varphi_0 - \varphi_\mathbf{k})$. Then, I use
\begin{equation}
\exp(iz \sin \varphi) = \sum_{m = - \infty}^\infty J_m (z)
\exp(im\varphi), \label{bessel}
\end{equation}

\noindent with $J_m (z)$ the Bessel function of the first kind, to
find
\begin{equation}
\phi \simeq \frac{1}{(2\pi)^3} \int d^3k\, \phi_\mathbf{k} \exp (
i\mathbf{k} \cdot \bR ) \sum_{m = - \infty}^\infty J_m (z) \exp [-
i m (\varphi_0 - \varphi_\mathbf{k})].
\end{equation}

\noindent Employing this expression, I obtain $\phiwig$ by
subtracting the average in $\varphi_0$ (component $m=0$), and I
find $\Phiwig$ by integrating $\phiwig$ over $\varphi_0$, giving
\begin{equation}
\Phiwig \simeq \frac{1}{(2\pi)^3} \int d^3k\, \phi_\mathbf{k} \exp
( i\mathbf{k} \cdot \bR ) \sum_{m \neq 0} \frac{i}{m} J_m (z) \exp
[- i m (\varphi_0 - \varphi_\mathbf{k})],
\end{equation}

\noindent where the summation includes every positive and negative
$m$ different from 0. To rewrite $\Phiwig$ as a function of $\br$,
$v_{||}$, $\mu_0$ and $\varphi_0$, I need the expression
\begin{equation}
\exp(i\mathbf{k} \cdot \bR) \simeq \exp(i\mathbf{k} \cdot \br)
\sum_{p = - \infty}^\infty J_p (z) \exp[ip(\varphi_0 -
\varphi_\mathbf{k})],
\end{equation}

\noindent deduced from $\bR \simeq \br + \Omega^{-1} \bv \times
\bun$ and \eq{bessel}. Then, I find
\begin{equation}
\Phiwig \simeq \frac{1}{(2\pi)^3} \int d^3k\, \phi_\mathbf{k} \exp
( i\mathbf{k} \cdot \br ) \sum_{m \neq 0, p} \frac{i}{m} J_m (z)
J_p (z) \exp [i (p - m) (\varphi_0 - \varphi_\mathbf{k})].
\end{equation}

\noindent Finally, integrating in $\varphi_0$, I obtain
\begin{equation}
\frac{1}{2\pi} \oint d\varphi_0\, \Phiwig = \frac{1}{(2\pi)^3}
\int d^3k\, \phi_\mathbf{k} \exp ( i\mathbf{k} \cdot \br ) \sum_{m
\neq 0} \frac{i}{m} [J_m (z)]^2 = 0
\end{equation}

\noindent since $J_{-m} (z) = (-1)^m J_m(z)$.

\chapter{Quasineutrality equation at long wavelengths \label{app_poisson_kL}}

In this Appendix, I obtain the gyrokinetic quasineutrality
equation at long wavelengths ($k_\bot \rho_i \ll 1$). In
section~\ref{sectapp_polarn_simple}, the polarization density
$n_{ip}$ in \eq{polar_n} is calculated for the intermediate scales
$\delta_i \ll k_\bot \rho_i \ll 1$. In
section~\ref{sectapp_poisson_kL}, the polarization density
$n_{ip}$ and the ion guiding center density $\hat{N}_i$ are
computed up to $O(\delta_i^2 n_e)$ for the extreme case of a
non-turbulent plasma.

\section{Polarization density at long wavelengths \label{sectapp_polarn_simple}}

For long wavelengths $\delta_i \ll k_\bot \rho_i \ll 1$, the
polarization density simplifies to give \eq{nip_krhosmall}. To
obtain this result, $\phiwig$ and hence $\phi$ must be obtained to
$O(\delta_i k_\bot \rho_i T_e/e)$. To this order, the potential is
\begin{equation}
\phi (\br) \simeq \phi (\bR - \bR_1) \simeq \phi(\bR) - \bR_1
\cdot \nabla_\bR \phi + \frac{1}{2} \bR_1 \bR_1 : \nabla_\bR
\nabla_\bR \phi, \label{phi_longwave_app}
\end{equation}

\noindent with $\bR_1 = \Omega_i^{-1} \bv \times \bun$. Here,
according to \eq{order_kbot}, $\nabla_\bR \phi \sim T_e/eL$ and
$\nabla_\bR \nabla_\bR \phi \sim k_\bot T_e/eL$. With the result
in \eq{phi_longwave_app}, the function $\phiwig$ is
\begin{equation}
\phiwig = \phi - \phiave \simeq - \bR_1 \cdot \nabla_\bR \phi +
\frac{1}{2} ( \bR_1 \bR_1 - \langle \bR_1 \bR_1 \rangle ) :
\nabla_\bR \nabla_\bR \phi, \label{phiwig_kL_1_app}
\end{equation}

\noindent where I neglect $\langle \bR_1 \rangle \sim \delta_i
\rho_i$. The gyroaverage of the first order correction $\bR_1$ is
not zero due to the difference between $\br$, $\mu_0$ and
$\varphi_0$ and the gyrokinetic variables $\bR$, $\mu$ and
$\varphi$. This difference will become important in
section~\ref{sectapp_poisson_kL}.

The integral $n_{ip}$ in \eq{polar_n} is performed over velocity
space holding $\br$ and $t$ fixed. Then, equation
\eq{phiwig_kL_1_app} has to be Taylor expanded around $\br$ to
obtain a function that depends on $\br$ and not $\bR$. To
$O(\delta_i k_\bot \rho_i T_e/e)$, the result is
\begin{equation}
\phiwig \simeq - \bR_1 \cdot \nabla \phi - \frac{1}{2} ( \bR_1
\bR_1 + \overline{\bR_1 \bR_1} ) : \nabla \nabla \phi,
\label{phiwig_kL_2_app}
\end{equation}

\noindent where I use $\nabla_\bR \phi \simeq \nabla \phi + \bR_1
\cdot \nabla \nabla \phi$. In the higher order terms, the
gyroaverage $\langle \ldots \rangle$ holding the gyrokinetics
variable fixed can be approximated by the gyroaverage
$\overline{(\ldots)}$ holding $\br$, $v_{||}$ and $v_\bot$ fixed.

Expression \eq{phiwig_kL_2_app} can be readily substituted into
\eq{polar_n} and integrated to give
\begin{equation}
n_{ip} \simeq \frac{cn_i}{B\Omega_i} (\matI - \bun \bun) : \nabla
\nabla \phi. \label{polarn_longwave_app}
\end{equation}

\noindent The final result in \eq{nip_krhosmall} is found by
realizing that $\nabla \cdot [ (cn_i/B\Omega_i) (\matI - \bun\bun)
] \cdot \nabla_\bot \phi$ is small by a factor $(k_\bot L)^{-1}$
compared to the term in \eq{polarn_longwave_app}.

\section{Quasineutrality equation for $k_\bot L \sim 1$ \label{sectapp_poisson_kL}}

In this section, I obtain the long wavelength quasineutrality
condition to order $\delta_i^2 n_e$. The derivation is not
applicable to turbulent plasmas because I will assume that neither
the potential nor the distribution function have short wavelength
pieces.

In quasineutrality, the ion distribution function must be written
in $\br$, $\bv$ variables. In a non-turbulent plasma, the ion
distribution function can be expanded around $\br$, $E_0$ and
$\mu_0$ up to $O(\delta_i^2 f_{Mi})$, giving
\begin{eqnarray}
f_i (\bR, E, \mu, t) = f_{i0} + (\bR_1 + \bR_2) \cdot \nablaave
f_{i0} + (E_1 + E_2) \frac{\partial f_{i0}}{\partial E_0} + \mu_1
\frac{\partial f_{i0}}{\partial \mu_0} \nonumber \\ + \frac{1}{2}
\bR_1 \bR_1 : \nablaave\, \nablaave f_{Mi} + E_1 \bR_1 \cdot
\nablaave \left ( \frac{\partial f_{Mi}}{\partial E_0} \right ) +
\frac{E_1^2}{2} \frac{\partial^2 f_{Mi}}{\partial E_0^2},
\label{Taylor_fi}
\end{eqnarray}

\noindent with $f_{i0} \equiv f_i (\br, E_0, \mu_0, t)$ and
$\nablaave$ the gradient holding $E_0$, $\mu_0$, $\varphi_0$ and
$t$ fixed. In $f_i$ I have not included the collisional gyrophase
dependent piece given in \eq{fwig_sol} because it will not
contribute to the density. In the higher order terms of the Taylor
expansion \eq{Taylor_fi}, the lowest order distribution function
$f_i \simeq f_{Mi}$ must be used.

To find the ion density, the ion distribution function is
integrated in velocity space. Some of the terms vanish because the
integral over gyrophase is zero; for example, $\int d^3v\; (\bR_1
+ \bR_2) \cdot \nablaave f_{i0} = 0 = \int d^3v\; E_2 (
\partial f_{i0} / \partial E_0)$. Then, the ion density becomes
\begin{equation}
n_i \simeq \hat{N}_i + \int d^3 v \, \frac{Ze\phiwig}{M} \left [
\frac{\partial f_{i0}}{\partial E_0} + \frac{1}{B} \frac{\partial
f_{i0}}{\partial \mu_0} + \frac{Ze\phiwig}{2M} \frac{\partial^2
f_{Mi}}{\partial E_0^2} + \bR_1 \cdot \nablaave \left (
\frac{\partial f_{Mi}}{\partial E_0} \right ) \right ],
\label{n_i_lk_1}
\end{equation}

\noindent where $\hat{N}_i ( \br, t )$ is the ion gyrocenter
density, defined as the portion of the ion density independent of
$\phiwig$ and given by
\begin{eqnarray}
\hat{N}_i = \int d^3 v \, f_{i0} - \int d^3 v \, \frac{v_{||}
v_\bot^2}{2 B \Omega_i} \bun \cdot \nabla \times \bun
\frac{\partial f_{i0}}{\partial \mu_0} + \int d^3 v \, \frac{\bR_1
\bR_1 }{2} : \nablaave\, \nablaave f_{Mi}. \label{n_hat_lk_1}
\end{eqnarray}

\noindent The formula for $\hat{N}_i$ can be simplified. The
second term in the right side of the equation is proportional to
$\int d^3 v \, (v_{||} v_\bot^2/2 B) (\partial f_{i0} / \partial
\mu_0)$. This integral is simplified by changing to the variables
$E_0 = v^2/2$, $\mu_0 = v_\bot^2/2B$ and $\varphi_0$ and
integrating by parts,
\begin{equation}
- \int d^3 v \, \frac{v_{||} v_\bot^2}{2 B} \frac{\partial
f_{i0}}{\partial \mu_0} = - B \sum_\sigma \int dE_0 \, d\mu_0 \,
d\varphi_0 \, \sigma \mu_0 \frac{\partial f_{i0}}{\partial \mu_0}
= \int d^3 v \, v_{||} f_{i0},
\end{equation}

\noindent where $\sigma = v_{||} / |v_{||}|$ is the sign of the
parallel velocity, the summation in front of the integral
indicates that the integral must be done for both signs of
$v_{||}$, and I have used the equality $d^3 v =
dE_0\,d\mu_0\,d\varphi_0 \, B/ |v_{||}|$. The third term in the
right side of \eq{n_hat_lk_1} is proportional to
\begin{equation}
M \int d^3 v \, ( \bv \times \bun ) ( \bv \times \bun ) :
\nablaave\, \nablaave f_{Mi} = ( \matrixtop{\mathbf{I}} - \bun
\bun ) : \nabla \nabla p_i.
\end{equation}

\noindent The final function $\hat{N}_i$ reduces to the result
shown in \eq{nhat_final_lk}.

In equation \eq{n_i_lk_1}, $\phiwig$ appears in several integrals.
The expression can be simplified by integrating first in the
gyrophase. Two integrals in \eq{n_i_lk_1} can be done by using the
lowest order result $\phiwig \simeq - \Omega_i^{-1} (\bv \times
\bun) \cdot \nabla \phi$. The integrals are
\begin{equation}
\int d^3 v \, \frac{Z^2e^2}{2M^2} \phiwig^2 \frac{\partial^2
f_{Mi}}{\partial E_0^2} = \frac{M c^2 n_i}{2T_i B^2} | \nabla_\bot
\phi |^2 \label{integ2_lk}
\end{equation}

\noindent and
\begin{equation}
\int d^3 v \, \frac{c}{B} \phiwig (\bv \times \bun) \cdot
\nablaave \left ( \frac{\partial f_{Mi}}{\partial E_0} \right ) =
\frac{c}{B \Omega_i} \nabla n_i \cdot \nabla_\bot \phi.
\label{integ3_lk}
\end{equation}

\noindent For $\int d^3 v \, \phiwig ( \partial f_{i0} / \partial
E_0 + B^{-1} \partial f_{i0} / \partial \mu_0 )$, only the
gyrophase integral $\int_0^{2\pi} \phiwig \, d \varphi_0$ is
needed, but $\phiwig$ must be written as a function of the
variables $\br$, $\bv$ to $O(\delta_i^2 T_e/e)$ to be consistent
with the order of the Taylor expansion. To do so, first I will
write $\phi ( \br, t)$ as a function of the gyrokinetic variables
by Taylor expansion to $O(\delta_i^2 T_e/e)$ [this Taylor
expansion is carried out to higher order than in
\eq{phi_longwave_app}]. The result is
\begin{equation}
\phi ( \br, t ) \simeq \phi ( \bR, t ) - \bR_1 \cdot \nabla_\bR
\phi + \frac{1}{2} \bR_1 \bR_1 : \nabla_\bR \nabla_\bR \phi -
\bR_2 \cdot \nabla_\bR \phi. \label{phi_rv2GK_lk}
\end{equation}

\noindent The second term in the right side of the equation needs
to be re-expanded in order to express $\phi$ as a self-consistent
function of the gyrokinetic variables to the right order. The
function $\bR_1$ is, to $O(\delta_i \rho_i)$,
\begin{equation}
\bR_1 = \frac{1}{\Omega_i} \bv \times \bun \simeq \frac{\sqrt{2\mu
B ( \bR )}}{\Omega_i ( \bR)} [ \eun_1 ( \bR ) \sin \varphi -
\eun_2 ( \bR ) \cos \varphi ] - \Delta \rhobf, \label{R1_o2_lk}
\end{equation}

\noindent with $\Delta \rhobf ( \bR, E, \mu, \varphi, t) \sim
\delta_i \rho_i$. The function $\Delta \rhobf$ is found by Taylor
expanding $\bR_1 (\br, \mu_0, \varphi_0)$ around $\bR$, $\mu$ and
$\varphi$ to $O(\delta_i \rho_i)$, giving
\begin{eqnarray}
\Delta \rhobf = - \frac{1}{2B\Omega_i^2} ( \bv \times \bun ) ( \bv
\times \bun ) \cdot \nabla B + \frac{Mc}{Ze v_\bot^2} \mu_1 (\bv
\times \bun) + \frac{1}{\Omega_i} \varphi_1 \bv_\bot \nonumber \\
+ \frac{v_\bot}{\Omega_i^2} ( \bv \times \bun ) \cdot ( \sin
\varphi_0 \nabla \eun_1 - \cos \varphi_0 \nabla \eun_2 ).
\label{Drho_lk_1}
\end{eqnarray}

\noindent Combining \eq{phi_rv2GK_lk} and \eq{R1_o2_lk}, $\phiwig$
is found to be
\begin{eqnarray}
\phiwig = \phi - \langle \phi \rangle \simeq - \frac{\sqrt{2\mu B
( \bR )}}{\Omega_i ( \bR)} [ \eun_1 ( \bR ) \sin \varphi - \eun_2
( \bR ) \cos \varphi ] \cdot \nabla_\bR \phi - \bR_2 \cdot
\nabla_\bR \phi \nonumber \\ + (\Delta \rhobf - \langle \Delta
\rhobf \rangle ) \cdot \nabla_\bR \phi + \frac{1}{2 \Omega_i^2} [
(\bv \times \bun) (\bv \times \bun) - \langle (\bv \times \bun)
(\bv \times \bun) \rangle ] : \nabla_\bR \nabla_\bR \phi.
\label{phiwig_lk}
\end{eqnarray}

\noindent The function $\phiwig$ must be written as a function of
$\br$, $\bv$ and $t$ since the integral in velocity space is done
for $\br$ and $t$ fixed. Taylor expanding the first term in
\eq{phiwig_lk} gives
\begin{eqnarray}
\phiwig \simeq - \frac{1}{\Omega_i} ( \bv \times \bun ) \cdot
\nabla \phi - \langle \Delta \rhobf \rangle \cdot \nabla \phi -
\bR_2 \cdot \nabla \phi - \frac{v_\bot^2}{4\Omega_i^2}
(\matrixtop{\mathbf{I}} - \bun \bun ) : \nabla \nabla \phi
\nonumber \\ - \frac{1}{2 \Omega_i^2} (\bv \times \bun) (\bv
\times \bun): \nabla \nabla \phi, \label{phiwig_o1}
\end{eqnarray}

\noindent where I use $\langle (\bv \times \bun) (\bv \times \bun)
\rangle = (v_\bot^2/2) (\matI - \bun \bun)$. Equation
\eq{phiwig_o1} is more complete than the approximation in
\eq{phiwig_kL_2_app}.

Gyroaveraging $\phiwig$ in \eq{phiwig_o1} holding $\br$, $v_{||}$
and $v_\bot$ fixed leads to
\begin{equation}
\frac{1}{2\pi} \int_0^{2\pi} \phiwig \, d\varphi_0 \simeq -
\langle \Delta \rhobf \rangle \cdot \nabla \phi -
\frac{v_\bot^2}{2\Omega_i^2} (\matrixtop{\mathbf{I}} - \bun \bun )
: \nabla \nabla \phi. \label{ave_phiwig_lk}
\end{equation}

\noindent with $\Delta \rhobf$ the function in \eq{Drho_lk_1}. To
simplify equation \eq{Drho_lk_1}, the gradients of the unit
vectors $\eun_1$ and $\eun_2$ are expressed as $\nabla \eun_1 = -
(\nabla \bun \cdot \eun_1) \bun - (\nabla \eun_2 \cdot \eun_1)
\eun_2$ and $\nabla \eun_2 = - (\nabla \bun \cdot \eun_2) \bun +
(\nabla \eun_2 \cdot \eun_1) \eun_1$, giving
\begin{eqnarray}
\Delta \rhobf = - \frac{1}{2B \Omega_i^2} ( \bv \times \bun ) (
\bv \times \bun ) \cdot \nabla B + \frac{Mc}{Ze v_\bot^2} \mu_1
(\bv \times \bun) + \frac{1}{\Omega_i} \varphi_1 \bv_\bot
\nonumber \\ - \frac{1}{\Omega_i^2} ( \bv \times \bun ) \cdot
\nabla \bun \cdot ( \bv \times \bun ) \bun - \frac{1}{\Omega_i^2}
\bv_\bot ( \bv \times \bun ) \cdot \nabla \eun_2 \cdot \eun_1,
\label{Drho_lk_2}
\end{eqnarray}

\noindent and its gyroaverage
\begin{equation}
\langle \Delta \rhobf \rangle = - \frac{c}{B\Omega_i} \nabla_\bot
\phi - \frac{v_\bot^2}{B\Omega_i^2} \nabla_\bot B -
\frac{v_{||}^2}{\Omega_i^2} \bun \cdot \nabla \bun -
\frac{v_\bot^2}{2 \Omega_i^2} (\nabla \cdot \bun) \bun,
\label{Drho_ave_lk_2}
\end{equation}

\noindent where I use
\begin{equation}
\langle \mu_1 ( \bv \times \bun ) \rangle = - \frac{c v_\bot^2}{2
B^2} \nabla_\bot \phi - \frac{v_\bot^4}{4 B^2 \Omega_i}
\nabla_\bot B - \frac{v_\bot^2 v_{||}^2}{2 B \Omega_i} \bun \cdot
\nabla \bun \label{mu1_vn_ave}
\end{equation}

\noindent and
\begin{equation}
\langle \varphi_1 \bv_\bot \rangle =  - \frac{c}{2B} \nabla_\bot
\phi - \frac{v_\bot^2}{2B\Omega_i} \nabla_\bot B -
\frac{v_{||}^2}{2\Omega_i} \bun \cdot \nabla \bun +
\frac{v_\bot^2}{2\Omega_i} \bun \times \nabla \eun_2 \cdot \eun_1.
\label{varphi1_vp_ave}
\end{equation}

\noindent These expressions are found by using the definitions of
$\mu_1$ and $\varphi_1$, given by \eq{correc_varphi_o1} and
\eq{correc_mu_o1}, and employing the lowest order expressions
$\phiwig \simeq - \Omega_i^{-1} ( \bv \times \bun ) \cdot \nabla
\phi$, $\Phiwig = \int^\varphi \phiwig \, d\varphi^\prime \simeq
\Omega_i^{-1} \bv_\bot \cdot \nabla \phi$ and $\partial \Phiwig /
\partial \mu \simeq (Mc/Ze v_\bot^2) \bv_\bot \cdot \nabla
\phi$.

Substituting \eq{Drho_ave_lk_2} in \eq{ave_phiwig_lk} gives
\begin{equation}
\frac{1}{2\pi} \int_0^{2\pi} \phiwig \, d \varphi_0 = -
\frac{v_\bot^2}{2} \nabla \cdot \left ( \frac{1}{\Omega_i^2}
\nabla_\bot \phi \right ) + \left ( v^2_{||} - \frac{v_\bot^2}{2}
\right ) \frac{1}{\Omega_i^2} \bun \cdot \nabla \bun \cdot \nabla
\phi + \frac{c}{B\Omega_i} | \nabla_\bot \phi |^2,
\end{equation}

\noindent where I have used
\begin{eqnarray}
\nabla \cdot \left ( \frac{1}{\Omega_i^2} \nabla_\bot \phi \right
) = \frac{1}{\Omega_i^2} ( \matrixtop{\mathbf{I}} - \bun \bun ) :
\nabla \nabla \phi - \frac{2}{\Omega_i^2} \nabla B \cdot
\nabla_\bot \phi - \frac{1}{\Omega_i^2} \bun \cdot \nabla \bun
\cdot \nabla \phi \nonumber \\ - \frac{1}{\Omega_i^2} (\bun \cdot
\nabla \phi) \nabla \cdot \bun.
\end{eqnarray}

\noindent Notice that the gyroaverage of $\phiwig$ is
$O(\delta_i^2 T_e/e)$, which means that the integral is
$O(\delta_i^2 n_i)$, and the lowest order distribution function,
$f_{Mi}$, can be used to write $\partial f_{i0} / \partial E_0
\simeq - (M / T_i) f_{Mi}$ and $\partial f_{i0} / \partial \mu_0
\simeq 0$. All these simplifications lead to the final result
\begin{equation}
\frac{Ze}{M} \int d^3 v \, \phiwig \left ( \frac{\partial
f_{i0}}{\partial E_0} + \frac{1}{B} \frac{\partial
f_{i0}}{\partial \mu_0} \right ) = n_i \nabla \cdot \left (
\frac{c}{B \Omega_i} \nabla_\bot \phi \right ) - \frac{M c^2
n_i}{T_i B^2} |\nabla_\bot \phi|^2, \label{integ1_lk_3}
\end{equation}

Using \eq{integ2_lk}, \eq{integ3_lk} and \eq{integ1_lk_3},
equation \eq{n_i_lk_1} becomes
\begin{equation}
n_i = \hat{N}_i + \nabla \cdot \left ( \frac{c n_i}{B \Omega_i}
\nabla_\bot \phi \right ) - \frac{M c^2 n_i}{2T_i B^2} |
\nabla_\bot \phi |^2,
\end{equation}

\noindent where $\hat{N}_i$ is given by \eq{nhat_final_lk}. Then
the quasineutrality condition is as shown in \eq{qn_highorder}.

\chapter{Gyrophase dependent piece of $f_i$ for $k_\bot L \sim 1$} \label{app_gyrodepend_kL}

In this Appendix, I show how to obtain the $O(\delta_i^2 f_{Mi})$
gyrophase dependent piece of the ion distribution function from
gyrokinetics in a non-turbulent plasma. The general gyrophase
dependent piece is found in section~\ref{sectapp_gyro_general}.
This result is useful because it can be compared to the drift
kinetic result \cite{simakov05}, proving that the higher order
gyrokinetic variables allow us to recover the higher order drift
kinetic results. In section~\ref{sectapp_gyro_tpinch}, the general
gyrophase dependent piece is specialized for the $\theta$-pinch,
and it is used in section~\ref{sect_thetapinch_2} to calculate the
radial electric field.

\section{General gyrophase dependent piece} \label{sectapp_gyro_general}

Part of the gyrophase dependence is in the corrections to the
gyrokinetic variables $\bR_1$, $\bR_2$, $E_1$, $E_2$ and $\mu_1$.
This gyrophase dependence can be extracted for $k_\bot L \sim 1$
by Taylor expanding $f_i (\bR, E, \mu, t)$ around $\br$, $E_0$ and
$\mu_0$, as already done in \eq{Taylor_fi}. The contribution of
the collisional piece $\fwig_i$, given by \eq{fwig_sol}, can be
always added later.

Employing \eq{Taylor_fi}, the gyrophase dependent part of $f_i$ is
found to be
\begin{eqnarray}
f_i - \overline{f}_i = \left ( \frac{1}{\Omega_i} \bv \times \bun
+ \bR_2 \right ) \cdot \nablaave f_{i0} + \left [ \frac{Ze}{M} (
\phiwig - \overline{\phiwig} ) + E_2 \right ] \frac{\partial
f_{i0}}{\partial E_0} \nonumber \\ + \frac{1}{4 \Omega_i^2} [ (\bv
\times \bun) (\bv \times \bun) - \bv_\bot \bv_\bot ]: \nablaave\,
\nablaave f_{Mi} + \frac{Z^2 e^2}{2M^2} ( \phiwig^2 -
\overline{\phiwig^2} ) \frac{\partial^2 f_{Mi}}{\partial E_0^2}
\nonumber \\ + \frac{c}{B} [\phiwig ( \bv \times \bun ) -
\overline{\phiwig ( \bv \times \bun )} ] \cdot \nablaave \left (
\frac{\partial f_{Mi}}{\partial E_0} \right ) + ( \mu_1 -
\overline{\mu}_1 ) \frac{\partial f_{i0}}{\partial \mu_0},
\label{f_gyrodepen}
\end{eqnarray}

\noindent with $f_{i0} \equiv f_i (\br, E_0, \mu_0, t)$. Here, I
have employed the lowest order distribution function $f_{Mi}$ in
the higher order results, and I have used \eq{v2wig} to rewrite
$(\bv \times \bun) ( \bv \times \bun ) - \overline{( \bv \times
\bun ) ( \bv \times \bun )}$.

The function $\phiwig$ must be written as a function of the $\br$,
$\bv$ variables. To do so, I use equation \eq{phiwig_o1} to find
\begin{eqnarray}
\phiwig - \overline{\phiwig} = - \frac{1}{\Omega_i} ( \bv \times
\bun ) \cdot \nabla \phi - \bR_2 \cdot \nabla \phi +
\frac{1}{4\Omega_i^2} [ \bv_\bot \bv_\bot - (\bv \times \bun) (\bv
\times \bun) ] : (\nabla \nabla \phi).
\end{eqnarray}

For the higher order terms in \eq{f_gyrodepen}, I can simply use
the lowest order result $\phiwig \simeq - \Omega_i^{-1} ( \bv
\times \bun ) \cdot \nabla \phi$, which leads to
\begin{equation}
\phiwig^2 - \overline{\phiwig^2} = - \frac{1}{2\Omega_i^2} [
\bv_\bot \bv_\bot - (\bv \times \bun) (\bv \times \bun) ] : (
\nabla \phi \nabla \phi )
\end{equation}

\noindent and
\begin{equation}
\phiwig ( \bv \times \bun ) - \overline{\phiwig ( \bv \times \bun
)} = \frac{1}{2\Omega_i} \nabla \phi \cdot [ \bv_\bot \bv_\bot -
(\bv \times \bun) (\bv \times \bun) ].
\end{equation}

Using these expressions, the gyrophase dependent part of the
distribution function becomes
\begin{eqnarray}
f_i - \overline{f}_i = \bv \cdot \mathbf{g}_\bot + (\mu_1 -
\overline{\mu}_1 ) \frac{\partial f_{i0}}{\partial \mu_0} + \bR_2
\cdot \mathbf{G} + E_2 \frac{\partial f_{i0}}{\partial E_0}
\nonumber \\ + \frac{1}{4 \Omega_i^2} [ ( \bv \times \bun ) ( \bv
\times \bun ) - \bv_\bot \bv_\bot ] : \left ( \nablaave \mathbf{G}
- \frac{Ze}{M} \nabla \phi \frac{\partial \mathbf{G}}{\partial
E_0} \right ), \label{f_gd_1}
\end{eqnarray}

\noindent with
\begin{equation}
\mathbf{g}_\bot = \frac{1}{\Omega_i} \bun \times \left ( \nablaave
f_{i0} - \frac{Ze}{M} \nabla \phi \frac{\partial f_{i0}}{\partial
E_0} \right ) \label{g_perp_sima}
\end{equation}

\noindent and
\begin{equation}
\mathbf{G} = \nablaave f_{i0} - \frac{Ze}{M} \nabla \phi
\frac{\partial f_{i0} }{\partial E_0}.
\end{equation}

\noindent Thus, $\mathbf{g}_\bot = \Omega_i^{-1} \bun \times
\mathbf{G}$. In the long wavelength limit, $\partial /
\partial t \ll v_i / L$, so $E_2$ as given in \eq{correc_E_o2}
is negligible since it contains a time derivative. Also, the
zeroth order Fokker-Planck equation for the ion distribution
function is
\begin{equation}
v_{||} \bun \cdot \mathbf{G} \equiv v_{||} \bun \cdot \left (
\nablaave f_{i0} - \frac{Ze}{M} \nabla \phi \frac{\partial
f_{i0}}{\partial E_0} \right ) = C \{ f_i \} = 0,
\label{parallel_condition}
\end{equation}

\noindent since the ion distribution function is assumed to be
Maxwellian to zeroth order. This condition is important in
\eq{f_gd_1} because it implies that the components of $\bR_2$ that
are parallel to the magnetic field do not enter $f_i -
\overline{f}_i$. Therefore, employing the definition of $\bR_2$ in
\eq{R_1_main} and using the fact that for long wavelengths $ (c/B
\Omega_i) \nabla_\bR \Phiwig \times \bun \sim \delta_i^2 k_\bot
\rho_i L$ is negligible, I obtain
\begin{eqnarray}
\bR_2 \cdot \mathbf{G} = \frac{1}{\Omega_i} \left [ \left ( v_{||}
\bun + \frac{\bv_\bot}{4} \right ) \bv\times \bun + \bv \times
\bun \left ( v_{||} \bun + \frac{\bv_\bot}{4} \right ) \right ] :
\left [ \nabla \left ( \frac{\bun}{\Omega_i} \right ) \times
\mathbf{G} \right ] \nonumber \\ + \frac{v_{||}}{\Omega_i^2}
\bv_\bot \cdot \nabla \bun \cdot \mathbf{G}. \label{R_2_gradgp}
\end{eqnarray}

\noindent Equation~\eq{R_2_gradgp} can be written in a more
recognizable manner by using
\begin{eqnarray}
[ \bun (\bv\times \bun) + (\bv\times \bun) \bun ] :
\matrixtop{\mathbf{h}} = - \frac{1}{\Omega_i} \bun \cdot \left (
\nablaave \mathbf{G} - \frac{Ze}{M} \nabla \phi \frac{ \partial
\mathbf{G} }{ \partial E_0 } \right ) \cdot \bv_\bot \nonumber \\
+ [ \bun (\bv\times \bun) + (\bv\times \bun) \bun ] : \left [
\nabla \left ( \frac{\bun}{\Omega_i} \right ) \times \mathbf{G}
\right ], \label{aux_viscos1}
\end{eqnarray}

\noindent where $\matrixtop{\mathbf{h}}$ is
\begin{equation}
\matrixtop{\mathbf{h}} = \nablaave \mathbf{g}_\bot - \frac{Ze}{M}
\nabla \phi \frac{\partial \mathbf{g}_\bot}{\partial E_0}.
\label{h_matrix_sima}
\end{equation}

\noindent The first term in the right side of \eq{aux_viscos1} can
be further simplified by using \eq{parallel_condition} to obtain
\begin{equation}
\bun \cdot \left ( \nablaave \mathbf{G} - \frac{Ze}{M} \nabla \phi
\frac{\partial \mathbf{G}}{\partial E_0} \right ) \cdot \bv_\bot =
\bv_\bot \cdot \nablaave \mathbf{G} \cdot \bun = - \bv_\bot \cdot
\nabla \bun \cdot \mathbf{G}.
\end{equation}

\noindent As a result, \eq{R_2_gradgp} becomes
\begin{eqnarray}
\bR_2 \cdot \mathbf{G} = \frac{v_{||}}{\Omega_i} [ \bun (\bv
\times \bun) + (\bv \times \bun) \bun ] : \matrixtop{\mathbf{h}}
\nonumber \\ + \frac{1}{4 \Omega_i} [ \bv_\bot (\bv\times \bun) +
(\bv\times \bun) \bv_\bot ] : \left [ \nabla \left (
\frac{\bun}{\Omega_i} \right ) \times \mathbf{G} \right ].
\end{eqnarray}

The gyrophase dependent part of the ion distribution function can
now be explicitly written as\begin{eqnarray} (f_i -
\overline{f}_i)_g = \bv \cdot \mathbf{g}_\bot - \left \{ \bv_d
\cdot \bv + \frac{v_{||}}{4 B \Omega_i}[ \bv_\bot (\bv \times
\bun) + (\bv \times \bun) \bv_\bot ] : \nabla \bun \right \}
\frac{\partial f_{i0}}{\partial \mu_0} \nonumber \\  +
\frac{1}{\Omega_i} \left [ \left ( v_{||} \bun +
\frac{\bv_\bot}{4} \right ) \bv \times \bun + \bv \times \bun
\left ( v_{||} \bun + \frac{\bv_\bot}{4} \right ) \right ] :
\matrixtop{\mathbf{h}}, \label{gp_dependent_sima}
\end{eqnarray}

\noindent where the subindex $g$ indicates the non-collisional
origin of this gyrophase dependence. Equation
\eq{gp_dependent_sima} is exactly the same gyrophase dependent
distribution function found in \cite{simakov05}.

\section{Gyrophase dependent piece in a $\theta$-pinch} \label{sectapp_gyro_tpinch}

The solution for $f_i$ found in \eq{solution_o2_theta} means that
for all the terms in \eq{gp_dependent_sima}, $f_{i0}$ is
approximately $f_{M0}$ from \eq{f_M0_tpinch}. Due to the geometry
in the $\theta$-pinch, the vector $\mathbf{g}_\bot$ defined in
\eq{g_perp_sima} is
\begin{equation}
\mathbf{g}_\bot = \frac{1}{\Omega_i} \hat{\thetabf} \left (
\frac{\partial f_{M0}}{\partial r} + \frac{Ze}{T_i} \frac{\partial
\phi}{\partial r} f_{M0} \right ),
\end{equation}

\noindent and the matrix $\matrixtop{\mathbf{h}}$ defined in
\eq{h_matrix_sima} is
\begin{eqnarray}
\matrixtop{\mathbf{h}} = \hat{\br} \hat{\thetabf} \left \{
\frac{\partial}{\partial r} \left [ \frac{1}{\Omega_i} \left (
\frac{\partial f_{M0}}{\partial r} + \frac{Ze}{T_i} \frac{\partial
\phi}{\partial r} f_{M0} \right ) \right ] + \frac{Ze}{\Omega_i}
\frac{\partial \phi}{\partial r} \left [ \frac{\partial}{\partial
r} \left ( \frac{f_{M0}}{T_i} \right ) + \frac{Ze}{T_i^2}
\frac{\partial \phi}{\partial r} f_{M0} \right ] \right \}
\nonumber \\ - \hat{\thetabf} \hat{\br} \left [ \frac{1}{r
\Omega_i} \left ( \frac{\partial f_{M0}}{\partial r} +
\frac{Ze}{T_i} \frac{\partial \phi}{\partial r} f_{M0} \right )
\right ],
\end{eqnarray}

\noindent where I use $\nabla \hat{\thetabf} = - \hat{\thetabf}
\hat{\br}/r$. Employing this results and taking into account that
$\partial f_{i0}/\partial \mu_0 \simeq 0$ in this case, I find
\eq{gp_tpinch}.

\chapter{Gyrokinetic equation in physical phase space \label{app_transportGK}}

This Appendix contains the details needed to write equation
\eq{eqGK_ini} as equation \eq{eqGK_final}. In
section~\ref{sectapp_drifts}, equation \eq{transf_phys} is
derived. Equation \eq{transf_phys} models the finite gyroradius
effects in the particle motion through the modified parallel
velocity $v_{||0}$ and the perpendicular drift $\tilde{\bv}_1$. In
section~\ref{sectapp_conserv}, equation \eq{transf_phys} is
written in conservative form, more convenient to obtain moment
equations.

\section{Gyrokinetic equation in $\br$, $E_0$, $\mu_0$ and
$\varphi_0$ variables} \label{sectapp_drifts}

In this section, I rewrite part of the gyrokinetic equation as a
function of the variables $\br$, $E_0$, $\mu_0$ and $\varphi_0$.
The gyrokinetic equation is only valid to $O(\delta_i f_{Mi}
v_i/L)$, and the expansions will be carried out only to that
order. In particular, I am interested in $\dot{\bR} \cdot
\nabla_\bR f_i + \dot{E} (\partial f_i / \partial E)$. In the term
$\dot{E} (\partial f_i / \partial E)$, I can make use of the
lowest order equality $\partial f_i / \partial E \simeq \partial
f_{Mi}/\partial E_0$. Then, employing $\dot{E}$ from
\eq{dE_dt_main} gives
\begin{equation}
\dot{\bR} \cdot \nabla_\bR f_i + \dot{E} \frac{\partial
f_i}{\partial E} \simeq \dot{\bR} \cdot \left ( \nabla_\bR f_i -
\frac{Ze}{M} \nabla_\bR \phiave \frac{\partial f_{Mi}}{\partial
E_0} \right ). \label{change_ini}
\end{equation}

\noindent For $\nabla_\bR f_i$, changing from $\bR$, $E$, $\mu$
and $\varphi$ to $\br$, $E_0$, $\mu_0$ and $\varphi_0$, I find
\begin{equation}
\nabla_{\bR} f_i \simeq \nabla_{\bR} \br \cdot \nablaave f_i +
\nabla_{\bR} E_0 \frac{\partial f_i}{\partial E_0} + \nabla_{\bR}
\mu_0 \frac{\partial f_i}{\partial \mu_0} + \nabla_{\bR} \varphi_0
\frac{\partial f_i}{\partial \varphi_0}. \label{gradchange}
\end{equation}

\noindent Here, $\partial f_i / \partial \mu_0$ and $\partial f_i
/ \partial \varphi_0$ are small because the zeroth order
distribution function is a stationary Maxwellian. The gradient of
$E_0$ is given by $0 = \nabla_{\bR} E \simeq \nabla_\bR (E_0 +
E_1) = \nabla_\bR E_0 + (Ze/M) \nabla_{\bR} \phiwig$. Similarly,
$\nabla_\bR \mu = 0 = \nabla_\bR \varphi$ give $\nabla_\bR \mu_0
\simeq - \nabla_\bR \mu_1$ and $\nabla_\bR \varphi_0 \simeq -
\nabla_\bR \varphi_1$. Then, to the required order
\begin{equation}
\nabla_\bR f_i \simeq \nabla_\bR \br \cdot \nablaave f_i -
\frac{Ze}{M} \nabla_\bR \phiwig \frac{\partial f_i}{\partial E_0}.
\end{equation}

\noindent Since $\partial f_i / \partial E \simeq (-M/T_i)
f_{Mi}$, and $\phiwig \, \nabla_{\bR} f_{Mi} \ll f_{Mi}
\nabla_{\bR} \phiwig$ because the perpendicular gradient of
$\phiwig$ is steeper and the parallel gradient of $f_{Mi}$ is
small, I find
\begin{equation}
\nabla_{\bR} f_i \simeq \nabla_\bR \br \cdot \nablaave \left ( f_i
+ \frac{Ze\phiwig}{T_i} f_{Mi} \right ) = \nabla_{\bR} \br \cdot
\nablaave f_{ig}, \label{gradfichange}
\end{equation}

\noindent where I have employed the lowest order result \eq{fi1}
and $\nabla_\bR \phiwig \simeq \nabla_\bR \br \cdot \nablaave
\phiwig$. To prove that $\nabla_\bR \phiwig \simeq \nabla_\bR \br
\cdot \nablaave \phiwig$ and $\nabla_\bR \phiave \simeq \nabla_\bR
\br \cdot \nablaave \phiave$, I follow a similar procedure to the
one used for $f_i$ in \eq{gradchange}. In this case, $\partial
\phiwig / \partial E_0$ and $\partial \phiave /
\partial E_0$ are small. Substituting equations \eq{gradfichange}
and $\nabla_\bR \phiave \simeq \nabla_\bR \br \cdot \nablaave
\phiave$ into \eq{change_ini}, I find equation \eq{transf_phys},
where to write $\nabla_\bR \br \simeq \nabla_{\bR_g} \br$ I have
used the fact that $\bR$ can be replaced by $\bR_g$ to lowest
order. The only coefficient left to evaluate is $\dot{\bR} \cdot
\nabla_{\bR_g} \br$. Since $\br = \bR_g - \Omega_i^{-1} \bv \times
\bun$, and $\bv \times \bun = \sqrt{2 \mu_0 B (\br)} [ \eun_1(\br)
\sin \varphi_0 - \eun_2 (\br) \cos \varphi_0 ]$, I find that
$\nabla_{\bR_g} \br \simeq \matI - \nablaave ( \Omega_i^{-1} \bv
\times \bun )$, where the gradient $\nablaave (\Omega_i^{-1} \bv
\times \bun)$ is evaluated holding $\mu_0$ and $\varphi_0$ fixed,
and it is given by
\begin{equation}
- \nablaave \left ( \frac{1}{\Omega_i} \bv \times \bun \right ) =
\frac{1}{\Omega_i} \left [ \frac{1}{2B}(\nabla B) (\bv \times
\bun) + \nabla \bun \cdot (\bv \times \bun) \bun + (\nabla \eun_2
\cdot \eun_1) \bv_\bot \right ]. \label{grad_rho_phys}
\end{equation}

\noindent In $\dot{\bR}$, given by \eq{dR_dt_main}, the terms
$\bv_M$ and $\bv_E$ are an order smaller than $u \bun (\bR)$ so I
can use $\nabla_{\bR_g} \br \simeq \matI$ for $\bv_M \cdot
\nabla_{\bR_g} \br$ and $\bv_E \cdot \nabla_{\bR_g} \br$ to find
the result in \eq{Rdot_gradRr}. In equation \eq{Rdot_gradRr}, I
have also used $u \bun (\bR) \simeq u \bun (\br) + (v_{||} /
\Omega_i) (\bv \times \bun) \cdot \nabla \bun$, $\bv_M \simeq
\bv_{M0}$, $\bv_E \simeq \bv_{E0}$ and
\begin{eqnarray}
 \nablaave \times \bv_\bot = (\bv \times \bun ) \cdot \nabla
\bun + \frac{1}{2B} \bv \times \bun (\bun \cdot \nabla B) +
\bv_\bot (\bun \cdot \nabla \eun_2 \cdot \eun_1) \nonumber \\ +
\bun \left [ \frac{1}{2B} ( \bv \times \bun ) \cdot \nabla B -
\bv_\bot \cdot \nabla \eun_2 \cdot \eun_1 \right ],
\label{curl_vbot}
\end{eqnarray}

\noindent where I employ $ \nablaave \times \bv_\bot = \nablaave
\times [ \bun \times (\bv \times \bun) ] = \nablaave \cdot [( \bv
\times \bun ) \bun] - \nablaave \cdot [ \bun ( \bv \times \bun
)]$. To find the result in equation \eq{vpar0}, notice that
$v_{||0} = u + (v_{||} / \Omega_i ) \bun \cdot \nabla \bun \cdot
(\bv \times \bun) - (v_{||}/\Omega_i) \bun \cdot \nablaave \times
\bv_\bot$, and the difference $u - v_{||}$ is given in
\eq{vpar_u_1}.

\section{Conservative form in $\br$, $E_0$, $\mu_0$ and $\varphi_0$ variables} \label{sectapp_conserv}

To obtain equation \eq{cons_Rdot} from \eq{transf_phys}, I just
need to prove that
\begin{eqnarray}
 \nablaave \cdot \left ( \frac{B}{v_{||}} \dot{\bR} \cdot
\nabla_{\bR_g} \br \right ) - \frac{\partial}{\partial \mu_0}
\left ( \mathbf{B} \cdot \nablaave \mu_{10} \right ) -
\frac{\partial}{\partial \varphi_0} \left ( \mathbf{B} \cdot
\nablaave \varphi_{10} \right ) \nonumber \\ -
\frac{\partial}{\partial E_0} \left ( \frac{B}{v_{||}}
\frac{Ze}{M} \dot{\bR} \cdot \nabla_{\bR_g} \br \cdot \nablaave
\phiave \right ) = 0. \label{div_phase}
\end{eqnarray}

\noindent Then, equation \eq{cons_Rdot} is found by using
$\partial f_{ig}/\partial E_0 \simeq \partial f_{Mi}/\partial
E_0$, $\partial f_{ig}/\partial \mu_0 \simeq 0$ and  $\partial
f_{ig}/\partial \varphi_0 \simeq 0$.

To prove \eq{div_phase}, I use the value of  $\dot{\bR} \cdot
\nabla_{\bR_g} \br$ from \eq{Rdot_gradRr} and the relations
$\partial ( \mathbf{B} \cdot \nablaave \mu_{10} )/\partial \mu_0 =
\nablaave \cdot [ \mathbf{B} (\partial \mu_{10}/\partial \mu_0) ]$
and $\partial ( \mathbf{B} \cdot \nablaave \varphi_{10} )/\partial
\varphi_0 = \nablaave \cdot [ \mathbf{B} (\partial
\varphi_{10}/\partial \varphi_0) ]$, to find
\begin{eqnarray}
\nablaave \cdot \left \{ \frac{B}{v_{||}} [ (v_{||0} - v_{||})
\bun + \bv_{M0} + \bv_{E0} ] - \mathbf{B} \left ( \frac{\partial
\mu_{10}}{\partial \mu_0} + \frac{\partial \varphi_{10}}{\partial
\varphi_0} \right ) \right \} \nonumber \\ -
\frac{\partial}{\partial E_0} \left ( \frac{B}{v_{||}}
\frac{Ze}{M} \bv_{M0} \right ) \cdot \nablaave \phiave = 0.
\label{div_phase2}
\end{eqnarray}

\noindent Here, I have also employed $\nabla \cdot \mathbf{B} =
0$, $\partial (\mathbf{B} \cdot \nablaave \phiave)/ \partial E =
0$, $\nablaave \cdot [ (B/v_{||}) \tilde{\bv}_1] = 0$ and
$\partial [ (B/v_{||}) \tilde{\bv}_1 ]/\partial E_0 = 0$ [these
last two expressions are easy to prove by using the definition of
$\tilde{\bv}_1$ from \eq{vtilde1}]. In the term $\dot{\bR} \cdot
\nabla_{\bR_g} \br \cdot \nablaave \phiave$, $(v_{||0} - v_{||})
\bun \cdot \nablaave \phiave$ is negligible because $\bun \cdot
\nablaave \phiave \simeq 0$ (the zeroth order potential is
constant along magnetic field lines). In equation \eq{div_phase2},
it is satisfied that
\begin{equation}
\nablaave \cdot \left ( \frac{B}{v_{||}} \bv_{E0} \right ) -
\frac{\partial}{\partial E_0} \left ( \frac{B}{v_{||}}
\frac{Ze}{M} \bv_{M0} \right ) \cdot \nablaave \phiave = 0,
\end{equation}

\noindent where I employ relation \eq{id_kappa} and $\bun \cdot
\nablaave \phiave \simeq 0$; and
\begin{equation}
\frac{v_{||0} - v_{||}}{v_{||}} - \frac{\partial
\mu_{10}}{\partial \mu_0} - \frac{\partial \varphi_{10}}{\partial
\varphi_0} = \frac{v_{||}}{\Omega_i} \bun \cdot \nabla \times
\bun,
\end{equation}

\noindent with $\partial \bv_\bot / \partial \mu_0 = (B/v_\bot^2)
\bv_\bot$, $\partial \bv_\bot / \partial \varphi_0 = - \bv \times
\bun$ and $v_{||0}$ defined by \eq{vpar0}. Using these relations,
equation \eq{div_phase2} becomes
\begin{eqnarray}
\frac{Mc}{Ze} \nablaave \cdot \left [ \frac{\mu_0}{v_{||}} \bun
\times \nabla B + v_{||} \nabla \times \bun \right ] =
\frac{Mc}{Ze} \nablaave \cdot [ \nablaave \times (v_{||} \bun)] =
0,
\end{eqnarray}

\noindent where the relation \eq{id_kappa} is used again.

\chapter{Details of the particle and momentum transport
calculation} \label{app_n_v}

This Appendix contains the details of the derivations of the
gyrokinetic particle conservation equation \eq{ntransp_krho} and
the momentum conservation equation \eq{vtransp_krho} from equation
\eq{div_GKtransp}.

\section{Details of the particle transport calculation} \label{sectapp_ntransp}

Equation \eq{div_GKtransp} with $G = 1$ leads to equation
\eq{ntransp_krho}, with the integral of the gyroaveraged collision
operator giving the term $\nabla \cdot (n_i \bV_{iC})$ as shown in
Appendix~\ref{app_collisiontransport}. The only other integral of
some difficulty is $\int d^3v\, f_{ig} \dot{\bR} \cdot
\nabla_{\bR_g} \br$. The integral of $v_{||} \bun$ is done
realizing that $f_i - f_{ig} = (-Ze\phiwig/T_i) f_{Mi}$ is even in
$v_{||}$ to write \eq{n_parV}. The integral of $( v_{||0} - v_{||}
) \bun$ is done by using $\nabla \cdot [ \int d^3v\, f_{ig} (
v_{||0} - v_{||} ) \bun ] \simeq \mathbf{B} \cdot \nabla [ \int
d^3v\, f_{ig} (v_{||0} - v_{||})/B ]$. Then $f_{ig}$ can be
replaced by $f_{Mi}$ because the gradient is along the magnetic
field line, and the slow parallel gradients make the small pieces
of the distribution function unimportant. From all the terms in
$v_{||0} - v_{||}$, only the gyrophase independent piece
$(v_\bot^2/2\Omega_i) \bun \cdot \nabla \times \bun$ gives a
non-vanishing contribution.

To perform the integral $\nabla \cdot (n_i \tilde{\bV}_i) = \nabla
\cdot (\int d^3v\, f_{ig} \tilde{\bv}_1) = \nabla \cdot [\int
d^3v\, f_{ig} (v_{||}/\Omega_i) \nablaave \times \bv_\bot]$, I use
the expression of $\nablaave \times \bv_\bot$ in equation
\eq{curl_vbot}. The contribution of the parallel component of
$\nablaave \times \bv_\bot$ is negligible because its divergence
only has parallel gradients, and they are small compared to the
perpendicular gradients. The integral of $(v_{||}/\Omega_i)
\bv_\bot (\bun \cdot \nabla \eun_2 \cdot \eun_1)$, on the other
hand, vanishes because its divergence becomes
\begin{equation}
\int d^3v\, \frac{v_{||}}{\Omega_i} \bun \cdot \nabla \eun_2 \cdot
\eun_1 \bv_\bot \cdot \nablaave f_{ig} = \int d^3v\, v_{||} \bun
\cdot \nabla \eun_2 \cdot \eun_1 \frac{\partial f_{ig}}{\partial
\varphi_0} = 0.
\end{equation}

\noindent Here, I neglect the gradients of any quantity that is
not $f_{ig}$ because they give contributions of order $\delta_i^2
k_\bot \rho_i n_e v_i/L$. To obtain $\partial f_{ig} /
\partial \varphi_0 \simeq \Omega_i^{-1} \bv_\bot \cdot \nablaave
f_{ig}$, I use that $f_{ig}$'s only dependence on $\varphi_0$ is
through $\bR_g$. The final result for $n_i \tilde{\bV}_i$ is
written in \eq{flowtilde_v2}.

\section{Details of the momentum transport calculation} \label{sectapp_vtransp}

Equation \eq{div_GKtransp} with $G = \bv$ gives equation
\eq{vtransp_krho}. The integral of the gyroaveraged collision
operator gives $\mathbf{F}_{iC}$. The details are given in
Appendix~\ref{app_collisiontransport}.

To simplify the integral $\int d^3v\, f_{ig} (\dot{\bR} \cdot
\nabla_{\bR_g} \br) M \bv$, I use that $\nabla \cdot [ \int d^3v\,
f_{ig} ( v_{||0} - v_{||} ) \bun \bv ] \simeq \mathbf{B} \cdot
\nabla [ \int d^3v\, f_{ig} \bv (v_{||0} - v_{||})/B ]$. Then
$f_{ig}$ can be replaced by $f_{Mi}$ because the gradient is along
the magnetic field line, and the next order corrections can be
neglected. The integral $\int d^3v\, f_{Mi} \bv (v_{||0} -
v_{||})/B$ vanishes because all the terms are either odd in
$v_{||}$ or odd in $\bv_\bot$. The final result is
\begin{equation}
\int d^3v\, f_{ig} \left ( \dot{\bR} \cdot \nabla_{\bR_g} \br
\right ) M \bv = p_{ig||} \bun \bun + \pibf_{ig||} \bun +
\matrixtop{\pibf}_{ig\times}, \label{flow_vGK}
\end{equation}

\noindent where I use the definitions of $p_{ig||} = \int d^3v\,
f_{ig} Mv_{||}^2$, $\pibf_{ig||}$ from \eq{pig_par}, and
$\matrixtop{\pibf}_{ig\times}$ from \eq{pig_perp}.

To find the integral $\int d^3 v\, M f_{ig} K \{ \bv \}$, with the
linear operator $K$ given in \eq{op_K}, I use $K\{ \bv \} = K\{
v_{||} \bun \} + K\{ \bv_\bot \}$. For $K\{v_{||} \bun \}$, I need
$\nablaave v_{||} = - \mu_0 \nabla B / v_{||}$, $\partial v_{||} /
\partial E_0 = v_{||}^{-1}$, $\partial v_{||} / \partial \mu_0 = -
B/v_{||}$ and $\partial v_{||} / \partial \varphi_0 = 0$. Then,
using the definitions of $p_{ig||}$ and $\pibf_{ig||}$ along with
\begin{equation}
(\bv_{M0} + \bv_{E0}) \cdot \left ( \frac{\mu_0 \nabla B}{v_{||}}
+ \frac{Ze}{Mv_{||}} \nablaave \phiave \right ) =
\frac{v_{||}}{\Omega_i} (\bun \times \kappabf) \cdot \left (\mu_0
\nabla B + \frac{Ze}{M} \nablaave \phiave \right ),
\end{equation}

\noindent I find
\begin{eqnarray}
\int d^3 v\, M f_{ig} K \{ v_{||} \bun \} = ( p_{ig||} \bun +
\pibf_{ig||} ) \cdot \nabla \bun - \int d^3 v\, f_{ig} M\mu_0 \bun
\bun \cdot \nabla B \nonumber \\ + \int d^3v\, f_{Mi}
\frac{Mv_\bot^2}{2\Omega_i} \bun (\bun \cdot \nabla \times \bun)
\bun \cdot \nablaave v_{||} + \int d^3 v\, f_{Mi} M \bun
(\mathbf{B} \cdot \nablaave \mu_{10}) \nonumber \\ - Ze \int d^3v
f_{ig} \bun ( \bun + \Omega_i^{-1} \nablaave \times \bv_\bot )
\cdot \nablaave \phiave. \label{force_par1}
\end{eqnarray}

\noindent Here, I have used that $(v_{||0} - v_{||}) \bun \cdot
\nabla \phiave \simeq 0$ and that, in the integrals that include
$(v_{||0} - v_{||}) \bun \cdot \nablaave (v_{||} \bun)$ and
$\tilde{\bv}_1 \cdot \nablaave v_{||}$, only the term
$(v_\bot^2/2\Omega_i) \bun (\bun \cdot \nabla \times \bun) \bun
\cdot \nablaave v_{||}$ gives a non-vanishing contribution. The
integral of $f_{Mi} (v_{||}/\Omega_i) (\bun \times \kappabf) \cdot
[\mu_0 \nabla B + (Ze/M) \nablaave \phiave ]$ vanishes because it
is odd in $v_{||}$. Equation \eq{force_par1} can be further
simplified by using
\begin{eqnarray}
\int d^3 v\, f_{Mi} \mathbf{B} \cdot \nablaave \mu_{10} = - \int
d^3 v\, f_{Mi} \mathbf{B} \cdot \nablaave \left (
\frac{v_{||}v_\bot^2}{2B\Omega_i} \bun \cdot \nabla \times \bun
\right ) = \nonumber \\ - \int d^3 v\, f_{Mi}
\frac{v_\bot^2}{2\Omega_i} (\bun \cdot \nabla \times \bun) \bun
\cdot \nablaave v_{||}, \label{int_fBnablamu10}
\end{eqnarray}

\noindent where I have used that in $\mu_{10}$ all the terms but
the gyrophase independent piece give vanishing integrals. The last
form of \eq{int_fBnablamu10} cancels with a term in
\eq{force_par1}. Using $p_{ig\bot} = \int d^3v\, f_{ig}
Mv_\bot^2/2$, equation \eq{force_par1} then becomes
\begin{eqnarray}
\int d^3 v\, M f_{ig} K \{ v_{||} \bun \} = - p_{ig\bot} \bun \bun
\cdot \nabla \ln B + ( p_{ig||} \bun + \pibf_{ig||} ) \cdot \nabla
\bun \nonumber \\ - Ze \int d^3v f_{ig} \bun ( \bun +
\Omega_i^{-1} \nablaave \times \bv_\bot ) \cdot \nablaave \phiave.
\label{force_par2}
\end{eqnarray}

The integral $\int d^3v\, f_{ig} K\{\bv_\bot\}$ is obtained using
$\nablaave \bv_\bot = (\nabla B/2B) \bv_\bot - \nabla \bun \cdot
\bv_\bot \bun + \nabla \eun_2 \cdot \eun_1 (\bv \times \bun)$,
$\partial \bv_\bot / \partial E_0 = 0$, $\partial \bv_\bot /
\partial \mu_0 = (2\mu_0)^{-1} \bv_\bot$ and $\partial \bv_\bot /
\partial \varphi_0 = - \bv \times \bun$. Then, I find
\begin{equation}
\int d^3v\, M f_{ig} K\{\bv_\bot\} = \int d^3v\, M f_{ig} v_{||}
\bun \cdot \nablaave \bv_\bot - \frac{Mc}{B} \int d^3v\, f_{Mi}
(\nablaave \phiave \times \bun) \cdot \nablaave \bv_\bot,
\end{equation}

\noindent where I employ that the integrals of $f_{Mi} [ (v_{||0}
- v_{||}) \bun + \bv_{M0} + \tilde{\bv}_1] \cdot \nablaave
\bv_\bot$, $f_{Mi} v_{||} \bun \cdot \nablaave \mu_{10} (\partial
\bv_\bot / \partial \mu_0)$ and $f_{Mi} v_{||} \bun \cdot
\nablaave \varphi_{10} (\partial \bv_\bot / \partial \varphi_0)$
vanish because the terms are either odd in $v_{||}$ or in
$\bv_\bot$. The integral that includes $\nablaave \phiave$ can be
rewritten by realizing that $\nablaave \phiave = \nabla \phi -
\nablaave \phiwig$, to find
\begin{equation}
\int d^3v\, M f_{ig} K\{\bv_\bot\} = \int d^3v\, M f_{ig} v_{||}
\bun \cdot \nablaave \bv_\bot + \frac{Mc}{B} \int d^3v\, f_{Mi}
(\nablaave \phiwig \times \bun) \cdot \nablaave \bv_\bot.
\label{force_perp}
\end{equation}

Combining equations \eq{force_par2} and \eq{force_perp}, I find
\begin{equation}
\int d^3v\, M f_{ig} K \{ \bv \} = p_{ig\bot} \bun (\nabla \cdot
\bun) + (p_{ig||} \bun + \pibf_{ig||}) \cdot \nabla \bun - Zen_i
\bun \bun \cdot \nabla \phi + \tilde{F}_{iE} \bun +
\mathbf{F}_{iB}, \label{force_par_perp}
\end{equation}

\noindent where I have used $\bun \cdot \nabla \ln B = -\nabla
\cdot \bun$, and the definitions of $\tilde{F}_{iE}$ from
\eq{force_E} and $\mathbf{F}_{iB}$ from \eq{force_B}.

Using \eq{flow_vGK} and \eq{force_par_perp} in \eq{div_GKtransp}
gives
\begin{eqnarray}
\frac{\partial}{\partial t} (n_i M \bV_{ig}) + \nabla \cdot (
p_{ig||} \bun \bun + \pibf_{ig||} \bun +
\matrixtop{\pibf}_{ig\times} ) = - Zen_i \bun \bun \cdot \nabla
\phi \nonumber \\ + p_{ig\bot} \bun (\nabla \cdot \bun) +
(p_{ig||} \bun + \pibf_{ig||}) \cdot \nabla \bun + \tilde{F}_{iE}
\bun + \mathbf{F}_{iB} + \mathbf{F}_{iC}.
\end{eqnarray}

\noindent Finally, employing $\nabla \cdot ( p_{ig||} \bun \bun )
= \bun ( \bun \cdot \nabla p_{ig||} + p_{ig||} \nabla \cdot \bun )
+ p_{ig||} \bun \cdot \nabla \bun$ and $\nabla \cdot (\pibf_{ig||}
\bun) = (\nabla \cdot \pibf_{ig||}) \bun + \pibf_{ig||} \cdot
\nabla \bun$, I am able to recover equation \eq{vtransp_krho}.
Multiplying equation \eq{vtransp_krho} by $\bun$ and taking into
account the cancellation of
\begin{equation}
\mathbf{F}_{iB} \cdot \bun = - M \int d^3v\, f_{ig} \left ( v_{||}
\bun + \frac{c}{B} \nablaave \phiwig \times \bun \right ) \cdot
\nabla \bun \cdot \bv_\bot
\end{equation}

\noindent and
\begin{equation}
(\nabla \cdot \matrixtop{\pibf}_{ig\times}) \cdot \bun \simeq - M
\int d^3v\, f_{ig} \left ( v_{||} \bun + \frac{c}{B} \nablaave
\phiwig \times \bun \right ) \cdot \nabla \bun \cdot \bv_\bot,
\label{par_gradpigperp}
\end{equation}

\noindent I find equation \eq{par_vtransp_krho}. To obtain
relation \eq{par_gradpigperp}, I have employed
$\matrixtop{\pibf}_{ig\times} \cdot \bun = 0$ and I have used the
lowest order distribution function $f_{Mi}$ for the higher order
terms. All the higher order terms, except for $(c/B) \nablaave
\phiave \times \bun$, cancel because they are either odd in
$v_{||}$ or $\bv_\bot$.

\chapter{Finite gyroradius effects in the like-collision operator} \label{app_collisiontransport}

In this Appendix, I show how to treat the gyroaveraged
like-collision operator, $\langle C\{f_i\} \rangle$. The
like-collision operator is
\begin{equation}
C \{ f_i \} = \gamma \nabla_v \cdot \left [ \int d^3v^\prime\,
\nabla_g \nabla_g g \cdot ( f_i^\prime \nabla_v f_i - f_i
\nabla_{v^\prime} f_i^\prime ) \right ],
\end{equation}

\noindent with $f_i = f_i(\bv)$, $f_i^\prime = f_i(\bv^\prime)$,
$\mathbf{g} = \bv - \bv^\prime$, $g = |\mathbf{g}|$, $\nabla_g
\nabla_g g = (g^2 \matI - \mathbf{g} \mathbf{g})/g^3$ and $\gamma
= 2\pi Z^4 e^4 \ln \Lambda/M^2$. Linearizing this equation for
$f_i = f_{Mi} + f_{i1}$, with $f_{i1} \ll f_{Mi}$, I find
\begin{equation}
C^{(\ell)} \{ f_{i1} \} = \gamma \nabla_v \cdot \Gammabf
\{f_{i1}\}, \label{linear_C}
\end{equation}

\noindent with
\begin{equation}
\Gammabf \{f_{i1}\} = \int d^3v^\prime\, f_{Mi} f_{Mi}^\prime
\nabla_g \nabla_g g \cdot \left [ \nabla_v \left (
\frac{f_{i1}}{f_{Mi}} \right ) - \nabla_{v^\prime} \left (
\frac{f_{i1}^\prime}{f_{Mi}^\prime} \right ) \right ].
\end{equation}

\noindent The vector $\Gammabf$ can also be written as in
\eq{Gamma_c} because $\Gammabf \{f_i\} = \Gammabf \{f_{Mi}\} +
\Gammabf \{f_{i1}\} = \Gammabf \{f_{i1}\}$. Using gyrokinetic
variables in equation \eq{linear_C} and gyroaveraging gives
\begin{equation}
\langle C^{(\ell)} \{ f_{i1} \} \rangle = \gamma \frac{u}{B} \Bigg
[ \frac{\partial}{\partial E} \left ( \frac{B}{u} \langle \Gammabf
\cdot \nabla_v E \rangle \right ) + \frac{\partial}{\partial \mu}
\left ( \frac{B}{u} \langle \Gammabf \cdot \nabla_v \mu \rangle
\right ) + \nabla_\bR \cdot \left ( \frac{B}{u} \langle \Gammabf
\cdot \nabla_v \bR \rangle \right ) \Bigg ]. \label{linear_C_GK}
\end{equation}

\noindent Here, $B/u \simeq \partial (\br, \bv)/\partial (\bR, E,
\mu, \varphi)$ is the approximate Jacobian, and I have used the
transformation rule for divergences from one reference system $\{
x_i \}$ to another $\{ y_j \}$:
\begin{equation}
\nabla_x \cdot \Gammabf = \sum_j \frac{1}{J_y}
\frac{\partial}{\partial y_j} (J_y \Gammabf \cdot \nabla_x y_j) =
\sum_j \frac{1}{J_y} \frac{\partial}{\partial y_j} (J_y \Gamma
_{y_j}) , \label{transf_diverg}
\end{equation}

\noindent where $J_y = \partial (x_i) / \partial (y_j)$ is the
Jacobian of the transformation, $\Gamma_{y_j} = \Gammabf \cdot
\nabla_x y_j$ and $\nabla_x$ is the gradient in the reference
system $\{x_i\}$. To rewrite equation \eq{linear_C_GK} in terms of
the variables $\br$, $E_0$, $\mu_0$ and $\varphi_0$, I need to use
\eq{transf_diverg} and the chain rule to find the transformation
between the two reference systems $\{y_j\}$ and $\{ z_k \}$
\begin{equation}
\frac{1}{J_y} \sum_j \frac{\partial}{\partial y_j}( J_y
\Gamma_{y_j} ) = \frac{1}{J_z} \sum_k \frac{\partial}{\partial
z_k} \left ( J_z \sum_j \Gamma_{y_j} \frac{\partial z_k}{\partial
y_j} \right ). \label{transf_diverg2}
\end{equation}

\noindent Employing this relation to write equation
\eq{linear_C_GK} as a function of $\br$, $E_0$, $\mu_0$ and
$\varphi_0$ gives
\begin{eqnarray}
\langle C^{(\ell)} \{ f_{i1} \} \rangle \simeq \gamma
\frac{v_{||}}{B} \Bigg \{ \frac{\partial}{\partial E_0} \left (
\frac{B}{v_{||}} \langle \Gammabf \cdot \nabla_v E \rangle \right
) + \frac{\partial}{\partial \mu_0} \left ( \frac{B}{v_{||}}
\langle \Gammabf \cdot \nabla_v \mu \rangle \right ) \nonumber \\
+ \nablaave \cdot \left [ \frac{B}{v_{||}} \left ( \langle
\Gammabf \cdot \nabla_v \mu \rangle \frac{\partial \br}{\partial
\mu} + \langle \Gammabf \cdot \nabla_v \bR \rangle \right ) \right
] \Bigg \}, \label{linear_C_phys}
\end{eqnarray}

\noindent where I have used the lowest order gyrokinetic variables
$\bR_g$, $E_0$, $\mu_0$ and $\varphi_0$. This approximation is
justified because the collision operator vanishes to lowest order,
and only the zeroth order definitions must be kept. Notice that I
keep the first order correction $\bR_1 = \Omega_i^{-1} \bv \times
\bun$ only within the spatial divergence because the spatial
gradients are steep. Employing $\nabla_v E \simeq \bv$, $\nabla_v
\mu \simeq \bv_\bot/B$, $\partial \br / \partial \mu \simeq -
\partial \bR_1 / \partial \mu \simeq - (2\mu_0\Omega_i)^{-1} \bv
\times \bun$ and $\nabla_v \bR \simeq \Omega_i^{-1} \matI \times
\bun$, I find
\begin{eqnarray}
 \langle C^{(\ell)} \{ f_{i1} \} \rangle \simeq \gamma
\frac{v_{||}}{B} \Bigg \{ \frac{\partial}{\partial E_0} \left (
\frac{B}{v_{||}} \langle \Gammabf \cdot \bv \rangle \right ) +
\frac{\partial}{\partial \mu_0} \left ( \frac{1}{v_{||}} \langle
\Gammabf \cdot \bv_\bot \rangle \right ) \nonumber \\ + \nablaave
\cdot \left [ \frac{Mc}{Zev_{||}} \left ( \langle \Gammabf \rangle
\times \bun - \frac{1}{v_\bot^2} \langle \Gammabf \cdot \bv_\bot
\rangle \bv \times \bun \right ) \right ] \Bigg \}.
\label{linear_C_final}
\end{eqnarray}

In the main text, there are two integrals that involve the
gyroaveraged collision operator, $\nabla \cdot (n_i \bV_{iC}) = -
\int d^3v\, \langle C\{ f_i \} \rangle$ and $\mathbf{F}_{iC} = M
\int d^3v\, \bv \langle C\{ f_i \} \rangle$. Using equation
\eq{linear_C_final}, I obtain equations \eq{nViC} and \eq{FiC}. To
find \eq{FiC}, I have integrated by parts using $\partial \bv /
\partial E_0 = v_{||}^{-1} \bun$ and $\partial \bv /\partial \mu_0 =
(B/v_\bot^2) \bv_\bot - (B/v_{||}) \bun$, and $\nablaave \bv$ has
been neglected because the gradient is of order $1/L$.

I can prove that the divergence of $n_i \bV_{iC}$, given in
\eq{nViC}, is of order $\delta_i (k_\bot \rho_i)^2 \nu_{ii} n_e$
rather than $\delta_i k_\bot \rho_i \nu_{ii} n_e$. For $k_\bot
\rho_i \ll 1$, the function $\Gammabf (\br, E_0, \mu_0,
\varphi_0)$ can be Taylor expanded around $\bR_g$ to find
$\Gammabf(\br, E_0, \mu_0, \varphi_0) \simeq \Gammabf(\bR_g, E_0,
\mu_0, \varphi_0) - \Omega_i^{-1} (\bv \times \bun) \cdot
\nabla_{\bR_g} \Gammabf$. Then, the gyroaverage $\langle \ldots
\rangle$ holding $\bR$, $E$, $\mu$ and $t$ fixed gives
\begin{equation}
\langle \Gammabf \rangle = \frac{1}{2\pi} \oint d\varphi_0\, \left
[ \Gammabf(\bR_g, E_0, \mu_0, \varphi_0) - \frac{1}{\Omega_i} (\bv
\times \bun) \cdot \nabla_{\bR_g} \Gammabf \right ],
\label{gammac_int_approx}
\end{equation}

\noindent where I have employed that to the order of interest
holding $\bR_g$, $E_0$ and $\mu_0$ fixed is approximately equal to
holding $\bR$, $E$ and $\mu$ fixed. To rewrite equation
\eq{gammac_int_approx} as a function of $\br$, $E_0$, $\mu_0$ and
$\varphi_0$, I Taylor expand $\Gammabf(\bR_g, E_0, \mu_0,
\varphi_0, t)$ around $\br$ to find
\begin{equation}
\frac{1}{2\pi} \oint d\varphi_0\, \Gammabf(\bR_g, E_0, \mu_0,
\varphi_0) \simeq \overline{\Gammabf} + \frac{1}{\Omega_i} (\bv
\times \bun) \cdot \nablaave\, \overline{\Gammabf},
\label{approx_ave_krho}
\end{equation}

\noindent with $\overline{\Gammabf} \equiv \overline{\Gammabf}
(\br, E_0, \mu_0) = \overline{\Gammabf(\br, E_0, \mu_0,
\varphi_0)}$ the gyroaverage holding $\br$, $E_0$, $\mu_0$ and $t$
fixed. The second term in the right side of \eq{gammac_int_approx}
is higher order and can be simply written as
\begin{equation}
- \frac{1}{2\pi} \oint d\varphi_0\, \frac{1}{\Omega_i} (\bv \times
\bun) \cdot \nabla_{\bR_g} \Gammabf \simeq - \frac{1}{\Omega_i}
\overline{(\bv \times \bun) \cdot \nablaave \Gammabf}.
\label{approx_ave_krho2}
\end{equation}

\noindent Employing equations \eq{approx_ave_krho} and
\eq{approx_ave_krho2} in equation \eq{gammac_int_approx}, I find
\begin{equation}
\langle \Gammabf \rangle = \overline{\Gammabf} +
\frac{1}{\Omega_i} (\bv \times \bun) \cdot \nablaave\,
\overline{\Gammabf} - \frac{1}{\Omega_i} \overline{(\bv \times
\bun) \cdot \nablaave \Gammabf}. \label{gammac_int1}
\end{equation}

\noindent Similarly, letting $\Gammabf \rightarrow \Gammabf \cdot
\bv_\bot$ in \eq{gammac_int1} and ignoring $\nablaave \bv_\bot$
corrections as small, I find
\begin{equation}
\langle \Gammabf \cdot \bv_\bot \rangle = \overline{\Gammabf \cdot
\bv_\bot} + \frac{1}{\Omega_i} (\bv \times \bun) \cdot \nablaave
\left ( \overline{\Gammabf \cdot \bv_\bot} \right ) -
\frac{1}{\Omega_i} \overline{(\bv \times \bun) \cdot \nablaave
\Gammabf \cdot \bv_\bot}. \label{gammac_int2}
\end{equation}

\noindent Using these results in \eq{nViC} the integral becomes
\begin{equation}
 \nabla \cdot ( n_i\bV_{iC} ) = - \nabla \cdot \left \{
\frac{\gamma}{\Omega_i} \int d^3 v\, \left [ \Gammabf \times \bun
- \frac{1}{\Omega_i} (\bv \times \bun) \cdot \nablaave \Gammabf
\times \bun - \frac{1}{2\Omega_i} \nablaave \Gammabf \cdot
\bv_\bot \right ] \right \},
\end{equation}

\noindent where I have used that $\int d^3v\,\overline{(\ldots)} =
\int d^3v\, (\ldots)$. The integral $\int d^3 v\, \Gammabf$ is
zero, as can be proven by exchanging the dummy integration
variables $\bv$ and $\bv^\prime$. The rest of the integral can be
written as
\begin{equation}
 \nabla \cdot (n_i \bV_{iC}) = \nabla \nabla : \left \{
\frac{\gamma}{\Omega_i^2} \int d^3 v\, \left [ (\bv \times \bun)
(\Gammabf \times \bun) + \frac{1}{2} (\matI - \bun\bun) (\Gammabf
\cdot \bv_\bot) \right ] \right \},
\end{equation}

\noindent where the spatial gradients of functions different from
$\Gammabf$ have been neglected. Using the definition of the
linearized collision operator \eq{linear_C} and employing $\int
d^3v\, [ (\bv \times \bun)(\bv \times \bun) + (v_\bot^2/2)(\matI -
\bun\bun)] \nabla_v \cdot \Gammabf = - \int d^3v\, [ (\Gammabf
\times \bun)(\bv \times \bun) + (\bv \times \bun)(\Gammabf \times
\bun) + (\Gammabf \cdot \bv_\bot)(\matI - \bun\bun)]$, I find
\begin{equation}
\nabla \cdot (n_i \bV_{iC}) = - \nabla \nabla : \left \{
\frac{1}{\Omega_i^2} \int d^3 v\, C^{(\ell)} \{ f_{i1} \} \left
[\frac{v_\bot^2}{4} (\matI - \bun\bun) + \frac{1}{2} (\bv \times
\bun) (\bv \times \bun) \right ] \right \},
\end{equation}

\noindent where I have also employed $\nablaave\, \nablaave :
[(\bv \times \bun) (\Gammabf \times \bun)] = \nablaave\, \nablaave
: [(\Gammabf \times \bun) (\bv \times \bun)]$. This integral, of
order $\delta_i (k_\bot \rho_i)^2 \nu_{ii} n_e$, can be simplified
by employing that, for $k_\bot \rho_i \ll 1$, the gyrophase
dependent part of the distribution function is proportional to
$\bv_\bot$ to zeroth order [see \eq{fwig_krho}]. Then, the
integral becomes
\begin{equation}
\nabla \cdot (n_i \bV_{iC}) = - \nabla \nabla : \left [
\frac{1}{\Omega_i^2} \int d^3 v\, C^{(\ell)} \{ f_{i1} \}
\frac{v_\bot^2}{2} (\matI - \bun\bun) \right ],
\label{intC_n_klong}
\end{equation}

\noindent where I used that $\overline{\bv_\bot (\bv \times
\bun)(\bv \times \bun)} = 0$ and $\overline{(\bv \times \bun)(\bv
\times \bun)} = (v_\bot^2/2) (\matI - \bun\bun)$.

Finally, I prove that $\mathbf{F}_{iC}$ from \eq{FiC} is of order
$\delta_i k_\bot \rho_i \nu_{ii} n_e M v_i$. Here, it is important
to realize that the first integral in \eq{FiC} must be carried to
the next order in $k_\bot \rho_i$, but the integrals in the
divergence only need the lowest order expressions. Then, using
expressions \eq{gammac_int1} and \eq{gammac_int2} in the first
integral of \eq{FiC}, and the lowest order expressions $\langle
\Gammabf \rangle \simeq \overline{\Gammabf}$ and $\langle \Gammabf
\cdot \bv_\bot \rangle \simeq \overline{\Gammabf \cdot \bv_\bot}$
for the second integral, I find that for $k_\bot \rho_i \ll 1$,
\begin{eqnarray}
\mathbf{F}_{iC} \simeq \nabla \cdot \Bigg \{
\frac{M\gamma}{\Omega_i} \int d^3 v\, \Bigg [ (\bv \times \bun)
(\Gammabf \cdot \bun) \bun + (\Gammabf \times \bun) v_{||} \bun +
(\Gammabf \cdot \bv_\bot) \matI \times \bun \Bigg ] \Bigg \} =
\nonumber \\ - \nabla \cdot \left \{ \frac{M}{\Omega_i} \int d^3
v\, C^{(\ell)} \{ f_{i1} \} \left [ (\bv \times \bun) v_{||} \bun
+ \frac{v_\bot^2}{2} \matI \times \bun \right ] \right \}.
\label{int_collmom}
\end{eqnarray}

\noindent To obtain the last result, I have employed that $\int
d^3 v\, [ (\bv \times \bun) v_{||} \bun + (v_\bot^2/2) \matI
\times \bun ] \nabla_v \cdot \Gammabf = - \int d^3 v\, [ (\Gammabf
\times \bun) v_{||} \bun + (\bv \times \bun) (\Gammabf \cdot \bun)
\bun + (\Gammabf \cdot \bv_\bot) \matI \times \bun ]$ and I have
used the definition of the linearized collision operator
\eq{linear_C}. Only the gyrophase dependent part of $f_{i1}$
contributes to the first part of integral \eq{int_collmom}, and
for $k_\bot \rho_i \ll 1$ the gyrophase dependent part is even in
$v_{||}$ [recall \eq{fwig_krho}] so this portion vanishes. As a
result, the integral becomes
\begin{equation}
\mathbf{F}_{iC} = - \nabla \cdot \left [ \frac{M}{\Omega_i} \int
d^3 v\, C^{(\ell)} \{ f_{i1} \} \frac{v_\bot^2}{2}\matI \times
\bun \right ], \label{intC_v_klong}
\end{equation}

\noindent of order $\delta_i k_\bot \rho_i \nu_{ii} n_e M v_i$.

\chapter{Gyrokinetic vorticity} \label{app_vortGK}

In this Appendix, I explain how to obtain the gyrokinetic
vorticity equation \eq{vort_type2} from equations
\eq{vtransp_krho} and \eq{vort_type1}.

Before adding equations \eq{vort_type1} and $\nabla \cdot \{ (c/B)
[ \mathrm{equation}\; \eq{vtransp_krho} ] \times \bun \}$, I
simplify the perpendicular component of the current density $Zen_i
\tilde{\bV}_i$. The perpendicular component of $\tilde{\bv}_1$,
defined in \eq{vtilde1}, is given by
\begin{equation}
\tilde{\bv}_{1\bot} \equiv \bun \times ( \tilde{\bv}_1 \times \bun
) = \frac{v_{||}}{\Omega_i} \bun \times ( \bun \cdot \nablaave
\bv_\bot + \nabla \bun \cdot \bv_\bot ),
\end{equation}

\noindent where I use that $(\nablaave \times \bv_\bot) \times
\bun = \bun \cdot \nablaave \bv_\bot - \nablaave \bv_\bot \cdot
\bun$ and $\nablaave \bv_\bot \cdot \bun = - \nabla \bun \cdot
\bv_\bot$. Employing $(\nabla \times \bun) \times \bv_\bot =
\bv_\bot \cdot \nabla \bun - \nabla \bun \cdot \bv_\bot$ and $\bun
\times [ (\nabla \times \bun) \times \bv_\bot ] = - \bv_\bot (\bun
\cdot \nabla \times \bun)$, I find
\begin{equation}
\tilde{\bv}_{1\bot} = \frac{v_{||}}{\Omega_i} \bun \times ( \bun
\cdot \nablaave \bv_\bot + \bv_\bot \cdot \nabla \bun ) +
\frac{v_{||}}{\Omega_i} \bv_\bot (\bun \cdot \nabla \times \bun).
\end{equation}

\noindent Then, the integral $n_i \tilde{\bV}_{i\bot}$ becomes
\begin{eqnarray}
 n_i \tilde{\bV}_{i\bot} = \frac{1}{\Omega_i} \bun \times \left
( \int d^3v\, f_{ig} v_{||} \bun \cdot \nablaave \bv_\bot + \int
d^3v\, f_{ig} v_{||} \bv_\bot \cdot \nabla \bun \right ) \nonumber
\\ + \frac{1}{\Omega_i} \bun \cdot \nabla \times \bun \int d^3v\,
f_{ig} v_{||} \bv_\bot. \label{relation_vtilde1}
\end{eqnarray}

Adding equations \eq{vort_type1} and $\nabla \cdot \{ (c/B) [
\mathrm{equation}\; \eq{vtransp_krho} ] \times \bun \}$, I find
\begin{eqnarray}
 \frac{\partial \varpi_G}{\partial t} = \nabla \cdot \Bigg [
J_{||} \bun + \bJ_{gd} + \tilde{\bJ}_i + Zen_i \tilde{\bV}_i +
Zen_i\bV_{iC} + \frac{c}{B} \bun \times (\nabla \cdot
\matrixtop{\pibf}_{ig\times}) \nonumber \\ - \frac{c}{B} \bun
\times \mathbf{F}_{iB} - \frac{c}{B} \bun \times \mathbf{F}_{iC}
\Bigg ], \label{vortGK_version1}
\end{eqnarray}

\noindent with $\varpi_G$ defined in \eq{vortGKdef}. In this
equation, adding $-(c/B) \bun \times \mathbf{F}_{iB}$ and the
expression in \eq{relation_vtilde1} for $Zen_i
\tilde{\bV}_{i\bot}$, two terms cancel to give
\begin{eqnarray}
Zen_i \tilde{\bV}_{i\bot} - \frac{c}{B} \bun \times
\mathbf{F}_{iB} = \frac{Mc}{B} \bun \times \int d^3v\, f_{ig}
v_{||} \bv_\bot \cdot \nabla \bun \nonumber \\ + \frac{Mc}{B} \bun
\cdot \nabla \times \bun \int d^3v\, f_{ig} v_{||} \bv_\bot -
\frac{Zec}{B\Omega_i} \bun \times \int d^3v\, f_{Mi} (\nablaave
\phiwig \times \bun ) \cdot \nablaave \bv_\bot.
\end{eqnarray}

\noindent The last term in this equation is absorbed in the
definition of $\tilde{\bJ}_{i\phi}$ in \eq{Jtilde_iphi}. I can
further simplify by realizing that $\bun \times \int d^3v\, f_{ig}
v_{||} \bv_\bot \cdot \nabla \bun = \bun \times [ \nabla \cdot (
\int d^3v\, f_{ig} \bv_\bot v_{||} \bun )]$, giving
\begin{eqnarray}
 \tilde{\bJ}_i + Zen_i \tilde{\bV}_{i\bot} - \frac{c}{B} \bun
\times \mathbf{F}_{iB} = \tilde{\bJ}_{i\phi} + \frac{c}{B} \bun
\times \left [ \nabla \cdot \left ( \int d^3v\, f_{ig} M \bv_\bot
v_{||} \bun \right ) \right ] \nonumber \\ + \frac{Mc}{B} \bun
\cdot \nabla \times \bun \int d^3v\, f_{ig} v_{||} \bv_\bot.
\end{eqnarray}

\noindent The integral $\int d^3v\, f_{ig} M \bv_\bot v_{||} \bun$
is part of the definition of $\matrixtop{\pibf}_{iG}$ in
\eq{piGfinal}, so I can finally write
\begin{eqnarray}
\tilde{\bJ}_i + Zen_i \tilde{\bV}_{i\bot} + \frac{c}{B} \bun
\times ( \nabla \cdot \matrixtop{\pibf}_{ig\times} ) - \frac{c}{B}
\bun \times \mathbf{F}_{iB} = \tilde{\bJ}_{i\phi} + \frac{c}{B}
\bun \times ( \nabla \cdot \matrixtop{\pibf}_{iG} ) \nonumber \\ +
\frac{Mc}{B} \bun \cdot \nabla \times \bun \int d^3v\, f_{ig}
v_{||} \bv_\bot.
\end{eqnarray}

\noindent Employing this result in equation \eq{vortGK_version1}
and using the fact that the divergence of $Zen_i \tilde{V}_{i||}
\bun$ and $(Mc/B) \bun \cdot \nabla \times \bun \int d^3v\, f_{ig}
v_{||} \bv_\bot$ is negligible, I recover equation
\eq{vort_type2}. The divergence of $Zen_i \tilde{V}_{i||} \bun
\sim \delta_i^2 k_\bot \rho_i e n_e v_i$ is small because the
parallel gradient is only order $1/L$, giving $\nabla \cdot (
Zen_i \tilde{V}_{i||} \bun ) \sim \delta_i^2 k_\bot \rho_i e n_e
v_i/L$; which is negligible with respect to the rest of the terms,
the smallest of which is order $\delta_i (k_\bot \rho_i)^2 e n_e
v_i/L$. The divergence of $(Mc/B) \bun \cdot \nabla \times \bun
\int d^3v\, f_{ig} v_{||} \bv_\bot \sim \delta_i^2 k_\bot \rho_i e
n_e v_i$ has only one term that is of order $\delta_i (k_\bot
\rho_i)^2 e n_e v_i / L$, given by
\begin{equation}
 \nabla \cdot \left ( \frac{Mc}{B} \bun \cdot \nabla \times
\bun \int d^3v\, f_{ig} v_{||} \bv_\bot \right )
\simeq\frac{Mc}{B} \bun \cdot \nabla \times \bun \int d^3v\,
v_{||} \bv_\bot \cdot \nablaave f_{ig}.
\end{equation}

\noindent Since the only dependence of $f_{ig}$ on $\varphi_0$ is
in $\bR_g = \br + \Omega_i^{-1} \bv \times \bun$, I find that
$\bv_\bot \cdot \nablaave f_{ig} = \Omega_i (\partial
f_{ig}/\partial \varphi_0)$. Thus, the divergence of $(Mc/B) \bun
\cdot \nabla \times \bun \int d^3v\, f_{ig} v_{||} \bv_\bot$
vanishes to the relevant order due to the gyrophase integration
\begin{equation}
 \nabla \cdot \left ( \frac{Mc}{B} \bun \cdot \nabla \times
\bun \int d^3v\, f_{ig} v_{||} \bv_\bot \right ) \simeq Ze \bun
\cdot \nabla \times \bun \int d^3v\, v_{||} \frac{\partial
f_{ig}}{\partial \varphi_0} = 0.
\end{equation}

\chapter{Flux surface averaged gyrokinetic vorticity equation} \label{app_tormom_long}

In this Appendix, I obtain the long wavelength limit of the flux
surface averaged vorticity equation \eq{tormomGK_1}. The long
wavelength limit of $\varpi_G$ is given by $\varpi_G \rightarrow
\varpi = \nabla \cdot [ (Ze/\Omega_i) n_i \bV_i \times \bun ]$, as
proven in \eq{vortGK_lw}. Then, upon using \eq{bxgradpsi} and
integrating once in $\psi$, equation \eq{tormomGK_1} becomes
\begin{eqnarray}
 - \frac{\partial}{\partial t} \langle c R n_i M \bV_i \cdot
\zun \rangle_\psi = \Bigg \langle \tilde{\bJ}_{i\phi} \cdot \nabla
\psi + \frac{cI}{B} (\nabla \cdot \pibf_{ig||} - \tilde{F}_{iE})
\nonumber \\ - \frac{c}{B} (\nabla \cdot \matrixtop{\pibf}_{iG})
\cdot (\bun \times \nabla \psi) + Zen_i \bV_{iC} \cdot \nabla \psi
- cR \mathbf{F}_{iC} \cdot \zun \Bigg \rangle_\psi.
\label{tormom_step1}
\end{eqnarray}

\noindent I will evaluate all the terms on the right side of
equation \eq{tormom_step1} to order $\delta_i^2 k_\bot \rho_i e
n_e v_i |\nabla \psi|$ for $k_\bot \rho_i \rightarrow 0$.

\section{Limit of $\langle (cI/B) \nabla \cdot \pibf_{ig||} \rangle_\psi$ for $k_\bot \rho_i \rightarrow 0$}

The term $\langle (cI/B) \nabla \cdot \pibf_{ig||} \rangle_\psi$
is written as
\begin{equation}
\left \langle \frac{cI}{B} \nabla \cdot \pibf_{ig||} \right
\rangle_\psi \simeq \left \langle \nabla \cdot \left (
\frac{cI}{B} \pibf_{ig||} \right ) \right \rangle_\psi =
\frac{1}{V^\prime} \frac{\partial}{\partial \psi} V^\prime \left
\langle \frac{cI}{B} \pibf_{ig||} \cdot \nabla \psi \right
\rangle_\psi. \label{pigpar_step1}
\end{equation}

\noindent The term $\pibf_{ig||} \cdot \nabla (cI/B)$ is neglected
because it is of order $\delta_i^3 e n_e v_i |\nabla \psi|$. In
the definition of $\pibf_{ig||}$ in \eq{pig_par}, one of the terms
is $- \int d^3v\, f_{ig} (c/B) (\nablaave \phiave \times \bun) M
v_{||}$. The difference between $f_{ig} (c/B) \nablaave \phiave
\times \bun$ and $f_i (c/B) \nabla \phi \times \bun$ gives rise to
the term
\begin{eqnarray}
\frac{cI}{B^2} \int d^3v\, Mv_{||} [f_{ig} (\nablaave \phiave
\times \bun) - f_i (\nabla \phi \times \bun) ] \cdot \nabla \psi =
\nonumber \\ \frac{cI}{B^2} \int d^3v\, Mv_{||} \left [
\frac{Ze\phiwig}{T_i} f_{Mi} (\nabla \phi \times \bun) - f_{ig}
\frac{c}{B} (\nablaave \phiwig \times \bun) \right ] \cdot \nabla
\psi \label{diff_ExB}
\end{eqnarray}

\noindent in $(I/B) \pibf_{ig||} \cdot \nabla \psi$, where I use
$f_i - f_{ig} = - (Ze\phiwig/T_i) f_{Mi}$ to obtain the second
equality. I will show below that the difference \eq{diff_ExB} is
of order $\delta_i^2 k_\bot \rho_i p_i R |\nabla \psi|$ and
therefore negligible compared to the other terms in $(I/B)
\pibf_{ig||} \cdot \nabla \psi$ that are of order $\delta_i^2 p_i
R |\nabla \psi|$. Then, in $(I/B) \pibf_{ig||} \cdot \nabla \psi$
the difference between $f_{ig} (c/B) \nablaave \phiave \times
\bun$ and $f_i (c/B) \nabla \phi \times \bun$ can be neglected to
write equation \eq{pigpar_step1} as
\begin{equation}
\left \langle \frac{cI}{B} \nabla \cdot \pibf_{ig||} \right
\rangle_\psi \simeq \frac{1}{V^\prime} \frac{\partial}{\partial
\psi} V^\prime \left \langle \frac{cI}{B} \pibf_{ig||}^\prime
\cdot \nabla \psi \right \rangle_\psi, \label{pigpar_final}
\end{equation}

\noindent with
\begin{equation}
\pibf_{ig||}^\prime = \int d^3v\, f_{ig} (\bv_{M0} +
\tilde{\bv}_1) Mv_{||} - \frac{Mc}{B} \nabla \phi \times \bun \int
d^3v\, f_i v_{||}. \label{pigpar_prime}
\end{equation}

\noindent Equation \eq{pigpar_final} is then seen to be of order
$\delta_i^2 k_\bot \rho_i e n_e v_i |\nabla \psi|$.

I will now prove that the difference \eq{diff_ExB} is of order
$\delta_i^2 k_\bot \rho_i p_i R |\nabla \psi|$. In equation
\eq{diff_ExB}, short wavelength components of $\phi$, $\phiwig$
and $f_{ig}$ can beat nonlinearly to give a long wavelength
component, and these functions cannot be Taylor expanded around
$\br$. However, the total long wavelength contribution
\eq{diff_ExB} can be expanded. The gyrophase dependence in $\bR_g$
then gives a contribution of order $\delta_i k_\bot \rho_i p_i R
|\nabla \psi|$ that can be ignored when integrating over velocity
space to order $\delta_i^2 p_i R |\nabla \psi|$. The result is
$(cI/B^2) \int d^3v (Ze\phiwig/T_i) f_{Mi} M v_{||} (\nablaave
\phiwig \times \bun) \cdot \nabla \psi$ because the rest of the
terms have vanishing gyroaverages to the order of interest. Since
both $f_{Mi}$ and $\phiwig$ are even in $v_{||}$, this integral
vanishes, and \eq{diff_ExB} is higher order than $\delta_i^2 p_i R
|\nabla \psi|$.

\section{Limit of $\langle \tilde{\bJ}_{i\phi} \cdot \nabla \psi - (c/B) (\nabla \cdot \matrixtop{\pibf}_{iG})
\cdot (\bun \times \nabla \psi) \rangle_\psi$ for $k_\bot \rho_i \rightarrow 0$}

I simplify the terms $\langle \tilde{\bJ}_{i\phi} \cdot \nabla
\psi \rangle_\psi$ and $- \langle (c/B) (\nabla \cdot
\matrixtop{\pibf}_{iG}) \cdot (\bun \times \nabla \psi)
\rangle_\psi$ by first calculating the divergence of
$\tilde{\bJ}_{i\phi} + (c/B) \bun \times (\nabla \cdot
\matrixtop{\pibf}_{iG})$. I employ the long wavelength result for
$\nabla \cdot \tilde{\bJ}_{i\phi}$ in \eq{div_Jtilde_iphi} to
obtain
\begin{equation}
\nabla \cdot \left [ \tilde{\bJ}_{i\phi} + \frac{c}{B} \bun \times
(\nabla \cdot \matrixtop{\pibf}_{iG}) \right ] = \nabla \cdot
\left [ \frac{c}{B} \bun \times ( \nabla \cdot
\matrixtop{\pibf}_{iG}^\prime + \nabla \cdot
\matrixtop{\pibf}_{iG}^{\prime\prime} ) \right ],
\label{Jiphi_piG}
\end{equation}

\noindent with
\begin{equation}
\matrixtop{\pibf}_{iG}^\prime = M \int d^3v\, f_{ig} v_{||} (\bun
\bv_\bot + \bv_\bot \bun) \sim \delta_i k_\bot \rho_i p_i
\label{pig_prime}
\end{equation}

\noindent and
\begin{equation}
\matrixtop{\pibf}_{iG}^{\prime\prime} = \int d^3v\, f_{ig}
(\bv_{M0} + \tilde{\bv}_1) M \bv_\bot -  \frac{Mc}{B} \nabla \phi
\times \bun \int d^3v\, f_i \bv_\bot \sim \delta_i^2 p_i.
\end{equation}

\noindent Flux surface averaging equation \eq{Jiphi_piG} and
integrating once in $\psi$, I find the order $\delta_i^2 k_\bot
\rho_i e n_e v_i |\nabla \psi|$ term
\begin{eqnarray}
\left \langle \tilde{\bJ}_{i\phi} \cdot \nabla \psi - \frac{c}{B}
(\nabla \cdot \matrixtop{\pibf}_{iG}) \cdot (\bun \times \nabla
\psi) \right \rangle_\psi = - \frac{1}{V^\prime}
\frac{\partial}{\partial \psi} V^\prime \left \langle \frac{c}{B}
\nabla \psi \cdot \matrixtop{\pibf}_{iG}^{\prime\prime} \cdot
(\bun \times \nabla \psi) \right \rangle_\psi,
\label{Jiphi_piG_final}
\end{eqnarray}

\noindent where I neglected $c
\matrixtop{\pibf}_{iG}^{\prime\prime}: \nabla [ (\bun \times
\nabla \psi)/B ] \sim \delta_i^3 e n_e v_i |\nabla \psi|$, I used
the definition of $\matrixtop{\pibf}_{iG}^\prime$ in
\eq{pig_prime} to obtain $\nabla \psi \cdot
\matrixtop{\pibf}_{iG}^\prime \cdot (\bun \times \nabla \psi) =
0$, and I will next prove that $\langle
\matrixtop{\pibf}_{iG}^\prime : \nabla [ (\bun \times \nabla
\psi)/B ] \rangle_\psi$ vanishes.

To see that $\langle \matrixtop{\pibf}_{iG}^\prime : \nabla [
(\bun \times \nabla \psi)/B ] \rangle_\psi = 0$, the velocity
integral $\int d^3v\, f_{ig} v_{||} \bv_\bot$ in
$\matrixtop{\pibf}_{iG}^\prime$ has to be found to order $\delta_i
k_\bot \rho_i n_e v_i^2$. This integral only depends on the
gyrophase dependent piece of $f_{ig}$, given to the required order
by
\begin{equation}
f_i - \overline{f}_i \simeq \frac{1}{\Omega_i} (\bv \times \bun)
\cdot \nablaave f_{i0} + \frac{1}{2\Omega_i^2}(\bv \times
\bun)(\bv \times \bun) : \nablaave\, \nablaave f_{i0},
\end{equation}

\noindent with $f_{i0} \equiv f_i (\br, E_0, \mu_0, \varphi_0)$ as
defined in \eq{fi0} and, thus, gyrophase independent. The integral
involving $\nablaave\, \nablaave f_{i0}$ vanishes, leaving
\begin{eqnarray}
\matrixtop{\pibf}_{iG}^\prime : \nabla \left ( \frac{\bun \times
\nabla \psi}{B} \right ) = \nonumber \\ \int d^3v\, \frac{M
v_{||}}{\Omega_i} [ (\bv \times \bun) \cdot \nablaave f_{i0} ]
(\bun \bv_\bot + \bv_\bot \bun): \nabla \left ( \frac{\bun \times
\nabla \psi}{B} \right ) \simeq  \nonumber \\ \nabla \cdot \left
[\int d^3v\, f_{i0} \frac{ M v_{||}}{\Omega_i} (\bv \times \bun)
(\bun \bv_\bot + \bv_\bot \bun): \nabla \left ( \frac{\bun \times
\nabla \psi}{B} \right ) \right ], \label{Jiphi_piG_piece2_1}
\end{eqnarray}

\noindent where terms of order $\delta_i^2 p_i$ are neglected.
Integrating in gyrophase and flux surface averaging, equation
\eq{Jiphi_piG_piece2_1} reduces to
\begin{eqnarray}
\left \langle \matrixtop{\pibf}_{iG}^\prime : \nabla \left (
\frac{\bun \times \nabla \psi}{B} \right ) \right \rangle_\psi =
\nonumber \\ \frac{1}{V^\prime} \frac{\partial}{\partial \psi}
V^\prime \left \langle M \int d^3v\, f_{i0} \frac{v_{||}
v_\bot^2}{2\Omega_i} [\bun (\bun \times \nabla \psi) + (\bun
\times \nabla \psi) \bun]: \nabla \left ( \frac{\bun \times \nabla
\psi}{B} \right ) \right \rangle_\psi.
\end{eqnarray}

\noindent This expression vanishes because $[\bun (\bun \times
\nabla \psi) + (\bun \times \nabla \psi) \bun]: \nabla [ (\bun
\times \nabla \psi)/B ] = 0$. To prove this, I employ equation
\eq{bxgradpsi} to write $\nabla [ (\bun \times \nabla \psi)/B] =
\nabla ( I\bun/B ) - \nabla (R\zun)$. The tensor $\nabla(R\zun) =
(\nabla R) \zun - \zun (\nabla R)$ gives zero contribution because
it is antisymmetric and it is multiplied by the symmetric tensor
$[\bun (\bun \times \nabla \psi) + (\bun \times \nabla \psi)
\bun]$. Then, I am only left with $\nabla (I\bun/B)$, giving
\begin{equation}
[\bun (\bun \times \nabla \psi) + (\bun \times \nabla \psi) \bun]:
\nabla \left ( \frac{\bun \times \nabla \psi}{B} \right ) = (\bun
\times \nabla \psi) \cdot \left [ \nabla \left ( \frac{I}{B}
\right ) + \frac{I}{B} \bun \cdot \nabla \bun \right ].
\end{equation}

\noindent To simplify, I use relation \eq{id_kappa}, with $\bun
\cdot \nabla \bun = \kappabf$, to write $\bun \cdot \nabla \bun
\cdot (\bun \times \nabla \psi) = - (\nabla \times \bun) \cdot
\nabla \psi = - \nabla \cdot ( \bun \times \nabla \psi)$. Finally,
I employ \eq{bxgradpsi} and $\zun \cdot \nabla (I/B) = 0 = \nabla
\cdot ( RB\zun )$ due to axisymmetry to obtain
\begin{equation}
 [\bun (\bun \times \nabla \psi) + (\bun \times \nabla \psi)
\bun]: \nabla \left ( \frac{\bun \times \nabla \psi}{B} \right ) =
I \bun \cdot \nabla \left ( \frac{I}{B} \right ) - \frac{I}{B}
\nabla \cdot ( I \bun ) = 0. \label{mag_property}
\end{equation}

\section{Limit of $\langle (cI/B) \tilde{F}_{iE} \rangle_\psi $ for $k_\bot \rho_i \rightarrow 0$}

The function $\tilde{F}_{iE}$, defined in \eq{force_E}, is written
as
\begin{equation}
\tilde{F}_{iE} = Ze \int dE_0\,d\mu_0\,d\varphi_0 \frac{\partial
v_{||}}{\partial E_0} B f_{Mi} \left ( \bun \cdot \nablaave
\phiwig + \frac{1}{\Omega_i} \nablaave \times \bv_\bot \cdot
\nablaave \phiwig \right ),
\end{equation}

\noindent where I use $d^3v = (B/v_{||}) dE_0\,d\mu_0\,d\varphi_0$
and $v_{||}^{-1} = \partial v_{||}/ \partial E_0$. Integrating by
parts in $E_0$, and making use of $\partial \phiwig / \partial E_0
= 0$, $\partial f_{Mi} /\partial E_0 = (-M/T_i) f_{Mi}$ and
$\tilde{\bv}_1 = (v_{||}/\Omega_i) \nablaave \times \bv_\bot$, I
find
\begin{equation}
\tilde{F}_{iE} = M \int dE_0\,d\mu_0\,d\varphi_0 B \frac{Ze}{T_i}
f_{Mi} ( v_{||} \bun + \tilde{\bv}_1 ) \cdot \nablaave \phiwig.
\label{FiE_step1}
\end{equation}

\noindent Multiplying equation \eq{FiE_step1} by $cI/B$ and
writing the result as a divergence give
\begin{equation}
\frac{cI}{B} \tilde{F}_{iE} \simeq \nabla \cdot \left (
\frac{cI}{B} \pibf_{iE}^\prime \right ) - \int d^3v\,
\frac{Z^2e^2\phiwig}{T_i} f_{Mi} v_{||} \bun \cdot \nablaave \left
( \frac{Iv_{||}}{\Omega_i} \right ), \label{FiE_step2}
\end{equation}

\noindent with $\pibf_{iE}^\prime = \int d^3v\, (Ze\phiwig/T_i)
f_{Mi} ( v_{||} \bun + \tilde{\bv}_1 ) Mv_{||}$. Here, I employ
$\bun \cdot \nablaave ( f_{Mi}/T_i ) = 0$, and neglect the
integral $Mc \int dE_0\,d\mu_0\,d\varphi_0 Ze \phiwig \nablaave
\cdot ( \tilde{\bv}_1  I f_{Mi} /T_i )$ because it is of order
$\delta_i^3 e n_e v_i |\nabla \psi|$. I will now consider the two
integrals in equation \eq{FiE_step2}. Upon using \eq{avepsi_div},
the flux surface average of the first integral gives
\begin{equation}
 \left \langle \nabla \cdot \left ( \frac{cI}{B}
\pibf_{iE}^\prime \right ) \right \rangle_\psi =
\frac{1}{V^\prime} \frac{\partial}{\partial \psi} V^\prime \left
\langle Mc \int d^3v\, \frac{Ze\phiwig}{T_i} f_{Mi}
\frac{Iv_{||}}{B} \tilde{\bv}_1 \cdot \nabla \psi \right
\rangle_\psi. \label{FiE_piece1_1}
\end{equation}

\noindent Multiplying equation \eq{bxgradpsi} by $\bv$, I find
$Iv_{||}/B = R \bv \cdot \zun + (\bv \times \bun) \cdot \nabla
\psi/B$. Substituting this result in equation \eq{FiE_piece1_1}, I
find that the integral of $\bv \times \bun$ vanishes because
$\phiwig$ and $f_{Mi}$ are even in $v_{||}$, and $\tilde{\bv}_1$
is odd. Thus, the first integral of equation \eq{FiE_step2} gives
\begin{equation}
\left \langle \nabla \cdot \left ( \frac{cI}{B} \pibf_{iE}^\prime
\right ) \right \rangle_\psi = \frac{1}{V^\prime}
\frac{\partial}{\partial \psi} V^\prime \left \langle c \int
d^3v\, \frac{Ze\phiwig}{T_i} f_{Mi} R M (\bv \cdot \zun)
\tilde{\bv}_1 \cdot \nabla \psi \right \rangle_\psi.
\label{FiE_piece1}
\end{equation}

In the second integral of \eq{FiE_step2}, I need to keep $\phiwig
(\bR_g, \mu_0, \varphi_0, t) \simeq \phiwig (\br, \mu_0,
\varphi_0, t) + \Omega_i^{-1} (\bv \times \bun) \cdot \nablaave
\phiwig$. Using $\oint d\varphi_0\, \phiwig (\br, \mu_0,
\varphi_0, t) = 0$ leaves
\begin{eqnarray}
\int d^3v\, \frac{Z^2 e^2\phiwig}{T_i} f_{Mi} v_{||} \bun \cdot
\nablaave \left ( \frac{Iv_{||}}{\Omega_i} \right ) \simeq
\nonumber \\ \frac{Ze}{\Omega_i} \int d^3v\, f_{Mi} \frac{Ze}{T_i}
(\bv \times \bun) \cdot \nablaave \phiwig \left [ v_{||} \bun
\cdot \nablaave \left ( \frac{Iv_{||}}{\Omega_i} \right ) \right ]
\simeq \nonumber \\ \nabla \cdot \left \{ \frac{Mc}{B} \int d^3v\,
\frac{Ze\phiwig}{T_i} f_{Mi} (\bv \times \bun) \left [ v_{||} \bun
\cdot \nablaave \left ( \frac{Iv_{||}}{\Omega_i} \right ) \right ]
\right \}, \label{FiE_piece2_1}
\end{eqnarray}

\noindent where terms of order $\delta_i^3 e n_e v_i |\nabla
\psi|$ are neglected to obtain the second equality. Using $v_{||}
\bun \cdot \nablaave (Iv_{||}/\Omega_i) = \bv_{M0} \cdot \nabla
\psi$ in equation \eq{FiE_piece2_1} and flux surface averaging, I
find
\begin{eqnarray}
\left \langle \int d^3v\, \frac{Z^2 e^2 \phiwig}{T_i} f_{Mi}
v_{||} \bun \cdot \nablaave \left ( \frac{Iv_{||}}{\Omega_i}
\right ) \right \rangle_\psi = \nonumber \\ \frac{1}{V^\prime}
\frac{\partial}{\partial \psi} V^\prime \left \langle Mc \int
d^3v\, \frac{Ze\phiwig}{T_i} f_{Mi} \frac{(\bv \times \bun) \cdot
\nabla \psi}{B} \bv_{M0} \cdot \nabla \psi \right \rangle_\psi.
\label{FiE_piece2_2}
\end{eqnarray}

\noindent Here, equation \eq{bxgradpsi} multiplied by $\bv$ gives
$(\bv \times \bun) \cdot \nabla \psi/B = Iv_{||}/B - R\bv \cdot
\zun$. The integral of $Iv_{||}/B$ vanishes because both $\phiwig$
and $f_{Mi}$ are even in $v_{||}$. Thus, equation
\eq{FiE_piece2_2} gives to relevant order
\begin{eqnarray}
\left \langle \int d^3v\, \frac{Z^2e^2 \phiwig}{T_i} f_{Mi} v_{||}
\bun \cdot \nablaave \left ( \frac{Iv_{||}}{\Omega_i} \right )
\right \rangle_\psi \simeq \nonumber \\ - \frac{1}{V^\prime}
\frac{\partial}{\partial \psi} V^\prime \left \langle c \int
d^3v\, \frac{Ze\phiwig}{T_i} f_{Mi} R M (\bv \cdot \zun) \bv_{M0}
\cdot \nabla \psi \right \rangle_\psi. \label{FiE_piece2}
\end{eqnarray}

Substituting equations \eq{FiE_piece1} and \eq{FiE_piece2} into
the flux surface average of equation \eq{FiE_step2}, I obtain
\begin{equation}
\left \langle \frac{cI}{B} \tilde{F}_{iE} \right \rangle_\psi
\simeq \frac{1}{V^\prime} \frac{\partial}{\partial \psi} V^\prime
\Bigg \langle c \int d^3v\, (f_{ig} - f_i) RM(\bv \cdot \zun)
(\bv_{M0} + \tilde{\bv}_1) \cdot \nabla \psi \Bigg \rangle_\psi,
\label{FiE_final}
\end{equation}

\noindent where I use $f_i - f_{ig} = - (Ze\phiwig/T_i) f_{Mi}$.

\section{Limit of $\langle Zen_i \bV_{iC} \cdot \nabla \psi - cR \mathbf{F}_{iC} \cdot \zun \rangle_\psi$ for $k_\bot \rho_i \rightarrow 0$}

This collisional combination vanishes. According to
\eq{intC_n_klong}, for $k_\bot \rho_i \ll 1$, $\langle Zen_i
\bV_{iC} \cdot \nabla \psi \rangle_\psi$ is given by
\begin{equation}
\langle Zen_i \bV_{iC} \cdot \nabla \psi \rangle_\psi = -
\frac{1}{V^\prime} \frac{\partial}{\partial \psi} V^\prime \left
\langle \frac{c}{B\Omega_i} \int d^3 v\, C \{ f_i \} |\nabla
\psi|^2 \frac{Mv_\bot^2}{2} \right \rangle_\psi.
\end{equation}

\noindent Equation \eq{intC_v_klong}, on the other hand, gives
\begin{equation}
\langle cR\mathbf{F}_{iC} \cdot \zun \rangle_\psi = -
\frac{1}{V^\prime} \frac{\partial}{\partial \psi} V^\prime \left
\langle \frac{c}{B \Omega_i} \int d^3 v\, C \{ f_i \} |\nabla
\psi|^2 \frac{M v_\bot^2}{2} \right \rangle_\psi,
\end{equation}

\noindent where I use equation \eq{B_def} to obtain $\nabla \psi
\cdot ( \bun \times \zun ) = |\nabla \psi|^2/RB$.

Finally, since the collisional piece vanishes to relevant order, I
just need to substitute equations \eq{pigpar_final},
\eq{Jiphi_piG_final} and \eq{FiE_final} into equation
\eq{tormom_step1} and employ \eq{bxgradpsi} to find
\eq{tormom_long}.

\chapter{Gyroviscosity in gyrokinetics} \label{app_gyrovisc}

In this Appendix, I show why the gyroviscosity must take the form
given in equation \eq{gyrovisc}, and later I simplify that
expression for up-down symmetric tokamaks by proving that the
collisional piece must vanish.

\section{Evaluation of equation \eq{gyrovisc}} \label{sectapp_gyrovisc}

To prove equation \eq{gyrovisc}, I employ that $\bv_{M0} \cdot
\nabla \psi = v_{||} \bun \cdot \nablaave (Iv_{||}/\Omega_i)$ and
$\tilde{\bv}_1 \cdot \nabla \psi = - v_{||} \bun \cdot \nablaave [
\Omega_i^{-1} (\bv \times \bun) \cdot \nabla \psi ]$, proven below
in \eq{v1_gradpsi}, to write
\begin{equation}
 (\bv_{M0} + \tilde{\bv}_1) \cdot \nabla \psi = \frac{Mc}{Ze}
v_{||} \bun \cdot \nablaave \left [ \frac{Iv_{||}}{B} - \frac{(\bv
\times \bun) \cdot \nabla \psi}{B} \right ] = \frac{Mc}{Ze} v_{||}
\bun \cdot \nablaave [ R (\bv \cdot \zun) ], \label{radial_drift}
\end{equation}

\noindent where I use equation \eq{bxgradpsi} to find the last
equality. Notice that in $\tilde{\bv}_1 \cdot \nabla \psi =
(v_{||}/\Omega_i) (\nablaave \times \bv_\bot) \cdot \nabla \psi =
(v_{||}/\Omega_i) \nablaave \cdot ( \bv_\bot \times \nabla \psi
)$, both $\bv_\bot$ and $\nabla \psi$ are perpendicular to $\bun$.
Then, $\bv_\bot \times \nabla \psi$ must be parallel to $\bun$,
giving $\bv_\bot \times \nabla \psi = - \bun [ (\bv \times \bun)
\cdot \nabla \psi ]$, and
\begin{equation}
\tilde{\bv}_1 \cdot \nabla \psi = - \frac{v_{||}}{\Omega_i}
\nablaave \cdot [ \bun (\bv \times \bun) \cdot \nabla \psi ] = -
v_{||} \bun \cdot \nablaave \left [ \frac{(\bv \times \bun) \cdot
\nabla \psi}{\Omega_i} \right ]. \label{v1_gradpsi}
\end{equation}

Substituting equation \eq{radial_drift} into equation
\eq{gyrovisc} gives
\begin{eqnarray}
\left \langle \int d^3v\, f_i R M(\bv \cdot \zun) (\bv_{M0} +
\tilde{\bv}_1 ) \cdot \nabla \psi \right \rangle_\psi = \nonumber
\\ \frac{M^2c}{2Ze} \left \langle \int d^3v\, f_i v_{||} \bun \cdot
\nablaave [R^2 (\bv \cdot \zun )^2] \right \rangle_\psi =
\nonumber \\ - \frac{M^2c}{2Ze} \left \langle \int d^3v\, [R^2
(\bv \cdot \zun )^2] v_{||} \bun \cdot \nablaave f_i \right
\rangle_\psi. \label{gyrovisc_1approx}
\end{eqnarray}

\noindent Only the axisymmetric, short radial wavelength pieces of
$f_i$ contribute to \eq{gyrovisc_1approx} because of the flux
surface average. To find this portion of $f_i$, I employ equations
\eq{VlasovGK}, \eq{varphidot_ddvarphi}, \eq{diff_ddt},
\eq{eqGK_ini}, \eq{transf_phys} and \eq{Rdot_gradRr} to obtain the
gyrokinetic equation
\begin{equation}
\left. \frac{\partial f_{ig}}{\partial t} \right |_{\br, \bv} +
(v_{||} \bun + \bv_{M0} + \tilde{\bv}_1 ) \cdot \left ( \nablaave
f_{ig} - \frac{Ze}{M} \nablaave \phiave \frac{\partial
f_{Mi}}{\partial E_0} \right ) + \bv_{E0} \cdot \nablaave f_{ig} =
\langle C \{ f_i \} \rangle. \label{GKeq_radial1}
\end{equation}

\noindent In this equation, I will only keep terms of order
$\delta_i f_{Mi} v_i/L$ or larger. These terms will give $R \zun
\cdot \matrixtop{\pibf}_i^{(0)} \cdot \nabla \psi \sim \delta_i^2
p_i R |\nabla \psi|$ as seen from \eq{gyrovisc_1approx}.

In the nonlinear term $\bv_{E0} \cdot \nablaave f_{ig}$, different
components of $\phiave$ and $f_{ig}$ beat together to give an
axisymmetric, short radial wavelength contribution to
\eq{GKeq_radial1}. The term $\bv_{E0} \cdot \nablaave f_{ig}$ can
be treated as the term $\bv_E \cdot \nabla_\bR f_i$ in the
axisymmetric piece of equation \eq{drift_kinetic_i}, where it was
neglected. In equation \eq{GKeq_radial1}, $\bv_{E0} \cdot
\nablaave f_{ig} \sim k_\bot \rho_i \delta_i f_{Mi} v_i/L \ll
\delta_i f_{Mi} v_i/L$ is negligible, too. This is another form of
the conclusion of section~\ref{sect_assumptions}, i.e., the long
wavelength, axisymmetric flows are neoclassical.

Neglecting $\bv_{E0} \cdot \nablaave f_{ig}$ and using $f_{ig} =
f_i + (Ze\phiwig/T_i) f_{Mi}$ in equation \eq{GKeq_radial1}, I
find that the axisymmetric, short radial wavelength portion of
$f_i$ must satisfy
\begin{eqnarray}
 \left. \frac{\partial f_i}{\partial t} \right |_{\br, \bv}
+ v_{||} \bun \cdot \left ( \nablaave f_i + \frac{Ze}{T_i} f_{Mi}
\nabla \phi \right ) + \left ( \frac{\partial f_i}{\partial \psi}
+ \frac{Ze}{T_i}f_{Mi} \frac{\partial \phi}{\partial \psi} \right
) ( \bv_{M0} + \tilde{\bv}_1 ) \cdot \nabla \psi \nonumber \\ =
\langle C \{ f_i \} \rangle. \label{GKeq_radial2}
\end{eqnarray}

\noindent Here, the terms that contain $(\bv_{M0} + \tilde{\bv}_1)
\cdot \nablaave \phiwig$ and $(\bv_{M0} + \tilde{\bv}_1) \cdot
\nablaave (\phiwig f_{Mi})$ are neglected because they are smaller
than the rest of the terms by a factor $k_\bot \rho_i$ since
$\phiwig \sim \delta_i T_e/e \sim k_\bot \rho_i \phiave$.
Additionally, I used $\nabla_\bot \simeq \nabla \psi (\partial /
\partial \psi)$ since only short radial wavelength effects matter
in \eq{GKeq_radial1}. Finally, substituting equation
\eq{radial_drift} into \eq{GKeq_radial2} gives
\begin{eqnarray}
 \left. \frac{\partial f_i}{\partial t} \right |_{\br, \bv}
+ v_{||} \bun \cdot \left ( \nablaave f_i + \frac{Ze}{T_i}f_{Mi}
\nabla \phi \right ) + \frac{Mc}{Ze} \left ( \frac{\partial
f_i}{\partial \psi} + \frac{Ze}{T_i}f_{Mi} \frac{\partial
\phi}{\partial \psi} \right ) v_{||} \bun \cdot \nablaave [R (\bv
\cdot \zun)] \nonumber \\ = \langle C \{ f_i \} \rangle.
\label{GKeq_radial3}
\end{eqnarray}

I substitute $v_{||} \bun \cdot \nablaave f_i$ from equation
\eq{GKeq_radial3} into equation \eq{gyrovisc_1approx} to find
\begin{eqnarray}
 \left \langle \int d^3v\, f_i R M(\bv \cdot \zun) (\bv_{M0} +
\tilde{\bv}_1 ) \cdot \nabla \psi \right \rangle_\psi = \nonumber
\\ \frac{M^2c}{2Ze} \left \langle \int d^3v\, [R^2 (\bv \cdot \zun
)^2] \left ( \left. \frac{\partial f_i}{\partial t} \right |_{\br,
\bv} - \langle C \{ f_i \} \rangle \right ) \right \rangle_\psi,
\label{gyrovisc_2approx}
\end{eqnarray}

\noindent where I use $(Ze/T_i) \bun \cdot \nabla \phi \int d^3v\,
f_{Mi} v_{||} [R^2 (\bv \cdot \zun)^2] = 0$ and
\begin{eqnarray}
 \frac{M^3 c^2}{2Z^2 e^2} \left \langle \int d^3v\, \left (
\frac{\partial f_i}{\partial \psi} + \frac{Ze}{T_i}f_{Mi}
\frac{\partial \phi}{\partial \psi} \right ) v_{||} \bun \cdot
\nablaave [ R ( \bv \cdot \zun )] R^2 ( \bv \cdot \zun )^2 \right
\rangle_\psi = \nonumber \\ \frac{M^3 c^2}{6Z^2 e^2} \left \langle
\int d^3v\, v_{||} \bun \cdot \nablaave \left [ \left (
\frac{\partial f_i}{\partial \psi} + \frac{Ze}{T_i}f_{Mi}
\frac{\partial \phi}{\partial \psi} \right ) R^3 ( \bv \cdot \zun
)^3 \right ] \right \rangle_\psi = 0. \label{int_no_gvisc}
\end{eqnarray}

\noindent To obtain the first equality in equation
\eq{int_no_gvisc}, terms that contain $\bun \cdot \nablaave
(\partial f_i / \partial \psi)$ and $\bun \cdot \nabla (\partial
\phi / \partial \psi)$ are neglected.

Finally, using $k_\bot \rho_i \ll 1$ in \eq{gyrovisc_2approx}, the
replacements $f_i \simeq \overline{f}_i$ and $\langle C \{ f_i \}
\rangle \simeq \overline{C \{ f_i \}}$ lead to \eq{gyrovisc}.

\section{Collisional piece in up-down symmetric tokamaks} \label{sectapp_collgyrovisc}

To prove that the collisional contribution to equation
\eq{gyrovisc} vanishes for up-down symmetric tokamaks I employ the
drift kinetic equation \eq{drift_kinetic_i} for the long
wavelength, axisymmetric component of $\overline{f}_i$. I already
proved that the term $\bv_E \cdot \nabla_\bR f_i$ can be safely
neglected. Additionally, I assume that the time derivative is
small once the statistical equilibrium is reached, and I split the
distribution function into $\overline{f}_i = f_{Mi} (\psi, E_0) +
\overline{h}_i$, with $\overline{h}_i (\psi, \theta, E_0, \mu_0,
t) \ll f_{Mi}$, giving
\begin{equation}
v_{||} \bun \cdot \nabla \theta \left [ \frac{\partial
\overline{h}_i}{\partial \theta} + \frac{\partial}{\partial
\theta} \left ( \frac{Iv_{||}}{\Omega_i} \frac{\partial
f_{Mi}}{\partial \psi} \right ) \right ] = C^{(\ell)} \{
\overline{h}_i \}, \label{eq_updown}
\end{equation}

\noindent where I use $\bv_{M0} \cdot \nabla \psi = v_{||} \bun
\cdot \nablaave ( Iv_{||}/\Omega_i )$. Since the tokamak is
up-down symmetric, $I v_{||}/\Omega_i$ is symmetric in $\theta$
and its derivative is antisymmetric.

In equation \eq{eq_updown}, replacing $\theta$ by $- \theta$,
$v_{||}$ by $ - v_{||}$ and $\overline{h}_i$ by $-\overline{h}_i$
does not change the equation. Then, $\overline{h}_i$ changes sign
if both $\theta$ and $v_{||}$ do. Due to this property, the
collisional integral in \eq{gyrovisc} is given by
\begin{equation}
\left \langle \frac{M}{2B\Omega_i} \int d^3v\, C^{(\ell)} \{
\overline{h}_i \} \left ( |\nabla \psi|^2 \frac{v_\bot^2}{2} + I^2
v_{||}^2 \right ) \right \rangle_\psi = 0.
\end{equation}

\noindent In the contributions to this integral, the piece of the
distribution function with positive $v_{||}$ in the upper half
($\theta > 0$) of the tokamak cancels the piece of the
distribution function with negative $v_{||}$ in the lower half
($\theta < 0$). Similarly, the piece with negative $v_{||}$ in the
upper half cancels the piece with positive $v_{||}$ in the lower
half.

\begin{singlespace}
\bibliography{main}
\bibliographystyle{unsrt}
\end{singlespace}

\end{document}